\documentclass[a4paper,11pt,oneside,american]{book}

\usepackage[latin1]{inputenc}
\usepackage{amsmath,amssymb,amsxtra,exscale} 

\usepackage[square,comma,numbers,sort&compress]{natbib}
\usepackage{babel}

\usepackage{fancyhdr} 

\fancypagestyle{plain}{%
  \fancyhf{}                       
  \cfoot{\thepage}                 
  
  }

\DeclareMathAlphabet{\matheurm}{U}{eur}{m}{n}  
\DeclareMathAlphabet{\matheubf}{U}{eur}{b}{n}  
\usepackage[mathscr]{eucal}                    

\usepackage{xspace} 

\newcommand{\eg}{e.g.\@\xspace}
\newcommand{\ie}{i.e.\@\xspace}
\newcommand{\cf}{cf.\@\xspace}
\newcommand{\etc}{etc.\@\xspace}
\newcommand{\rhs}{r.h.s.\@\xspace}
\newcommand{\lhs}{l.h.s.\@\xspace}
\newcommand{\eom}{e.o.m.\@\xspace}

\newcommand{\Cha}{Chapter~}
\newcommand{\Sec}{Section~}
\newcommand{\App}{Appendix~}
\newcommand{\Tab}{Table~}

\newcommand{\susy}{\breve}

\newcommand{\id}{\operatorname{id}}
\newcommand{\1}{1\hspace{-0.243em}\text{l}} 
\newcommand{\diag}{\operatorname{diag}} 

\newcommand{\R}{\mathbb{R}} 

\newcommand{\half}{\frac{1}{2}}

\newcommand{\const}{\operatorname{const}}

\newcommand{\sign}{\operatorname{sign}}
\newcommand{\body}{\operatorname{body}}

\newcommand{\sdet}{\operatorname{sdet}}
\newcommand{\tr}{\operatorname{tr}}

\newcommand{\cycl}{\operatorname{cycl}}
\newcommand{\gcycl}{\operatorname{gcycl}}

\newcommand{\Hodge}{\star} 
\newcommand{\Lie}{\mathscr{L}} 
\newcommand{\D}{D} 
\newcommand{\SD}{\mathcal{D}} 

\newcommand{\s}{\hat} 
\newcommand{\spartial}{\s\partial} 

\newcommand{\maps}{\text{:}\;}

\newcommand{\LC}{\Omega} 
\newcommand{\tor}{\tau} 
\newcommand{\RS}{\psi} 

\newcommand{\BMf}{\mathcal{M}}
\newcommand{\TSp}{\mathcal{N}}
\newcommand{\Poisson}{\mathcal{P}}
\newcommand{\Casimir}{C}
\newcommand{\casimir}{c}
\newcommand{\Lagrange}{\mathcal{L}}
\newcommand{\Action}{L}
\newcommand{\N}{N}
\newcommand{\detv}{\Delta}
\newcommand{\derv}{\nabla}

\DeclareMathAlphabet{\matheurm}{U}{eur}{m}{n}  
\newcommand{\q}[1]{\matheurm{#1}}

\renewcommand{\^}{{}^}
\renewcommand{\_}{\!{}_}

\newcommand{\eqnsplit}{\notag \\ {} &\hspace{1.35em} {}}
\newcommand{\eqnspl}{\notag \\ {} &\, {}}

\newcommand{\lvec}[1]{\overset{\raisebox{-0.5ex}{$\scriptscriptstyle
      \leftarrow$}}{#1}{}}
\newcommand{\rvec}[1]{\overset{\,
    \raisebox{-0.5ex}{$\scriptscriptstyle \rightarrow$}}{#1}{}}

\newcommand{\lpartial}{\lvec{\partial}}
\newcommand{\rpartial}{\rvec{\partial}}

\newcommand{\clmn}[2]{\left(
    \begin{array}{rr}
      #1 \\
      #2 
    \end{array}\right)
  }

\newcommand{\mtrx}[4]{\left(
    \begin{array}{cc}
      #1 & #2 \\
      #3 & #4
    \end{array}\right)
  }

\newcommand{\bibJBD}{%
  jordan55,Dicke:1957RM,Jordan:1959eg,Fierz:1956,Brans:1961sx}
\newcommand{\bibQuint}{%
  Wang:1999fa,Diaz-Rivera:1999wd,Matos:1999et,Coley:1999yq,%
  Bertolami:1999dp,Sokolowski:1995dk}
\newcommand{\bibQuintObserv}{%
  Perlmutter:1997hx,Perlmutter:1998np,Riess:1998cb,%
  Garnavich:1998nb,Garnavich:1998th,Schmidt:1998}
\newcommand{\bibTele}{%
  Hayashi:1977jd,Hayashi:1979qx,hehl79,kopczynski82,Kopczynski:1990af,%
  Muller-Hoissen:1983vc,mueller-hoissen85,Bakler:1988ub,%
  Mielke:1990my,mielke92,nester89,DeAndrade:2000sf}

\newcommand{\bibString}{green87,luest89,polchinski98}

\newcommand{\bibSUSY}{Wess:1977fn,Wess:1978bu,Golfand:1971iw,Volkov:1973ix}
\newcommand{\bibSUGRA}{%
Freedman:1976xh,Freedman:1976py,Deser:1976eh,Deser:1976rb,Grimm:1978kp}
\newcommand{\bibSGeom}{Vladimirov:1984zj,Vladimirov:1985ZJ}

\newcommand{\bibSBH}{%
  Aichelburg:1978fn,Aichelburg:1980eb,Aichelburg:1983ux,%
  Rosenbaum:1986cn,Knutt-Wehlau:1998gq}
\newcommand{\bibSuperfield}{%
  berezin66,gates83,dewitt84,%
  Vladimirov:1984zj,Vladimirov:1985ZJ,constantinescu94}
\newcommand{\bibGaugeSUGRA}{VanNieuwenhuizen:1981ae}

\newcommand{\bibSAlg}{%
  Kac:1977qb,Scheunert:1976uf,Scheunert:1976ug,Frappat:1996pb}

\newcommand{\bibGrF}{Schmidt:1999wb,Obukhov:1997uc,Strobl:1999Habil}
\newcommand{\bibKtor}{Katanaev:1993fu,Katanaev:1997je}
\newcommand{\bibKV}{Katanaev:1986wk,Katanaev:1990qm}
\newcommand{\bibGrFermi}{Solodukhin:1993xs,Solodukhin:1995ux}
\newcommand{\bibFOG}{Schaller:1994np,Schaller:1994es}
\newcommand{\bibKlStr}{Klosch:1996fi,Klosch:1996qv,Kloesch:1997fi}
\newcommand{\bibDil}{%
  Banks:1991mk,Odintsov:1991qu,%
  Louis-Martinez:1994eh,Gegenberg:1995pv,%
  Klosch:1996fi,Klosch:1996qv,Kloesch:1997fi,Klosch:1998yh}
\newcommand{\bibJT}{%
  Barbashov:1979bm,Teitelboim:1983ux,teitelboim84,%
  jackiw84,Jackiw:1985je}
\newcommand{\bibDilKKL}{Katanaev:1996bh,Katanaev:1997ni}
\newcommand{\bibGrRed}{%
  Thomi:1984na,Hajicek:1984mz,Schmidt:1997mq,Schmidt:1998ih}
\newcommand{\bibSI}{Verlinde:1991rf,Cangemi:1992bj,Cangemi:1993sd}
\newcommand{\bibDBH}{Mandal:1991tz,Elitzur:1991cb,Witten:1991yr}
\newcommand{\bibDBHmatter}{%
  Callan:1992rs,%
  Mikovic:1992id,Mikovic:1993vy,Mikovic:1995ub,Mikovic:1997de,%
  Kuchar:1997zm,Varadarajan:1998qz,%
  Cangemi:1996yz,Benedict:1996qy}
\newcommand{\bibQGr}{
  Kummer:1997hy,Kummer:1997jj,Kummer:1998zs}
\newcommand{\bibPSM}{Schaller:1994es,Schaller:1994uj,Schaller:1995xk}
\newcommand{\bibNLGT}{Ikeda:1993qz,Ikeda:1993nk,Ikeda:1993aj}
\newcommand{\bibPSMstar}{Cattaneo:1999fm,Cattaneo:2000iw}
\newcommand{\bibPoisson}{Weinstein:1983MM,choquet-bruhat89}
\newcommand{\bibGrMatterC}{Kummer:1995qv,Grumiller:1999rz}

\newcommand{\bibHoweOther}{%
  Brown:1979ma,Martinec:1983um,Rocek:1986iz,Ertl:1997ib}
\newcommand{\bibSJTsuperfield}{Chamseddine:1991fg}
\newcommand{\bibSJTgauge}{Rivelles:1994xs,Cangemi:1994mj}
\newcommand{\bibNLSGT}{Ikeda:1994dr,Ikeda:1994fh}
\newcommand{\bibIzq}{%
  Ikeda:1994dr,Ikeda:1994fh,Izquierdo:1998hg,Strobl:1999zz}

\begin{document}

\frontmatter

\fancyhf{}

\begin{titlepage}


\thispagestyle{empty}

    \centering

    \vspace*{4.5\baselineskip}

    {\huge
      \textbf{DISSERTATION}}
        
    \vspace{3\baselineskip}

    {\Huge
      \textbf{Supergravity in}

      \medskip
    
      \textbf{Two Spacetime Dimensions}}
    
    \vspace{4\baselineskip}

    A Thesis \\
    Presented to the Faculty of Science and Informatics \\
    Vienna University of Technology

    \vspace{1.5\baselineskip}
        
    Under the Supervision of Prof.\ Wolfgang Kummer \\
    Institute for Theoretical Physics

    \vspace{1.5\baselineskip}
        
    In Partial Fulfillment \\
    Of the Requirements for the Degree \\
    Doctor of Technical Sciences
    
    \vspace{1.5\baselineskip}
    
    By

    \vspace{1.5\baselineskip}
    
    \textbf{Dipl.-Ing.\ Martin Franz Ertl} \\
    Streitwiesen 13, 3653 Weiten, Austria \\
    E-mail: \texttt{ertl@tph.tuwien.ac.at}
  
    \vspace{\fill}
  
    \leftline{Vienna, January 31, 2001}

\end{titlepage}


\chapter*{Acknowledgements}
\label{cha:ackn}

\begin{samepage} \thispagestyle{empty}


I am most grateful to my parents for rendering my studies possible and
their invaluable support during this time. Further thanks go to my
sister Eva and my brother in law Fritz, as well as to my grandmother
and my aunts Maria and Gertrud.

Let me express my gratitude to Prof.\ Wolfgang Kummer who was the
supervisor of both my diploma thesis and my PhD thesis, and whose FWF
projects established the financial background for them. He advised and
assisted me in all problems encountered during this work.

I gratefully acknowledge the profit I gained from collaborations with
Michael Katanaev (Steklov Institute Moskau) und Thomas Strobl
(currently at Friedrich-Schiller University Jena). Furthermore, I also
thank Manfred Schweda and Maximilian Kreuzer for financial support.

Finally, I am also very grateful to all the members of the institute
for constantly encouraging me and for the nice atmosphere. Special
thanks go to Herbert Balasin, Daniel Grumiller, Dragica Kahlina,
Alexander Kling, Thomas Kl\"osch, Herbert Liebl, Mahmoud
Nikbakht-Tehrani, Anton Rebhan, Axel Schwarz and Dominik Schwarz.

This work was supported by projects \mbox{P\,10.221-PHY} and
\mbox{P\,12.815-TPH} and in its final stage by project
\mbox{P\,13.126-PHY} of the FWF (\"Osterreichischer Fonds zur
F\"orderung der wissenschaftlichen Forschung).

\end{samepage}


\chapter*{Abstract}
\label{cha:abs}


\thispagestyle{empty}

Supersymmetry is an essential component of all modern theories such as
strings, branes and supergravity, because it is considered as an
indispensable ingredient in the search for a unified field theory.
Lower dimensional models are of particular importance for the
investigation of physical phenomena in a somewhat simplified context.
Therefore, two-dimensional supergravity is discussed in the present
work.

The constraints of the superfield method are adapted to allow for
supergravity with bosonic torsion. As the analysis of the Bianchi
identities reveals, a new vector superfield is encountered besides the
well-known scalar one. The constraints are solved both with
superfields using a special decomposition of the supervielbein, and
explicitly in terms of component fields in a Wess-Zumino gauge.

The graded Poisson Sigma Model (gPSM) is the alternative method used
to construct supersymmetric gravity theories.  In this context the
graded Jacobi identity is solved algebraically for general cases. Some
of the Poisson algebras obtained are singular, or several potentials
contained in them are restricted. This is discussed for a selection of
representative algebras. It is found, that the gPSM is far more
flexible and it shows the inherent ambiguity of the supersymmetric
extension more clearly than the superfield method. Among the various
models spherically reduced Einstein gravity and gravity with torsion
are treated. Also the Legendre transformation to eliminate auxiliary
fields, superdilaton theories and the explicit solution of the gPSM
equations of motion for a typical model are presented.

Furthermore, the PSM field equations are analyzed in detail, leading
to the so called "symplectic extension". Thereby, the Poisson tensor
is extended to become regular by adding new coordinates to the target
space. For gravity models this is achieved with one additional
coordinate.

Finally, the relation of the gPSM to the superfield method is
established by extending the base manifold to become a supermanifold.

\thispagestyle{empty}


\lhead{\slshape \contentsname}                       \rhead{\thepage}
\tableofcontents
\clearpage

\mainmatter

\lhead{\slshape \leftmark}                           \rhead{\thepage}
\renewcommand{\chaptermark}[1]{
  \markboth{\chaptername\ \thechapter.\ #1}{}}

\chapter{Introduction}
\label{cha:intro}



The study of diffeomorphism invariant theories in $1+1$ dimensions has
for quite some time been a fertile ground for acquiring some insight
into the unsolved problems of quantum gravity in higher dimensions.
Indeed, the whole field of spherically symmetric gravity belongs to
this class, from $d$-dimensional Einstein theory to extended theories
like the Jordan-Brans-Dicke theory \cite{\bibJBD} or `quintessence'
\cite{\bibQuint} which may now seem to obtain observational support
\cite{\bibQuintObserv}.  Also, equivalent formulations for 4d Einstein
theory with nonvanishing torsion (`teleparallelism' \cite{\bibTele})
and alternative theories including curvature and torsion
\cite{Hehl:1995ue} are receiving increasing attention.

On the other hand, supersymmetric extensions of gravity
\cite{\bibSUGRA} are believed to be a necessary ingredient for a
consistent solution of the problem of quantizing gravity, especially
within the framework of string/brane theory \cite{\bibString}. These
extensions so far are based upon bosonic theories with vanishing
torsion.


Despite the fact that so far no tangible direct evidence for
supersymmetry has been discovered in nature, supersymmetry
\cite{\bibSUSY} managed to retain continual interest within the aim to
arrive at a fundamental `theory of everything' ever since its
discovery: first in supergravity \cite{\bibSUGRA} in $d = 4$, then in
generalizations to higher dimensions of higher $\N$
\cite{Vafa:1996xn}, and finally incorporated as a low energy limit of
superstrings \cite{Witten:1995ex} or of even more fundamental theories
\cite{Sezgin:1997cj} in 11 dimensions.

Even before the advent of strings and superstrings the importance of
studies in $1 + 1$ `spacetime' had been emphasized \cite{Howe:1979ia}
in connection with the study of possible superspace formulations
\cite{Fayet:1977yc}. To the best of our knowledge, however, to this
day no attempt has been made to generalize the supergravity
formulation of (trivial) Einstein-gravity in $d = 2$ to the
consideration of two-dimensional $(1, 1)$ supermanifolds for which the
condition of vanishing (bosonic) torsion is removed. Only attempts to
formulate theories with higher powers of curvature (at vanishing
torsion) seem to exist \cite{Hindawi:1996fy}. There seem to be only
very few exact solutions of supergravity in $d = 4$ as well
\cite{\bibSBH}.

Especially at times when the number of arguments in favour of the
existence of an, as yet undiscovered, fundamental theory increase
\cite{Vafa:1996xn} it may seem appropriate to also exploit---if
possible---\emph{all} (super-)geometrical generalizations of the
two-dimensional stringy world sheet. Actually, such an undertaking can
be (and indeed is) successful, as suggested by the recent much
improved insight, attained for all (non-supersymmetric)
two-dimensional diffeomorphism invariant theories, including dilaton
theory, and also comprising torsion besides curvature \cite{\bibKV} in
the most general manner
\cite{\bibFOG,Strobl:1994eu,\bibKlStr,\bibDilKKL}. In the absence of
matter-fields (non-geometrical degrees of freedom) all these models
are integrable at the classical level and admit the analysis of all
global solutions \cite{\bibKtor,\bibKlStr}.  Integrability of
two-dimensional gravity coupled to chiral fermions was demonstrated in
\cite{kummer92,\bibGrFermi}. Even the general aspects of quantization
of any such theory now seem to be well understood
\cite{Kummer:1992rt,\bibFOG,Strobl:1994eu,Haider:1992cw,Kummer:1997jj}.
By contrast, in the presence of matter and if singularities like black
holes occur in such models, integrable solutions are known only for
very few cases. These include interactions with fermions of one
chirality \cite{kummer92} and, if scalar fields are present, only the
dilaton black hole \cite{\bibDBH,Callan:1992rs} and models which have
asymptotical Rindler behaviour \cite{Fabbri:1996bz}.  Therefore, a
supersymmetric extension of the matterless case suggests that the
solvability may carry over, in general. Then `matter' could be
represented by superpartners of the geometric bosonic field variables.


In view of this situation it seems surprising that the following
problem has not been solved so far:

Given a general geometric action of pure gravity in two spacetime
dimensions of the form (\cf \cite{\bibGrF} and references therein)
\begin{equation}
  \Action^{\mathrm{gr}} = \int d^2\!x \sqrt{-g} F(R,\tor^2), \label{grav}
\end{equation}
what are its possible supersymmetric generalizations?

In two dimensions there are only two algebraic-geometric invariants of
curvature and torsion, denoted by $R$ and $\tor^2$, respectively, and
$F$ is some sufficiently well-behaved function of these invariants.
For the case that $F$ does not depend on its second argument, $R$ is
understood to be the Ricci scalar of the torsion-free Levi-Civita
connection.

As a prototype of a theory with dynamical torsion we may consider the
specification of (\ref{grav}) to the Katanaev-Volovich (KV) model
\cite{\bibKV}, quadratic in curvature and torsion.  Even for this
relatively simple particular case of (\ref{grav}), a supergravity
generalization has not been presented to this day.

The bosonic theory (\ref{grav}) may be reformulated as a first order
gravity action (FOG) by introducing auxiliary fields $\phi$ and $X^a$
(the standard momenta in a Hamiltonian reformulation of the model; \cf
\cite{\bibFOG} for particular cases and \cite{Strobl:1999Habil} for
the general discussion)
\begin{equation}
  \Action^\mathrm{FOG} = \int_\BMf \phi d\omega + X_a De^a + \epsilon
  v(\phi,Y) \label{FOG}
\end{equation}
where $Y=X^2/2 \equiv X^a X_a/2$ and $v$ is some two-argument function
of the indicated variables.  In (\ref{FOG}) $e^a$ is the zweibein and
$\omega_{ab} = \omega\epsilon_{ab}$ the Lorentz or spin connection,
both 1-form valued, and $\epsilon = \frac{1}{2} e^a \wedge e^b
\epsilon_{ba} = e d^2\!x$ is the two-dimensional volume form ($e =
\det(e_m\^a)$). The torsion 2-form is $De^a = de^a + e^b \wedge \omega
\epsilon_b\^a$.

The function $v(\phi,Y)$ is the Legendre transform
\cite{Strobl:1999Habil} of $F(R,\tor^2)$ with respect to the
\emph{three} arguments $R$ and $\tor^a$, or, if $F$ depends on the
Levi-Civita curvature $R$ only, with respect to this single variable
($v$ depending only on $\phi$ then). In view of its close relation to
the corresponding quantity in generalized dilaton theories which we
recall below, we shall call $\phi$ the `dilaton' also within the
action (\ref{FOG}).\footnote{In the literature also $\Phi =
  -\frac{1}{2} \ln \, \phi$ carries this name. This definition is
  useful when $\phi$ is restricted to $\R_+$ only, as it is often the
  case for specific models.}  The equivalence between (\ref{grav}) and
the action (\ref{FOG}) holds at a global level, if there is a globally
well-defined Legendre transform of $F$. Prototypes are provided by
quadratic actions, \ie $R^2$-gravity and the model of \cite{\bibKV}.
Otherwise, in the generic case one still has local (patchwise)
equivalence.  Only theories for which $v$ or $F$ do not have a
Legendre transform even locally, are not at all covered by the
respective other formulation. In any case, as far as the
supersymmetrization of 2d gravity theories is concerned, we will
henceforth focus on the family of actions given by (\ref{FOG}).

The FOG formulation (\ref{FOG}) also covers general dilaton theories
in two dimensions \cite{\bibDil} ($\widetilde{R}$ is the torsion free
curvature scalar),
\begin{equation}
  \label{dil}
  \Action^{\mathrm{dil}} = \int d^2\!x \sqrt{-g}
  \left[
    \frac{\widetilde{R}}{2} \phi - \half Z(\phi) (\partial^n \phi)
    (\partial_n \phi) + V(\phi)
  \right].
\end{equation}
Indeed, by eliminating $X^a$ and the torsion-dependent part of
$\omega$ in (\ref{FOG}) by their algebraic equations of motion, for
regular 2d spacetimes ($e = \sqrt{-g} \neq 0$) the theories
(\ref{dil}) and (\ref{FOG}) are locally and globally equivalent if in
(\ref{FOG}) the `potential' is chosen as \cite{\bibDilKKL} (\cf
  also \Sec\ref{sec:sdil} below for some details as well as
  \cite{Obukhov:1997uc} for a related approach)
\begin{equation}
  \label{vdil}
  v^{\mathrm{dil}}(\phi,Y) = Y Z(\phi) + V(\phi).
\end{equation}

There is also an alternative method for describing dilaton gravity by
means of an action of the form (\ref{FOG}), namely by using the
variables $e^a$ as a zweibein for a metric $\bar g$, related to $g$ in
(\ref{dil}) according to $g_{mn} = \Omega(\phi) \bar{g}_{mn}$ for a
suitable choice of the function $\Omega$ (it is chosen in such a way
that after transition from the Einstein-Cartan variables in
(\ref{FOG}) and upon elimination of $X^a$ one is left with an action
for $\bar g$ of the form (\ref{dil}) with $\bar Z = 0$, \ie without
kinetic term for the dilaton, \cf \eg
\cite{Verlinde:1991rf,Banks:1991mk,Louis-Martinez:1994eh}).\footnote{Some
  details on the two approaches to general dilaton gravity may also be
  found in \cite{Strobl:1999Habil}.}  This formulation has the
advantage that the resulting potential $v$ depends on $\phi$ only. It
has to be noted, however, that due to a possibly singular behavior of
$\Omega$ (or $1/\Omega$) the global structures of the resulting
spacetimes (maximally extended with respect to $\bar g$ versus $g$) in
part are quite different.  Moreover, the change of variables in a path
integral corresponding to the `torsion' description of dilaton
theories ($Y$-dependent potential (\ref{vdil})) seems advantageous
over the one in the `conformal' description. In this description even
interactions with (scalar) matter can be included in a systematic
perturbation theory, starting from the (trivially) exact path integral
for the geometric part (\ref{FOG}) \cite{\bibQGr}.  Therefore when
describing dilaton theories within the present thesis we will
primarily focus on potentials (\ref{dil}), linear in $Y$.

In any case there is thus a huge number of 2d gravity theories
included in the present framework. We select only a few for
illustrative purposes. One of these is spherically reduced Einstein
gravity (SRG) from $d$ dimensions \cite{\bibGrRed}
\begin{equation}
  \label{SRG}
  Z_{\mathrm{SRG}} = -\frac{d-3}{(d-2)\phi}, \qquad
  V_{\mathrm{SRG}} = -\lambda^2 \phi^{\frac{d-4}{d-2}} \, ,
\end{equation}
where $\lambda$ is some constant. In the `conformal approach'
mentioned above the respective potentials become
\begin{equation}
  \label{EBH}
  Z_{\overline{\mathrm{SRG}}} = 0, \qquad
  V_{\overline{\mathrm{SRG}}} =
  -\frac{\lambda^2}{\phi^{\frac{1}{d-2}}}.
\end{equation}
The KV-model, already referred to above, results upon 
\begin{equation}
  \label{KV}
  Z_{\mathrm{KV}} = \alpha, \qquad
  V_{\mathrm{KV}} = \frac{\beta}{2} \phi^2 - \Lambda,
\end{equation}
where $\Lambda$, $\alpha$ and $\beta$ are constant.
 Two other particular examples are the
so-called Jackiw-Teitelboim (JT) model \cite{\bibJT} with vanishing
torsion in (\ref{FOG}) and no kinetic term of $\phi$ in (\ref{dil}),
\begin{equation}
  \label{JT}
  Z_{\mathrm{JT}} = 0, \qquad V_{\mathrm{JT}} = -\Lambda \phi,
\end{equation}
and the string inspired dilaton black hole (DBH) \cite{\bibDBH} (\cf
also \cite{\bibDBHmatter})
\begin{equation}
  \label{DBH}
  Z_{\mathrm{DBH}} = -\frac{1}{\phi}, \qquad
  V_{\mathrm{DBH}} = -\lambda^2 \phi,
\end{equation}
which, incidentally, may also be interpreted as the formal limit $d
\rightarrow \infty$ of (\ref{SRG}).

For later purposes it will be crucial that (\ref{FOG}) may be
formulated as a Poisson Sigma Model (PSM) \cite{\bibPSM,Klosch:1996fi}
(\cf also \cite{\bibNLGT,\bibPSMstar}).  Collecting zero form and
one-form fields within (\ref{FOG}) as
\begin{equation}
  \label{ident1}
  (X^i) := (\phi,X^a), \qquad
  (A_i) = (dx^m A_{mi}(x)) := (\omega,e_a),
\end{equation}
and after a partial integration, the action (\ref{FOG}) may be
rewritten identically as
\begin{equation}
  \label{PSM}
  \Action^{\mathrm{PSM}} = \int_\BMf dX^i \wedge A_i + \half
  \Poisson^{ij} A_j \wedge A_i,
\end{equation}
where the matrix $\Poisson^{ij}$ may be read off by direct comparison.
The basic observation in this framework is that this matrix defines a
Poisson bracket on the space spanned by coordinates $X^i$, which is
then identified with the target space of a Sigma Model. In the
present context this bracket $\{ X^i , X^j \} := \Poisson^{ij}$ has
the form
\begin{align}
  \{ X^a, \phi \} &= X^b \epsilon_b\^a, \label{LorentzB} \\
  \{ X^a, X^b \} &= v(\phi,Y) \epsilon^{ab},  \label{PB}
\end{align} 
where (throughout this thesis) $Y \equiv \half X^a X_a$. This bracket
may be verified to obey the Jacobi identity.

The gravitational origin of the underlying model is reflected by the
first set of brackets: It shows that $\phi$ is the generator of
Lorentz transformations (with respect to the bracket) on the target
space $\R^3$. The form of the second set of brackets is already
completely determined by the following: Antisymmetry of the bracket
leads to proportionality to the $\epsilon$-tensor, while the Jacobi
identity for the bracket requires $v$ to be a function of the Lorentz
invariant quantities $\phi$ and $X^2$ only.

Inspection of the local symmetries of a general PSM,
\begin{equation}
  \label{PSM-symms}
  \delta X^i = \Poisson^{ij} \epsilon_j, \qquad
  \delta A_i = -d\epsilon_i - (\partial_i \Poisson^{jk}) \epsilon_k
  A_j,
\end{equation}
shows that the Lorentz symmetry of the bracket gives rise to the
\emph{local} Lorentz symmetry of the gravity action (\ref{FOG})
(specialization of (\ref{PSM-symms}) to an $\epsilon$ with only
nonzero $\phi$ component, using the identification (\ref{ident1})).
On the other hand, the second necessary ingredient for the
construction of a gravity action, diffeomorphism invariance, is
automatically respected by an action of the form (\ref{PSM}). (It may
be seen that the diffeomorphism invariance is also encoded
\emph{on-shell} by the remaining two local symmetries
(\ref{PSM-symms}), \cf \eg \cite{Klosch:1996fi,Strobl:1994PhD}).

PSMs relevant for 2d gravity theories (without further gauge field
interactions) possess one `Casimir function' $\casimir(X)$ which is
characterized by the vanishing of the Poisson brackets $\{ X^i,
\casimir\}$.  Different constant values of $\casimir$ characterize
symplectic leaves \cite{\bibPoisson}. In the language of gravity
theories, for models with asymptotic Minkowski behavior, $\casimir$ is
proportional to the ADM mass of the system.\footnote{We remark that an
  analogous conservation law may be established also in the presence
  of additional matter fields \cite{\bibGrMatterC}.}

To summarize, the gravity models (\ref{PSM}), and thus implicitly also
any action of the form (\ref{FOG}) and hence generically of
(\ref{grav}), may be obtained from the construction of a Lorentz
invariant bracket on the two-dimensional Minkowski space $\R^2$
spanned by $X^a$, with $\phi$ entering as an additional
parameter.\footnote{Actually, this point of view was already used in
  \cite{Schaller:1994es} so as to arrive at (\ref{PSM}), but without
  fully realizing the relation to (\ref{grav}) at that time.  Let us
  remark on this occasion that in principle one might also consider
  theories (\ref{PSM}) with $X^a$ replaced \eg by $X^a \cdot
  f(\phi,X^2)$. For a nonvanishing function $f$ this again yields a
  PSM after a suitable reparametrization of the target space. Also,
  the identification of the gauge fields $A_i$ in (\ref{ident1}) could
  be modified in a similar manner. Hence we do not have to cover this
  possibility explicitly in what follows. Nevertheless, it could be
  advantageous to derive by this means a more complicated gravity
  model from a simpler PSM structure.}  The resulting bracket as well
as the corresponding models are seen to be parametrized by one
two-argument function $v$ in the above way.

The Einstein-Cartan formulation of 2d gravities as in (\ref{FOG}) or,
even more so, in the PSM form (\ref{PSM}) will turn out to be
particularly convenient for obtaining the most general supergravities
in $d=2$. While the metrical formulation of gravity due to Einstein in
$d=4$ appeared very cumbersome for a supersymmetric generalization,
the Einstein-Cartan approach appeared to be best suited for the needs
of introducing additional fermionic degrees of freedom to pure gravity
\cite{\bibSUGRA}.

We now briefly digress to the corresponding strategy of constructing a
supersymmetric extension of a gravity theory in a spacetime of general
dimension $d$. By adding to the vielbein $e_m\^a$ and the Lorentz
connection $\omega_{ma}\^b$ appropriate terms containing a fermionic
spin-vector $\psi_m\^\alpha$, the Rarita-Schwinger field, an action
invariant under local supersymmetry can be constructed, where
$\psi_m\^\alpha$ plays the role of the gauge field for that symmetry.
In this formulation the generic local infinitesimal supersymmetry
transformations are of the form
\begin{equation}
  \label{susytrafo}
  \delta e_m\^a = -2i (\epsilon \gamma^a \psi_m), \qquad
  \delta \psi_m\^\alpha = -D_m \epsilon^\alpha + \cdots
\end{equation}
with $\epsilon = \epsilon(x)$ arbitrary.

In the course of time various methods were developed to make the
construction of supergravity actions more systematical. One of these
approaches, relying on superfields \cite{\bibSuperfield}, extends the
Einstein-Cartan formalism by adding anticommuting coordinates to the
spacetime manifold, thus making it a supermanifold, and,
simultaneously, by enlarging the structure group with a spinorial
representation of the Lorentz group. This method adds many auxiliary
fields to the theory, which can be eliminated by choosing appropriate
constraints on supertorsion and supercurvature and by choosing a
Wess-Zumino type gauge. It will be the approach used in the first part
of this thesis (\Cha\ref{cha:sfield}).

The other systematic approach to construct supergravity models for
general $d$ is based on the similarity of gravity to a gauge theory.
The vielbein and the Lorentz-connection are treated as gauge fields on
a similar footing as gauge fields of possibly additional gauge groups.
Curvature and torsion appear as particular components of the total
field strength.\footnote{Note that nevertheless standard gravity
  theories cannot be just reformulated as YM gauge theories with all
  symmetries being incorporated in a principle fiber bundle
  description; one still has to deal with the infinite-dimensional
  diffeomorphism group (\cf also \cite{Strobl:1993xt} for an
  illustration).}  By adding fermionic symmetries to the gravity gauge
group, usually taken as the Poincar\'e, de~Sitter or conformal group,
one obtains the corresponding supergravity theories
\cite{\bibGaugeSUGRA}.

In the two-dimensional case, the supergravity multiplet was first
constructed using the superfield approach \cite{Howe:1979ia}. Based on
that formalism, it was straightforward to \emph{formulate} a
supersymmetric generalization of the dilaton theory (\ref{dil}), \cf
\cite{Park:1993sd}.  Before that the supersymmetric generalization of
the particular case of the Jackiw-Teitelboim or de~Sitter model
\cite{\bibJT} had been achieved within this framework in
\cite{\bibSJTsuperfield}.  Up to global issues, this solved implicitly
also the problem of a supersymmetrization of the theories (\ref{grav})
in the \emph{torsion-free} case.

Still, the supergravity multiplet obtained from the set of constraints
used in \cite{Howe:1979ia} consists of the vielbein, the
Rarita-Schwinger field and an auxiliary scalar field, but the
Lorentz-connection is lost as independent field. It is expressed in
terms of the vielbein and the Rarita-Schwinger field. Without a
formalism using an independent Lorentz-connection the construction of
supersymmetric versions of general theories of the $F(R, \tor^2)$-type
is impossible. A straightforward approach consists of a repetition of
the calculation of \cite{Howe:1979ia} while relaxing the original
superspace constraints to allow for a nonvanishing bosonic torsion.

As in higher dimensional theories, the gauge theoretic approach
provides a much simpler method for supersymmetrization than the
superfield approach. However, it is restricted to relatively simple
Lagrangians such as the one of the Jackiw-Teitelboim model (\ref{JT})
\cite{\bibSJTgauge}. The generic model (\ref{FOG}) or also (\ref{dil})
cannot be treated in this manner.

On the other hand, first attempts showed that super dilaton theories
may fit into the framework of `nonlinear' supergauge theories
\cite{\bibNLSGT}, and the action for a super dilaton theory was
obtained (without superfields) by a nonlinear deformation of the
graded de~Sitter group using free differential algebras in
\cite{Izquierdo:1998hg}.

Recently, it turned out \cite{Strobl:1999zz} (but \cf also
\cite{\bibNLSGT}) that the framework of PSMs \cite{\bibPSM}, now with
a graded target space, represents a very direct formalism to deal with
super dilaton theories. In particular, it allowed for a simple
derivation of the general solution of the corresponding field
equations, and in this process yielded the somewhat surprising result
that, \emph{in the absence of additional matter fields}, the
supersymmetrization of the dilaton theories (\ref{dil}) is on-shell
trivial. By this we mean that, up to the choice of a gauge, in the
general solution to the field equations all fermionic fields can be
made to vanish identically by an appropriate choice of gauge while the
bosonic fields satisfy the field equations of the purely bosonic
theory and are still subject to the symmetries of the latter.  This
local on-shell triviality of the supersymmetric extension may be
interpreted superficially to be yet another consequence of the fact
that, from the Hamiltonian point of view, the `dynamics' of
(\ref{PSM}) is described by just one variable (the Casimir function)
which does not change when fermionic fields are added. This type of
triviality will cease to prevail in the case of additional matter
fields (as is already obvious from a simple counting of the fields and
local symmetries involved). Furthermore, the supersymmetrization may
be used \cite{Park:1993sd} as a technical device to prove positive
energy theorems for supersymmetric \emph{and} non-supersymmetric
dilaton theories. Thus, the (local) on-shell triviality of pure 2d
supergravity theories by no means implicitly demolishes all the
possible interest in their supersymmetric generalizations. This
applies similarly to the $F(R,\tor^2)$-theories and to the FOG
formulation (\ref{FOG}) in which we are primarily interested.

Graded PSMs (gPSMs) turn out also to provide a unifying and most
efficient framework for the \emph{construction} of supersymmetric
extensions of a two-dimensional gravity theory, at least as far as
theories of the initially mentioned type (\ref{grav}) are considered.
This route, sketched already briefly in \cite{Strobl:1999zz}, will be
followed in detail within the present thesis.

The main idea of this approach will be outlined in
\Sec\ref{sec:outline}. It will be seen that within this framework the
problem for a supersymmetric extension of a gravity theory
(\ref{grav}) is reduced to a finite dimensional problem: Given a
Lorentz invariant Poisson bracket on a two-dimensional Minkowski space
(which in addition depends also on the `dilaton' or, equivalently, on
the generator of Lorentz transformations $\phi$), one has to extend
this bracket consistently and in a Lorentz covariant manner to the
corresponding superspace.

In spirit this is closely related to the analogous extension of Lie
algebras to superalgebras \cite{\bibSAlg}. In fact, in the particular
case of a linear dependence of $v$ in (\ref{PB}) the original Poisson
bracket corresponds to a three-dimensional Lie algebra, and likewise
any \emph{linear} extension of this Poisson algebra corresponds to a
superalgebra.  Here we are dealing with general nonlinear Poisson
algebras, particular cases of which can be interpreted as finite
$W$-algebras (\cf \cite{deBoer:1996nu}). Due to that nonlinearity the
analysis necessary for the fermionic extension is much more involved
and there is a higher ambiguity in the extension (except if one
considers this only modulo arbitrary (super)diffeomorphisms).
Therefore we mainly focus on an $\N=1$ extension within this thesis.


\Sec\ref{sec:sgeom} is devoted to the general definitions of
superspace used in our present work. A decomposition of the
supervielbein useful for solving the supergravity constraints is
presented in \Sec\ref{sec:svb}. In \Sec\ref{sec:sugra-howe} the
supergravity model of Howe \cite{Howe:1979ia} is given. The new
supergravity constraints for a supergravity model in terms of
superfields with independent Lorentz connection are derived in
\Sec\ref{sec:susy-cov} and the constraints are solved in
\Sec\ref{sec:sugra-ertl}.

In \Cha\ref{cha:PSM} the PSM approach is presented for the bosonic
case. Beside the known parts (\Sec\ref{sec:gravity-psm} and
\ref{sec:C}) we present a general method to solve the PSM field
equations (\Sec\ref{sec:eoms}). We also show a new `symplectic
extension'
(\Sec\ref{sec:target-space-ext}--\ref{sec:symplectic-gravity}).

After recapitulating some material on gPSMs in \Sec\ref{sec:gPSM2} and
also setting our notation and conventions, the solution of the
$\phi$-components of the Jacobi identities is given in
\Sec\ref{sec:ans-gP} simply by writing down the most general Lorentz
covariant ansatz for the Poisson tensor.  In \Sec\ref{sec:remJac} the
remaining Jacobi identities are solved in full generality for
nondegenerate and degenerate $\N=1$ fermionic extensions.

The observation that a large degree of arbitrariness is present in
these extensions is underlined also by the study of target space
diffeomorphisms in \Sec\ref{sec:diffeo}. We also point out the
advantages of this method in the quest for new algebras and
corresponding gravity theories.

In \Sec\ref{sec:Poisson} we shall consider particular examples of the
general result. This turns out to be superior to performing a general
abstract discussion of the results of \Sec\ref{sec:Jacobi}.  The more
so, because fermionic extensions of specific bosonic 2d gravity
theories, which have been discussed already in the literature, can be
investigated. Supersymmetric extensions of the KV-model (\ref{KV}) as
compared to SRG (\ref{SRG}) will serve for illustrative purposes.

The corresponding actions and their relation to the initial problem
(\ref{grav}) are given in \Sec\ref{sec:models}.  Also the general
relation to the supersymmetric dilatonic theories (\ref{dil}) will be
made explicit using the results of \Sec\ref{sec:action}. Several
different supersymmetrizations (one of which is even parity violating)
for the example of SRG are compared to the one provided previously in
the literature \cite{Park:1993sd}.  For each model the corresponding
supersymmetry is given explicitly.

In \Sec\ref{sec:sdil-sol} the explicit solution for a supergravity
theory with the bosonic part corresponding to $v^{\mathrm{dil}}$ in
(\ref{vdil}) is given.

In \Cha\ref{cha:SPSM} we discuss the relations between the superfield
approach of \Cha\ref{cha:sfield} and the gPSM supergravity theories of
\Cha\ref{cha:gPSM}.

In the final \Cha\ref{cha:concl} we will summarize our findings and
comment on possible further investigations.

\App\ref{app:gravity} and \ref{app:spinors} define notations and
summarize useful identities.


\chapter{Supergravity with Superfields}
\label{cha:sfield}


Supergravity can be formulated in superspace, where the
two-dimensional $x$-space of pure gravity is enlarged by fermionic
(anticommutative) $\theta$-variables. On this underlying superspace a
gravity theory comprising Einstein-Cartan variables $E_M\^A(x,\theta)$
and $\Omega_{MA}\^B(x,\theta)$ is established. The basic properties of
supergeometry are given in \Sec\ref{sec:sgeom}. As target space group
the direct sum of a vector and a spinor representation of the Lorentz
group is chosen (\cf (\ref{elorcs}) and (\ref{eantis}) below). In
order to reduce the large number of component fields, the coefficients
in the $\theta$-expansion of the superfields, constraints in
superspace are imposed. The choice of the supergravity constraints is
a novel one. We start with the original constraints of the $\N = 1$
supergravity model of Howe \cite{Howe:1979ia}, shortly reviewed in
\Sec\ref{sec:sugra-howe}, but in order to retain an independent
Lorentz connection $\omega_m(x)$ these constraints have to be
modified. A short derivation leading to the new constraints is given
in \Sec\ref{sec:susy-cov}. Then, in \Sec\ref{sec:sugra-ertl}, the new
supergravity model based on the new set of constraints is
investigated. The solution in terms of superfields as well as the
component field expansion in a Wess-Zumino type gauge are given. It
turns out that the more general new supergravity model contains in
addition to the well-known supergravity multiplet $\{ e_m\^a,
\RS_m\^\alpha, A \}$, where $e_m\^a$ is the zweibein, $\RS_m\^\alpha$
the Rarita-Schwinger field and $A$ the auxiliary field, a new
connection multiplet $\{ k^a, \varphi_m\^\alpha, \omega_m \}$
consisting of a vector field $k^a$, a further spin-vector
$\varphi_m\^\alpha$ and the Lorentz connection $\omega_m$.

\section{Supergeometry}
\label{sec:sgeom}

Although the formulae of supergeometry often look quite similar to the
ones of pure bosonic geometry, some attention has to be devoted to a
convenient definition \eg of the signs. The difference to ordinary
gravity becomes obvious when Einstein-Cartan variables are extended to
superspace.  As the structure group a direct product of a vector and a
spinor representation of the Lorentz group is needed. This leads to a
rich structure for supertorsion and supercurvature components and for
the corresponding Bianchi identities.

\subsection{Superfields}
\label{sec:sfield}

In $d = 2$ we consider a superspace with two commuting (bosonic) and
two anticommuting (Grassmann or spinor) coordinates
$z^M=\{x^m,\theta^\mu\}$ where lower case Latin ($ m=0,1$) and Greek
indices ($\mu=1,2$) denote commuting and anticommuting coordinates,
respectively:
\begin{equation}
  z^Mz^N=z^Nz^M(-1)^{MN}.
\end{equation}
Within our conventions for Majorana spinors (\cf
\App\ref{app:spinors}) the first anticommuting element of the
Grassmann algebra is supposed to be real, $(\theta^+)^* = \theta^+$,
while the second one is purely imaginary, $(\theta^-)^* = -\theta^-$
(\cf \App\ref{app:spinors}).

Our construction is based on differential geometry of superspace.  We
shall not deal with subtle mathematical definitions
\cite{\bibSGeom}. We just set our basic conventions.

In superspace right and left derivatives have to be distinguished. The
relation between the partial derivatives
\begin{equation}
  \label{rl-der}
  \rpartial_M \equiv \frac{\rpartial}{\partial z^M}, \qquad
  \lpartial_M \equiv \frac{\lpartial}{\partial z^M},
\end{equation}
which act to right and to the left, respectively, becomes
\begin{equation}
  \label{lr-partial}
  \rpartial_M f = f\, \lpartial_M (-1)^{M(f+1)},
\end{equation}
where in the exponent $M$ and $f$ are $1$ for anticommuting quantities
and $0$ otherwise. For our purpose it is sufficient to follow one
simple working rule allowing to generalize ordinary formulae of
differential geometry to superspace.
Any vectorfield in superspace
\begin{equation}
  \rvec V = V^M \rpartial_M
\end{equation}
is invariant under arbitrary nondegenerate coordinate changes
$z^M\rightarrow \bar{z}\^{M}(z)$:
\begin{equation}
  \label{ecooch}
  V^M \rpartial_M = \bar{V}^{M} \frac{\rpartial z^L}{\partial \bar{z}^{M}}
  \frac{\rpartial \bar{z}^{N}}{\partial z^L} \rpartial_{\bar{N}}
\end{equation}
Summation over repeated indices is assumed, and derivatives are always
supposed to act \emph{to the right} from now. So we drop the arrows in
the sequel. From (\ref{ecooch}) follows our simple basic rule: Any
formula of differential geometry in ordinary space can be taken over
to superspace if the summation is always performed from the upper left
corner to the lower right one with no indices in between (`ten to
four'), and the order of the indices in each term of the expression
must be the same. Otherwise an appropriate factor $(-1)$ must be
included.

The components of differential superforms of degree $p$ are defined by
\begin{equation}
  \label{sform}
  \Phi = \frac{1}{p!} dz^{M_p} \wedge \cdots \wedge dz^{M_1}
  \Phi_{M_1\cdots M_p}
\end{equation}
and the exterior derivative by
\begin{equation}
  \label{sform-d}
  d\Phi = \frac{1}{p!} dz^{M_p} \wedge \cdots \wedge dz^{M_1} \wedge
  dz^N \partial_N \Phi_{M_1\cdots M_p}.
\end{equation}
A simple calculation shows that as a consequence of the Leibniz rule
for the partial derivative and of (\ref{sform-d}) the rule of the
exterior differential acting on a product of a $q$-superform $\Psi$
and a $p$-superform $\Phi$ becomes
\begin{equation}
  \label{sform-Leibnitz}
  d(\Psi \wedge \Phi) = \Psi \wedge d\Phi + (-1)^{p} d\Psi
  \wedge \Phi.
\end{equation}
Thus we arrive at the simple prescription that $d$ effectively acts
\emph{from the right}. This should not be confused with the
\emph{partial derivative} in our convention acting to the right.

Each superfield $S(x,\theta)$ can be expanded in the anticommutative
variable ($\theta$-expansion)
\begin{equation}
  \label{sf-theta-exp}
  S(x,\theta) = s(x) + \theta^\lambda s_\lambda(x) + \half \theta^2
  s_2(x),
\end{equation}
so that the coefficients $s$, $s_\lambda$ and $s_2$ are functions of
the commutative variable $x^m$ only. For the zeroth order in $\theta$
the shorthand $S| = s$ is utilized. If $s$ is invertible, then
\begin{align}
  \label{sf-inv}
  S^{-1} = \frac{1}{s} - \frac{1}{s^2}\, \theta^\lambda s_\lambda -
  \half \theta^2\, \left(\frac{1}{s^3} s^\lambda s_\lambda +
    \frac{1}{s^2} s_2 \right)
\end{align}
is the inverse of the superfield $S$. 

For a matrix-valued superfield
\begin{equation}
  \label{sf-matrix}
  A = a + \theta^\lambda a_\lambda + \half \theta^2 a_2,
\end{equation}
where the coefficients $a$, $a_\lambda$ and $a_2$ are matrices of
equal dimensions, $a^{-1}$ exists and $A$ has definite parity $p(A) =
p(a) = p(a_\lambda) + 1 = p(a_2)$, the inverse reads
\begin{equation}
  \label{sf-matrix-inv}
  A^{-1} = a^{-1} - a^{-1} (\theta^\lambda a_\lambda) a^{-1} - \half
  \theta^2\, (a^{-1} a^\lambda a^{-1} a_\lambda a^{-1} + a^{-1} a_2
  a^{-1}).
\end{equation}

The Taylor expansion in terms of the $\theta$-variables of a function
$V(S)$, where $S$ is of the form (\ref{sf-theta-exp}), is given by
\begin{equation}
  \label{sf-Taylor}
  V(S) = V(s) + \theta^\lambda s_\lambda V'(s) + \half \theta^2
  \left[ s_2 V'(s) - \half s^\lambda s_\lambda V''(s) \right].
\end{equation}

\subsection{Superspace Metric}
\label{sec:smetric}

The invariant interval reads
\begin{equation}
  ds^2=dz^M\! \otimes dz^N G_{NM} = dz^M\! \otimes dz^N G_{MN}
  (-1)^{MN},
\end{equation}
where $G_{MN}$ is the superspace metric. This metric can be used to
lower indices of a vector field,
\begin{equation}
  \label{elowin}
  V_M=V^NG_{NM}=G_{MN}V^N(-1)^N.
\end{equation}
The generalization to an arbitrary tensor is obvious.  Defining the
inverse metric according to the rule
\begin{equation}
  V^M=G^{MN}V_N=V_NG^{NM}(-1)^N,
\end{equation}
and demanding that sequential lowering and raising indices shall be
the identical operation yields the main property of the inverse metric
\begin{equation}
  G^{MN} G_{NP} = \delta_P\^M (-1)^{MP} = \delta_P\^M (-1)^M = \delta_P\^M
  (-1)^P.
\end{equation}
The last identities follow from the diagonality of the Kronecker
symbol $\delta_P\^M = \delta_P^M$. Thus the inverse metric is not an
inverse matrix in the usual sense.  From (\ref{elowin}) the quantity
\begin{equation}
  V^2=V^MV_M=V_MV^M(-1)^M
\end{equation}
is a scalar, (but \eg $V_M V^M$ is not!).

\subsection{Linear Superconnection}
\label{sec:sconn}

We assume that our superspace is equipped with a Riemann-Cartan
geometry that is with a metric and with a metrical connection
$\Gamma_{MN}\^P$. The latter defines the covariant derivative of a
tensor field. Covariant derivatives of a vector $V^N$ and covector
$V_N$ read as
\begin{align}
  \nabla_M V^N &= \partial_M V^N + V^P \Gamma_{MP}\^N (-1)^{PM}, \\
  \nabla_M V_N &= \partial_M V_N - \Gamma_{MN}\^P V_P.
\end{align}
The metricity condition for the metric is
\begin{equation}                                          
  \label{emecom}
  \nabla_MG_{NP} = \partial_M G_{NP} - \Gamma_{MN}\^R G_{RP} -
  \Gamma_{MP}\^R G_{RN} (-1)^{NP} = 0.
\end{equation}
The action of an (anti)commutator of covariant derivatives,
\begin{equation}
  [\nabla_M,\nabla_N\} = \nabla_M\nabla_N-\nabla_N\nabla_M(-1)^{MN}
\end{equation}
on a vector field (\ref{elowin}),
\begin{equation}
  [\nabla_M,\nabla_N\}V_P=-R_{MNP}\^RV_R-T_{MN}\^R\nabla_RV_P,
\end{equation}
is defined in terms of curvature and torsion:
\begin{align}                                        
  R_{MNP}\^R &= \partial_M \Gamma_{NP}\^R - \Gamma_{MP}\^S
  \Gamma_{NS}\^R
  (-1)^{N(S+P)} - (M \leftrightarrow N) (-1)^{MN}, \label{ecurvs} \\
  T_{MN}\^R &= \Gamma_{MN}\^R - \Gamma_{NM}\^R (-1)^{MN}
  \label{etorss}
\end{align}

\subsection{Einstein-Cartan Variables}
\label{sec:cartan}

In our construction we use Cartan variables: the superspace vierbein
$E_M\^A$ and the superconnection $\LC_{MA}\^B$.  Capital Latin indices
from the beginning of the alphabet $(A={a,\alpha})$ transform under
the Lorentz group as a vector $(a=0,1)$ and spinor $(\alpha=1,2)$,
respectively. Cartan variables are defined by
\begin{equation}
  \label{eviers}
  G_{MN}=E_M\^AE_N\^B\eta_{BA}(-1)^{AN},
\end{equation}
and the metricity condition is
\begin{equation}
 \label{elocos}
 \nabla_ME_N\^A=\partial_ME_N\^A-\Gamma_{MN}\^PE_P\^A
 +E_N\^B\LC_{MB}\^A(-1)^{M(B+N)}=0.
\end{equation}
Raising and lowering of the anholonomic indices $(A,B,\dots)$ is
performed by the superspace Minkowski metric
\begin{equation}
  \label{eminms}
  \eta_{AB} = \mtrx{\eta_{ab}}{0}{0}{\epsilon_{\alpha\beta}}, \quad
  \eta^{AB} = \mtrx{\eta^{ab}}{0}{0}{\epsilon^{\alpha\beta}},
\end{equation}
consisting of the two-dimensional Minkowskian metric
$\eta_{ab}=\eta^{ab}=\diag(+-)$ and $\epsilon_{\alpha\beta}$, the
antisymmetric (Levi-Civita) tensor defined in \App\ref{app:spinors}.
The Minkowski metric and its inverse (\ref{eminms}) in superspace obey
\begin{equation}
  \eta_{AB}=\eta_{BA}(-1)^A, \quad \eta^{AB}\eta_{BC}=\delta_C\^A(-1)^A.
\end{equation}
The metric (\ref{eminms}) is invariant under the Lorentz group acting
on tensor indices from the beginning of the alphabet. In fact
(\ref{eminms}) is not unique in this respect because
$\epsilon_{\alpha\beta}$ may be multiplied by an arbitrary nonzero
factor. This may represent a freedom to generalize our present
approach. In fact, in order to have a correct dimension of all terms
in the line element of superspace, that factor should carry the
dimension of length. A specific choice for it presents a freedom in
approaches to supersymmetry. In the following this factor will be
suppressed. Therefore any apparent differences in dimensions between
terms below are not relevant.

The transformation of anholonomic indices $(A,B,\dots)$ into holonomic
ones $(M,N,\dots)$ and vice versa is performed using the supervierbein
and its inverse $E_A\^M$ defined as
\begin{equation}
  \label{einvsv}
  E_A\^ME_M\^B=\delta_A\^B, \quad E_M\^AE_A\^N=\delta_M\^N.
\end{equation}

To calculate the superdeterminant parts of the supervielbein and its
inverse are needed:
\begin{equation}
  \label{sdetE}
  \sdet(E_M\^A) = \det(E_m\^a) \det(E_\alpha\^\mu).
\end{equation}

In the calculations below we have found it extremely convenient to
work directly in the anholonomic basis
\begin{displaymath}
  \spartial_A := E_A\^M \partial_M,
\end{displaymath}
defined by the inverse supervierbein (for the ordinary zweibein the
notation $\partial_a = e_a\^m \partial_m$ is used).

The anholonomicity coefficients, defined by $[\spartial_A,
\spartial_B] = C_{AB}\^C \spartial_C$, or in terms of differential
forms by $d E^A = -C^A$, can be calculated from the supervierbein and
its inverse
\begin{equation}
  \label{anholC}
  C_{AB}\^C = (E_A\^N \partial_N E_B\^M - (-1)^{AB} E_B\^N \partial_N
  E_A\^M) E_M\^C.
\end{equation}

The metricity condition (\ref{elocos}) formally establishes a
one-to-one correspondence between the metrical connection
$\Gamma_{MN}\^P$ and the superconnection $\LC_{MA}\^B$.  Together with
(\ref{emecom}) it implies
\begin{equation}
  \nabla_M \eta_{AB} = 0
\end{equation}
and the symmetry property
\begin{equation}
  \label{esupco}
  \LC_{MAB} + \LC_{MBA} (-1)^{AB} = 0.
\end{equation}
In general, the superconnection $\LC_{MA}\^B$ is not related to
Lorentz transformations alone. The Lorentz connection in superspace
must have a specific form and is defined (in $d = 2$) by $\LC_M$, a
superfield with one vector index,
\begin{equation}
  \label{elorcs}
  \LC_{MA}\^B = \LC_M L_A\^B,
\end{equation}
where
\begin{equation}
  \label{eantis}
  L_A\^B = \mtrx{\epsilon_a\^b}{0}{0}{-\frac12 \gamma^3\_\alpha\^\beta}
\end{equation}
contains the Lorentz generators in the bosonic and fermionic sectors.
Here the factor in front of $\gamma^3$ is fixed by the requirement
that under Lorentz transformations $\gamma$-matrices are invariant
under simultaneous rotations of vector and spinor indices. Definition
and properties of $\gamma$-matrices are given in
\App\ref{app:spinors}.  $L_A\^B$ has the properties
\begin{equation}
  L_{AB} = -L_{BA} (-1)^A, \quad L_A\^B L_B\^C =
  \mtrx{\delta_a\^c}{0}{0}{\frac{1}{4} \delta_{\alpha}\^{\gamma}},
  \quad \nabla_M L_A\^B = 0.
\end{equation}
The superconnection $\LC_{MA}\^B$ in the form (\ref{elorcs}) is very
restricted because the original 32 independent superfield components
for the Lorentz superconnection reduce to 4. As a consequence,
(\ref{elocos}) with (\ref{elorcs}) also entails restrictions on the
metric connection $\Gamma_{MN}\^P$.

In terms of the connection (\ref{elorcs}) covariant derivatives of a
Lorentz supervector read
\begin{align}
  \nabla_M V^A &= \partial_M V^A + \LC_M V^B L_B\^A, \\
  \nabla_M V_A &= \partial_M V_A - \LC_M L_A\^B V_B.
\end{align}

\subsection{Supercurvature and Supertorsion}
\label{sec:scurtor}

The (anti)commutator of covariant derivatives,
\begin{equation}
  [\nabla_M,\nabla_N\}V_A =-R_{MNA}\^BV_B-T_{MN}\^P\nabla_PV_A,
\end{equation}
is defined by the same expressions for curvature and torsion as given
by (\ref{ecurvs}) and (\ref{etorss}), which in Cartan variables become
\begin{align}
  R_{MNA}\^B &= \partial_M \LC_{NA}\^B - \LC_{MA}\^C \LC_{NC}\^B
  (-1)^{N(A+C)} - (M \leftrightarrow N) (-1)^{MN}, \label{ecarcu} \\
  T_{MN}\^A &= \partial_M E_N\^A + E_N\^B \LC_{MB}\^A (-1)^{M(B+N)} -
  (M \leftrightarrow N) (-1)^{MN}. \label{ecarto}
\end{align}
In the anholonomic basis (\ref{ecarto}) turns into
\begin{equation}
  \label{stor}
  T_{AB}\^C = -C_{AB}\^C + \Omega_A L_B\^C - (-1)^{AB} \Omega_B L_A\^C,
\end{equation}
which due to the special form of the Lorentz connection (\ref{elorcs})
yields for the various components
\begin{align}
  T_{\alpha\beta}\^\gamma &= -C_{\alpha\beta}\^\gamma - \half
  \Omega_\alpha \gamma^3\_\beta\^\gamma - \half \Omega_\beta
  \gamma^3\_\alpha\^\gamma, \label{T-fff} \\
  T_{\alpha\beta}\^c &= -C_{\alpha\beta}\^c, \label{T-ffb} \\
  T_{a\beta}\^\gamma &= -C_{a\beta}\^\gamma - \half \Omega_a
  \gamma^3\_\beta\^\gamma, \label{T-bff} \\
  T_{\alpha b}\^c &= -C_{\alpha b}\^c + \Omega_\alpha \epsilon_b\^c,
  \label{T-fbb} \\
  T_{ab}\^\gamma &= -C_{ab}\^\gamma, \label{T-bbf} \\
  T_{ab}\^c &= -C_{ab}\^c + \Omega_a \epsilon_b\^c - \Omega_b
  \epsilon_a\^c. \label{T-bbb}
\end{align}
In terms of the Lorentz connection (\ref{elorcs}) the curvature does
not contain quadratic terms
\begin{equation}
  \label{ecurts}
  R_{MNA}\^B = \left(\partial_M \LC_N - \partial_N \LC_M (-1)^{MN}\right)
  L_A\^B = F_{MN} L_A\^B,
\end{equation}
where $F_{AB}$ can be calculated in the anholonomic basis by the
formulae
\begin{align}
  F_{AB} &= \spartial_A \Omega_B - (-1)^{AB} \spartial_B \Omega_A -
  C_{AB}\^C \Omega_C, \\
  &= \nabla_A \Omega_B - (-1)^{AB} \nabla_B \Omega_A + T_{AB}\^C
  \Omega_C. \label{F-cov}
\end{align}
Ricci tensor and scalar curvature of the manifold are
\begin{align}
  R_{AB} &= R_{CAB}\^C (-1)^{C(A+B+C)} = L_B\^C F_{CA} (-1)^{AB},
  \label{esrict} \\
  R &= R_A\^A (-1)^A = L^{AB} F_{BA}. \label{esuscu}
\end{align}

\subsection{Bianchi Identities}
\label{sec:bianchi}

The first Bianchi identity $\s\D T^A = E^B \wedge R_B\^A$ reads in
component form
\begin{equation}
  \label{bianchi1}
  \Delta_{ABC}\^D = R_{[ABC]}\^D,
\end{equation}
where
\begin{multline}
  \Delta_{ABC}\^D := \nabla_{[A} T_{BC]}\^D + T_{[AB|}\^E T_{E|C]}\^D
  \\
  = \nabla_A T_{BC}\^D + \nabla_B T_{CA}\^D (-1)^{A(B+C)} + \nabla_C
  T_{AB}\^D (-1)^{C(A+B)} \\ + T_{AB}\^E T_{EC}\^D + T_{BC}\^E
  T_{EA}\^D (-1)^{A(B+C)} + T_{CA}\^E T_{EB}\^D (-1)^{C(A+B)}
\end{multline}
and
\begin{equation}
  R_{[ABC]}\^D = F_{AB} L_C\^D + F_{BC} L_A\^D (-1)^{A(B+C)} + F_{CA}
  L_B\^D (-1)^{C(A+B)}.
\end{equation}
Due to the restricted form of the Lorentz connection (\ref{elorcs})
the bosonic and spinorial parts of $R_{[ABC]}\^D$ are given by
\begin{align}
  R_{[\alpha\beta\gamma]}\^d &= 0, \label{R-fffb} \\
  R_{[\alpha\beta\gamma]}\^\delta &= -\half F_{\alpha\beta}
  \gamma^3\_\gamma\^\delta - \half F_{\beta\gamma}
  \gamma^3\_\alpha\^\delta - \half F_{\gamma\alpha}
  \gamma^3\_\beta\^\delta, \\
  R_{[\alpha\beta c]}\^d &= F_{\alpha\beta} \epsilon_c\^d, \\
  R_{[a\beta\gamma]}\^\delta &= -\half F_{a\beta}
  \gamma^3\_\gamma\^\delta - \half F_{a\gamma} \gamma^3\_\beta\^\delta,
  \\
  R_{[a\beta c]}\^d &= F_{a\beta} \epsilon_c\^d - F_{c\beta}
  \epsilon_a\^d, \\
  R_{[ab\gamma]}\^\delta &= -\half F_{ab} \gamma^3\_\gamma\^\delta, \\
  R_{[abc]}\^d &= F_{ab} \epsilon_c\^d + F_{bc} \epsilon_a\^d + F_{ca}
  \epsilon_b\^d, \\
  R_{[abc]}\^\delta &= 0. \label{R-bbbf}
\end{align}

The second Bianchi identity $\s\D R_A\^B = 0$, where the various
components are denoted by the symbol $\Delta_{ABC}$,
\begin{align}
  \Delta_{ABC} &:= \nabla_{[A} F_{BC]} + T_{[AB|}\^D F_{D|C]} \\
  &= \nabla_A F_{BC} + \nabla_B F_{CA} (-1)^{A(B+C)} + \nabla_C F_{AB}
  (-1)^{C(A+B)} \eqnsplit + T_{AB}\^D F_{DC} + T_{BC}\^D F_{DA}
  (-1)^{A(B+C)} + T_{CA}\^D F_{DB} (-1)^{C(A+B)},
\end{align}
is stated in component form
\begin{equation}
  \label{bianchi2}
  \Delta_{ABC} = 0.
\end{equation}

\section{Decomposition of the Supervielbein}
\label{sec:svb}

For the solution of the supergravity constraints it will be very
useful to decompose the supervierbein and its inverse in terms of the
new superfields $B_m\^a$, $B_\mu\^\alpha$, $\Phi_\mu\^m$ and
$\Psi_m\^\mu$:
\begin{align}
  \label{SVdecomp}
  E_M\^A &= \mtrx{B_m\^a}{\Psi_m\^\nu B_\nu\^\alpha}{\Phi_\mu\^n
    B_n\^a}{B_\mu\^\alpha + \Phi_\mu\^n \Psi_n\^\nu B_\nu\^\alpha} \\
  E_A\^M &= \mtrx{B_a\^m + B_a\^n \Psi_n\^\nu \Phi_\nu\^m}{-B_a\^n
    \Psi_n\^\mu}{-B_\alpha\^\nu
    \Phi_\nu\^m}{B_\alpha\^\mu} \label{ISVdecomp}
\end{align}
The superfields $B_a\^m$ and $B_\alpha\^\mu$ are the inverse of
$B_m\^a$ and $B_\mu\^\alpha$, respectively:
\begin{equation}
  \label{svb-B-inv}
  B_a\^m B_m\^b = \delta_a\^b, \qquad B_\alpha\^\mu B_\mu\^\beta =
  \delta_\alpha\^\beta.
\end{equation}
To shorten notation the frequently occurring products of $\Phi_\mu\^m$
and $\Psi_m\^\mu$ with the $B$-fields are abbreviated by
\begin{alignat}{3}
  \Phi_\nu\^a &= \Phi_\nu\^m B_m\^a, &\qquad
  \Phi_\alpha\^m &= B_\alpha\^\nu \Phi_\nu\^m, &\qquad
  \Phi_\alpha\^a &= B_\alpha\^\nu \Phi_\nu\^m B_m\^a, \label{svb-Phi} \\
  \Psi_a\^\nu &= B_a\^m \Psi_m\^\nu, &\qquad
  \Psi_m\^\alpha &= \Psi_m\^\nu B_\nu\^\alpha, &\qquad
  \Psi_a\^\alpha &= B_a\^m \Psi_m\^\nu B_\nu\^\alpha. \label{svb-Psi}
\end{alignat}
The superdeterminant of the supervielbein (\ref{sdetE}) expressed in
terms of $B_m\^a$ and $B_\mu\^\alpha$ reads
\begin{equation}
  \label{svb-sdet}
  \sdet(E_M\^A) = \frac{\det(B_m\^a)}{\det(B_\mu\^\alpha)}.
\end{equation}

\newcommand{\fermB}{B_\alpha\^\mu B_\beta\^\nu + B_\beta\^\mu
  B_\alpha\^\nu} \newcommand{\bosB}{B_a\^m B_b\^n - B_b\^m B_a\^n}

Now the anholonomicity coefficients (\ref{anholC}) in terms of their
decomposition (\ref{SVdecomp}) become
\begin{equation}
  C_{\alpha\beta}\^c = (\fermB) (\Phi_\mu\^l \partial_l
  \Phi_\nu\^n - \partial_\mu \Phi_\nu\^n) B_n\^c, \label{C-ffb}
\end{equation}
\begin{multline}
  C_{\alpha\beta}\^\gamma =  (\fermB) (\Phi_\mu\^l \partial_l
  \Phi_\nu\^n - \partial_\mu \Phi_\nu\^n) \Psi_n\^\gamma \\
  + (\fermB) (\Phi_\mu\^l \partial_l B_\nu\^\gamma -
  \partial_\mu B_\nu\^\gamma), \label{C-fff}
\end{multline}
\begin{multline}
  C_{a\beta}\^c = -\Psi_a\^\alpha (\fermB) (\Phi_\mu\^l \partial_l
  \Phi_\nu\^n - \partial_\mu \Phi_\nu\^n) B_n\^c \\
  - B_a\^n B_\beta\^\mu ((\partial_n \Phi_\mu\^l) B_l\^c +
  \Phi_\mu\^l \partial_l B_n\^c - \partial_\mu B_n\^c), \label{C-bfb}
\end{multline}
\begin{multline}
  C_{a\beta}\^\gamma = -\Psi_a\^\alpha (\fermB) (\Phi_\mu\^l
  \partial_l \Phi_\nu\^n - \partial_\mu \Phi_\nu\^n) \Psi_n\^\gamma
  \\
  - \Psi_a\^\alpha (\fermB) (\Phi_\mu\^l \partial_l
  B_\nu\^\gamma - \partial_\mu B_\nu\^\gamma) \\
  - B_a\^n B_\beta\^\mu ((\partial_n \Phi_\mu\^l) \Psi_l\^\gamma
  + \Phi_\mu\^l \partial_l \Psi_n\^\gamma - \partial_\mu
  \Psi_n\^\gamma + \partial_n B_\mu\^\gamma), \label{C-bff}
\end{multline}

\begin{multline}
  C_{ab}\^c = \Psi_b\^\beta \Psi_a\^\alpha (\fermB) (\Phi_\mu\^l
  \partial_l \Phi_\nu\^n - \partial_\mu \Phi_\nu\^n) B_n\^c \\
  - (\bosB) \Psi_m\^\mu ((\partial_n \Phi_\mu\^l) B_l\^c +
  \Phi_\mu\^l \partial_l B_n\^c - \partial_\mu B_n\^c) \\
  + (B_a\^m \partial_m B_b\^n - B_b\^m \partial_m B_a\^n) B_n\^c,
  \label{C-bbb}
\end{multline}
\begin{multline}
  C_{ab}\^\gamma = \Psi_b\^\beta \Psi_a\^\alpha (\fermB)
  (\Phi_\mu\^l \partial_l \Phi_\nu\^n - \partial_\mu \Phi_\nu\^n)
  \Psi_n\^\gamma \\
  - (\bosB) \Psi_m\^\mu ((\partial_n \Phi_\mu\^l) \Psi_l\^\gamma
  + \Phi_\mu\^l \partial_l \Psi_n\^\gamma - \partial_\mu
  \Psi_n\^\gamma + \partial_n B_\mu\^\gamma) \\
  - (\bosB) (\partial_m \Psi_n\^\gamma). \label{C-bbf}
\end{multline}

\section{Supergravity Model of Howe}
\label{sec:sugra-howe}

Our next task is the recalculation of the two-dimensional supergravity
model originally found in \cite{Howe:1979ia} and also in
\cite{\bibHoweOther}. It is not our intention to review the steps of
calculation in this section, we merely give a summary of the results
for reference purposes. Details of the calculation can be found in
\cite{Ertl:1997ib} and in \Sec\ref{sec:sugra-ertl} below, where a more
general supergravity model with torsion is considered.

The original supergravity constraints chosen by Howe were
\begin{equation}
  \label{constr-Howe}
  T_{\alpha\beta}\^\gamma = 0, \qquad
  T_{\alpha\beta}\^c = 2i \gamma^c\_{\alpha\beta}, \qquad
  T_{ab}\^c = 0.
\end{equation}
An equivalent set of constraints is
\begin{equation}
  \label{constr-Howe2}
  T_{\alpha\beta}\^\gamma = 0, \qquad
  T_{\alpha\beta}\^c = 2i \gamma^c\_{\alpha\beta}, \qquad
  \gamma_a\^{\beta\alpha} F_{\alpha\beta} = 0,
\end{equation}
as can be seen when inspecting the Bianchi identities.

\subsection{Component Fields}
\label{sec:Howe-comp}

The constraints (\ref{constr-Howe}) are identically fulfilled with the
expressions for the supervielbein and the Lorentz superconnection
below. These expression are obtained using a Wess-Zumino type gauge;
details can be found in \cite{Ertl:1997ib}. We also specify the
superfields of the decomposition of the supervielbein (\ref{SVdecomp}),
which is needed to solve the supergravity constraints with superfield
methods (\cf \Sec\ref{sec:ertl-constr} below).

As physical $x$-space variables one obtains the vielbein $e_m\^a$, the
Rarita-Schwinger field $\RS_m\^\alpha$ and the auxiliary scalar field
$A$. They constitute the supergravity multiplet $\{ e_m\^a,
\RS_m\^\alpha, A \}$. Although the multiplet was derived using
Einstein-Cartan variables in superspace, in the residual $x$-space no
Einstein-Cartan variables were left over. There is no independent
Lorentz connection. As a consequence of the constraints
(\ref{constr-Howe}) it is eliminated by the condition $t_{ab}\^c = 0$,
where
\begin{equation}
  \label{Howe-tor}
  t_{ab}\^c = -c_{ab}\^c + \omega_a \epsilon_b\^c - \omega_b
  \epsilon_a\^c - 2i (\RS_a \gamma^c \RS_b),
\end{equation}
yielding $\omega_a = \susy\omega_a$ with
\begin{align}
  \label{Howe-om}
  \susy\omega_a &:= \tilde\omega_a - i \epsilon^{bc} (\RS_c \gamma_a
  \RS_b) \\ &\phantom{:}= \tilde\omega_a - 4i (\RS \gamma^3 \lambda_c).
\end{align}
Here $\tilde\omega^a = \epsilon^{nm} (\partial_m e_n\^a)$ is the usual
torsion free connection of pure bosonic gravity, and for the second
line (\ref{rs-ga-rs}) was used. The term quadratic in $\RS_m\^\alpha$
is necessary to make $\susy\omega_a$ covariant with respect to
supersymmetry transformations \ie no derivative of the supersymmetry
parameter $\epsilon^\alpha$ shows up in its transformation rule.

A similar argument applies to the derivative of the Rarita-Schwinger
field
\begin{align}
  \susy\sigma_\mu &:= \epsilon^{mn} \left( \partial_n \RS_{m\mu} + \half
    \susy\omega_n (\gamma^3 \RS_m)_\mu - \frac{i}{2} A (\gamma_n
    \RS_m)_\mu \right), \label{Howe-si} \\
  &\phantom{:}= \epsilon^{nm} \susy\D_m \RS_{n\mu} + i A
  (\gamma^3\RS)_\mu,
\end{align}
which is obviously covariant with respect to Lorentz transformations,
and also with respect to supersymmetry.

We obtain for the supervielbein
\begin{align}
  E_m\^a &= e_m\^a + 2i (\theta \gamma^a \RS_m) + \half \theta^2
  \left[ A e_m\^a \right], \label{Howe-E-bb} \\
  E_m\^\alpha &= \RS_m\^\alpha - \half \susy\omega_m (\theta
  \gamma^3)^\alpha + \frac{i}{2} A (\theta \gamma_m)^\alpha
  - \half \theta^2 \left[ \frac{3}{2} A \RS_m\^\alpha + i
    (\susy\sigma \gamma_m \gamma^3)^\alpha \right], \label{Howe-E-bf} \\
  E_\mu\^a &= i (\theta \gamma^a)_\mu, \label{Howe-E-fb} \\
  E_\mu\^\alpha &= \delta_\mu\^\alpha + \half \theta^2 \left[ -\half A
    \delta_\mu\^\alpha \right], \label{Howe-E-ff}
\end{align}
and for its inverse
\begin{align}
  E_a\^m &= e_a\^m - i (\theta \gamma^m \RS_a) + \half \theta^2
  \left[ -2 (\RS_a \lambda^m) \right], \\
  E_a\^\mu &= -\RS_a\^\mu + i (\theta \gamma^b \RS_a) \RS_b\^\mu +
  \half \susy\omega_a (\theta \gamma^3)^\mu - \frac{i}{2} A (\theta
  \gamma_a)^\mu \eqnsplit + \half \theta^2 \left[ 2 (\RS_a
    \lambda^b) \RS_b\^\mu - \frac{i}{2} \susy\omega_b (\RS_a \gamma^b
    \gamma^3)^\mu + i (\susy\sigma \gamma_a \gamma^3)^\mu \right], \\
  E_\alpha\^m &= -i (\theta \gamma^m)_\alpha + \half \theta^2 \left[ -2
    \lambda^m\_\alpha \right], \\
  E_\alpha\^\mu &= \delta_\alpha\^\mu + i (\theta \gamma^b)_\alpha
  \RS_b\^\mu + \half \theta^2 \left[ 2 \lambda^b\_\alpha \RS_b\^\mu
    - \frac{i}{2} \susy\omega_b (\gamma^b \gamma^3)_\alpha\^\mu - \half A
    \delta_\alpha\^\mu \right].
\end{align}

The above expressions for the supervielbein are derived from the
decomposition (\ref{SVdecomp}). The $B_a\^m$ superfield and its
inverse with the zweibein at zeroth order read
\begin{align}
  B_a\^m &= e_a\^m - 2i (\theta \gamma^m \RS_a) + \half \theta^2
  \left[ -8 (\RS_a \lambda^m) - A e_a\^m \right], \\
  B_m\^a &= e_m\^a + 2i (\theta \gamma^a \RS_m) + \half \theta^2
  \left[ A e_m\^a \right].
\end{align}
The $B_\alpha\^\mu$ and its inverse are given by
\begin{align}
  B_\alpha\^\mu &= \delta_\alpha\^\mu + i (\theta \gamma^b)_\alpha
  \RS_b\^\mu + \half \theta^2 \left[ 2 \lambda^b\_\alpha \RS_b\^\mu
    - \frac{i}{2} \susy\omega_b (\gamma^b \gamma^3)_\alpha\^\mu - \half A
    \delta_\alpha\^\mu \right], \\
  B_\mu\^\alpha &= \delta_\mu\^\alpha - i (\theta \gamma^b)_\mu
  \RS_b\^\alpha + \half \theta^2 \left[ -4 \lambda^b\_\mu
    \RS_b\^\alpha + \frac{i}{2} \susy\omega_b (\gamma^b
    \gamma^3)_\mu\^\alpha + \half A \delta_\mu\^\alpha \right].
\end{align}
We note that the lowest order $\delta_\alpha\^\mu$ is responsible for
the promotion of the coordinates $\theta^\mu$ to Lorentz spinors. In
the superfield $\Phi_\mu\^m$, the first term inherently carries the
superspace structure,
\begin{equation}
  \Phi_\mu\^m = i (\theta \gamma^m)_\mu + \half \theta^2 \left[ 4
    \lambda^m\_\mu \right],
\end{equation}
and $\Psi_m\^\mu$, whose zeroth component is the Rarita-Schwinger
field, reads
\begin{align}
  \Psi_m\^\mu &= \RS_m\^\mu + i (\theta \gamma^b \RS_m) \RS_b\^\mu
  - \half \susy\omega_m (\theta \gamma^3)^\mu + \frac{i}{2} A (\theta
  \gamma_m)^\mu \eqnsplit + \half \theta^2 \left[ 2 (\RS_m \lambda^b)
    \RS_b\^\mu - \frac{i}{2} \susy\omega_b (\RS_m \gamma^b \gamma^3)^\mu
    - A \RS_m\^\mu - i (\susy\sigma \gamma_m \gamma^3)^\mu \right].
\end{align}

The equally complicated result for the Lorentz superconnection becomes
\begin{align}
  \Omega_a &= \susy\omega_a - i (\theta \gamma^b \RS_a) \susy\omega_b + A
  (\theta \gamma^3 \RS_a) + 2i (\theta \gamma_a \susy\sigma) \eqnsplit +
  \half \theta^2 \left[ -2 (\RS_a \lambda^b) \susy\omega_b + 2i A (\RS
    \gamma^3 \lambda_a) + 4 (\RS \gamma_a \susy\sigma) + \epsilon_a\^b
    (\partial_b A) \right], \\
  \Omega_\alpha &= -i (\theta \gamma^b)_\alpha \susy\omega_b + A (\theta
  \gamma^3)_\alpha + \half \theta^2 \left[ -2 \lambda^b\_\alpha
    \susy\omega_b - 2i A (\gamma^3 \RS)_\alpha + 4 \susy\sigma_\alpha
  \right],
\end{align}
and also for world indices
\begin{align}
  \Omega_m &= \susy\omega_m + 2 A (\theta \gamma^3 \RS_m) + 2i (\theta
  \gamma_m \susy\sigma) \eqnsplit + \half \theta^2 \left[ A
    \susy\omega_m -4 (\lambda_m \susy\sigma) + \epsilon_m\^n
    (\partial_n A) \right], \label{Howe-Om-b} \\
  \Omega_\mu &= A (\theta \gamma^3)_\mu. \label{Howe-Om-f}
\end{align}


Finally, the superdeterminant can be obtained from (\ref{sdetE}) or
(\ref{svb-sdet}):
\begin{equation}
  \label{Howe-E}
  E = e \left( 1 - 2i (\theta \RS) + \half \theta^2 \left[ A + 2
      \RS^2 + \lambda^2 \right] \right)
\end{equation}

The formulae in this section were cross-checked against a computer
calculation for which a symbolic computer algebra program was adapted
\cite{Ertl:Index-0.14.2}.

\subsection{Symmetry Transformations}
\label{sec:Howe-symms}

Under a superdiffeomorphism $\delta z^M = -\xi^M(z)$ and a local
Lorentz transformation $\delta V^A = V^B L(z) L_B\^A$ the
supervielbein transforms according to
\begin{equation}
  \label{svb-trafo}
  \delta E_M\^A = -\xi^N (\partial_N E_M\^A) - (\partial_M \xi^N)
  E_N\^A + L E_M\^B L_B\^A.
\end{equation}
There are $16+4=20$ $x$-space transformation parameters in the
superfields $\xi^M(x,\theta)$ and $L(x,\theta)$. The $10+5=15$ gauge
fixing conditions (\ref{gf-0}) and (\ref{gf-1}) reduce this number to
$5$.\footnote{The gauge conditions are indeed the same as the one in
  \Sec\ref{sec:ertl-gf} below.} The remaining symmetries are the
$x$-space diffeomorphism with parameter $\eta^m(x)$, the local Lorentz
transformation with parameter $l(x)$ and the local supersymmetry
parametrized with $\epsilon^\alpha(x)$. The superfields for $x$-space
diffeomorphism are characterized by
\begin{equation}
  \xi^m = \eta^m, \qquad \xi^\mu = 0, \qquad L = 0,
\end{equation}
and local Lorentz transformations by
\begin{equation}
  \label{Howe-symm-l}
  \xi^m = 0, \qquad
  \xi^\mu = -\half l (\theta \gamma^3)^\mu, \qquad
  L = l.
\end{equation}
In addition to the frame rotation $L = l$ there is also a rotation of
the coordinate $\theta^\mu$ as expressed by the term for $\xi^\mu$ in
(\ref{Howe-symm-l}). This is a consequence of the gauge condition
$E_\mu\^\alpha| = \delta_\mu\^\alpha$ (\cf (\ref{gf-0})) and leads to
the identification of the anticommutative coordinate $\theta^\mu$ as a
Lorentz spinor. The superfield transformation parameters of local
supersymmetry are a bit more complicated:
\begin{align}
  \xi^m &= -i (\epsilon \gamma^m \theta) + \half \theta^2 \left[ 2
    (\epsilon \lambda^m) \right], \\
  \xi^\mu &= \epsilon^\mu + i (\epsilon \gamma^b \theta)
  \RS_b\^\mu + \half \theta^2 \left[ -2 (\epsilon \lambda^b)
    \RS_b\^\mu + \frac{i}{2} \susy\omega_b (\epsilon \gamma^b
    \gamma^3)^\mu \right], \\
  L &= i (\epsilon \gamma^b \theta) \susy\omega_b - A (\epsilon
  \gamma^3 \theta) + \half \theta^2 \left[ -2 (\epsilon \lambda^b)
    \susy\omega_b - 2i A (\epsilon \gamma^3 \RS) + 2 (\epsilon
    \susy\sigma) \right].
\end{align}
These expressions are derived by $\theta$-expansion of
(\ref{svb-trafo}), and by taking the gauge conditions (\ref{gf-0}) and
(\ref{gf-1}) into account. Details of the calculation can be found in
\cite{Ertl:1997ib}.

Similarly, from (\ref{svb-trafo}) the transformation laws of the
physical $x$-space fields
\begin{align}
  \delta e_a\^m &= 2i (\epsilon \gamma^m \RS_a), \qquad \delta
  e_m\^a = -2i (\epsilon \gamma^a \RS_m), \\
  \delta \RS_m\^\mu &= -\left( \partial_m \epsilon^\mu - \half
    \susy\omega_m (\epsilon \gamma^3)^\mu + \frac{i}{2} A (\epsilon
    \gamma_m)^\mu \right), \\
  \delta A &= -2 (\epsilon \gamma^3 \susy\sigma)
\end{align}
follow. Also the variation of the dependent Lorentz connection
$\susy\omega_m$ defined in (\ref{Howe-om}) might be of interest:
\begin{equation}
  \delta \susy\omega_m =  -2i (\epsilon \gamma_m \susy\sigma) - 2 A
  (\epsilon \gamma^3 \RS_m).
\end{equation}

\subsection{Supertorsion and Supercurvature}
\label{sec:Howe-torcur}

The Bianchi identities (\ref{bianchi1}) and (\ref{bianchi2}) show that
as a consequence of the supergravity constraints (\ref{constr-Howe})
all components of supertorsion and supercurvature depend on only one
single superfield $S$:
\begin{alignat}{2}
  T_{\alpha\beta}\^c &= 2i \gamma^c\_{\alpha\beta}, &\qquad
  T_{\alpha\beta}\^\gamma &= 0, \\
  T_{\alpha b}\^c &= 0, &\qquad
  T_{a\beta}\^\gamma &= -\frac{i}{2} S \gamma_{a\beta}\^\gamma, \\
  T_{ab}\^c &= 0, &\qquad
  T_{ab}\^\gamma &= \half \epsilon_{ab} \gamma^3\^{\gamma\delta}
  \nabla_\delta S,
\end{alignat}
and
\begin{align}
  F_{\alpha\beta} &= 2 S \gamma^3\_{\alpha\beta}, \label{Howe-F-ff} \\
  F_{a\beta} &= -i (\gamma_a \gamma^3)_\beta\^\gamma \nabla_\gamma S,
  \\
  F_{ab} &= \epsilon_{ab} \left( S^2 - \half \nabla^\alpha
    \nabla_\alpha S \right).
\end{align}
Using the component field expansion of supervielbein and
supercurvature stated in \Sec\ref{sec:Howe-comp} one finds
\begin{equation}
  \label{Howe-S-field}
  S = A + 2 (\theta \gamma^3 \susy\sigma) + \half \theta^2 \left[
    \epsilon^{mn} (\partial_n \susy\omega_m) - A (A + 2 \RS^2 + \lambda^2) - 
    4i (\RS \gamma^3 \susy\sigma) \right].
\end{equation}
Here $\susy\omega_m$ and $\susy\sigma_\mu$ are the expressions defined
in (\ref{Howe-om}) and (\ref{Howe-si}).

The superfield $S$ is the only quantity that can be used to build
action functionals invariant with respect to local supersymmetry. A
general action in superspace invariant with respect to
superdiffeomorphisms takes the form
\begin{equation}
  \label{sf-action}
  \Action = \int d^2\!x \underline{d}^2\!\theta\, E\, F(x,\theta),
\end{equation}
where $E = \sdet(E_M\^A)$ is the superdeterminant of the supervielbein
and $F(x,\theta)$ is an arbitrary scalar superfield. For an
explanation of the Berezin integration (\ref{sf-action}) we refer to
\cite{\bibSuperfield}. The superspace actions
\begin{equation}
  \Action_n = \int d^2\!x \underline{d}^2\!\theta\, E\, S^n
\end{equation}
constructed with $E$ and $S$ from above (\cf (\ref{Howe-E}) and
(\ref{Howe-S-field})) result in
\begin{align}
  \Action_0 &= \int d^2\!x\, e \left( A + 2 \RS^2 + \lambda^2 \right), \\
  \Action_1 &= \int d^2\!x\, \frac{e}{2} \susy r, \\
  \Action_2 &= \int d^2\!x\, e \left( A \susy r + 4 \susy\sigma^2 -
    A^2 (A + 2 \RS^2 + \lambda^2) \right),
\end{align}
where the $\theta$-integration was done, and where $\susy r := 2
\epsilon^{mn} (\partial_n \susy\omega_m)$.

\section{Lorentz Covariant Supersymmetry}
\label{sec:susy-cov}

As a preparation for the `new' supergravity in this section we
consider a theory with covariantly constant supersymmetry.

Starting from the supervielbein and superconnection of rigid SUSY we
separate the coordinate space and the tangent space by the
introduction of the zweibein $e_a\^m(x)$. This makes the theory
covariant with respect to $x$-space coordinate transformations given
by the superspace parameters $\xi^m(x, \theta) = \eta^m(x)$,
$\xi^\mu(x, \theta) = 0$ and $L(x, \theta) = 0$. The local Lorentz
transformation in the tangent space is given by the superspace
parameter $L(x, \theta) = l(x)$. The gauge fixing condition
$B_\alpha\^\mu| = \delta_\alpha\^\mu$ shows that this local Lorentz
transformation is accompanied by a superdiffeomorphism with parameters
$\xi^m(x, \theta) = 0$ and $\xi^\mu(x, \theta) = -\half l(x) (\theta
\gamma^3)^\mu$. When one calculates the transformations under this
local Lorentz boost of the various component fields of the
supervielbein and superconnection one finds that some components
transform into a derivative of the transformation parameter. In order
to isolate this inhomogeneous term we introduce a connection field
$\omega_m(x)$ with the Lorentz transformation property $\delta
\omega_m = - \partial_m l$. In this way the superfields listed below
are derived:
\begin{align}
  E_M\^A &= \mtrx{e_m\^a}{-\half \omega_m (\theta \gamma^3)^\alpha}{i
    (\theta \gamma^a)_\mu}{\delta_\mu\^\alpha} \\
  E_A\^M &= \mtrx{e_a\^m}{\half \omega_a (\theta \gamma^3)^\mu}{-i
    (\theta \gamma^m)_\alpha}{\delta_\alpha\^\mu - \frac{i}{4}
    \theta^2 \omega_n (\gamma^n \gamma^3)_\alpha\^\mu}
  \label{susy-svb-inv}
\end{align}
\begin{alignat}{2}
  B_a\^m &= e_a\^m &\qquad
  B_m\^a &= e_m\^a \\
  B_\alpha\^\mu &= \delta_\alpha\^\mu - \frac{i}{4} \theta^2 \omega_n
  (\gamma^n \gamma^3)_\alpha\^\mu &\qquad
  B_\mu\^\alpha &= \delta_\mu\^\alpha + \frac{i}{4} \theta^2 \omega_n
  (\gamma^n \gamma^3)_\mu\^\alpha \\
  \Phi_\nu\^n &= i (\theta \gamma^n)_\nu &\qquad
  \Psi_n\^\nu &= -\half \omega_n (\theta \gamma^3)^\nu  
\end{alignat}
\begin{alignat}{2}
  \Omega_\alpha &= -i\, \omega_n (\theta \gamma^n)_\alpha &\qquad
  \Omega_a &= \omega_a \\
  \Omega_\mu &= 0 &\qquad \Omega_m &= \omega_m
\end{alignat}

From (\ref{susy-svb-inv}) the superspace covariant derivatives
$\spartial_A = E_A\^M \partial_M$, now also covariant with respect to
local Lorentz transformation,
\begin{align}
  \spartial_\alpha &= \delta_\alpha\^\mu \partial_\mu - i (\theta
  \gamma^m)_\alpha \partial_m - \frac{i}{4} \theta^2 \omega_n
  (\gamma^n \gamma^3)_\alpha\^\mu \partial_\mu, \\
  \spartial_a &= e_a\^m \partial_m + \half \omega_a (\theta
  \gamma^3)^\mu \partial_\mu
\end{align}
are obtained.

The anholonomicity coefficients are derived from (\ref{anholC})
yielding
\begin{align}
  C_{\alpha\beta}\^c &= -2i \gamma^c\_{\alpha\beta} - \theta^2\,
  \tor^c \gamma^3\_{\alpha\beta}, \\
  C_{\alpha\beta}\^\gamma &= \frac{i}{2} \omega_n (\theta
  \gamma^n)_\alpha \gamma^3\_\beta\^\gamma + \frac{i}{2} \omega_n
  (\theta \gamma^n)_\beta \gamma^3\_\alpha\^\gamma, \\
  C_{a\beta}\^c &= i (\theta \gamma^b)_\beta\, (-c_{ab}\^c + \omega_a
  \epsilon_b\^c), \\
  C_{a\beta}\^\gamma &= -\half \omega_a \gamma^3\_\beta\^\gamma +
  \frac{i}{8} \theta^2\, r\, \gamma_{a\beta}\^\gamma, \\
  C_{ab}\^c &= c_{ab}\^c, \\
  C_{ab}\^\gamma &= \half f_{ab} (\theta \gamma^3)^\gamma.
\end{align}
The calculation of the supertorsion is based upon the formulae
(\ref{stor}),
\begin{align}
  T_{\alpha\beta}\^c &= 2i \gamma^c\_{\alpha\beta} + \theta^2\,
  \tor^c \gamma^3\_{\alpha\beta}, \label{LCSUSY-struc} \\
  T_{\alpha\beta}\^\gamma &= 0, \\
  T_{a\beta}\^c &= -i (\theta \gamma^b)_\beta\, t_{ab}\^c = +i
  (\theta \gamma_a \gamma^3)_\beta \tor^c, \\
  T_{a\beta}\^\gamma &= -\frac{i}{8} \theta^2\, r\,
  \gamma_{a\beta}\^\gamma, \\
  T_{ab}\^c &= t_{ab}\^c, \label{LCSUSY-tor} \\
  T_{ab}\^\gamma &= -\half f_{ab} (\theta \gamma^3)^\gamma,
\end{align}
and from (\ref{F-cov}) the supercurvature components
\begin{align}
  F_{\alpha\beta} &= \half \theta^2\, r\, \gamma^3\_{\alpha\beta},
  \label{LCSUSY-r} \\
  F_{a\beta} &= -i f_{ab} (\theta \gamma^b)_\beta = \frac{i}{2} r
  (\theta \gamma_a \gamma^3)_\beta, \\
  F_{ab} &= f_{ab} = \half \epsilon_{ab} r
\end{align}
are obtained. For the definition of bosonic torsion and curvature
$t_{ab}\^c$ (or $\tor^c$) and $f_{ab}$ (or $r$) we refer to
\App\ref{app:gravity}.

The verification of the first and second Bianchi identities can be
done by direct calculation and indeed they do not restrict the
connection $\omega_m$. The following remarkable formulae are
consequences of the Bianchi identities:
\begin{alignat}{2}
  T_{ab}\^c &= -\frac{i}{2} \gamma_b\^{\beta\alpha} \nabla_\alpha
  T_{a\beta}\^c &\qquad F_{ab} &= -\frac{i}{2} \gamma_b\^{\beta\alpha}
  \nabla_\alpha F_{a\beta} \\
  T_{a\beta}\^c &= \frac{i}{4} (\gamma_a \gamma^3)_\beta\^\alpha
  \nabla_\alpha (\gamma^3\^{\delta\gamma} T_{\gamma\delta}\^c) &\qquad
  F_{a\beta} &= \frac{i}{4} (\gamma_a \gamma^3)_\beta\^\alpha
  \nabla_\alpha (\gamma^3\^{\delta\gamma} F_{\gamma\delta}) \\
  T_{ab}\^c &= \frac{1}{8} \epsilon_{ab} \epsilon^{\beta\alpha}
  \nabla_\alpha \nabla_\beta (\gamma^3\^{\delta\gamma}
  T_{\gamma\delta}\^c) &\qquad F_{ab} &= \frac{1}{8} \epsilon_{ab}
  \epsilon^{\beta\alpha} \nabla_\alpha \nabla_\beta
  (\gamma^3\^{\delta\gamma} F_{\gamma\delta})
\end{alignat}

In this way we get a theory which is covariant with respect to
$x$-space coordinate transformations and local Lorentz
transformations. The model also has a further symmetry, namely
supersymmetry, given by the superfield parameters $\xi^m(x, \theta) =
0$, $\xi^\mu(x, \theta) = \epsilon^\mu(x)$ and $L(x, \theta) = 0$, but
the parameter $\epsilon^\mu(x)$ must be covariantly constant,
$\nabla_m \epsilon^\mu = 0$. The main purpose of the present section
was the derivation of the supertorsion- and supercurvature components
listed above. They are very helpful in finding constraints for the
supergeometry of the most general supergravity, where the $x$-space
torsion $t_{ab}\^c$ or curvature $f_{ab}$ are not restricted. We will
also see that the original two-dimensional supergravity constraints
(\ref{constr-Howe}) of Howe \cite{Howe:1979ia} set the bosonic torsion
to zero from the very beginning, and why our first attempt
\cite{Ertl:1997ib} to find a supergravity model with independent
connection $\omega_m$ could not have been but partially successful.

\section{New Supergravity}
\label{sec:sugra-ertl}

In our generalization we relax the original supergravity constraints
(\ref{constr-Howe}) in order to obtain an independent bosonic Lorentz
connection $\omega_a$ in the superfield Einstein-Cartan variables
$E_A\^M$ and $\Omega_A$. The first constraint in (\ref{constr-Howe})
can be left untouched, but we cannot use the third constraint
$T_{ab}\^c = 0$, because this would lead to $t_{ab}\^c = 0$ as can be
seen from (\ref{LCSUSY-tor}). This observation was the basis of the
work in \cite{Ertl:1997ib}, where that third constraint was omitted,
leading to an additional Lorentz connection supermultiplet $\Omega_a$,
in which the bosonic Lorentz connection was the zeroth component of
the $\theta$-expansion, $\omega_a = \Omega_a|$. The drawback of that
set of supergravity constraints was the appearance of a second Lorentz
connection, dependent on zweibein and Rarita-Schwinger field. Here we
choose $\gamma_a\^{\beta\alpha} F_{\alpha\beta} = 0$ from the
alternative set (\ref{constr-Howe2}).  This equalizes the Lorentz
connection terms in $\Omega_a$ and the other superfields $E_a\^m$ and
$\Omega_\alpha$, but does not force the bosonic torsion to zero (\cf
also (\ref{LCSUSY-r})). Also the second constraint in
(\ref{constr-Howe}) cannot be maintained. The considerations of the
former section showed that this would again lead to a constraint on
the bosonic torsion appearing in the $\theta^2$-component of
$T_{\alpha\beta}\^c$ (\cf (\ref{LCSUSY-struc})), so we have to weaken
that condition and choose the new set of supergravity constraints
\begin{equation}
  \label{constr-new}
  T_{\alpha\beta}\^\gamma = 0, \qquad
  \gamma_a\^{\beta\alpha} T_{\alpha\beta}\^c = -4i \delta_a\^c, \qquad
  \gamma_a\^{\beta\alpha} F_{\alpha\beta} = 0.
\end{equation}
The terms proportional to $\gamma^3\_{\alpha\beta}$ of
$T_{\alpha\beta}\^c$ and $F_{\alpha\beta}$ are a vector superfield
$K^c$ and a scalar superfield $S$, defined according to the ansatz
\begin{equation}
  \label{ansatz-TF}
  T_{\alpha\beta}\^c = 2i \gamma^c\_{\alpha\beta} + 2 K^c
  \gamma^3\_{\alpha\beta}, \qquad 
  F_{\alpha\beta} = 2 S \gamma^3\_{\alpha\beta}.
\end{equation}
Whereas the scalar superfield $S$ was already present in the original
supergravity model of Howe where the bosonic curvature $r$ was part of
the $\theta^2$-component of $S$ (\cf (\ref{Howe-F-ff}) and
(\ref{Howe-S-field})) the vector superfield $K^c$ enters as a new
superfield playing a similar role for the bosonic torsion $\tor^c$.

It should be stressed that the superspace constraints do not break the
symmetries of the theory such as superdiffeomorphisms or Lorentz
supertransformations. The constraints merely reduce the number of
independent components of the supervielbein $E_A\^M$ and the Lorentz
superconnection $\Omega_A$ as can be seen in \Sec\ref{sec:ertl-constr}
below.

\subsection{Bianchi Identities for New Constraints}
\label{sec:ertl-bianchi}

The Bianchi identities $\Delta_{ABC}\^D = R_{[ABC]}\^D$ (\cf
(\ref{bianchi1})) and $\Delta_{ABC} = 0$ (\cf (\ref{bianchi2})) are
relations between the components of supertorsion and supercurvature
and their derivatives.  In the presence of constraints they can be
used to determine an independent set of superfields from which all
components of supertorsion and supercurvature can be derived. In our
case this set turns out to be formed by the superfields $S$ and $K^c$
of ansatz (\ref{ansatz-TF}).

The decompositions similar to the one of the Rarita-Schwinger field
(\ref{ederas}) of \App\ref{app:spin-tensor} are very useful for the
derivation of the formulae below. For present needs we employ
\begin{align}
  F_{a\beta} &= F^-\_{a\beta} - \gamma_{a\beta}\^\alpha F^+\_\alpha,
  \label{decomp-F-bf} \\
  T_{a\beta}\^c &= T^{-}\_{a\beta}\^c - \gamma_{a\beta}\^\alpha
  T^{+}\_\alpha\^c, \label{decomp-T-bfb} \\
  T_{a\beta}\^\gamma &= T^{-}\_{a\beta}\^\gamma -
  \gamma_{a\beta}\^\alpha T^{+}\_\alpha\^\gamma, \label{decomp-T-bff}
\end{align}
where $F^-$ and $T^-$ are $\gamma^a$ traceless, \eg we have
$\gamma^a\_\alpha\^\beta F^-\_{a\beta} = 0$ in conformance with
(\ref{equlam}). Further helpful formulae to derive the results below
are again found in \App\ref{app:spin-tensor} in particular the
relations (\ref{W-fff})--(\ref{V-fff}).

The Bianchi identity $\Delta_{\alpha\beta\gamma} = 0$ yields, for the
parts of $F_{a\beta}$ as defined by (\ref{decomp-F-bf}),
\begin{equation}
  \label{bianchi-F-bf-1}
  F^-\_{a\beta} = 0, \qquad
  \nabla_\alpha S = -i K_\alpha\^\beta F^+\_\beta.
\end{equation}
Here the superfield operator
\begin{equation}
  \label{gamma-K}
  K_\alpha\^\beta := \gamma^3\_\alpha\^\beta + i K^a
  \gamma_{a\alpha}\^\beta
\end{equation}
was introduced because this particular shift of $\gamma^3$ by the
vector superfield $K^a$ is encountered frequently. Due to
\begin{equation}
  K_\alpha\^\beta K_\beta\^\gamma = (1 - K^a K_a)\, \delta_\alpha\^\gamma
\end{equation}
$K_\alpha\^\beta$ can be inverted if $\body(K^a K_a) \neq 1$:
\begin{equation}
  \label{gamma-K-inv}
  K^{-1}\_\alpha\^\beta = (1 - K^a K_a)^{-1} K\_\alpha\^\beta.
\end{equation}
Therefore, from (\ref{bianchi-F-bf-1})
\begin{equation}
  \label{bianchi-F-bf-2}
  F_{a\beta} = -(\gamma_a F^+)_\beta, \qquad F^+\_\alpha = i
  K^{-1}\_\alpha\^\beta \nabla_\beta S
\end{equation}
is obtained.

In the same way the Bianchi identity $\Delta_{\alpha\beta\gamma}\^d =
R_{\alpha\beta\gamma}\^d$ yields for the separate parts of the
decomposition of $T_{a\beta}\^c$ (\cf (\ref{decomp-T-bfb}))
\begin{equation}
  \label{bianchi-T-bfb-1}
  T^-\_{a\beta}\^c = 0, \qquad
  \nabla_\alpha K^c = -i K\_\alpha\^\beta T^+\_\beta\^c.
\end{equation}

From $\Delta_{\alpha\beta\gamma}\^\delta =
R_{\alpha\beta\gamma}\^\delta$ we get for $T_{a\beta}\^\gamma$ in its
decomposition (\ref{decomp-T-bff})
\begin{equation}
  \label{bianchi-T-bff-1}
  T^{-}\_{a\beta}\^\gamma = 0, \qquad
  K_\beta\^\alpha T^+\_\alpha\^\gamma = \frac{i}{2} S
  \gamma^3\_\beta\^\gamma.
\end{equation}
The second equation of (\ref{bianchi-T-bff-1}) can be rewritten using
a rescaled superfield $S'$
\begin{equation}
  S = (1 - K^a K_a) S', \qquad
  T^{+}\_\alpha\^\beta = \frac{i}{2} S' (K \gamma^3)_\alpha\^\beta,
\end{equation}
where $S'$ can be easily calculated from $S' = -i
T^{+}\_\alpha\^\alpha$.

Further relations are derived from the Bianchi identity
$\Delta_{a\beta\gamma} = 0$. The symmetrized contraction
$\gamma_{(b}\^{\gamma\beta} \Delta_{a)\beta\gamma} = 0$ yields
\begin{equation}
  \label{bianchi-F-exp}
  \epsilon^{\beta\alpha} \nabla_\alpha F^{+}\_\beta = (T^{+}\^a
  \gamma_a F^{+}),
\end{equation}
and the antisymmetric part $\gamma_{[b}\^{\gamma\beta}
\Delta_{a]\beta\gamma} = 0$ gives
\begin{equation}
  \label{bianchi-F-bb-1}
  \gamma^3\^{\beta\alpha} \nabla_\alpha F^+\_{\beta} = -(T^+\^a
  \gamma_a \gamma^3 F^+) - 2i S' S + i \epsilon^{ba} F_{ab}.
\end{equation}
The contraction $\gamma^3\^{\gamma\beta} \Delta_{a\beta\gamma} = 0$
leads to the same relations (\ref{bianchi-F-exp}) and
(\ref{bianchi-F-bb-1}), but needs more tedious calculations.

Similar formulae are derived from the Bianchi identity
$\Delta_{\alpha\beta c}\^d = R_{[\alpha\beta c]}\^d$. From the
symmetrized contraction $\gamma_{(b|}\^{\alpha\beta}
\Delta_{\alpha\beta |c)}\^d$ one gets the relation
\begin{equation}
  \label{bianchi-T-exp}
  \epsilon^{\beta\alpha} \nabla_\alpha T^{+}\_\beta\^d = (T^{+}\^a
  \gamma_a T^{+}\^d) + 2i S' K^a \epsilon_a\^d,
\end{equation}
and from the antisymmetrized contraction $\gamma_{[b|}\^{\alpha\beta}
\Delta_{\alpha\beta |c]}\^d$
\begin{equation}
  \label{bianchi-T-bbb-1}
  \gamma^3\^{\beta\alpha} \nabla_\alpha T^+\_{\beta}\^d = -(T^+\^a
  \gamma_a \gamma^3 T^+\^d) - 2i S' K^d + i \epsilon^{ba} T_{ab}\^d.
\end{equation}

From the Bianchi identity $\Delta_{\alpha\beta c}\^\delta =
R_{[\alpha\beta c]}\^\delta$ also an expression for $T_{ab}\^\gamma$
could be obtained.

Finally we summarize the expressions for the supercurvature and
supertorsion components in terms of the independent and unconstrained
superfields $S$ and $K^a$; the latter is also used to build the matrix
$K$ and its inverse $K^{-1}$ (\cf (\ref{gamma-K}) and
(\ref{gamma-K-inv})):
\begin{alignat}{2}
  F_{\alpha\beta} &= 2 S \gamma^3\_{\alpha\beta}, &\qquad
  T_{\alpha\beta}\^c &= 2i \gamma^c\_{\alpha\beta} + 2 K^c
  \gamma^3\_{\alpha\beta}, \label{ertl-F-ff-T} \\
  F_{a\beta} &= -i (\gamma_a K^{-1})_\beta\^\alpha (\nabla_\alpha S),
  &\qquad
  T_{a\beta}\^c &= -i (\gamma_a K^{-1})_\beta\^\alpha (\nabla_\alpha
  K^c), \label{ertl-F-bf-T} \\
  & & T_{a\beta}\^\gamma &= -\frac{i}{2} S (\gamma_a K^{-1}
  \gamma^3)_\beta\^\gamma, \label{ertl-T-bff}
\end{alignat}
and the more complicated expressions
\begin{align}
  \half \epsilon^{ba} F_{ab} &= \half \gamma^3\^{\beta\alpha}
  \nabla_\alpha (K^{-1}\_\beta\^\gamma \nabla_\gamma S) \eqnsplit +
  \frac{i}{2} (K^{-1} \gamma_a \gamma^3 K^{-1})^{\beta\alpha}
  (\nabla_\alpha K^a) (\nabla_\beta S) + \frac{S^2}{1-K^a K_a},
  \label{ertl-F-bb} \\
  \half \epsilon^{ba} T_{ab}\^c &= \half \gamma^3\^{\beta\alpha}
  \nabla_\alpha (K^{-1}\_\beta\^\gamma \nabla_\gamma K^c) \eqnsplit +
  \frac{i}{2} (K^{-1} \gamma_a \gamma^3 K^{-1})^{\beta\alpha}
  (\nabla_\alpha K^a) (\nabla_\beta K^c) + \frac{S}{1-K^a K_a}
  K^c. \label{ertl-T-bb}
\end{align}

\subsection{Solution of the Constraints}
\label{sec:ertl-constr}

The solution of the new supergravity constraints (\ref{constr-new})
can be derived solely with superfield methods. To achieve that goal
the formulae for the supertorsion components
(\ref{T-fff})--(\ref{T-bbb}) and for the supercurvature components
(\ref{R-fffb})--(\ref{R-bbbf}), in each case consequences of the
restricted tangent space group (\ref{eantis}), are employed. However,
the decomposition of the supervielbein in terms of the superfields
$B_m\^a$, $\Psi_m\^a$, $\Phi_\mu\^a$ and $B_\mu\^\alpha$ (\cf
(\ref{SVdecomp})) and the ensuing formulae for the anholonomicity
coefficients (\ref{C-ffb})--(\ref{C-bbf}) turn out to be essential.
The method for solving the supergravity constraints in terms of
superfields employed here is similar to the one developed in
\cite{Rocek:1986iz}.

We start with the ansatz (\ref{ansatz-TF}) for $T_{\alpha\beta}\^c$.
Contraction with $\gamma_a\^{\beta\alpha}$ leads to the second
constraint $\gamma_a\^{\beta\alpha} T_{\alpha\beta}\^c = -4i
\delta_a\^c$ (\cf (\ref{constr-new})).  Looking into the expression
for the supertorsion (\ref{T-ffb}) we recognize that the
superconnection drops out there.  The remaining anholonomicity
coefficient, calculated according to (\ref{C-ffb}), allows to express
$B_a\^n$ in terms of $B_\alpha\^\mu$ and $\Phi_\nu\^n$:
\begin{equation}
  \label{constr-B-bb}
  B_a\^n = -\frac{i}{2} \gamma_a\^{\beta\alpha} B_\alpha\^\mu
  B_\beta\^\nu \left( \Phi_\mu\^l \partial_l \Phi_\nu\^n -
    \partial_\mu \Phi_\nu\^n \right).
\end{equation}
The contraction with $\gamma^3\^{\beta\alpha}$ of $T_{\alpha\beta}\^c$
(\cf (\ref{ansatz-TF})) yields $K^c = -\frac{1}{4}
\gamma^3\^{\beta\alpha} T_{\alpha\beta}\^c$.  Again using
(\ref{T-ffb}) and (\ref{C-ffb}) together with $K^n := K^c B_c\^n$
\begin{equation}
  \label{constr-K-b}
  K^n = \half \gamma^3\^{\beta\alpha} B_\alpha\^\mu B_\beta\^\nu
  (\Phi_\mu\^l \partial_l \Phi_\nu\^n - \partial_\mu \Phi_\nu\^n)
\end{equation}
is obtained, thus expressing $K^n$ in terms of $B_\alpha\^\mu$ and
$\Phi_\nu\^n$.

Next we turn to the investigation of the constraint
$T_{\alpha\beta}\^\gamma = 0$. Contracting (\ref{T-fff}) with
$\gamma_a\^{\beta\alpha}$ and $\gamma^3\^{\beta\alpha}$ yields
\begin{align}
  -\gamma_a\^{\beta\alpha} C_{\alpha\beta}\^\gamma + \Omega^\alpha
  (\gamma_a \gamma^3)_\alpha\^\gamma &= 0, \\
  -\gamma^3\^{\beta\alpha} C_{\alpha\beta}\^\gamma + \Omega^\gamma &=
  0.
\end{align}
These two equations can be used to eliminate $\Psi_a\^\nu$ and
$\Omega^\alpha$: Indeed, using formula (\ref{C-fff}) to calculate the
anholonomicity coefficients and observing that parts of that formula
are of the form (\ref{constr-B-bb}) and (\ref{constr-K-b}) we derive
\begin{gather}
  \Psi_a\^\nu = -\frac{i}{2} \gamma_a\^{\beta\alpha} B_\alpha\^\mu
  \left( \Phi_\mu\^l \partial_l B_\beta\^\nu - \partial_\mu
    B_\beta\^\nu \right) - \frac{i}{4} \Omega^\alpha (\gamma_a
  \gamma^3)_\alpha\^\beta B_\beta\^\nu, \label{constr-Psi-bf} \\
  -4 K^n \Psi_n\^\gamma + 2 \gamma^3\^{\beta\alpha} B_\alpha\^\mu
  \left( \Phi_\mu\^m \partial_m B_\beta\^\lambda - \partial_\mu
    B_\beta\^\lambda \right) B_\lambda\^\gamma + \Omega^\gamma =
  0. \label{constr-Om-f-0}
\end{gather}
The first equation expresses $\Psi_a\^\nu$ in terms of
$B_\alpha\^\mu$, $\Phi_\nu\^n$ and $\Omega^\alpha$, whereas an
appropriate combination of both equations gives
\begin{equation}
  (\Omega K \gamma^3)^\gamma = -2 K^{\beta\alpha} B_\alpha\^\mu \left(
    \Phi_\mu\^l \partial_l B_\beta\^\nu - \partial_\mu B_\beta\^\nu
  \right) B_\nu\^\gamma,
\end{equation}
so that $\Omega^\alpha$ can be eliminated in favour of
$B_\alpha\^\mu$, $\Phi_\nu\^n$ and $K^c$:
\begin{equation}
  \label{constr-Om-f}
  \Omega^\delta = -2 K^{\beta\alpha} B_\alpha\^\mu \left( \Phi_\mu\^l
    \partial_l B_\beta\^\nu - \partial_\mu B_\beta\^\nu \right)
  B_\nu\^\gamma (\gamma^3 K^{-1})_\gamma\^\delta.
\end{equation}
We recall that $B_\nu\^\gamma$ is the inverse of $B_\alpha\^\mu$
defined in (\ref{svb-B-inv}) and that the matrix $K_\beta\^\alpha$ and
its inverse were given in (\ref{gamma-K}) and (\ref{gamma-K-inv}).

The third constraint $\gamma_a\^{\beta\alpha} F_{\alpha\beta} = 0$ of
(\ref{constr-new}) allows to eliminate the superconnection component
$\Omega_a$. Calculating the supercurvature according to (\ref{F-cov})
the expression
\begin{equation}
    2 \gamma_a\^{\beta\alpha} (\nabla_\alpha \Omega_\beta) +
  \gamma_a\^{\beta\alpha} T_{\alpha\beta}\^\gamma \Omega_\gamma +
  \gamma_a\^{\beta\alpha} T_{\alpha\beta}\^c \Omega_c = 0
\end{equation}
is found. Then the constraints on the supertorsion (\cf
(\ref{constr-new})) immediately lead to
\begin{equation}
  \label{constr-Om-b}
  \Omega_a = -\frac{i}{2} \gamma_a\^{\beta\alpha} (\nabla_\alpha
    \Omega_\beta)
    = \frac{i}{2} \gamma_a\^{\beta\alpha} B_\alpha\^\mu \left(
      \Phi_\mu\^l \partial_l \Omega_\beta - \partial_\mu \Omega_\beta
    \right).
\end{equation}

This gives a complete solution of the new supergravity constraints
(\ref{constr-new}) in terms of the unconstrained and independent
superfields $B_\alpha\^\mu$ and $\Phi_\nu\^n$.  Due to the
supergravity constraints the superfields $B_a\^n$ (\ref{constr-B-bb})
and $\Psi_a\^\nu$ (\ref{constr-Psi-bf}), member of the supervielbein,
as well as the whole superconnection $\Omega_\alpha$
(\ref{constr-Om-f}) and $\Omega_a$ (\ref{constr-Om-b}) were
eliminated. The vector superfield $K^a$ appeared as the special
combination (\ref{constr-K-b}).

\subsection{Physical Fields and Gauge Fixing}
\label{sec:ertl-gf}

Although the superfield solution given in the previous section looks
quite pleasant, calculating the $\theta$-expansion to recover the
physical $x$-space fields is cumbersome. Let us have a quick look at
the number of components that remain: The $6+4+2=12$ superspace
constraints (\ref{constr-new}) reduce the original $16+4=20$
superfields formed by $E_M\^A$ and $\Omega_M$ down to $4+4=8$
superfields contained in $B_\alpha\^\mu$ and $\Phi_\nu\^n$.  Thus, the
remaining number of $x$-space fields is $4 \times 8=32$. This is still
more than the $4+4+2=10$ components of the zweibein $e_m\^a$, the
Rarita-Schwinger field $\RS_m\^\alpha$ and the Lorentz connection
$\omega_m$ which we want to accommodate within the superfields.

An additional complication is that the identification of the physical
$x$-space fields is not made within the independent superfields
$B_\alpha\^\mu$ and $\Phi_\nu\^n$, but with respect to the already
eliminated superfields to lowest order in $\theta$. The identification
is expected to be
\begin{equation}
  \label{ident-fields}
  E_m\^a| = e_m\^a, \qquad
  E_m\^\alpha| = \RS_m\^\alpha, \qquad
  \Omega_m| = \omega_m.
\end{equation}
Whereas the first two relations will be found to hold, the
identification of $\omega_m$ in the explicit formulae below will be
different. It will be made at first order in $\theta$ of
(\ref{ertl-Om-f}) leading to the result (\ref{ertl-Om-b}) for
$\Omega_m$. A simple redefinition of $\omega_m$ will fix that
disagreement.

In order to reduce the $32$ independent $x$-space fields a Wess-Zumino
type gauge fixing is chosen. It is the same as in the original model
of Howe \cite{Howe:1979ia}. In particular to zeroth order in $\theta$
there are the $4+4+2=10$ conditions
\begin{equation}
  \label{gf-0}
  E_\mu\^\alpha| = \delta_\mu\^\alpha, \qquad
  E_\mu\^a = 0|, \qquad
  \Omega_\mu| = 0
\end{equation}
and to first order the $2+2+1=5$ conditions
\begin{equation}
  \label{gf-1}
  \partial_{[\nu} E_{\mu]}\^\alpha| = 0, \qquad
  \partial_{[\nu} E_{\mu]}\^a| = 0, \qquad
  \partial_{[\nu} \Omega_{\mu]}| = 0,
\end{equation}
which reduce the number of independent component fields down to $17$.
A corresponding reduction also follows for the number of symmetries
contained within superdiffeomorphism and Lorentz
supertransformation. The same considerations as in
\Sec\ref{sec:Howe-symms} apply, for a detailed analysis \cf
\cite{Ertl:1997ib}.


The remaining $17$ degrees of freedom are constituted by the $10$
components of zweibein, Rarita-Schwinger field and Lorentz connection,
and by additional $7$ components consisting of the well-known
auxiliary field $A$ and the newly introduced fields $k^a$ and
$\varphi_m\^\alpha$.  The identification for $k^a$ is
\begin{equation}
  \label{ident-k-b}
  K^a| = k^a,
\end{equation}
the others at the end of the calculation are to be identified with
zeroth components of supertorsion and supercurvature.

\subsection{Component Fields of New Supergravity}
\label{sec:ertl-cf}

To recover the component fields the superfield expressions
(\ref{constr-B-bb}), (\ref{constr-K-b}), (\ref{constr-Psi-bf}),
(\ref{constr-Om-f-0}) and (\ref{constr-Om-b}) have to be worked out
order by order in the anticommutative coordinate $\theta$. For the
decomposition of spin-tensors we refer to \App\ref{app:spin-tensor}.
The reader should especially consult (\ref{ederas}) where the
decomposition of the Rarita-Schwinger field is given.

In the calculations new $\Gamma$-matrices, dependent on the vector
field $k^a$ (\cf (\ref{ident-k-b})), were encountered frequently:
\begin{equation}
  \label{Gamma}
  \Gamma^a := \gamma^a - i k^a \gamma^3, \qquad
  \Gamma^3 := \gamma^3 + i k^a \gamma_a.
\end{equation}
Their (anti-)commutator algebra and other properties as well as
formulae used in the calculations can be found in \App\ref{app:Gamma}.

In the $\theta$-expansion some covariant derivatives of $x$-space
fields are encountered. These are, respectively, the torsion, the
curvature, the covariant derivative of the Rarita-Schwinger field
\begin{align}
  t_{ab}\^c &= -c_{ab}\^c + \omega_a \epsilon_b\^c - \omega_b
  \epsilon_a\^c - A k_a \epsilon_b\^c + A k_b \epsilon_a\^c \eqnsplit
  - 2i (\RS_a \Gamma^c \RS_b) - 4i \epsilon_{ab} (\RS
  \gamma^3 \varphi^c), \label{ertl-tor} \\
  f_{mn} &= \partial_m \omega_n - \partial_n \omega_m - \partial_m (A
  k_n) + \partial_n (A k_m) \eqnsplit - 2 (1-k^2) A (\RS_m \gamma^3
  \RS_n) - 4i \epsilon_{mn} (\RS \gamma^3 \Gamma^3 \gamma^3 \sigma)
  \eqnsplit - 4 A \epsilon_{mn} \epsilon^{dc} (\RS \gamma^3 \Gamma_c
  \varphi_d) + 4i A \epsilon_{mn} (\RS \gamma^3 \varphi^c) k_c,
  \label{ertl-cur} \\
  \sigma_{mn}\^\gamma &= \partial_m \RS_n\^\gamma - \half \omega_m (\RS_n
  \gamma^3)^\gamma - \frac{i}{2} A \epsilon_m\^l (\RS_n \Gamma_l
  \gamma^3)^\gamma - (m \leftrightarrow n), \label{ertl-si}
\end{align}
and the covariant derivative of $k^b$
\begin{equation}
  \label{ertl-derk}
  \SD_a k^b = \partial_a k^b + \omega_a k^c \epsilon_c\^b - A k_a k^c
  \epsilon_c\^b - 2 (\RS_a \Gamma^3 \varphi^b).
\end{equation}
They are (obviously) covariant with respect to Lorentz
transformations, but they are also covariant with respect to
supersymmetry transformations. For the Hodge duals $\tor^c = \half
\epsilon^{ba} t_{ab}\^c$ and $\sigma^\gamma = \half \epsilon^{nm}
\sigma_{mn}\^\gamma$ one obtains (\cf (\ref{HodgeDual}) and also
(\ref{omT}))
\begin{align}
  \tor^c &= \tilde{\omega}^c - \omega^c + A k^c - i \epsilon^{ba}
  (\RS_a \Gamma^c \RS_b) - 4i (\RS \gamma^3 \varphi^c), \\
  \sigma^\gamma &= \epsilon^{nm} \partial_m \RS_n\^\gamma - \half
  \epsilon^{nm} \omega_m (\RS_n \gamma^3)^\gamma - \frac{i}{2} A
  (\RS_n \Gamma^n \gamma^3)^\gamma.
\end{align}
Here it becomes obvious that the identification of $\omega_m$ in
(\ref{ertl-Om-b}) was not the best choice. This could be fixed by the
replacement $\omega_a \rightarrow \omega'_a = \omega_a + A k_a$.

First we summarize the results for the decomposition superfields of
the supervielbein. For $B_a\^m$ and its inverse $B_m\^a$ (\cf
(\ref{svb-B-inv})), where the zweibein $e_a\^m$ and $e_m\^a$ are the
zeroth components, we obtain
\begin{align}
  B_a\^m &= e_a\^m - 2i (\theta \Gamma^m \RS_a) - 2i (\theta \gamma_a
  \varphi^m) \eqnspl + \half \theta^2 \left[ -\epsilon_a\^b (\SD_b
    k^c) e_c\^m + k_a \tor^m - \omega_a k^m - A k_a k^m - A e_a\^m -
    k^2 A e_a\^m \right] \eqnsplit + \half \theta^2 \left[ -4 (\RS_a
    \Gamma^b \Gamma^m \RS_b) - 6 (\RS_a \varphi^m) - 4 (\RS_b
    \Gamma^m \gamma_a \varphi^b) \right], \\
  B_m\^a &= e_m\^a + 2i (\theta \Gamma^a \RS_m) + 2i (\theta \gamma_m
  \varphi^a) \eqnspl + \half \theta^2 \left[ \epsilon_m\^n (\SD_n
    k^a) - k_m \tor^a + \omega_m k^a + A k_m k^a + A e_m\^a + k^2 A
    e_m\^a \right] \eqnsplit + \half \theta^2 \left[ -2 (\RS_m
    \varphi^a) + 4i k^b (\RS_m \gamma^3 \gamma_b \varphi^a) - 4
    (\varphi^b \gamma_m \gamma_b \varphi^a) \right].
\end{align}
The component fields in $B_\alpha\^\mu$ and its inverse
$B_\mu\^\alpha$ (\cf again (\ref{svb-B-inv})) are
\begin{align}
  B_\alpha\^\mu &= \delta_\alpha\^\mu + i (\theta \Gamma^b)_\alpha
  \RS_b\^\mu \eqnspl + \half \theta^2 \left[ - \frac{i}{2} \omega_b
    (\Gamma^b \gamma^3)_\alpha\^\mu - \half A (\Gamma^3
    \gamma^3)_\alpha\^\mu + (\Gamma^b \Gamma^c \RS_b)_\alpha
    \RS_c\^\mu + 4 \varphi^c\_\alpha \RS_c\^\mu \right], \\
  B_\mu\^\alpha &= \delta_\mu\^\alpha - i (\theta \Gamma^b)_\mu
  \RS_b\^\alpha \eqnspl + \half \theta^2 \left[ \frac{i}{2}
    \omega_b (\Gamma^b \gamma^3)_\mu\^\alpha + \half A (\Gamma^3
    \gamma^3)_\mu\^\alpha -2 (\Gamma^b \Gamma^c \RS_b)_\mu
    \RS_c\^\alpha - 4 \varphi^c\_\mu \RS_c\^\alpha \right].
\end{align}
The various contractions with $\Phi_\mu\^m$ (\cf (\ref{svb-Phi})) are
given by
\begin{align}
  \Phi_\mu\^m &= i (\theta \Gamma^m)_\mu + \half \theta^2 \left[ 4
    \varphi^m\_\mu + 2 (\Gamma^b \Gamma^m \RS_b)_\mu \right], \\
  \Phi_\mu\^a &= i (\theta \Gamma^a)_\mu + \half \theta^2 \left[ 2i
    k^b (\gamma^3 \gamma_b \varphi^a)_\mu \right], \\
  \Phi_\alpha\^a &= i (\theta \Gamma^a)_\alpha + \half \theta^2 \left[
    2i k^b (\gamma^3 \gamma_b \varphi^a)_\alpha - (\Gamma^b \Gamma^a
    \RS_b)_\alpha \right], \\
  \Phi_\alpha\^m &= i (\theta \Gamma^m)_\alpha + \half \theta^2 \left[
    4 \varphi^m\_\alpha + (\Gamma^b \Gamma^m \RS_b)_\alpha \right].
\end{align}
The superfield $\Psi_m\^\alpha$ with the Rarita-Schwinger
$\RS_m\^\alpha$ field at lowest order and its various contractions
(\cf (\ref{svb-Psi})) read
\begin{align}
  \Psi_m\^\alpha &= \RS_m\^\alpha - \half \omega_m (\theta
  \gamma^3)^\alpha - \frac{i}{2} A \epsilon_m\^n (\theta \Gamma_n
  \gamma^3)^\alpha \eqnspl + \half \theta^2 \left[ -\frac{3}{2} A
    (\RS_m \Gamma^3 \gamma^3)^\alpha - 2 A \epsilon_m\^n (\varphi_n
    \gamma^3)^\alpha - i \sigma_{mn}\^\beta (\gamma^3 \Gamma^n
    \gamma^3)_\beta\^\alpha \right], \\
  \Psi_m\^\mu &= \RS_m\^\mu - \half \omega_m (\theta \gamma^3)^\mu -
  \frac{i}{2} A \epsilon_m\^n (\theta \Gamma_n \gamma^3)^\mu + i
  (\theta \Gamma^n \RS_m) \RS_n\^\mu + \half \theta^2
  \left[ \cdots \right],
\end{align}
\begin{align}
  \Psi_a\^\alpha &= \RS_a\^\alpha - \half \omega_a (\theta
  \gamma^3)^\alpha - \frac{i}{2} A \epsilon_a\^b (\theta \Gamma_b
  \gamma^3)^\alpha - 2i (\theta \Gamma^b \RS_a) \RS_b\^\alpha - 2i
  (\theta \gamma_a \varphi^b) \RS_b\^\alpha \eqnspl + \half
  \theta^2 \left[ -\epsilon_a\^b (\SD_b k^c) \RS_c\^\alpha + k_a
    \tor^b \RS_b\^\alpha - i \sigma_{ab}\^\beta (\gamma^3 \Gamma^b
    \gamma^3)_\beta\^\alpha \right] \eqnspl + \half \theta^2 \left[
    i \omega_b (\RS_a \Gamma^b \gamma^3)^\alpha - \omega_a k^b
    \RS_b\^\alpha + i \omega_b (\varphi^b \gamma_a \gamma^3)^\alpha
  \right] \eqnspl + \half \theta^2 \left[ - \half A (\RS_a \gamma^3
    \Gamma^3)^\alpha - A k_a k^b \RS_b\^\alpha - k^2 A \RS_a\^\alpha
    - A \epsilon^{cb} (\varphi_b \Gamma_c \gamma_a \gamma^3)^\alpha
  \right] \eqnsplit + \half \theta^2 \left[ - 4 (\RS_a \Gamma^b
    \Gamma^c \RS_b) \RS_c\^\alpha - 6 (\RS_a \varphi^b)
    \RS_b\^\alpha - 4 (\RS_b \Gamma^c \gamma_a \varphi^b)
    \RS_c\^\alpha \right],
\end{align}
\begin{align}
  \Psi_a\^\mu &= \RS_a\^\mu - \half \omega_a (\theta \gamma^3)^\mu -
  \frac{i}{2} A \epsilon_a\^b (\theta \Gamma_b \gamma^3)^\mu - i
  (\theta \Gamma^b \RS_a) \RS_b\^\mu - 2i (\theta \gamma_a
  \varphi^b) \RS_b\^\mu \eqnspl + \half \theta^2 \left[
    -\epsilon_a\^b (\SD_b k^c) \RS_c\^\mu + k_a \tor^b \RS_b\^\mu - i
    \sigma_{ab}\^\beta (\gamma^3 \Gamma^b \gamma^3)_\beta\^\mu \right]
  \eqnspl + \half \theta^2 \left[ \frac{i}{2} \omega_b (\RS_a
    \Gamma^b \gamma^3)^\mu + i \omega_b (\varphi^b \gamma_a
    \gamma^3)^\mu \right] \eqnspl + \half \theta^2 \left[ - A k_a
    k^b \RS_b\^\mu - A k^2 \RS_a\^\mu - A \epsilon^{cb} (\varphi_b
    \Gamma_c \gamma_a \gamma^3)^\mu \right] \eqnsplit + \half \theta^2
  \left[ - (\RS_a \Gamma^b \Gamma^c \RS_b) \RS_c\^\mu - 2 (\RS_a
    \varphi^b) \RS_b\^\mu - 2 (\RS_b \Gamma^c \gamma_a \varphi^b)
    \RS_c\^\mu \right].
\end{align}

The $K^a$ vector superfield constituting a new multiplet of fields for
supergravity with torsion and the contraction $K^m = K^a B_a\^m$ are
given by
\begin{align}
  K^a &= k^a + 2 (\theta \Gamma^3 \varphi^a) \eqnsplit + \half
  \theta^2 \left[ (1 - k^2) \left( \tor^a - 2 A k^a \right) + k^c
    \epsilon_c\^b (\SD_b k^a) + 4i (\varphi^b \Gamma^3 \gamma_b
    \varphi^a) \right], \label{ertl-K-b} \\
  K^m &= k^m + 2 (\theta \gamma^3 \varphi^m) - 2i k^b (\theta \Gamma^m
  \RS_b) \eqnspl + \half \theta^2 \left[ \tor^m - k^b \omega_b k^m
    - 3 A k^m - 4 k^c (\RS_c \Gamma^b \Gamma^m \RS_b) \right]
  \eqnsplit + \half \theta^2 \left[ -6 k^c (\RS_c \varphi^m) - 4i
    (\RS_b \Gamma^m \gamma^3 \varphi^b) \right].
\end{align}
Here the vector $k^a$ appears at zeroth order (as was already
mentioned in (\ref{ident-k-b})), the spin-vector $\varphi_m\^\alpha$
at first order and the torsion $\tor^a$ at second order in $\theta$,
constituting a new multiplet of $x$-space fields $\{ k^a,
\varphi_m\^\alpha, \tor^a \}$, where the torsion $\tor^a$ could also
be replaced by the Lorentz connection $\omega_a$.

The inverse supervielbein $E_A\^M$ can be derived from the above
decomposition superfields according to (\ref{ISVdecomp})
\begin{align}
  E_a\^m &= e_a\^m - i (\theta \Gamma^m \RS_a) - 2i (\theta \gamma_a
  \varphi^m) \eqnspl + \half \theta^2 \left[ - \epsilon_a\^b (\SD_b
    k^c) e_c\^m + k_a \tor^m - A k_a k^m - A k^2 e_a\^m \right]
  \eqnsplit + \half \theta^2 \left[ - (\RS_a \Gamma^n \Gamma^m
    \RS_n) - 2 (\RS_a \varphi^m) - 2 (\RS_b \Gamma^m \gamma_a
    \varphi^b) \right], \\
  E_a\^\mu &= -\RS_a\^\mu + \half \omega_a (\theta \gamma^3)^\mu +
  \frac{i}{2} A \epsilon_a\^b (\theta \Gamma_b \gamma^3)^\mu + i
  (\theta \Gamma^b \RS_a) \RS_b\^\mu + 2i (\theta \gamma_a
  \varphi^b) \RS_b\^\mu \eqnspl + \half \theta^2 \left[
    \epsilon_a\^b (\SD_b k^c) \RS_c\^\mu - k_a \tor^b \RS_b\^\mu + i
    \sigma_{ab}\^\beta (\gamma^3 \Gamma^b \gamma^3)_\beta\^\mu \right]
  \eqnspl + \half \theta^2 \left[ -\frac{i}{2} \omega_b (\RS_a
    \Gamma^b \gamma^3)^\mu - i \omega_b (\varphi^b \gamma_a
    \gamma^3)^\mu \right] \eqnspl + \half \theta^2 \left[ A k_a
    k^b \RS_b\^\mu + A k^2 \RS_a\^\mu + A \epsilon^{cb} (\varphi_b
    \Gamma_c \gamma_a \gamma^3)^\mu \right] \eqnsplit + \half \theta^2
  \left[ (\RS_a \Gamma^b \Gamma^c \RS_b) \RS_c\^\mu + 2 (\RS_a
    \varphi^b) \RS_b\^\mu + 2 (\RS_b \Gamma^c \gamma_a \varphi^b)
    \RS_c\^\mu \right], \\
  E_\alpha\^m &= -i (\theta \Gamma^m)_\alpha + \half \theta^2 \left[
    -4 \varphi^m\_\alpha - (\Gamma^b \Gamma^m \RS_b)_\alpha \right],
  \\
  E_\alpha\^\mu &= \delta_\alpha\^\mu + i (\theta \Gamma^b)_\alpha
  \RS_b\^\mu \eqnspl + \half \theta^2 \left[ - \frac{i}{2} \omega_b
    (\Gamma^b \gamma^3)_\alpha\^\mu - \half A (\Gamma^3
    \gamma^3)_\alpha\^\mu + (\Gamma^b \Gamma^c \RS_b)_\alpha
    \RS_c\^\mu + 4 \varphi^c\_\alpha \RS_c\^\mu \right].
\end{align}

For the $\theta$-expansion of the supervielbein $E_M\^A$ now applying
the decomposition (\ref{SVdecomp}) we arrive at
\begin{align}
  E_m\^a &= e_m\^a + 2i (\theta \Gamma^a \RS_m) + 2i (\theta \gamma_m
  \varphi^a) \eqnspl + \half \theta^2 \left[ \epsilon_m\^n (\SD_n
    k^a) - k_m \tor^a + \omega_m k^a + A k_m k^a + A e_m\^a + k^2 A
    e_m\^a \right] \eqnsplit + \half \theta^2 \left[ -2 (\RS_m
    \varphi^a) + 4i k^b (\RS_m \gamma^3 \gamma_b \varphi^a) - 4
    (\varphi^b \gamma_m \gamma_b \varphi^a) \right], \\
  E_m\^\alpha &= \RS_m\^\alpha - \half \omega_m (\theta
  \gamma^3)^\alpha - \frac{i}{2} A \epsilon_m\^n (\theta \Gamma_n
  \gamma^3)^\alpha \eqnspl + \half \theta^2 \left[ -\frac{3}{2} A
    (\RS_m \Gamma^3 \gamma^3)^\alpha - 2 A \epsilon_m\^n (\varphi_n
    \gamma^3)^\alpha - i \sigma_{mn}\^\beta (\gamma^3 \Gamma^n
    \gamma^3)_\beta\^\alpha \right], \\
  E_\mu\^a &= i (\theta \Gamma^a)_\mu + \half \theta^2 \left[ 2i k^b
    (\gamma^3 \gamma_b \varphi^a)_\mu \right], \\
  E_\mu\^\alpha &= \delta_\mu\^\alpha + \half \theta^2 \left[ - \half
    A (\Gamma^3 \gamma^3)_\mu\^\alpha \right].
\end{align}

Finally the Lorentz superconnection $\Omega_A$ reads
\begin{align}
  \Omega_a &= \omega_a - A k_a - i (\theta \Gamma^b \RS_a) \omega_b +
  A (\theta \Gamma^3 \RS_a) - 2i (\theta \gamma_a \varphi^b) \omega_b
  + 2 A \epsilon^{cb} (\theta \gamma_a \Gamma_b \varphi_c) \eqnsplit +
  2i (\theta \gamma_a \Gamma^3 \gamma^3 \sigma) + \half \theta^2
  \left[ \cdots \right], \\
  \Omega_\alpha &= -i (\theta \Gamma^b)_\alpha \omega_b + A (\theta
  \Gamma^3)_\alpha + \half \theta^2 \left[ -(\Gamma^c \Gamma^b
    \RS_c)_\alpha \omega_b + i A (\Gamma^3 \Gamma^b \RS_b)_\alpha
  \right] \eqnsplit + \half \theta^2 \left[ -4 \varphi^b\_\alpha
    \omega_b - 4i A \epsilon^{cb} (\Gamma_b \varphi_c)_\alpha + 4
    (\Gamma^3 \gamma^3 \sigma)_\alpha \right], \label{ertl-Om-f}
\end{align}
and for $\Omega_M = E_M\^A \Omega_A$ we obtain
\begin{align}
  \Omega_m &= \omega_m - A k_m + 2 (1-k^2) A (\theta \gamma^3 \RS_m)
  + 2 A \epsilon^{cb} (\theta \gamma_m \Gamma_b \varphi_c) - 2i A
  (\theta \gamma_m \varphi^b) k_b \eqnsplit + 2i (\theta \gamma_m
  \Gamma^3 \gamma^3 \sigma) + \half \left[ \cdots \right],
  \label{ertl-Om-b} \\
  \Omega_\mu &= (1-k^2) A (\theta \gamma^3)_\mu + \half \theta^2
  \left[ \cdots \right].
\end{align}

When terms were omitted in the formulae above there is no difficulty
of principle to work them out, but it is tedious to do so.

\subsection{Supertorsion and Supercurvature}
\label{sec:ertl-torcur}

Finally we calculate the supertorsion and supercurvature components to
zeroth order in $\theta$. There the auxiliary fields $A$, $k^a$ and
$\varphi_m\^\alpha$ can be detected. For $T_{\alpha\beta}\^c$ we refer
to ansatz (\ref{ansatz-TF}) and to (\ref{ertl-K-b}). With
$\sigma_{ab}\^\gamma$ and $t_{ab}\^c$ given by (\ref{ertl-si}) and
(\ref{ertl-tor}) the supertorsion components read (\cf also
decomposition (\ref{decomp-T-bfb}) and (\ref{decomp-T-bff}))
\begin{alignat}{2}
  T_{ab}\^\gamma| &= \sigma_{ab}\^\gamma, &\qquad
  T_{ab}\^c| &= t_{ab}\^c, \\
  T^+\_\beta\^\gamma| &= \frac{i}{2} A (\Gamma^3
  \gamma^3)_\beta\^\gamma, &\qquad
  T^+\_\beta\^c| &= 2i \varphi^c\_\beta,
\end{alignat}
the superfield $S$ of ansatz (\ref{ansatz-TF}) for $F_{\alpha\beta}$
to lowest order is
\begin{equation}
  \label{ertl-S-field}
  S| = (1-k^2) A,
\end{equation}
and for the supercurvature components with $f_{ab}$ given by
(\ref{ertl-cur}) (\cf also (\ref{decomp-F-bf}))
\begin{align}
  F_{ab}| &= f_{ab}, \\
  F^+\_\alpha| &= 2 A \epsilon^{cb} (\Gamma_b \varphi_c)_\alpha - 2i A
  \varphi^b\_\alpha k_b + 2i (\Gamma^3 \gamma^3 \sigma)_\alpha
\end{align}
is obtained.

The calculation, simplification and analysis of the model is outside
the scope of our present work. The complexity of these results
suggested the treatment of supergravity along a different path (\cf
\Cha\ref{cha:gPSM} below). We note that in
(\ref{ertl-tor})--(\ref{ertl-derk}) always the combination $\omega_a -
A k_a$ appears. Actually this combination will be found to be the
important one below in the PSM approach (\Cha\ref{cha:gPSM}).


\chapter{PSM Gravity and its Symplectic Extension}
\label{cha:PSM}


\newcommand{\sbi}[1]{\matheurm{#1}} 

The PSM in general and its relation to two-dimensional gravity is
presented. The method to solve the \eom{}s of general PSMs is derived
and the symplectic extension of the two-dimensional gravity PSM is
constructed.

\section{PSM Gravity}
\label{sec:gravity-psm}

A large class of gravity models in two dimensions can be written in
first order form
\begin{equation}
  \label{eq:action-first-order}
  \Action = \int_{\BMf} (DX^a) e_a + (d\phi) \omega + \half V
  \epsilon^{ba} e_a e_b,
\end{equation}
where the potential $V = V(\phi, Y)$ is a function of $\phi$ and $Y$
with $Y = \half X^a X_a$, and the covariant exterior derivative is
given by
\begin{equation}
  \label{eq:cov-ext-der}
  DX^a = dX^a + X^b \omega \epsilon_b\^a.
\end{equation}
Particular choices of $V$ yield various gravity models, among which
spherically reduced gravity is an important physical example.

This action is of the form of a Poisson Sigma Model
\cite{\bibPSM,Klosch:1996fi,Strobl:1999Habil} (\cf also
\cite{\bibNLGT,\bibPSMstar}),
\begin{equation}
  \label{eq:action-psm}
  \Action = \int_\BMf dX^i A_i + \half \Poisson^{ij} A_j A_i.
\end{equation}
The coordinates on the target space $\TSp$ are denoted by $X^i = (X^a,
\phi)$. The same symbols are used to denote the mapping from $\BMf$ to
$\TSp$, therefore $X^i = X^i(x^m)$. In this sloppy notation the $dX^i$
in the integral above stand for the pull-back of the target space
differentials $dX^i = dx^m \partial_m X^i$, and $A_i$ are 1-forms on
$\BMf$ with values in the cotangent space of $\TSp$. The Poisson
tensor $\Poisson^{ij}$ is an antisymmetric bi-vector field on the
target space $\TSp$, which fulfills the Jacobi identity
\begin{equation}
  \label{Jacobi}
  J^{ijk} = \Poisson^{il} \partial_l \Poisson^{jk} + \cycl(ijk) = 0.
\end{equation}
For the gravity model (\ref{eq:action-first-order}) we obtain
\begin{equation}
  \Poisson^{ab} = V \epsilon^{ab}, \qquad
  \Poisson^{a\phi} = X^b \epsilon_b\^a.\label{eq:poisson-tensor-gr}
\end{equation}
The equations of motion are
\begin{gather}
  dX^i + \Poisson^{ij} A_j = 0, \label{eq:eomX} \\
  dA_i + \half (\partial_i \Poisson^{jk}) A_k A_j = 0, \label{eq:eomA}
\end{gather}
and the symmetries of the action are found to be
\begin{equation}
  \label{eq:psm_symmetries}
  \delta X^i = \Poisson^{ij} \epsilon_j, \qquad
  \delta A_i = -d\epsilon_i - (\partial_i \Poisson^{jk}) \epsilon_k A_j.
\end{equation}
Note that $\epsilon'\_j = \epsilon_j + \partial_j \Casimir$ leads to
the same transformations of $X^i$.  Under the local symmetries
$\epsilon_i = \epsilon_i(x^m)$ the action transforms into a total
derivative, $\delta \Action = \int d(dX^i \epsilon_i)$. We get Lorentz
transformations with parameter $l = l(x^m)$ when making the particular
choice $\epsilon_i = (\epsilon_a, \epsilon_\phi) = (0, l)$. We can
also represent any infinitesimal diffeomorphism on $\BMf$, given by
$\delta x^m = -\xi^m$, by symmetry transformations with parameter
$\epsilon_i = \xi^m A_{m i} = (\xi^m e_{m a}, \xi^m \omega_m)$. These
symmetries yield the usual transformation rules for the fields back,
$\delta X^i = -\xi^m \partial_m X^i$ and $\delta A_{m i} = -\xi^n
\partial_n A_{m i} - (\partial_m \xi^n) A_{n i}$, when going on shell.
The transformations with parameter $\epsilon_i = (\epsilon_a, 0)$
reveals the zweibeins $e_a$ to be the corresponding gauge fields,
$\delta e_a = -d\epsilon_a + \cdots$, leading us to the notion of
`local translations' for this symmetry.

\section{Conserved Quantity}
\label{sec:C}

If $\Poisson^{ij}$ is not of full rank, then we have functions
$\Casimir$ so that
\begin{equation}
  \label{eq:casimir-function}
  \{X^i, \Casimir\} = \Poisson^{ij} \partial_j \Casimir = 0.
\end{equation}
In the case of PSM gravity there is only one such function $\Casimir =
\Casimir(\phi, Y)$ given by its defining equation
(\ref{eq:casimir-function})
\begin{equation}
  \label{eq:C}
  \Casimir' - V \dot \Casimir = 0,
\end{equation}
where $\Casimir' = \partial_\phi \Casimir$ and $\dot \Casimir =
\partial_Y \Casimir$.
\begin{table}
  \begin{displaymath}
    \begin{array}{|l|l|}
      \hline
      V = 0 & \Casimir = f\left( Y \right) \\
      V = v_0(\phi) & \Casimir = f\left( Y + \int_{0}^{\phi} v_0(x) dx
      \right) \\
      V = v_0(\phi) + v_1 Y & \Casimir = f\left( Y e^{v_1 \phi} +
        \int_{0}^{\phi} e^{v_1 x} v_0(x) dx \right) \\
      V = v_0(\phi) + v_1(\phi) Y & \Casimir = f\left( Y
        e^{\int_{0}^{\phi} v_1(x') dx'} + \int_{0}^{\phi}
        e^{\int_{0}^{x} v_1(x') dx'} v_0(x) dx \right) \\
      \hline
    \end{array}
  \end{displaymath}
  \caption{Conserved Quantity $\Casimir$}
  \label{tab:conserved-quantitiy-C}
\end{table}

From 
\begin{equation}
  d\Casimir = dX^i \partial_i \Casimir = - \Poisson^{ij} A_j
  \partial_i \Casimir = 0,
\end{equation}
where the PSM \eom (\ref{eq:eomX}) and condition
(\ref{eq:casimir-function}) were employed, we derive that $d\Casimir =
0$ on shell.  Another view is to take a particular linear combination
of (\ref{eq:eomX}),
\begin{equation}
  \label{eq:eomC}
  (dX^i + \Poisson^{ij} A_j) \partial_i \Casimir = dX^i \partial_i
  \Casimir = d\Casimir,
\end{equation}
which immediately shows that $d\Casimir = 0$ is an equation of motion.
This result can also be derived using the second Noether theorem and
making a local transformation with parameter $\epsilon_C$,
\begin{equation}
  \label{eq:symm-local-C}
  \delta X^i = 0, \qquad
  \delta A_i = -(\partial_i \Casimir)\, d\epsilon_C,
\end{equation}
for which we get $\delta \Action = -\int d(d\Casimir \epsilon_C)$.
This transformation differs from the symmetries
(\ref{eq:psm_symmetries}) with parameter $\epsilon_i = (\partial_i
\Casimir) \epsilon_C$ only by a term proportional to the equations of
motion. One can also use the first Noether theorem to establish the
same results.  Although the symmetry which yields the conserved
quantity is already a local one, because $d\Casimir = 0$ is an \eom,
one can use a trick and perform a rigid transformation instead. The
transformation
\begin{equation}
  \label{eq:symm-rigid-C}
  \delta X^i = 0, \qquad
  \delta A_i = -(\partial_i \Casimir)\, df \epsilon
\end{equation}
with an arbitrary function $f = f(\phi, Y)$ and a rigid transformation
parameter $\epsilon$ leaves the action (\ref{eq:action-psm})
invariant.  Furthermore, the \eom{}s transform in a total derivative
$d\Casimir$, stating that the conserved charge is $\Casimir$
\cite{Kummer:1994ur}. We will give a more geometrical meaning to this
symmetry later on. Note that this is a symmetry which acts on the one
forms $A_i$, but has no influence on the coordinates $X^i$.

\section{Solving the Equations of Motion}
\label{sec:eoms}

Assuming for simplicity that there is only one function
$\Casimir(X^i)$, then the equation $\Casimir(X^i) =
\underline{\Casimir} \in \R$ solves one of the differential equations
(\ref{eq:eomX}), but how can we solve the remaining ones?  This is
easily done by making the coordinate transformation $(X^i) = (X^{++},
X^\sbi{i}) \rightarrow (\underline{\Casimir}, \underline{X}^\sbi{i})$,
\begin{align}
  \underline{\Casimir} &= \Casimir(X^{++}, X^\sbi{l}), \\
  \underline{X}^\sbi{i} &= X^\sbi{i}.
\end{align}
Notice the difference between the index $i$ and $\sbi{i}$ and that
$X^\sbi{i}$ stands for $(X^{--}, \phi)$. For $A_i = (A_{++},
A_\sbi{i})$, using $A_i = \frac{\partial \underline{X}^j}{\partial
  X^i} \underline{A}_j$, we get
\begin{equation}
  \label{eq:Aorig}
  A_{++} = (\partial_{++} \Casimir) \underline{A}_C, \qquad
  A_\sbi{i} = \underline{A}_\sbi{i} + (\partial_\sbi{i} \Casimir)
  \underline{A}_C,
\end{equation}
and for the Poisson tensor in the new coordinate system we have
\begin{equation}
  \underline{\Poisson}^{C\sbi{j}} = 0, \qquad
  \underline{\Poisson}^{\sbi{i}\sbi{j}} = \Poisson^{\sbi{i}\sbi{j}}.
\end{equation}
The reduced Poisson tensor of the subspace spanned by $(X^\sbi{i})$ is
invertible. The inverse defined by
$\underline{\Poisson}^{\sbi{l}\sbi{j}}
\underline{\Omega}_{\sbi{l}\sbi{i}} = \delta_\sbi{i}\^\sbi{j}$
supplies this subspace with the symplectic two form
$\underline{\Omega} = \half d\underline{X}^\sbi{i}
d\underline{X}^\sbi{j} \underline{\Omega}_{\sbi{j}\sbi{i}}$, which, as
a consequence of the Jacobi identity of the Poisson tensor, is closed
on shell, $d\underline{\Omega} = d\underline{\Casimir} \gamma$
(evaluated on the target space, where it is nontrivial). The action
reads in the $(\underline{\Casimir}, \underline{X}^\sbi{i})$
coordinate system of the target space
\begin{equation}
  \label{eq:action-psm-C}
  \Action = \int_\BMf d\underline{\Casimir}\, \underline{A}_C +
  d\underline{X}^\sbi{i} \underline{A}_\sbi{i} + \half
  \underline{\Poisson}^{\sbi{j}\sbi{i}} \underline{A}_\sbi{i}
  \underline{A}_\sbi{j}.
\end{equation}
The \eom{}s by the variations $\delta \underline{A}_C$ and $\delta
\underline{A}_\sbi{i}$,
\begin{equation}
  d\underline{\Casimir} = 0, \qquad d\underline{X}^\sbi{i} +
  \underline{\Poisson}^{\sbi{i}\sbi{j}} \underline{A}_j = 0
\end{equation}
are then easily solved by $\underline{\Casimir} = \const$, as we
already knew, and by
\begin{equation}
  \label{eq:Asbi}
  \underline{A}_\sbi{i} = -d\underline{X}^\sbi{j}
  \underline{\Omega}_{\sbi{j}\sbi{i}}.
\end{equation}
In order to investigate the remaining \eom{}s, following from the
variations $\delta \underline{\Casimir}$ and $\delta
\underline{X}^\sbi{i}$, we first insert the solutions for the
$\underline{A}_\sbi{i}$ back into the action, yielding
\begin{equation}
  \label{eq:actionCOm}
  \Action = \int_\BMf d\underline{\Casimir}\, \underline{A}_C - \half
  d\underline{X}^\sbi{i} d\underline{X}^\sbi{j}
  \underline{\Omega}_{\sbi{j}\sbi{i}},
\end{equation}
and then derive the equations
\begin{alignat}{2}
  \delta \underline{A}_C &\text{:}\qquad & d\underline{\Casimir} &= 0,
  \\
  \delta \underline{\Casimir} &\text{:}\qquad & d\underline{A}_C -
  \half d\underline{X}^\sbi{i} d\underline{X}^\sbi{j}\,
  (\underline{\partial}_C \underline{\Omega}_{\sbi{j}\sbi{i}})  &= 0,
  \label{eq:AC} \\
  \delta \underline{X}^\sbi{i} &\text{:}\qquad &
  \underline{\partial}_\sbi{i} \underline{\Omega}_{\sbi{j}\sbi{k}} +
  \cycl(\sbi{i}\sbi{j}\sbi{k}) &= 0.
\end{alignat}
We see (\ref{eq:AC}) is the only differential equation which remains.
In order to see how it has to be treated we first stick to PSM gravity
and take a closer look at the $(\underline{\Casimir}, \underline{X}^{--},
\underline{\phi})$, the $(\underline{X}^{++}, \underline{\Casimir},
\underline{\phi})$ and the $(\underline{X}^{++}, \underline{X}^{--},
\underline{\Casimir})$ coordinate systems in turn.

In the $(\underline{\Casimir}, \underline{X}^{--}, \underline{\phi})$
coordinate system the Poisson tensor of PSM gravity is
\begin{equation}
  \label{eq:poisson-gr1}
  \underline{\Poisson}^{ij} = \left(
  \begin{array}{ccc}
    0 & 0 & 0 \\
    0 & 0 & \underline{X}^{--} \\
    0 & -\underline{X}^{--} & 0
  \end{array}
  \right).
\end{equation}
The restriction of it to the subspace $\underline{X^\sbi{i}} =
(\underline{X}^{--}, \underline{\phi})$ is invertible, as long as
$\underline{X}^{--} \neq 0$, giving this subspace a symplectic
structure:
\begin{equation}
  \label{eq:symplectic-gr1}
  \underline{\Poisson}^{\sbi{i}\sbi{j}} = \left(
  \begin{array}{cc}
    0 & \underline{X}^{--} \\
    -\underline{X}^{--} & 0
  \end{array}
  \right), \qquad
  \underline{\Omega}_{\sbi{i}\sbi{j}} = \left(
  \begin{array}{cc}
    0 & \frac{1}{\underline{X}^{--}} \\
    -\frac{1}{\underline{X}^{--}} & 0
  \end{array}
  \right).
\end{equation}
The solution of the one forms $\underline{A}_\sbi{i}$ follow
immediately from (\ref{eq:Asbi}), $\underline{A}_{--} =
d\underline{\phi}/\underline{X}^{--}$ and $\underline{A}_{\phi} =
-d\underline{X}^{--}/\underline{X}^{--}$. The remaining equation
(\ref{eq:AC}) for $\underline{A}_C$ is particularly simple, due to
$\underline{\partial}_C \underline{\Omega}_{\sbi{i}\sbi{j}} = 0$ in
this special coordinate system, yielding $d\underline{A}_C = 0$. This
is solved by $\underline{A}_C = -d\underline{F}$, where $\underline{F}
= \underline{F}(x)$. Going back to the original $(X^{++}, X^{--},
\phi)$ coordinate system using (\ref{eq:Aorig}) we get
\begin{align}
  e_{++} &= -X_{++} \dot{\Casimir} dF, \\
  e_{--} &= \frac{d\phi}{X^{--}} - X_{--} \dot{\Casimir} dF, \\
  \omega &= -\frac{dX^{--}}{X^{--}} - \Casimir' dF.
\end{align}

The generalization of the PSM to the graded case (\Cha\ref{cha:gPSM})
will be the basis of a very direct way to obtain 2d supergravity
theories. Also the way to obtain the solution for a particular model
(\Sec\ref{sec:sdil-sol}) will essentially follow the procedure
described here.

\section{Symplectic Extension of the PSM}
\label{sec:target-space-ext}

The solution of the PSM \eom{}s in the $(\underline{X}^{++},
\underline{\Casimir}, \underline{\phi})$ coordinate system provides no
new insight, it is similar to the one before, but in the
$(\underline{X}^{++}, \underline{X}^{--}, \underline{\Casimir})$
system things turn out to be more difficult. For the Poisson tensor we
obtain
\begin{equation}
  \label{eq:poisson-gr3}
  \underline{\Poisson}^{ij} = \left(
  \begin{array}{ccc}
    0 & \underline{V} & 0 \\
    -\underline{V} & 0 & 0 \\
    0 & 0 & 0
  \end{array}
  \right).
\end{equation}
The restriction to the subspace $(\underline{X}^{++},
\underline{X}^{--})$ gives
\begin{equation}
  \label{eq:symplectic-gr3}
  \underline{\Poisson}^{\sbi{i}\sbi{j}} = \left(
  \begin{array}{cc}
    0 & \underline{V} \\
    -\underline{V} & 0
  \end{array}
  \right), \qquad
  \underline{\Omega}_{\sbi{i}\sbi{j}} = \left(
  \begin{array}{cc}
    0 & \frac{1}{\underline{V}} \\
    -\frac{1}{\underline{V}} & 0
  \end{array}
  \right).
\end{equation}
The difficulties stem from the fact that $\underline{\partial}_C
\underline{\Omega}_{\sbi{i}\sbi{j}} \neq 0$, therefore yielding for
(\ref{eq:AC}) the equation $d\underline{A}_C - d\underline{X}^{--}
d\underline{X}^{++} \underline{\partial}_C (\frac{1}{\underline{V}}) =
0$. Finding a solution for this equation in an elegant way is the
intention of a symplectic extension of the PSM where the Poisson
tensor is no longer singular.

\subsection{Simple Case}
\label{sec:target-space-ext-simple}

Let us first treat the more simple case $\underline{\partial}_C
\underline{\Omega}_{\sbi{i}\sbi{j}} = 0$ using the coordinate system
$(\underline{\Casimir}, \underline{X}^{--}, \underline{\phi})$ as an
example. The solution of (\ref{eq:AC}) is $\underline{A}_C =
-d\underline{F}$, the minus sign is conventional.  Inserting this
solution back into action (\ref{eq:actionCOm}) we get
\begin{equation}
  \Action = \int_\BMf -d\underline{\Casimir} d\underline{F} - \half
  d\underline{X}^\sbi{i} d\underline{X}^\sbi{j}
  \underline{\Omega}_{\sbi{j}\sbi{i}}.
\end{equation}
This directly suggests to extend the target space by adding a new
target space coordinate $\underline{F}$. Using $\underline{X}^I =
(\underline{F}, \underline{\Casimir}, \underline{X}^{--},
\underline{\phi})$ as coordinate system we write the action in the
form
\begin{equation}
  \Action = \int_\BMf -\half d\underline{X}^I
  d\underline{X}^J \underline{\Omega}_{JI},
\end{equation}
from which we can immediately read off the extended symplectic matrix
and calculate its inverse,
\begin{equation}
  \label{eq:symplectic-ext1}
  \underline{\Omega}_{IJ} =
  \left(
    \begin{array}{cccc}
      0 & 1 & 0 & 0 \\
      -1 & 0 & 0 & 0 \\
      0 & 0 & 0 & \frac{1}{\underline{X}^{--}} \\
      0 & 0 & -\frac{1}{\underline{X}^{--}} & 0 \\
    \end{array}
  \right), \quad
  \underline{\Poisson}^{IJ} =
    \left(
    \begin{array}{cccc}
      0 & 1 & 0 & 0 \\
      -1 & 0 & 0 & 0 \\
      0 & 0 & 0 & \underline{X}^{--} \\
      0 & 0 & -\underline{X}^{--} & 0 \\
    \end{array}
  \right).
\end{equation}
Now let's take a look at the corresponding action in PSM form. When
using $\underline{F}$ as target space coordinate it is also necessary
to introduce its corresponding one form $\underline{A}_F$, therefore
with $\underline{A}_{--}$ and $\underline{A}_\phi$ already eliminated
we get the equivalent action
\begin{equation}
  \Action = \int_\BMf d\underline{F}\, \underline{A}_F +
  d\underline{\Casimir}\, \underline{A}_C - \underline{A}_F \underline{A}_C -
  \half d\underline{X}^\sbi{i} d\underline{X}^\sbi{j}
  \underline{\Omega}_{\sbi{j}\sbi{i}},
\end{equation}
from which we immediately derive the \eom{}s
\begin{alignat}{2}
  \delta \underline{A}_C &\text{:}\qquad & d\underline{\Casimir} -
  \underline{A}_F &= 0, \\
  \delta \underline{A}_F &\text{:}\qquad & d\underline{F} +
  \underline{A}_C &= 0, \label{eq:AC-ext} \\
  \delta \underline{\Casimir} &\text{:}\qquad & d\underline{A}_C &= 0,
  \label{eq:dAC-ext} \\
  \delta \underline{F} &\text{:}\qquad & d\underline{A}_F &= 0.
\end{alignat}
The general solution in the symplectic case can be immediately written
down,
\begin{equation}
  \underline{A}_I = -d\underline{X}^J \underline{\Omega}_{JI}.
\end{equation}
We see, gauge-fixing $\underline{A}_F = 0$ yields our original model
back.  Of course, field equations (\ref{eq:AC-ext}) and
(\ref{eq:dAC-ext}) only fit when $\underline{\partial}_C
\underline{\Omega}_{\sbi{i}\sbi{j}} = 0$. If this is not so,
(\ref{eq:dAC-ext}) would read $d\underline{A}_C - \half
d\underline{X}^\sbi{i} d\underline{X}^\sbi{j} \underline{\partial}_C
\underline{\Omega}_{\sbi{j}\sbi{i}} = 0$ and (\ref{eq:AC-ext}) has to
be extended by adding some more terms, which means
$\underline{\Poisson}^{iF} \neq 0$, or when inverting the Poisson
tensor $\underline{\Omega}_{Ci} \neq 0$. Note that the extended
Poisson tensor (\ref{eq:symplectic-ext1}) already fulfills the Jacobi
identity $\underline{\Poisson}^{IL} \underline{\partial}_L
\underline{\Poisson}^{JK} + \cycl(IJK) = 0$ and that its inverse, the
symplectic matrix, obeys $\underline{\partial}_I
\underline{\Omega}_{JK} + \cycl(IJK) = 0$, which is just the statement
that the symplectic form on the extended target space $\Omega = \half
d\underline{X}^I d\underline{X}^J \underline{\Omega}_{JI}$ is closed
on the target space, $d\Omega = 0$.  And it is just the closure of the
symplectic form, or, when considering the dual problem, the Jacobi
identity, which determines $\underline{\Omega}_{Ci}$ and/or
$\underline{\Poisson}^{iF}$ in the case $\underline{\partial}_C
\underline{\Omega}_{\sbi{i}\sbi{j}} \neq 0$, as we demonstrate in a
minute.

Next we will investigate the new symmetry transformations we have
implicitly added to the extended action. From the transformations
$\delta \underline{X}^I = \underline{\Poisson}^{IJ} \underline{\epsilon}_J$
and $\delta \underline{A}_I = -d\underline{\epsilon}_I +
\underline{\partial}_I \underline{\Poisson}^{JK} \underline{A}_K
\underline{\epsilon}_J$ we first look at the one with parameter
$\underline{\epsilon}_C$,
\begin{equation}
  \delta \underline{F} = \underline{\epsilon}_C, \qquad
  \delta \underline{A}_C = -d\underline{\epsilon}_C.
\end{equation}
Although this symmetry was originally available in the gauge potential
sector only, $\delta \underline{A}_C = -d\underline{\epsilon}_C$, it
can now be interpreted as a transformation in the extended target
space, $\delta \underline{F} = \underline{\epsilon}_C$.  Transforming
back to the coordinate system $(F, X^{++}, X^{--}, \phi)$, which is the
original one extended by $F = \underline{F}$, using (\ref{eq:Aorig})
shows that
\begin{equation}
  \delta A_i = -(\partial_i \Casimir) d\underline{\epsilon}_C,
\end{equation}
is indeed the gauge symmetry of the extended action which is the local
version of (\ref{eq:symm-rigid-C}) corresponding to the conserved
quantity $\Casimir$ of the original action. Furthermore, any departure from
$\underline{A}_F = 0$ can be made by a symmetry transformation with
parameter $\underline{\epsilon}_F$,
\begin{equation}
  \delta \underline{\Casimir} = -\underline{\epsilon}_F, \qquad
  \delta \underline{A}_F = -d\underline{\epsilon}_F.
\end{equation}

For completeness we write down the extension in the $(\underline{F},
\underline{X}^{++}, \underline{\Casimir}, \underline{\phi})$
coordinate system too:
\begin{equation}
  \label{eq:symplectic-ext2}
  \underline{\Omega}_{IJ} =
  \left(
    \begin{array}{cccc}
      0 & 0 & 1 & 0 \\
      0 & 0 & 0 & -\frac{1}{\underline{X}^{++}} \\
      -1 & 0 & 0 & 0 \\
      0 & \frac{1}{\underline{X}^{++}} & 0 & 0 \\
    \end{array}
  \right), \quad
  \underline{\Poisson}^{IJ} =
    \left(
    \begin{array}{cccc}
      0 & 0 & 1 & 0 \\
      0 & 0 & 0 & -\underline{X}^{++} \\
      -1 & 0 & 0 & 0 \\
      0 & \underline{X}^{++} & 0 & 0 \\
    \end{array}
  \right).
\end{equation}

\subsection{Generic Case}
\label{sec:target-space-ext-generic}

As already promised, we are going to solve (\ref{eq:AC}) in the
$(\underline{F}, \underline{X}^{++}, \underline{X}^{--},
\underline{\Casimir})$ coordinate system by extending the target space
and looking for an appropriate Poisson tensor. The above
considerations suggest the ansatz
\begin{equation}
  \underline{\Poisson}^{IJ} =
    \left(
    \begin{array}{cccc}
      0 & \underline{\Poisson}^{F++} & \underline{\Poisson}^{F--} & 1 \\
      -\underline{\Poisson}^{F++} & 0 & V & 0 \\
      -\underline{\Poisson}^{F--} & -V & 0 & 0 \\
      -1 & 0 & 0 & 0 \\
    \end{array}
  \right),
\end{equation}
or equivalently for the symplectic form
\begin{equation}
  \underline{\Omega}_{IJ} =
  \left(
    \begin{array}{cccc}
      0 & 0 & 0 & 1 \\
      0 & 0 & \frac{1}{V} & -\underline{\Omega}_{C++} \\
      0 & -\frac{1}{V} & 0 & -\underline{\Omega}_{C--} \\
      -1 & \underline{\Omega}_{C++} & \underline{\Omega}_{C--} & 0 \\
    \end{array}
  \right).
\end{equation}
The quantities $\underline{\Poisson}^{Fb}$ and
$\underline{\Omega}_{Cb}$ should not depend on the coordinate
$\underline{F}$. We can further expand $\underline{\Omega}_{Cb}$ in a
Lorentz covariant way, $\underline{\Omega}_{Cb} = \half A
\underline{X}_b + \half B \underline{X}^a \epsilon_{ab}$. The closure
relation for $\Omega$ by $\underline{J}_{IJK} =
\underline{\partial}_{[I} \underline{\Omega}_{JK]}$ yields
\begin{equation}
  -\half \epsilon^{ba} \underline{J}_{Cab} = \underline{Y}
  (\underline{\partial}_Y B) + B + \underline{\partial}_C (\frac{1}{V}) 
  = 0.
\end{equation}
Then, after inserting $B = \frac{G}{\underline{Y}}$, we arrive at
\begin{equation}
  \underline{\partial}_Y G + \underline{\partial}_C (\frac{1}{V}) 
  = 0.
\end{equation}
This equation is solved by
\begin{equation}
  G = \frac{1}{\Casimir'} + g(\underline{\Casimir}),
\end{equation}
as can be seen easily by going back to our original coordinate system,
thus using $\underline{\partial}_Y = \partial_Y - \frac{1}{V}
\partial_\phi$, $\underline{\partial}_C = \frac{1}{\Casimir'}
\partial_\phi$ and $V = \frac{\Casimir'}{\dot \Casimir}$. We set $A =
0$, which can be reached by a coordinate transformation of the type
$\underline{F} \rightarrow \underline{F} + f(\underline{X}^i,
\underline{\Casimir})$ as we see later, and calculate
$\underline{\Poisson}^{Fb}$ using $\underline{\Omega}_{Cb}$,
\begin{equation}
  \underline{\Omega}_{Cb} = \frac{1}{2 \underline{Y}}\, G
  \underline{X}^a \epsilon_{ab}, \qquad
  \underline{\Poisson}^{Fb} = \frac{1}{2 \underline{Y}}\, G V
  \underline{X}^b,
\end{equation}
and then go back to our original coordinate system, in which we have,
in addition to (\ref{eq:poisson-tensor-gr}),
\begin{equation}
  \Poisson^{F\phi} = 0, \qquad \Poisson^{Fb} = \frac{1}{2 Y}\, G V X^b.
\end{equation}
The homogeneous part in the solution $G$ can also be set to zero, thus
we finally have the extended action
\begin{equation}
  \label{eq:action-ext}
  \Action^{\mathrm{ext}} = \int_\BMf (dF) A_F + (DX^a) e_a + (d\phi)
  \omega + \half V \epsilon^{ba} e_a e_b + \frac{1}{X^a X_a
    \Dot{\Casimir}} X^b e_b A_F.
\end{equation}
This is a generalization of 2d gravity with $U(1)$ gauge field $A$;
there $V$ is a function of $\phi$, $Y$ and $F$ but the last term is
not present \cite{Kummer:1995qv}. The original model
(\ref{eq:action-first-order}) is found again when choosing the gauge
$A_F = 0$. Of course the gauge $A_F = 0$ restricts the target space to
the surface $\Casimir(X^i) = \const$, as can be seen immediately from
the solution of the \eom{}s of model (\ref{eq:action-ext}), given by $A_I
= - dX^J \Omega_{JI}$, where $\Omega_{JI}$ is the inverse of the
Poisson tensor,
\begin{equation}
  \Omega_{Fi} = \partial_i \Casimir, \qquad \Omega_{\phi b} =
  \frac{1}{2 Y} X^a \epsilon_{ab}, \qquad \Omega_{ab} = 0,
\end{equation}
which is given by
\begin{align}
  A_F &= d\Casimir, \\
  e_a &= -dF X_a \Dot{\Casimir} - d\phi \frac{1}{2Y} X^b
  \epsilon_{ba}, \\
  \omega &= -dF \Casimir' + \frac{1}{2Y} dX^a X^b \epsilon_{ba}.
\end{align}

\subsection{Uniqueness of the Extension}
\label{sec:target-space-ext-uniqueness}

We now have a look at the various possibilities one has in extending
the action. We will see, by using Casimir-Darboux coordinates, that
all these possibilities are connected by coordinate transformations of
particular types.  The simplest case of a symplectic extension is the
extension of a model in Casimir-Darboux coordinates. Denoting the
coordinates by $X^i = (\Casimir, Q, P)$ we have
\begin{equation}
  \label{eq:action-CD}
  \Action^{\mathrm{CD}} = \int_\BMf d\Casimir A_C + dQ A_Q + dP A_P -
  A_Q A_P
\end{equation}
and the Poisson tensor reads
\begin{equation}
  \label{eq:poisson-CD}
  \Poisson^{ij} = \left(
  \begin{array}{ccc}
    0 & 0 & 0 \\
    0 & 0 & 1 \\
    0 & -1 & 0
  \end{array}
  \right).
\end{equation}
The \eom{}s for $A_Q$ and $A_P$ can be read off immediately. The
remaining equations are $d\Casimir = 0$ and $dA_C = 0$, where the
latter is solved by $A_C = -dF$. We also see that the same equations
can be derived from the extended action
\begin{equation}
  \label{eq:action-CD-ext}
  \Action^{\mathrm{ext}} = \int_\BMf dF A_F + d\Casimir A_C - A_F A_C
  + dQ A_Q + dP A_P - A_Q A_P
\end{equation}
as long as we restrict the solution to the surface $\Casimir = \const$ 
or alternatively gauge $A_F = 0$.

The extended target space is uniquely determined by the conditions
$\partial_F \Poisson^{IJ} = 0$, and the Jacobi identity $J^{IJK} = 0$.
One can easily see this by making the ansatz in the extended target
space $X^I = (F, \Casimir, Q, P)$
\begin{equation}
  \label{eq:poisson-CD-ext}
  \Poisson^{IJ} =
    \left(
    \begin{array}{cccc}
      0 & \Poisson^{FC} & \Poisson^{FQ} & \Poisson^{FP} \\
      -\Poisson^{FC} & 0 & 0 & 0 \\
      -\Poisson^{FQ} & 0 & 0 & 1 \\
      -\Poisson^{FP} & 0 & -1 & 0 \\
    \end{array}
  \right).
\end{equation}
From the Jacobi identities $J^{FCQ} = J^{FCP} = 0$ we get
$\Poisson^{FC} = \kappa(\Casimir)$. Using a coordinate transformation
of the form $\underline{\Casimir} = g(\Casimir)$ we can set
$\Poisson^{FC} = 1$. This can be done because any redefinition of
$\Casimir$ does not change the original model.  The remaining Jacobi
identity $J^{FQP} = 0$ yields $\partial_Q \Poisson^{FQ} + \partial_P
\Poisson^{FP} = 0$, but any solution of this equation can be put to
zero by a coordinate transformation $\underline{F} = F + f(\Casimir,
Q, P)$. The one forms $A_C$, $A_Q$ and $A_F$ get then contributions of
$A_F$, but on the surface $\Casimir = \const$ we have $A_F$ = 0,
therefore this doesn't change the model too. We conclude that the
conditions $\partial_F \Poisson^{IJ} = 0$ and $J^{IJK} = 0$ provide a
coordinate independent way to extend the target space and to supply it
with a symplectic structure.

\section{Symplectic Geometry}
\label{sec:symplectic-geometry}

Symplectic form and Poisson tensor read
\begin{equation}
  \Omega = \half dX^I dX^J \Omega_{JI}, \qquad
  \Poisson = \half \Poisson^{IJ} \partial_J \partial_I,
\end{equation}
where $\Omega_{IJ} = -\Omega_{JI}$, $\Poisson^{IJ} = -\Poisson^{JI}$
and $\Poisson^{IJ} \Omega_{IK} = \delta_K\^J$. The related Poisson
bracket of functions becomes
\begin{equation}
  \{f, g\} = \Poisson^{IJ} (\partial_J g) (\partial_I f).
\end{equation}

The symplectic form defines an isomorphism between vectors an
covectors. Let $v = v^I \partial_I$ be a vector field, then it's
correspondent 1-form $A_v$ is
\begin{equation}
  A_v = v^\flat = v \rfloor \Omega = dX^I v^J \Omega_{JI}.
\end{equation}
The vector corresponding to a 1-form $\alpha = dX^I \alpha_I$ is
\begin{equation}
  \alpha^\sharp = \Poisson^{IJ} \alpha_J \partial_I.
\end{equation}
We have $(v^\flat)^\sharp = v$ and $(\alpha^\sharp)^\flat = \alpha$ of
course. Now the Poisson bracket of the 1-forms $\alpha$ and $\beta$
can be defined by
\begin{equation}
  \{\alpha, \beta\} = [\alpha^\sharp, \beta^\sharp]^\flat,
\end{equation}
and the antisymmetric scalar product of vectors reads
\begin{equation}
  [u|v] = u^I v^J \Omega_{JI}.
\end{equation}

Hamiltonian vector fields become
\begin{equation}
  \label{eq:hamiltonian-vectorfield}
  V_f = (df)^\sharp = \{\,.\,, f\} = \Poisson^{IJ} (\partial_J f)
  \partial_I.
\end{equation}
(Note: $V_f \rfloor \Omega = df$). The commutator of Hamiltonian
vector fields is
\begin{equation}
  \label{eq:hamilton-vec-commutator}
  [V_f, V_g] = V_{\{g, f\}},
\end{equation}
and the antisymmetric scalar product of Hamiltonian vector fields
reads
\begin{equation}
  [V_f | V_g] = \{g, f\}.
\end{equation}
On the basis of these formulae the Lie derivative of functions in the
direction of Hamiltonian vector fields becomes
\begin{equation}
  \Lie_{V_f}(g) = V_f(g) = \{g, f\} = -V_g(f) = -\Lie_{V_g}(f),
\end{equation}
and the Lie derivative of vectors in the direction of Hamiltonian
vector fields
\begin{equation}
  \Lie_{V_f}(u) = [V_f, u].
\end{equation}
Symplectic form and Poisson tensor obey
\begin{equation}
  \Lie_{V_f}(\Omega) = \Lie_{V_f}(\Poisson) = 0.
\end{equation}

\section{Symplectic Gravity}
\label{sec:symplectic-gravity}

In terms of the coordinates $X^I = (F, X^a, \phi)$ the new components
of the Poisson tensor are $\Poisson^{Fb} = \frac{1}{2Y \dot{\Casimir}}
X^b$ and $\Poisson^{F\phi} = 0$. Its inverse, the symplectic form,
consists of the components $\Omega_{Fi} = (\partial_i \Casimir)$ and
$\Omega_{\phi b} = \frac{1}{2Y} X^c \epsilon_{cb}$.
\begin{alignat}{2}
  \Poisson^{IJ} &=
  \left(
    \begin{array}{c|cc}
      0 & \Poisson^{Fb} & \Poisson^{F\phi} \\
      \hline
      \Poisson^{aF} & \Poisson^{ab} & \Poisson^{a\phi} \\
      \Poisson^{\phi F} & \Poisson^{\phi b} & 0
    \end{array}
  \right) &&=
  \left(
    \begin{array}{c|cc}
      0 & \frac{1}{2Y\dot{\Casimir}} X^b & 0 \\
      \hline
      -\frac{1}{2Y\dot{\Casimir}} X^a & V \epsilon^{ab} & X^c
      \epsilon_c\^a \\
      0 & -X^c \epsilon_c\^b & 0
    \end{array}
  \right), \label{eq:poisson-ext} \\
  \Omega_{IJ} &=
  \left(
    \begin{array}{c|cc}
      0 & \Omega_{Fb} & \Omega_{F\phi} \\
      \hline
      \Omega_{aF} & \Omega_{ab} & \Omega_{a\phi} \\
      \Omega_{\phi F} & \Omega_{\phi b} & 0
    \end{array}
  \right) &&=
  \left(
    \begin{array}{c|cc}
      0 & \partial_b \Casimir & \partial_\phi \Casimir \\
      \hline
      -\partial_a \Casimir & 0 & -\frac{1}{2Y} X^c \epsilon_{ca} \\
      -\partial_\phi \Casimir & \frac{1}{2Y} X^c \epsilon_{cb} & 0
    \end{array}
  \right). \label{eq:symplectic-matrix}
\end{alignat}
\begin{align}
  \Poisson^{IJ} &=
  \left(
    \begin{array}{c|ccc}
      0 & \frac{1}{2 X^{--} \dot{\Casimir}} & \frac{1}{2 X^{++}
        \dot{\Casimir}} & 0 \\
      \hline
      -\frac{1}{2 X^{--} \dot{\Casimir}} & 0 & V & -X^{++} \\
      -\frac{1}{2 X^{++} \dot{\Casimir}} & -V & 0 & X^{--} \\
      0 & X^{++} & -X^{--} & 0
    \end{array}
  \right), \label{eq:poisson-matrix-ext} \\
  \Omega_{IJ} &=
  \left(
    \begin{array}{c|ccc}
      0 & X^{--} \dot{\Casimir} & X^{++} \dot{\Casimir} & \Casimir'
      \\
      \hline
      -X^{--} \dot{\Casimir} & 0 & 0 & -\frac{1}{2 X^{++}} \\
      -X^{++} \dot{\Casimir} & 0 & 0 & \frac{1}{2 X^{--}} \\
      -\Casimir' & \frac{1}{2 X^{++}} & -\frac{1}{2 X^{--}} & 0
    \end{array}
  \right). \label{eq:symplectic-matrix-ext}
\end{align}
\begin{equation}
  \det(\Omega_{IJ}) = \dot{\Casimir}^2, \qquad
  \det(\Poisson^{IJ}) = \frac{1}{\det(\Omega_{IJ})} =
  \frac{1}{\dot{\Casimir}^2}.
\end{equation}

Let $\Phi \maps \BMf \to \TSp$ and $(x^m)$ be a coordinate system on
$\BMf$. We have the relations for the 1-forms $A_I$:
\begin{alignat}{2}
  A_I &= A_{\partial_I} = (\partial_I)^\flat = \partial_I \rfloor
  \Omega = -dX^J \Omega_{JI} \quad &&\in \Lambda^1(\TSp), \\
  \Phi^{*} A_I &= -dx^m (\partial_m \Phi^J) \Omega_{JI} = dx^m
  A_{m I} \quad &&\in \Lambda^1(\BMf), \\
  A_{\Phi_{*} \partial_m} &= dX^I (\partial_m \Phi^J) \Omega_{JI} = 
  - dX^I A_{m I} \quad &&\in \Lambda^1(\TSp), \\
  A_{m I} &= -\partial_I \rfloor (A_{\Phi_{*} \partial_m})  =
  -(\partial_m \Phi^J) \Omega_{JI}.
\end{alignat}
The PSM Lagrangian can be written as
\begin{equation}
  \Lagrange = \Phi^{*}\left( dX^I \wedge (\partial_I \rfloor \Omega) + \half
    \Poisson^{IJ} (\partial_J \rfloor \Omega) \wedge (\partial_I \rfloor
    \Omega) \right),
\end{equation}
and the Hamiltonian flux of $\Casimir$ is
\begin{equation}
  V_C = (d\Casimir)^\sharp = \partial_F.
\end{equation}
Let $\Phi \maps \R \to \TSp$ be the integral curve $\Phi(\tau)$ of the
Hamiltonian vector field $V_C$. It satisfies the condition $\Phi_{*}
\partial_\tau = V_C$. When using the coordinate representation the
Hamiltonian equations
\begin{equation}
  \frac{dX^I}{d\tau} = \Poisson^{IJ} \frac{\partial \Casimir}{\partial
    X^J}
\end{equation}
follow, thus $\frac{dF}{d\tau} = 1$ and $\frac{dX^i}{d\tau} = 0$. This
suggests to use $(x^m) = (F, x^1)$ as coordinate system on $\BMf$,
because the $X^i$ are then functions of $x^1$ only, $X^i = X^i(x^1)$.
The induced symplectic form on the world sheet $\BMf$ given by $\omega
= \Phi^{*} \Omega$ in that coordinates reads $\omega = -dF dx^1
(\partial_1 X^i) (\partial_i \Casimir) = -dF dx^1
\frac{d\Casimir}{dx^1}$, which is zero on the surface $\Casimir =
\text{const}$.

The surface $\Casimir = \const$ in parameter representation using $(F,
x^1)$ as coordinate system on $\BMf$ follows from $\Phi^{*}(d\Casimir)
= dx^1 ((\partial_1 Y) \dot{\Casimir} + (\partial_1 \phi) \Casimir') =
0$, thus yielding the differential equation
\begin{equation}
  \frac{\partial_1 Y}{\partial_1 \phi} = -V(\phi, Y).
\end{equation}
This immediately leads to the $(x^m) = (F, \phi)$ system on
$\BMf$, where we have to solve
\begin{equation}
  \frac{dY}{d\phi} = -V(\phi, Y(\phi)).
\end{equation}

\subsection{Symmetry Adapted Coordinate Systems}
\label{sec:symplectic-gravity-coordinates}

The coordinates $(X^a) = (X^{++}, X^{--})$ transform under Lorentz
transformations with infinitesimal parameter $l$ according to $\delta
X^a = l X^b \epsilon_b\^a$. It is natural to replace $(X^{++},
X^{--})$ by new coordinates, one of these being the invariant $Y =
\half X^a X_a = X^{++} X^{--}$. In order to determine the second new
coordinate we calculate the finite Lorentz transformations by solving
$\frac{dX^a(\lambda)}{d\lambda}|_{\lambda=0} = X^b_{(0)}
\epsilon_b\^a$, for which we get
\begin{equation}
  \left\{
    \begin{aligned}
      X^{++}(\lambda) &= e^{-\lambda}\, X^{++}_{(0)} \\
      X^{--}(\lambda) &= e^{\lambda}\, X^{--}_{(0)}
    \end{aligned}
  \right..
\end{equation}
Now we see that $\lambda = -\half \ln(\frac{X^{++}}{X^{--}})$, or
$\lambda = -\half \ln(-\frac{X^{++}}{X^{--}})$ if
$\frac{X^{++}}{X^{--}}$ is negative, is an appropriate choice for the
remaining new coordinate. This coordinate is best suited, because any
Lorentz transformation is a translation in the $\lambda$-space.

Therefore, we change from the original coordinates $(X^{++}, X^{--})$
to symmetry adapted coordinates $(Y, \lambda)$, where $Y$ is an
invariant of the symmetry transformation and $\lambda$ is in a sense
isomorphic to the transformation group, by incorporating the
coordinate transformations:
\begin{eqnarray}
  X^{++} X^{--} > 0: &&\qquad
  \left\{
    \begin{aligned}
      Y &= X^{++} X^{--} \\
      \lambda &= -\half \ln\left( \frac{X^{++}}{X^{--}} \right)
    \end{aligned}
  \right. \\
  X^{++} X^{--} < 0: &&\qquad
  \left\{
    \begin{aligned}
      Y &= X^{++} X^{--} \\
      \lambda &= -\half \ln\left( -\frac{X^{++}}{X^{--}} \right)
    \end{aligned}
  \right. \\
  X^{++} = 0,\, X^{--} \neq 0: &&\qquad
  \left\{
    \begin{aligned}
      Y &= 0 \\
      \lambda &= \ln\left( X^{--} \right)
    \end{aligned}
  \right. \\
  X^{++} \neq 0,\, X^{--} = 0: &&\qquad
  \left\{
    \begin{aligned}
      Y &= 0 \\
      \lambda &= -\ln\left( X^{++} \right)
    \end{aligned}
  \right.
\end{eqnarray}
The point $X^{++} = 0$, $X^{--} = 0$ is non-sensitive to Lorentz
transformations and therefore cannot be represented in this way.

A short calculation yields the transformed components of the Poisson
tensor (\ref{eq:poisson-matrix-ext}) and the symplectic form
(\ref{eq:symplectic-matrix-ext}). Going from the original coordinates
$X^I = (F, X^{++}, X^{--}, \phi)$ to the Lorentz symmetry adapted
system $\underline{X}^I = (F, Y, \lambda, \phi)$ with $Y = X^{++}
X^{--}$ and $\lambda = -\half \ln \frac{X^{++}}{X^{--}}$ using
$\underline{\Poisson}^{IJ} = \Poisson^{KL} (\partial_L
\underline{X}^J) (\partial_K \underline{X}^I)$ we get
\begin{align}
  \underline{\Poisson}^{IJ} &=
  \left(
    \begin{array}{cccc}
      0 & \frac{1}{\dot{\Casimir}} & 0 & 0
      \\
      -\frac{1}{\dot{\Casimir}} & 0 & V & 0 \\
      0 & -V & 0 & 1 \\
      0 & 0 & -1 & 0
    \end{array}
  \right), \label{eq:poisson-matrix-ext1} \\
  \underline{\Omega}_{IJ} &=
  \left(
    \begin{array}{cccc}
      0 & \dot{\Casimir} & 0 & \Casimir'
      \\
      -\dot{\Casimir} & 0 & 0 & 0 \\
      0 & 0 & 0 & 1 \\
      -\Casimir' & 0 & -1 & 0
    \end{array}
  \right). \label{eq:symplectic-matrix-ext1}
\end{align}
\begin{equation}
  \det(\underline{\Omega}_{IJ}) = \dot{\Casimir}^2, \qquad
  \det(\underline{\Poisson}^{IJ}) = \frac{1}{\dot{\Casimir}^2}.
\end{equation}

In flat space where $V = 0$ we have $\Casimir = Y$ and the Poisson
tensor (\ref{eq:poisson-matrix-ext1}) is already in Darboux form. In
this case $Y$ is not only Lorentz invariant, but also invariant under
spacetime translations. This suggests to use an invariant for both
spacetime as well as Lorentz transformations as new coordinate
instead of $Y$, but this is the quantity $\Casimir$. Using $(F,
\Casimir, \lambda, \phi)$ as coordinates of the target space, we
arrive at the Darboux form in the curved case too, where the conjugate
pairs are
\begin{equation}
  \label{eq:poisson-bracket-ext}
  \{F, \Casimir\} = 1, \qquad \{\lambda, \phi\} = 1,
\end{equation}
and the symplectic form reads
\begin{equation}
  \label{eq:symplectic-form-ext}
  \Omega = d\Casimir dF + d\phi d\lambda.
\end{equation}
The closure of the symplectic form is now evident, and we can
immediately write down the Cartan 1-form $\theta$, which is the
potential for the symplectic form, $\Omega = d\theta$, determined up
to an exact form:
\begin{equation}
  \label{eq:cartan-form}
  \theta = -\Casimir dF + \lambda d\phi
\end{equation}


\chapter{Supergravity from Poisson Superalgebras}
\label{cha:gPSM}


The method presented in this chapter is able to provide the geometric
actions for most general $\N=1$ supergravity in two spacetime
dimensions. Our construction implies the possibility of an extension
to arbitrary $\N$.  This provides a supersymmetrization of any
generalized dilaton gravity theory or of any theory with an action
being an (essentially) arbitrary function of curvature and torsion.

Technically we proceed as follows: The bosonic part of any of these
theories may be characterized by a generically nonlinear Poisson
bracket on a three-dimensional target space. In analogy to the given
ordinary Lie algebra, we derive all possible $\N=1$ extensions of any
of the given Poisson (or $W$-) algebras. Using the concept of graded
Poisson Sigma Models, any extension of the algebra yields a possible
supergravity extension of the original theory, local Lorentz and
super-diffeomorphism invariance follow by construction. Our procedure
automatically restricts the fermionic extension to the minimal one;
thus local supersymmetry is realized on-shell. By avoiding a
superfield approach of \Cha\ref{cha:sfield} we are also able to
circumvent in this way the introduction of constraints and their
solution. Instead, we solve the Jacobi identities of the graded
Poisson algebra. The rank associated with the fermionic extension
determines the number of arbitrary parameter functions in the
solution. In this way for many well-known dilaton theories different
supergravity extensions are derived.  It turns out that these
extensions also may yield restrictions on the range of the bosonic
variables.

\section{Graded Poisson Sigma Model}
\label{sec:gPSM}

\subsection{Outline of the Approach}
\label{sec:outline}

The PSM formulation of gravity theories allows a direct
generalization, yielding possible supergravity theories. Indeed, from
this perspective it is suggestive to replace the Minkowski space with
its linear coordinates $X^a$ by its superspace analogue, spanned by
$X^a$ and (real, \ie Majorana) spinorial and (one or more)
Grassmann-valued coordinates $\chi^{\q i\alpha}$ (where $\q i =
1,\ldots,\N$).  In the purely bosonic case we required that $\phi$
generates Lorentz transformations on Minkowski space. We now extend
this so that $\phi$ is the generator of Lorentz transformations on
superspace. This implies in particular that besides (\ref{LorentzB})
now also
\begin{equation}
  \label{LorentzF}
  \{ \chi^{\q i\alpha}, \phi \} = 
  -\half\chi^{\q i\beta} \gamma^3\_\beta\^\alpha,
\end{equation} 
has to hold, where $-\half \gamma^3\_\beta\^\alpha$ is the generator
of Lorentz transformations in the spinorial representation. For the
choice of the $\gamma$-matrices and further details on notation and
suitable identities we refer to \App\ref{app:spinors}.

Within the present work we first focus merely on a consistent
extension of the original bosonic Poisson algebra to the total
superspace. This superspace can be built upon $\N$ pairs of
coordinates obeying (\ref{LorentzF}). Given such a graded Poisson
algebra, the corresponding Sigma Model provides a possible
$\N$-supergravity extension of the original gravity model
corresponding to the purely bosonic sigma model. We shall mainly focus
on the construction of a graded Poisson tensor $\Poisson^{IJ}$ for the
simplest supersymmetric extension $\N=1$, \ie on a (`warped') product
of the above superspace and the linear space spanned by the generator
$\phi$.  Upon restriction to the bosonic submanifold $\chi^\alpha=0$,
the bracket will be required to coincide with the bracket
(\ref{LorentzB}) and (\ref{PB}) corresponding to the bosonic theory
(\ref{PSM}). Just as the framework of PSMs turns out to provide a
fully satisfactory and consistent 2d gravity theory with all the
essential symmetries for any given (Lorentz invariant) Poisson bracket
(\ref{LorentzB}) and (\ref{PB}), the framework of graded Poisson Sigma
Models (gPSM) will provide possible generalizations for any of the
brackets $\Poisson^{ij}$ with a local `supersymmetry' of the generic
type (\ref{susytrafo}). In particular, by construction of the general
theory (cf \cite{Strobl:1999zz} or \Sec\ref{sec:Jacobi} below) and
upon an identification which is a straightforward extension of
(\ref{ident1}), the resulting gravity theory will be invariant
automatically with respect to local Lorentz transformations, spacetime
diffeomorphisms \emph{and} local supersymmetry transformations. In
particular, the Rarita-Schwinger field $\psi_\alpha$ (or $\psi_{\q
  i\alpha}$, $\q i = 1,\ldots,\N$ in the more general case) is seen to
enter naturally as the fermionic component of the one-form valued
multiplet $A_I$.  Likewise, specializing the local symmetries
(\ref{PSM-symms}) (or rather their generalization to the graded case
provided in (\ref{gPSM-symms}) below) to the spinorial part
$\epsilon_\alpha$, local supersymmetry transformations of the form
(\ref{susytrafo}) are found, which, by construction, are symmetries of
the action (In fact, it is here where the graded Jacobi identity for
$\Poisson$ enters as an essential ingredient!). Finally, by
construction, the bosonic part of the action of the gPSM corresponding
to the bracket $\Poisson^{IJ}$ will coincide with (\ref{PSM}).  Thus,
for any such a bracket $\Poisson^{IJ}$, the resulting model should
allow the interpretation as a permissible supersymmetric
generalization of the original bosonic starting point.

The relations (\ref{LorentzB}), (\ref{LorentzF}) fix the $\phi$
components of the sought for (graded) Poisson tensor $\Poisson^{IJ}$.
We are thus left with determining the remaining components
$\Poisson^{AB}$, $A$ and $B$ being indices in the four-dimensional
superspace with $X^A = (X^a,\chi^\alpha)$. As will be recapitulated in
\Sec\ref{sec:gPSM2}, besides the graded symmetry of the tensor
$\Poisson^{IJ}$, the only other requirement it has to fulfill by
definition is the graded Jacobi identity. This requires the vanishing
of a 3-tensor $J^{IJK}$ (\cf (\ref{gJacobi}) below), which may be
expressed also as the Schouten-Nijenhuis bracket $[ \cdot , \cdot
]_{SN}$ of $(\Poisson^{IJ})$ with itself. In this formulation
$(\Poisson^{IJ})$ is meant to be the Poisson tensor itself and not its
components (abstract indices).  It is straightforward to verify (\cf
also \cite{Strobl:1999zz}) that the relations $J^{IJK}=0$ with at
least one of the indices coinciding with the one corresponding to
$\phi$ are satisfied, \emph{iff} $(\Poisson^{AB})$ is a Lorentz
covariant 2-tensor,
\begin{equation} \label{Pinv} \mathcal{L}_{(\Poisson^{A \phi})}
(\Poisson^{AB}) = 0,
\end{equation}
\ie depending on $X^a$, $\chi^\alpha$ and also on the Lorentz invariant
quantity $\phi$ in a covariant way as determined by its indices.  Thus
one is left with finding the general solution of $J^{ABC}=0$ starting
from a Lorentz covariant ansatz for $(\Poisson^{AB})$.

Let us note on this occasion that the above considerations do
\emph{not} imply that $(\Poisson^{AB})$ forms a bracket on the
Super-Minkowski space, a subspace of the target space under
consideration. The reason is that the equations $J^{ABC}=0$ contain
also derivatives of $\Poisson^{AB}$ with respect to $\phi$: in terms
of the Schouten-Nijenhuis bracket, the remaining equations become
\begin{equation}
  \label{Schouten}
  [(\Poisson^{AB}),(\Poisson^{AB})]_{SN} = (\Poisson^{A
    \phi}) \wedge (\partial_\phi \Poisson^{AB}),
\end{equation} 
where the components of the supervector $(\Poisson^{A \phi})$ are
given implicitly by eqs.~(\ref{LorentzB}) and (\ref{LorentzF}) above.
So $(\Poisson^{AB})$ defines a graded Poisson bracket for the $X^A$
only if it is independent of $\phi$.  However, in the present context
$\phi$-independent Poisson tensors are uninteresting in view of our
discussion of actions of the form (\ref{FOG}).

It should be remarked that given a particular bosonic model and its
corresponding bracket, there is by no means a unique graded extension,
even for fixed $\N$. Clearly, any (super-)diffeomorphism leaving
invariant the bosonic sector as well as the brackets (\ref{LorentzF})
applied to a solution of the (graded) Jacobi identities yields another
solution.  This induces an ambiguity for the choice of a
superextension of a given gravity model (\ref{dil}) or also
(\ref{grav}). This is in contrast to the direct application of, say,
the superfield formalism of Howe \cite{Howe:1979ia}, which when
applied to the (necessarily torsionfree) theory (\ref{dil})
\cite{Park:1993sd}, yields one particular superextension. This now turns
out as just one possible choice within an infinite dimensional space
of admissible extensions.  {}From one or the other perspective,
however, different extensions (for a given $\N$) may be regarded also
as effectively equivalent. We shall come back to these issues below.

A final observation concerns the relation of our supersymmetric
extensions to `ordinary' supergravity. {}From the point of view of the
seminal work on the 2d analogue \cite{Deser:1976rb,Brink:1977vg} of 4d
supergravity our supergravity algebra is `deformed' by the presence of
a dilaton field. Such a feature is known also from the dimensional
reduction of supergravity theories in higher dimensions, where one or
more dilaton fields arise from the compactification.

\subsection{Details of the gPSM}
\label{sec:gPSM2}

In this Section we recollect for completeness some general and
elementary facts about graded Poisson brackets and the corresponding
Sigma Models. This Section (\cf also \Sec\ref{sec:sfield} and
\App\ref{app:gravity}, \ref{app:spinors}) also sets the conventions
about signs \etc used within the present work, which are adapted to
those of \cite{Ertl:1997ib} and which differ on various instances from
those used in \cite{Strobl:1999zz}.

For the construction of the gPSM we take a 2-dimensional base manifold
$\BMf$, also called world sheet or spacetime manifold, with purely
bosonic (commutative) coordinates $x^m$, and the target space $\TSp$
with coordinates $X^I = (\phi, X^A) = (\phi, X^a, \chi^\alpha)$,
$\phi$ and $X^a$ being bosonic and $\chi^\alpha$ fermionic
(anticommutative), promoting $\TSp$ to a supermanifold.  The
restriction to one Majorana spinor means that only the case $\N=1$ is
implied in what follows.  To the coordinate functions $X^I$ correspond
gauge fields $A_I$ which we identify with the usual Lorentz-connection
1-form $\omega$ and the vielbein 1-form $e_a$ of the Einstein-Cartan
formalism of gravity and with the Rarita-Schwinger 1-form
$\psi_\alpha$ of supergravity according to $A_I = (\omega, e_A) =
(\omega, e_a, \psi_\alpha)$.  They can be viewed as 1-forms on the
base manifold $\BMf$ with values in the cotangential space of $\TSp$
and may be collected in the total 1-1-form $A = dX^I A_I = dX^I dx^m
A_{mI}$.

As the main structure of the model we choose a Poisson tensor
$\Poisson^{IJ} = \Poisson^{IJ}(X)$ on $\TSp$, which encodes the
desired symmetries and the dynamics of the theory to be constructed.
Due to the grading of the coordinates of $\TSp$ it is graded
antisymmetric $\Poisson^{IJ} = -(-1)^{IJ} \Poisson^{JI}$ and is
assumed to fulfill the graded Jacobi identity ($\rpartial_I
=\rpartial/\partial X^I$ is the right derivative) of which we list
also a convenient alternative version
\begin{align}
  J^{IJK} &= \Poisson^{IL} \rpartial_L \Poisson^{JK} + \gcycl(IJK)
  \label{gJacobi} \\
  &= \Poisson^{IL} \rpartial_L \Poisson^{JK}
  + \Poisson^{JL} \rpartial_L \Poisson^{KI} (-1)^{I(J+K)}
  + \Poisson^{KL} \rpartial_L \Poisson^{IJ} (-1)^{K(I+J)} \\
  &= 3 \Poisson^{I]L} \rpartial_L \Poisson^{[JK} = 0.
\end{align}
The relation between the right partial derivative $\rpartial_I$ and
the left partial derivative $\lpartial_I$ for the graded case formally
is the same as in (\ref{lr-partial}) ($M \rightarrow I$).  The Poisson
tensor defines the Poisson bracket of functions $f$, $g$ on $\TSp$,
\begin{equation}
  \label{gPB}
  \{f, g\} = (f \lpartial_J) \Poisson^{JI} (\rpartial_I g),
\end{equation}
implying for the coordinate functions $\{X^I, X^J\} = \Poisson^{IJ}$.
With (\ref{lr-partial}) the Poisson bracket (\ref{gPB}) may be written
also as
\begin{equation}
  \label{gPBv2}
  \{f, g\} = \Poisson^{JI} (\rpartial_I g) (\rpartial_J f) (-1)^{g(f+J)}.
\end{equation}
This bracket is graded anticommutative,
\begin{equation}
  \label{gPB-symm}
  \{f, g\} = -(-1)^{fg} \{g, f\},
\end{equation}
and fulfills the graded Jacobi identity
\begin{multline}
  \label{gPB-Jacobi}
  \{X^I, \{X^J, X^K\}\} (-1)^{IK} + \{X^J, \{X^K, X^I\}\} (-1)^{JI} \\
  + \{X^K, \{X^I, X^J\}\} (-1)^{KJ} = 0,
\end{multline}
which is equivalent to the graded derivation property
\begin{equation}
  \label{gPB-der}
  \{X^I, \{X^J, X^K\}\} = \{\{X^I, X^J\}, X^K\} + (-1)^{IJ} \{X^J, \{X^I,
  X^K\}\}.
\end{equation}

The PSM action (\ref{PSM}) generalizes to
\begin{equation}
  \label{gPSM}
  L^\mathrm{gPSM} = \int_\BMf dX^I A_I + \half \Poisson^{IJ} A_J A_I,
\end{equation}
where in the graded case the sequence of the indices is important.
The functions $X^I(x)$ represent a map from the base manifold to the
target space in the chosen coordinate systems of $\BMf$ and $\TSp$,
and $dX^I$ is the shorthand notation for the derivatives
$d^\BMf\!X^I(x) = dx^m \partial_m X^I(x)$ of these functions. The
reader may notice the overloading of the symbols $X^I$ which sometimes
are used to denote the map from the base manifold to the target space
and sometimes, as in the paragraph above, stand for target space
coordinates. This carries over to other expressions like $dX^I$ which
denote the coordinate differentials $d^\TSp\!X^I$ on $\TSp$ and, on
other occasions, as in the action (\ref{gPSM}), the derivative of the
map from $\BMf$ to $\TSp$.

The variation of $A_I$ and $X^I$ in (\ref{gPSM}) yields the gPSM field
equations
\begin{gather}
  dX^I + \Poisson^{IJ} A_J = 0, \label{gPSM-eomX} \\
  dA_I + \half (\rpartial_I \Poisson^{JK}) A_K A_J = 0.
  \label{gPSM-eomA}
\end{gather}
These are first order differential equations of the fields $X^I(x)$
and $A_{mI}(x)$ and the Jacobi identity (\ref{gJacobi}) of the Poisson
tensor ensures the closure of (\ref{gPSM-eomX}) and (\ref{gPSM-eomA}).
As a consequence of (\ref{gJacobi}) the action exhibits the symmetries
\begin{equation}
  \label{gPSM-symms}
  \delta X^I = \Poisson^{IJ} \epsilon_J, \qquad
  \delta A_I = -d\epsilon_I - (\rpartial_I \Poisson^{JK}) \epsilon_K A_J,
\end{equation}
where corresponding to each gauge field $A_I$ we have a symmetry
parameter $\epsilon_I(x)$ with the same grading which is a function of
$x$ only. In general, when calculating the commutator of these
symmetries, parameters depending on both $x$ and $X$ are obtained.
For two parameters $\epsilon_{1I}(x,X)$ and $\epsilon_{2I}(x,X)$
\begin{align}
  \left( \delta_1 \delta_2 - \delta_2 \delta_1 \right) X^I &= \delta_3
  X^I, \label{gCommX} \\
  \left( \delta_1 \delta_2 - \delta_2 \delta_1 \right) A_I &= \delta_3
  A_I + \left( dX^J + \Poisson^{JK} A_K \right) (\rpartial_J
  \rpartial_I \Poisson^{RS}) \epsilon_{1S} \epsilon_{2R}
  \label{gCommA}
\end{align}
follows, where $\epsilon_{3I}(x,X)$ of the resulting variation
$\delta_3$ are given by the Poisson bracket (or Koszul-Lie bracket) of
the 1-forms $\epsilon_1 = dX^I \epsilon_{1I}$ and $\epsilon_2 = dX^I
\epsilon_{2I}$, defined according to
\begin{equation}
  \label{gCommEp3}
  \epsilon_{3I} = \{ \epsilon_2, \epsilon_1 \}_I
  := (\rpartial_I \Poisson^{JK}) \epsilon_{1K} \epsilon_{2J} +
  \Poisson^{JK} \left( \epsilon_{1K} \rpartial_J \epsilon_{2I} -
    \epsilon_{2K} \rpartial_J \epsilon_{1I} \right).
\end{equation}
Note, that the commutator of the PSM symmetries closes if the Poisson
tensor is linear, for non-linear Poisson tensors the algebra closes
only on-shell (\ref{gCommA}).

Right and left Hamiltonian vector fields are defined by $\rvec{T}^I =
\{X^I, \cdot\}$ and $\lvec{T}^I = \{\cdot, X^I\}$, respectively, \ie
by
\begin{equation}
  \label{gHVF}
  \rvec{T}^I \cdot f = \{X^I, f\} = \Poisson^{IJ} (\rpartial_J f),
  \qquad
  f \cdot \lvec{T}^I = \{f, X^I\} = (f \lpartial_J) \Poisson^{JI}.
\end{equation}
The vector fields $\lvec{T}^I$ are the generators of the symmetries,
$\delta X^I = X^I \cdot \lvec{T}^J \epsilon_J$. From their commutator
the algebra
\begin{equation}
  \label{gStrucFunc}
  [\lvec{T}^I, \lvec{T}^J] = \lvec{T}^K f_K\^{IJ}(X)
\end{equation}
follows with the structure functions $f_K\^{IJ} = (\rpartial_K
\Poisson^{IJ})$.  Structure constants and therefore Lie algebras are
obtained when the Poisson tensor depends only linearly on the
coordinates, which is true for Yang-Mills gauge theory and simple
gravity models like (anti-)de~Sitter gravity.

As in the purely bosonic case the kernel of the graded Poisson
algebra determines the so-called Casimir functions $\Casimir$ obeying
$\{ \Casimir, X^I\} = 0$. When the co-rank of the bosonic
theory---with one Casimir function---is not changed we shall call this
case non-degenerate. Then $\Poisson^{\alpha\beta}|$, the bosonic part
of the fermionic extension, must be of full rank. For $\N=1$
supergravity and thus one target space Majorana spinor $\chi^\alpha$,
the expansion of $\Casimir$ in $\chi^\alpha$ reads ($\chi^2 =
\chi^\alpha \chi_\alpha$, \cf\App\ref{app:spinors})
\begin{equation}
  \label{Casimir}
  \Casimir = \casimir + \half \chi^2 \casimir_2,
\end{equation}
where $\casimir$ and $\casimir_2$ are functions of $\phi$ and $Y
\equiv \half X^a X_a$ only.  This assures that the Poisson bracket
$\{\phi, \Casimir\}$ is zero.  From the bracket $\{X^a, \Casimir\} =
0$, to zeroth order in $\chi^\alpha$, the defining equation of the
Casimir function for pure bosonic gravity PSMs becomes
\begin{equation}
  \label{c}
  \derv \casimir :=
  \left( \partial_\phi - v \partial_Y \right) \casimir = 0.
\end{equation}
This is the well-known partial differential equation of that quantity
\cite{Klosch:1996fi,Strobl:1999Habil}. The solution of (\ref{c}) for
bosonic potentials relevant for kinetic dilaton theories (\ref{vdil})
can be given by ordinary integration,
\begin{alignat}{2}
  \casimir(\phi,Y) &= Y e^{Q(\phi)} + W(\phi), \label{c-sol} \\
  Q(\phi) &= \int_{\phi_1}^{\phi} Z(\varphi) d\varphi, &\qquad
  W(\phi) &= \int_{\phi_0}^{\phi} e^{Q(\varphi)} V(\varphi)
  d\varphi. \label{QW}
\end{alignat}
The new component $\casimir_2$ is derived by considering the terms
proportional to $\chi^\beta$ in the bracket $\{\chi^\alpha, \Casimir\}
= 0$. Thus $\casimir_2$ will depend on the specific fermionic
extension. In the degenerate case, when $\Poisson^{\alpha\beta}$ is
not of full rank, there will be more than one Casimir function,
including purely Grassmann valued ones (see \Sec\ref{sec:DFS-P} and
\ref{sec:CPA-P}).

\section{Solution of the Jacobi Identities}
\label{sec:Jacobi}

As mentioned above, in order to obtain the general solution of the
graded Jacobi identities a suitable starting point is the use of
Lorentz symmetry in a most general ansatz for $\Poisson^{IJ}$.
Alternatively, one could use a simple $\Poisson^{IJ}_{(0)}$ which
trivially fulfills (\ref{gJacobi}). Then the most general
$\Poisson^{IJ}$ may be obtained by a general diffeomorphism in target
space. The first route will be followed within this section. We will
comment upon the second one in \Sec\ref{sec:diffeo}.

\subsection{Lorentz-Covariant Ansatz for the Poisson-Tensor}
\label{sec:ans-gP}

Lorentz symmetry determines the mixed components $\Poisson^{A\phi}$ of
$\Poisson^{IJ}$,
\begin{equation}
  \label{P-Lorentz}
  \Poisson^{a\phi} = X^b \epsilon_b\^a, \qquad
  \Poisson^{\alpha\phi} = - \half \chi^\beta \gamma^3\_\beta\^\alpha.
\end{equation}
All other components of the Poisson tensor must be Lorentz-covariant
(\cf the discussion around (\ref{Pinv})).  Expanding them in
terms of invariant tensors $\eta^{ab}$, $\epsilon^{ab}$,
$\epsilon^{\alpha\beta}$ and $\gamma$-matrices yields
\begin{align}
  \Poisson^{ab} &= V \epsilon^{ab}, \label{ans-P-bb} \\
  \Poisson^{\alpha b} &= \chi^\beta F^b\_\beta\^\alpha,
  \label{ans-P-fb} \\
  \Poisson^{\alpha\beta} &= U \gamma^3\^{\alpha\beta} + i \widetilde U X^c
  \gamma_c\^{\alpha\beta} + i \widehat U X^c \epsilon_c\^d
  \gamma_d\^{\alpha\beta}. \label{ans-P-ff}
\end{align}
The quantities $V$, $U$, $\widetilde U$ and $\widehat U$ are functions
of $\phi$, $Y$ and $\chi^2$.  Due to the anticommutativity of
$\chi^\alpha$ the dependence on $\chi^2$ is at most linear. Therefore
\begin{equation}
  \label{ans-V}
  V = v(\phi, Y) + \half \chi^2\, v_2(\phi, Y)
\end{equation}
depends on two Lorentz-invariant functions $v$ and $v_2$ of $\phi$ and
$Y$. An analogous notation will be implied for $U, \widetilde U$ and
$\widehat{U}$, using the respective lower case letter for the
$\chi$-independent component of the superfield and an additional index
2 for the respective $\chi^2$-component.  The component
(\ref{ans-P-fb}) contains the spinor matrix $F^a\_\beta\^\gamma$,
which may be first expanded in terms of the linearly independent
$\gamma$-matrices,
\begin{equation}
  \label{ans-f-bff}
  F^a\_\beta\^\gamma = f_{(1)}^a \delta_\beta\^\gamma + i f^{ab}
  \gamma_{b\beta}\^\gamma + f_{(3)}^a \gamma^3\_\beta\^\gamma.
\end{equation}
The Lorentz-covariant coefficient functions in (\ref{ans-f-bff}) are
further decomposed according to
\begin{align}
  f_{(1)}^a &= f_{(11)} X^a - f_{(12)} X^b \epsilon_b\^a,
  \label{ans-f1-b} \\
  f_{(3)}^a &= f_{(31)} X^a - f_{(32)} X^b \epsilon_b\^a,
  \label{ans-f3-b} \\
  f^{ab} &= f_{(s)} \eta^{ab} + f_{(t)} X^a X^b - f_{(h)} X^c
  \epsilon_c\^a X^b + f_{(a)} \epsilon^{ab}. \label{ans-f-bb}
\end{align}
The eight Lorentz-invariant coefficients $f_{(11)}$, $f_{(12)}$,
$f_{(31)}$, $f_{(32)}$, $f_{(s)}$, $f_{(t)}$, $f_{(h)}$ and $f_{(a)}$
are functions of $\phi$ and $Y$ only. The linearity in $\chi^\alpha$
of (\ref{ans-P-fb}) precludes any $\chi^2$ term in (\ref{ans-f-bff}).

Below it will turn out to be convenient to use a combined notation for
the bosonic and the $\chi^2$-dependent part of
$\Poisson^{\alpha\beta}$,
\begin{equation}
  \label{v-ff}
  \Poisson^{\alpha\beta} = v^{\alpha\beta} + \half \chi^2
  v_2^{\alpha\beta},
\end{equation}
where $v^{\alpha\beta}$ and $v_2^{\alpha\beta}$ are particular
matrix-valued functions of $\phi$ and $X^a$, namely, in the notation
above (\cf also \App\ref{app:spin-comp} for the definition of $X^{++}$
and $X^{--}$),
\begin{equation}
  \label{matrix}
  v^{\alpha\beta} = \mtrx{\sqrt{2}X^{++}(\tilde u - \hat u)}{-u}
  {-u}{\sqrt{2}X^{--}(\tilde u + \hat u)},
\end{equation}
and likewise with suffix 2. Note that the symmetric $2 \times 2$
matrix $v^{\alpha\beta}$ still depends on three arbitrary real
functions; as a consequence of Lorentz invariance, however, they are
functions of $\phi$ and $Y$ only. A similar explicit matrix
representation may be given also for $F^{\pm \pm}\_\alpha\^\beta$.

\subsection{Remaining Jacobi Identities}
\label{sec:remJac}

The Jacobi identities $J^{\phi BC}=0$ have been taken care of
automatically by the Lorentz covariant parametrization introduced in
\Sec\ref{sec:ans-gP}. In terms of these functions we write the
remaining identities as
\begin{align}
  J^{\alpha\beta\gamma} &= \rvec T^\alpha(\Poisson^{\beta\gamma}) + \cycl(\alpha\beta\gamma)
  = 0, \label{J-fff} \\
  J^{\alpha\beta c} &= \rvec T^c(\Poisson^{\alpha\beta}) + \rvec
  T^\alpha(\chi F^c)^\beta + \rvec T^\beta(\chi F^c)^\alpha = 0,
  \label{J-ffb} \\
  \half J^{\alpha bc} \epsilon_{cb} &= \rvec T^\alpha(V) - \rvec
  T^b(\chi F^c)^\alpha \epsilon_{cb} = 0. \label{J-fbb}
\end{align}
Here $\rvec T^a$ and $\rvec T^\alpha$ are Hamiltonian vector fields
introduced in (\ref{gHVF}), yielding ($\partial_\phi =
\frac{\partial}{\partial\phi}$, $\partial_a = \frac{\partial}{\partial
  X^a}$, $\partial_\alpha = \frac{\partial}{\partial \chi^\alpha}$)
\begin{align}
  \rvec T^a &= X^b \epsilon_b\^a \partial_\phi + \left( v +
    \half\chi^2 v_2 \right) \epsilon^{ab}
  \partial_b - (\chi F^a)^\beta \partial_\beta, \label{T-b} \\
  \rvec T^\alpha &= -\half (\chi\gamma^3)^\alpha \partial_\phi + (\chi
  F^b)^\alpha \partial_b + \left( v^{\alpha\beta} + \half\chi^2
    v_2\^{\alpha\beta} \right) \partial_\beta.
  \label{T-f}
\end{align}
To find the solution of (\ref{J-fff})--(\ref{J-fbb}) it is necessary
to expand in terms of the anticommutative coordinate $\chi^\alpha$.
Therefore, it is convenient to split off any dependence on
$\chi^\alpha$ and its derivative also in (\ref{T-b}) and (\ref{T-f}),
using instead the special Lorentz vector and spinor matrix valued
derivatives\footnote{When (\ref{derv-b}) acts on an invariant function
  of $\phi$ and $Y$, $\derv^c$ essentially reduces to the `scalar'
  derivative, introduced in (\ref{c}).}
\begin{align}
  \derv^c &:= X^d \epsilon_d\^c \partial_\phi + v \epsilon^{cd}
  \partial_d, \label{derv-b} \\
  \derv_\delta\^\alpha &:= -\half \gamma^3\_\delta\^\alpha
  \partial_\phi + F^d\_\delta\^\alpha \partial_d. \label{derv-ff}
\end{align}
Then the Jacobi identities, arranged in the order $J^{\alpha\beta
  c}|$, $J^{\alpha\beta\gamma}|_{\chi}$, $J^{\alpha bc}|_{\chi}$ and
$J^{\alpha\beta c}|_{\chi^2}$, that is the order of increasing
complexity best adapted for our further analysis, read
\begin{gather}
  v^{\alpha)\gamma} F^c\_\gamma\^{(\beta} + \half \derv^c v^{\alpha\beta}
  = 0, \label{J-ffb-0} \\
  v_\delta\^\alpha v_2^{\beta\gamma} - \derv_\delta\^\alpha
  v^{\beta\gamma} + \cycl(\alpha\beta\gamma) = 0, \label{J-fff-1}
  \\
  v_\delta\^\alpha v_2 - \derv_\delta\^\alpha v + \derv^c
  F^b\_\delta\^\alpha \epsilon_{bc} - (F^c F^b)_\delta\^\alpha
  \epsilon_{bc} = 0 , \label{J-fbb-1} \\
  \derv^c v_2^{\alpha\beta} - F^c\_\delta\^\delta v_2^{\alpha\beta} +
  v_2 \epsilon^{cd} \partial_d v^{\alpha\beta} + 2
  \derv^{\delta(\alpha|} F^c\_\delta\^{|\beta)} + 2
  v_2^{\alpha)\delta} F^c\_\delta\^{(\beta} = 0. \label{J-ffb-2}
\end{gather}

All known solutions for $d=2$ supergravity models found in the
literature have the remarkable property that the Poisson tensor has
(almost everywhere, \ie except for isolated points) constant rank
four, implying exactly one conserved Casimir function $\Casimir$
\cite{Strobl:1999zz}. Since the purely bosonic Poisson tensor has
(almost everywhere) a maximum rank of two, this implies that the
respective fermionic bracket $\Poisson^{\alpha\beta}$ (or,
equivalently, its $\chi$-independent part $v^{\alpha\beta}$) must be
of full rank if only one Casimir function is present in the fermionic
extension. In the following subsection we will consider this case, \ie
we will restrict our attention to (regions in the target space with)
invertible $\Poisson^{\alpha\beta}$. For describing the rank we
introduce the notation $(B|F)$. Here $B$ denotes the rank of the
bosonic body of the algebra, $F$ the one of the extension. In this
language the nondegenerate case has rank $(2|2)$. The remaining
degenerate cases with rank $(2|0)$ and $(2|1)$ will be analyzed in a
second step (\Sec\ref{sec:DFS-P} and \Sec\ref{sec:CPA-P}).

\subsubsection{Nondegenerate Fermionic Sector}
\label{sec:sol-gP}

When the matrix $v^{\alpha\beta}$ in (\ref{v-ff}) is nondegenerate,
\ie when its determinant
\begin{equation}
  \label{detv}
  \detv := \det(v^{\alpha\beta}) = \half v^{\alpha\beta}
  v_{\beta\alpha}
\end{equation}
is nonzero, for a given bosonic bracket this yields all supersymmetric
extensions of maximal total rank. We note in parenthesis that due to
the two-dimensionality of the spinor space (and the symmetry of
$v^{\alpha\beta}$) the inverse matrix to $v^{\alpha\beta}$ is nothing
else but $v_{\alpha\beta}/\detv$, which is used in several intermediary steps
below.

The starting point of our analysis of the remaining Jacobi identities
$J^{ABC}=0$ will always be a certain ansatz, usually for
$v^{\alpha\beta}$. Therefore, it will be essential to proceed in a
convenient sequence so as to obtain the restrictions on the remaining
coefficient functions in the Poisson tensor with the least effort.
This is also important because it turns out that several of these
equations are redundant. This sequence has been anticipated in
(\ref{J-ffb-0})--(\ref{J-ffb-2}). There are already redundancies
contained in the second and third step (eqs.\ (\ref{J-fff-1}) and
(\ref{J-fbb-1})), while the $\chi^2$-part of $J^{\alpha\beta c} = 0$
(eq.\ (\ref{J-ffb-2})) turns out to be satisfied identically because
of the other equations.  It should be noted, though, that this
peculiar property of the Jacobi identities is \emph{not} a general
feature, resulting \eg from some hidden symmetry, it holds true only
in the case of a nondegenerate $\Poisson^{\alpha\beta}$ (\cf the discussion
of the degenerate cases below).

For fixed (nondegenerate) $v^{\alpha\beta}$, all solutions of
(\ref{J-ffb-0}) are parametrized by a Lorentz vector field $f^a$ on
the coordinate space $(\phi, X^a)$:
\begin{equation}
  \label{sol-f-bff}
  F^c\_\alpha\^\beta =  \left[ f^c \epsilon^{\gamma\beta} - \derv^c
    v^{\gamma\beta} \right] \frac{v_{\gamma\alpha}}{2 \detv}
\end{equation}
Eq.\ (\ref{J-fff-1}) can be solved to determine $v_2^{\alpha\beta}$
in terms of $v^{\alpha\beta}$:
\begin{align}
  v_2\^{\alpha\beta} &= -\frac{1}{4\detv} v_\gamma\^\delta \left[
    \derv_\delta\^\gamma v^{\alpha\beta} + \cycl(\alpha\beta\gamma)
  \right] \label{sol-v2-ff}
\end{align}
Multiplying (\ref{J-fbb-1}) by $v_\beta\^\gamma$ yields
\begin{equation}
  \detv \delta_\beta\^\alpha v_2 =  v_\beta\^\delta \left[ -
    \derv_\delta\^\alpha v + \derv^c F^b\_\delta\^\alpha \epsilon_{bc}
    - (F^c F^b)_\delta\^\alpha \epsilon_{bc} \right]. \label{sol-v2}
\end{equation}
The trace of (\ref{sol-v2}) determines $v_2$, which is thus seen to
depend also on the original bosonic potential $v$ of (\ref{PB}).

Neither the vanishing
traces of (\ref{sol-v2}) multiplied with $\gamma^3$ or with
$\gamma_a$, nor the identity (\ref{J-ffb-2}) provide new restrictions
in the present case. This has been checked by extensive computer
calculations \cite{Ertl:Index-0.14.2}, based upon the explicit
parametrization (\ref{P-Lorentz})--(\ref{ans-f-bb}), which were
necessary because of the extreme algebraic complexity of this problem.
It is a remarkable feature of (\ref{sol-f-bff}), (\ref{sol-v2-ff}) and
(\ref{sol-v2}) that the solution of the Jacobi identities for the
nondegenerate case can be obtained from algebraic equations only.

As explained at the end of \Sec\ref{sec:gPSM2} the fermionic extension
of the bosonic Casimir function $\casimir$ can be derived from $\{
\chi^\alpha, \Casimir \} = 0$. The general result for the
nondegenerate case we note here for later reference
\begin{equation}
  \label{c2}
  \casimir_2 = -\frac{1}{2\detv} v_\alpha\^\beta \left( -\half
    \gamma^3\_\beta\^\alpha \partial_\phi + F^d\_\beta\^\alpha
    \partial_d \right) \casimir.
\end{equation}

The algebra of full rank $(2|2)$ with the above solution for
$F^c\_\alpha\^\beta$, $v_2\^{\alpha\beta}$ and $v_2$ depends on 6
independent functions $v$, $v^{\alpha\beta}$ and $f^a$ and their
derivatives.  The original bosonic model determines the `potential'
$v$ in (\ref{FOG}) or (\ref{PB}). Thus the arbitrariness of
$v^{\alpha\beta}$ and $f^a$ indicates that the supersymmetric
extensions, obtained by fermionic extension from the PSM, are far from
unique. This has been mentioned already in the previous section and we
will further illuminate it in the following one.

\subsubsection[Degenerate Fermionic Sector, Rank $(2|0)$]{Degenerate
  Fermionic Sector, Rank \mathversion{bold}$(2|0)$}
\label{sec:DFS-P}

For vanishing rank of $\Poisson^{\alpha\beta}|$, \ie $v^{\alpha\beta}
= 0$, the identities (\ref{J-ffb-0}) and (\ref{J-fff-1}) hold
trivially whereas the other Jacobi identities become complicated
differential equations relating $F^a$, $v$ and $v_2^{\alpha\beta}$.
However, these equations can again be reduced to algebraic ones for
these functions when the information on additional Casimir functions
is employed, which appear in this case. These have to be of fermionic
type with the general ans\"a{}tze
\begin{align}
  \Casimir^{(+)} &= \chi^+ \left| \frac{X^{--}}{X^{++}}
  \right|^{\frac{1}{4}} \casimir_{(+)}, \label{DFS-Cf+} \\
  \Casimir^{(-)} &= \chi^- \left| \frac{X^{--}}{X^{++}}
  \right|^{-\frac{1}{4}} \casimir_{(-)}. \label{DFS-Cf-}
\end{align}
The quotients $X^{--}/X^{++}$ assure that $\casimir_{(\pm)}$ are
Lorentz invariant functions of $\phi$ and $Y$.  This is possible
because the Lorentz boosts in two dimensions do not mix chiral
components and the light cone coordinates $X^{\pm\pm}$.

Taking a Lorentz covariant ansatz for the Poisson tensor as specified
in \Sec\ref{sec:ans-gP}, $\Casimir^{(+)}$ and $\Casimir^{(-)}$ must
obey $\{ X^a, \Casimir^{(+)} \} = \{ X^a, \Casimir^{(-)} \} = 0$. Both
expressions are linear in $\chi^\alpha$, therefore, the coefficients
of $\chi^\alpha$ have to vanish separately. This leads to $F^a\_-\^+
= 0$ and $F^a\_+\^- = 0$ immediately. With the chosen representation
of the $\gamma$-matrices (\cf \App\ref{app:spinors}) it is seen that
(\ref{ans-f-bff}) is restricted to $f^{ab}=0$, \ie the potentials
$f_{(s)}$, $f_{(t)}$, $f_{(h)}$ and $f_{(a)}$ have to vanish. A
further reduction of the system of equations reveals the further
conditions $f_{(11)} = 0$\footnote{In fact $f_{(11)}$ vanishes in all
  cases, \ie also for rank $(2|2)$ and $(2|1)$.} and
\begin{equation}
  \label{DFS-v}
  v = 4Y f_{(31)}.
\end{equation}
This leaves the differential equations for $\casimir_{(+)}$ and
$\casimir_{(-)}$
\begin{align}
  \left( \derv + f_{(12)} + f_{(32)} \right) \casimir_{(+)} &= 0,
  \label{DFS-c+} \\
  \left( \derv + f_{(12)} - f_{(32)} \right) \casimir_{(-)} &=
  0. \label{DFS-c-}
\end{align}
The brackets $\{ \chi^+, \Casimir^{(+)} \}$ and $\{ \chi^-,
\Casimir^{(-)} \}$ are proportional to $\chi^2$; the resulting
equations require $\tilde{u}_2 = \hat{u}_2 = 0$. The only surviving
term $u_2$ of $\Poisson^{\alpha\beta}$ is related to $F^a$ via $u_2 =
-f_{(12)}$ as can be derived from $\{ \chi^-, \Casimir^{(+)} \} = 0$
as well as from $\{ \chi^+, \Casimir^{(-)} \} = 0$, which are
equations of order $\chi^2$ too.

Thus the existence of the fermionic Casimir functions (\ref{DFS-Cf+})
and (\ref{DFS-Cf-}) has lead us to a set of \emph{algebraic} equations
among the potentials of the Lorentz covariant ansatz for the Poisson
tensor, and the number of independent potentials has been reduced
drastically.  The final question, whether the Jacobi identities are
already fulfilled with the relations found so far finds a positive
answer, and the general Poisson tensor with degenerate fermionic
sector, depending on four parameter functions $v(\phi,Y)$,
$v_2(\phi,Y)$, $f_{(12)}(\phi,Y)$ and $f_{(32)}(\phi,Y)$ reads
\begin{align}
  \Poisson^{ab} &= \left( v + \half\chi^2 v_2 \right)
  \epsilon^{ab}, \label{DFS-P-bb} \\
  \Poisson^{\alpha b} &= \frac{v}{4Y} X^b (\chi\gamma^3)^\alpha -
  f_{(32)} X^c \epsilon_c\^b (\chi\gamma^3)^\alpha - f_{(12)} X^c
  \epsilon_c\^b \chi^\alpha, \label{DFS-P-fb} \\
  \Poisson^{\alpha\beta} &= -\half\chi^2 f_{(12)} \gamma^3\^{\alpha\beta}.
  \label{DFS-P-ff}
\end{align}
This Poisson tensor possesses three Casimir functions: two fermionic
ones defined in regions $Y \neq 0$ according to (\ref{DFS-Cf+}) and
(\ref{DFS-Cf-}), where $\casimir_{(+)}$ and $\casimir_{(-)}$ have to
fulfill the first order differential equations (\ref{DFS-c+}) and
(\ref{DFS-c-}), respectively, and one bosonic Casimir function
$\Casimir$ of the form (\ref{Casimir}), where $\casimir$ is a solution
of the bosonic differential equation (\ref{c})---note the definition
of $\derv$ therein---and where $\casimir_2$ has to obey
\begin{equation}
  \left( \derv + 2 f_{(12)} \right) \casimir_2 = v_2 \partial_Y
  \casimir.
\end{equation}

Let us finally emphasize that it was decisive within this subsection
to use the information on the \emph{existence} of Casimir functions.
This follows from the property of the bivector $\Poisson^{IJ}$ to be
surface-forming, which in turn is a consequence of the (graded) Jacobi
identity satisfied by the bivector. However, the inverse does not hold
in general: Not any surface-forming bivector satisfies the Jacobi
identities.  Therefore, it was necessary to check their validity in a
final step.

\subsubsection[Degenerate Fermionic Sector, Rank $(2|1)$]{Degenerate
  Fermionic Sector, Rank \mathversion{bold}$(2|1)$}
\label{sec:CPA-P}

When the fermionic sector has maximal rank one, again the existence of
a fermionic Casimir function is very convenient. We start with
`positive chirality'\footnote{`Positive chirality' refers to the
  structure of (\ref{CPA-v-ff}). It does not preclude the coupling to
  the negative chirality component $\chi^-$ in other terms. A genuine
  chiral algebra (similar to $\N=(1,0)$ supergravity) is a special
  case to be discussed below in \Sec\ref{sec:CPA}}. We choose
  the ansatz (\cf \App\ref{app:spinors})
\begin{equation}
  \label{CPA-v-ff}
  \Poisson^{\alpha\beta}| = v^{\alpha\beta} = i\tilde{u} X^c (\gamma_c
  P_{+})^{\alpha\beta} = \mtrx{\sqrt{2}\tilde{u} X^{++}}{0}{0}{0}.
\end{equation}
The most general case of rank $(2|1)$ can be reduced to
  (\ref{CPA-v-ff}) by a (target space) transformation of the spinors.
Negative chirality where $P_{+}$ is replaced with $P_{-}$ is
considered below. Testing the ans\"a{}tze (\ref{DFS-Cf+}) and
(\ref{DFS-Cf-}) reveals that $\Casimir^{(-)}$ now again is a Casimir
function, but $\Casimir^{(+)}$ is not.  Indeed $\{ \chi^+,
\Casimir^{(+)} \}| \propto \tilde{u} \casimir_{(+)} \neq 0$ in
general, whereas $\{ \chi^+, \Casimir^{(-)} \}| \equiv 0$ shows that
the fermionic Casimir function for positive chirality is
$\Casimir^{(-)}$, where $\casimir_{(-)} = \casimir_{(-)}(\phi,Y)$ has
to fulfill a certain differential equation, to be determined below.

The existence of $\Casimir^{(-)}$ can be used to obtain information
about the unknown components of $\Poisson^{AB}$. Indeed an
investigation of $\{ X^A, \Casimir^{(-)} \} = 0$ turns out to be much
simpler than trying to get that information directly from the Jacobi
identities. The bracket $\{ X^a, \Casimir^{(-)} \} = 0$ results in
$F^a\_+\^- = 0$ and from $\{ \chi^\alpha, \Casimir^{(-)} \} = 0$ the
relation $v_2\^{--} = 0$ can be derived. This is the reason why the
ansatz (\ref{ans-f-bff}) and (\ref{ans-f-bb}), retaining
(\ref{ans-f1-b}) and (\ref{ans-f3-b}), attains the simpler form
\begin{align}
  F^a &= f_{(1)}^a \1 + i f^{ab} (\gamma_b P_{+}) + f_{(3)}^a \gamma^3,
  \label{CPA-F-bff} \\
  f^{ab} &= f_{(s)} \eta^{ab} + f_{(t)} X^a X^b. \label{CPA-f-bb}
\end{align}
Likewise for the $\chi^2$-component of $\Poisson^{\alpha\beta}$ we set
\begin{equation}
  \label{CPA-v2-ff}
  v_2\^{\alpha\beta} = i \tilde{u}_2 X^c (\gamma_c P_{+})^{\alpha\beta}
  + u_2 \gamma^3\^{\alpha\beta}.
\end{equation}

Not all information provided by the existence of $\Casimir^{(-)}$ has
been introduced at this point. Indeed using the chiral ansatz
(\ref{CPA-v-ff}) together with (\ref{CPA-F-bff})--(\ref{CPA-v2-ff})
the calculation of $\{ X^a, \Casimir^{(-)} \} = 0$ in conjunction with
the Jacobi identities $J^{\alpha\beta c}| = 0$ (\cf
(\ref{J-ffb-0})) requires $f_{(11)} = 0$ and
\begin{equation}
  \label{CPA-v}
  v = 4Y f_{(31)}.
\end{equation}
It should be noted that the results $f_{(11)} = 0$ and (\ref{CPA-v})
follow from $\{ \chi^\alpha, \Casimir \} = 0$ too, where $\Casimir$ is
a bosonic Casimir function. The remaining equation in $\{ X^a,
\Casimir^{(-)} \} = 0$ together with $\{ \chi^\alpha,\Casimir^{(-)} \}
= 0$ yields $u_2 = -f_{(12)}$. With the solution obtained so far any
calculation of $\{ X^A, \Casimir^{(-)} \} = 0$ leads to one and only
one differential equation (\ref{DFS-c-}) which must be satisfied in
order that (\ref{DFS-Cf-}) is a Casimir function.

We now turn our attention to the Jacobi identities.  The inspection of
$J^{++c}|=0$ (\ref{J-ffb-0}), $J^{+++}|_{\chi}=0$ (\ref{J-fff-1}) and
$J^{+bc}|_{\chi}=0$ (\ref{J-fbb-1}) leads to the conditions
\begin{align}
  f_{(32)} &= \half (\derv\ln|\tilde{u}|) - f_{(12)} - \frac{v}{4Y},
  \label{CPA-f32+} \\
  \tilde{u}_2 &= f (\partial_Y \ln|\tilde{u}|) + f_{(t)},
  \label{CPA-uT2} \\
  v_2 &= \left( \derv + 2 f_{(12)} + (\partial_Y v) \right)
  \frac{f}{\tilde{u}}, \label{CPA-v2}
\end{align}
respectively. In order to simplify the notation we introduced 
\begin{equation}
  \label{CPA-f}
  f = f_{(s)} + 2 Y f_{(t)}.
\end{equation}
All other components of the Jacobi tensor are found to vanish
identically.

The construction of graded Poisson tensors with `negative
chirality', \ie with fermionic sector of the form
\begin{equation}
  \Poisson^{\alpha\beta}| = v^{\alpha\beta} = i\tilde{u} X^c (\gamma_c
  P_{-})^{\alpha\beta} = \mtrx{0}{0}{0}{\sqrt{2}\tilde{u} X^{--}},
\end{equation}
proceeds by the same steps as for positive chirality. Of course, the
relevant fermionic Casimir function is now $\Casimir^{(+)}$ of
(\ref{DFS-Cf+}) and $P_{+}$ in (\ref{CPA-F-bff}) and (\ref{CPA-v2-ff})
has to be replaced by $P_{-}$.  The results $f_{(11)}=0$,
(\ref{CPA-v}), (\ref{CPA-uT2}) and (\ref{CPA-v2}) remain the same,
only $f_{(32)}$ acquires an overall minus sign,
\begin{equation}
  \label{CPA-f32-}
  f_{(32)} = -\half (\derv\ln|\tilde{u}|) + f_{(12)} + \frac{v}{4Y},
\end{equation}
to be inserted in the differential equation (\ref{DFS-c+}) for
  $\casimir_{(+)}$.

The results for graded Poisson tensors of both chiralities can be
summarized as (\cf (\ref{CPA-f}))
\begin{align}
  \Poisson^{ab} &= \left( v + \half\chi^2 \left[ \derv + 2 f_{(12)} +
      (\partial_Y v) \right] \frac{f}{\tilde{u}} \right) \epsilon^{ab},
  \label{CPA-P-bb} \\
  \Poisson^{\alpha b} &= (\chi F^b)^\alpha \label{CPA-P-fb} \\
  \Poisson^{\alpha\beta} &= i \left( \tilde{u} + \half\chi^2 \left[ f
      (\partial_Y \ln|\tilde{u}|) + f_{(t)} \right] \right) X^c
  (\gamma_c P_{\pm})^{\alpha\beta} - \half\chi^2 f_{(12)}
  \gamma^3\^{\alpha\beta}.
  \label{CPA-P-ff}
\end{align}
Eq.\ (\ref{CPA-P-fb}) reads explicitly
\begin{multline}
  \label{CPA-F-b}
  F^b = \frac{v}{4Y} (X^b \pm X^c \epsilon_c\^b) \gamma^3 - 2 f_{(12)}
  X^c \epsilon_c\^b P_{\mp} \\
  + i f_{(s)} (\gamma^b P_{\pm}) + i f_{(t)} X^b X^c (\gamma_c P_{\pm})
  \mp \half (\derv\ln|\tilde{u}|) X^c \epsilon_c\^b \gamma^3.
\end{multline}
Eqs.\ (\ref{CPA-P-bb})--(\ref{CPA-F-b}) represent the generic solution
of the graded $\N=1$ Poisson algebra of rank $(2|1)$. In addition to
$v(\gamma,Y)$ it depends on four parameter functions $\tilde{u}$,
$f_{(12)}$, $f_{(s)}$ and $f_{(t)}$, all depending on $\phi$ and $Y$.

Each chiral type possesses a bosonic Casimir function $\Casimir =
\casimir + \half\chi^2 \casimir_2$, where $\casimir(\phi,Y)$ and
$\casimir_2(\phi,Y)$ are determined by $\derv \casimir = 0$ and
\begin{equation}
  \label{CPA-c2}
  \casimir_2 = \frac{f \partial_Y \casimir}{\tilde{u}}.
\end{equation}
The fermionic Casimir function for positive chirality is
$\Casimir^{(-)}$ and for negative chirality $\Casimir^{(+)}$ (\cf
(\ref{DFS-Cf-}) and (\ref{DFS-Cf+})), where $\casimir_{(\mp)}(\phi,Y)$
are bosonic scalar functions solving the same differential equation in
both cases when eliminating $f_{(32)}$,
\begin{equation}
  \label{CPA-c1}
  \left( \derv + 2 f_{(12)} + \frac{v}{4Y} - \half(\derv\ln|\tilde{u}|)
  \right) \casimir_{(\mp)} = 0,
\end{equation}
derived from (\ref{DFS-c-}) with (\ref{CPA-f32+}) and from
(\ref{DFS-c+}) with (\ref{CPA-f32-}).

\section{Target space diffeomorphisms}
\label{sec:diffeo}

When subjecting the Poisson tensor of the action (\ref{gPSM}) to a
diffeomorphism
\begin{equation}
  \label{diffeo}
  X^I \to \bar{X}^I = \bar{X}^I(X)
\end{equation}
on the target space $\TSp$, another action of gPSM form is generated
with the new Poisson tensor
\begin{equation}
  \label{tr-P}
  \bar{\Poisson}^{IJ} = (\bar{X}^I \lpartial_K ) \Poisson^{KL}
  (\rpartial_L \bar{X}^J).
\end{equation}
It must be emphasized that in this manner a \emph{different} model is
created with---in the case of 2d gravity theories and their fermionic
extensions---in general different bosonic `body' (and global
topology). Therefore, such transformations are a powerful tool to
create new models from available ones. This is important, because---as
shown in \Sec\ref{sec:Jacobi} above---the solution of the Jacobi
identities as a rule represents a formidable computational problem.
This problem could be circumvented by starting from a simple
$\bar{\Poisson}^{IJ}(\bar{X})$, whose Jacobi identities have been
solved rather trivially. As a next step a transformation
(\ref{diffeo}) is applied. The most general Poisson tensor can be
generated by calculating the inverse of the Jacobi matrices
\begin{alignat}{2}
  J_I\^{\bar J}(X) &= \rpartial_I \bar{X}^J, &\qquad
  J_I\^{\bar K} (J^{-1})_{\bar K}\^J &= \delta_I\^J, \\
  I^{\bar I}\_J(X) &= \bar{X}^I \lpartial_J, &\qquad
  (I^{-1})^I\_{\bar K} I^{\bar K}\_J &= \delta_I\^J.
\end{alignat}
According to
\begin{equation}
  \label{generalP}
  P^{IJ}(X) = (I^{-1})^I\_{\bar K} \bar{P}^{KL}|_{\bar{X}(X)}
  (J^{-1})_{\bar L}\^J
\end{equation}
the components $\Poisson^{IJ}$ of the transformed Poisson tensor are
expressed in terms of the coordinates $X^I$ without the need to invert
(\ref{diffeo}).

The drawback of this argument comes from the fact that in our problem
the (bosonic) part of the `final' algebra is given, and the inverted
version of the procedure described here turns out to be very difficult
to implement.

Nevertheless, we construct explicitly the diffeomorphisms connecting
the dilaton prepotential superalgebra given in \Sec\ref{sec:Izq-P}
with a prototype Poisson tensor in its simplest form, \ie with a
Poisson tensor with constant components.  Coordinates where the
nonzero components take the values $\pm 1$ are called Casimir-Darboux
coordinates. This immediately provides the explicit solution of the
corresponding gPSM too; for details \cf \Sec\ref{sec:sdil-sol}.

In addition, we have found target space diffeomorphisms very useful to
incorporate \eg bosonic models related by conformal transformations.
An example of that will be given in \Sec\ref{sec:cIzq-P} where an
algebra referring to models without bosonic torsion---the just
mentioned dilaton prepotential algebra---can be transformed quite
simply to one depending quadratically on torsion and thus representing
a dilaton theory with kinetic term ($Z\neq 0$ in (\ref{dil})) in its
dilaton version.  There the identification $A_I=(\omega, e_a,
\psi_\alpha)$ with `physical' Cartan variables is used to determine
the solution of the latter theory ($Z \neq 0$) from the simpler model
($\bar{Z}=0$) with PSM variables $(\bar{X}^I, \bar{A}_I)$ by
\begin{equation}
  \label{tr-A}
  A_I = (\rpartial_I \bar{X}^J) \bar{A}_J.
\end{equation}

Interesting information regarding the arbitrariness to obtain
supersymmetric extensions of bosonic models as found in the general
solutions of \Sec\ref{sec:Jacobi} can also be collected from target
space diffeomorphisms. Imagine that a certain gPSM has been found,
solving the Jacobi identities with a particular ansatz. A natural
question would be to find out which other models have the same bosonic
body. For this purpose we single out at first all transformations
(\ref{diffeo}) which leave the components $\Poisson^{A\phi}$ form
invariant as given by (\ref{LorentzB}) and (\ref{LorentzF}):
\begin{equation}
  \bar{\phi} = \phi, \qquad
  \bar{X}^a = X^b C_b\^a, \qquad
  \bar{\chi}^\alpha = \chi^\beta h_\beta\^\alpha.
\end{equation}
Here $C_b\^a$ and $h_\beta\^\alpha$ are Lorentz covariant functions
(resp.\ spinor matrices)
\begin{align}
  C_b\^a &= L \delta_b\^a + M \epsilon_b\^a = c_b\^a + \half\chi^2
  (c_2)_b\^a, \label{tr-bb} \\
  h_\beta\^\alpha &= \left[ h_{(1)} \1 + h_{(2)} \gamma^3 + i h_{(3)}
    X^c \gamma_c + i h_{(4)} X^d \epsilon_d\^c \gamma_c
  \right]_\beta\^\alpha, \label{tr-ff}
\end{align}
when expressed in terms of $\chi^2$ ($L = l + \half\chi^2 l_2$ and
similar for $M$) and in terms of $\phi$ and $Y$ ($l$, $l_2$, $m$,
$m_2$, $h_{(i)}$).

The `stabilisator' ($\bar{\Poisson}^{ab} = \Poisson^{ab}$) of the
bosonic component $v(\phi,Y) = v(\bar{\phi},\bar{Y})$ of a graded
Poisson tensor will be given by the restriction of $c_b\^a$ to a
Lorentz transformation on the target space $\TSp$ with $l^2 - m^2 = 1$
in (\ref{tr-bb}). Furthermore from the two parameters $h_{(1)}$ and
$h_{(2)}$ a Lorentz transformation can be used to reduce them to one
independent parameter. Thus no less than five arbitrary two argument
functions are found to keep the bosonic part of $\Poisson^{ab}$
unchanged, but produce different fermionic extensions with
supersymmetries different from the algebra we started from. This
number for rank $(2|2)$ exactly coincides with the number of arbitrary
invariant functions found in \Sec\ref{sec:sol-gP}. For rank $(2|1)$
in the degenerate case a certain `chiral' combination of $h_{(3)}$
and $h_{(4)}$ in (\ref{tr-ff}) must be kept fixed, reducing that
number to four---again in agreement with \Sec\ref{sec:CPA-P}. In a
similar way also the appearance of just three arbitrary functions in
\Sec\ref{sec:DFS-P} for rank $(2|0)$ can be understood.

\section{Particular Poisson Superalgebras}
\label{sec:Poisson}

The compact formulae of the last sections do not seem suitable for a
general discussion, especially in view of the large arbitrariness of
gPSMs. We, therefore, elucidate the main features in special models of
increasing complexity. The corresponding actions and their relations
to the alternative formulations (\ref{grav}) and, or the dilaton
theory form (\ref{dil}) will be discussed in \Sec\ref{sec:models}.

\subsection{Block Diagonal Algebra}
\label{sec:BDA}

The most simple ansatz which, nevertheless, already shows the generic
features appearing in fermionic extensions, consists in setting the
mixed components $\Poisson^{\alpha b}=0$ so that the nontrivial
fermionic brackets are restricted to the block
$\Poisson^{\alpha\beta}$. Then (\ref{J-ffb-0})--(\ref{J-ffb-2}) reduce
to
\begin{gather}
  \derv^c v^{\alpha\beta} = 0, \label{BDA-J-ffb-0} \\
  v_\delta\^\alpha v_2\^{\beta\gamma} + \half \gamma^3\_\delta\^\alpha
  \partial_\phi v^{\beta\gamma} + \cycl(\alpha\beta\gamma) = 0,
  \label{BDA-J-fff-1} \\
  v_\delta\^\alpha v_2 + \half \gamma^3\_\delta\^\alpha \partial_\phi
  v = 0, \label{BDA-J-fbb-1} \\
  v_2 \epsilon^{cd} \partial_d v^{\alpha\beta} + \derv^c
  v_2^{\alpha\beta} = 0. \label{BDA-J-ffb-2}
\end{gather}
Eq.\ (\ref{BDA-J-fbb-1}) implies the spinorial structure
\begin{equation}
  v^{\alpha\beta} = u \gamma^3\^{\alpha\beta}.
\end{equation}
The trace of (\ref{BDA-J-ffb-0}) with $\gamma^3$ leads to the
condition (\ref{c}) for $u$, \ie $u = u(\casimir(\phi, Y))$
depends on the combination of $\phi$ and $Y$ as determined by the
bosonic Casimir function.

For $u \neq 0$ the remaining equations (\ref{BDA-J-fff-1}),
(\ref{BDA-J-fbb-1}) and (\ref{BDA-J-ffb-2}) are fulfilled by
\begin{equation}
  \label{BDA-v2}
  v_2^{\alpha\beta} = -\frac{\partial_\phi u}{2u}
  \gamma^3\^{\alpha\beta}, \qquad
  v_2 = -\frac{\partial_\phi v}{2u}.
\end{equation}
For the present case according to (\ref{c2}) the Casimir function is
\begin{equation}
  \label{BDA-C}
  \Casimir = \casimir - \half\chi^2 \frac{\partial_\phi\casimir}{2
    u(\casimir)}.
\end{equation}
It is verified easily that
\begin{equation}
  \label{BDA-U}
  U = u(\Casimir) = u(\casimir) + \half\chi^2 u_2, \qquad u_2 =
  -\frac{\partial_\phi\casimir}{2u(\casimir)} \frac{du}{d\casimir}.
\end{equation}

Already in this case we observe that in the fermionic extension
$\detv^{-1}$, $u^{-1}$ from the inverse of $v^{\alpha\beta}$ may
introduce singularities. It should be emphasized that $u=u(c)$ is an
arbitrary function of $c(\phi,Y)$. Except for $u=u_0=\const$ (see
below) any generic choice of the arbitrary function $u(\casimir)$ by
the factors $u^{-1}$ in (\ref{BDA-v2}), thus may introduce
restrictions on the allowed range of $\phi$ and $Y$ or new
singularities on a certain surface where $u(\casimir(\phi,Y))$
vanishes, not present in the purely bosonic bracket. Indeed, these
obstructions in certain fermionic extensions are a generic feature of
gPSMs. The singularities are seen to be caused here by $\detv^{-1}$,
the inverse of the determinant (\ref{detv}), except for cases with
$\detv = \const$ or when special cancellation mechanisms are invoked.
Another source for the same phenomenon will appear below in connection
with the appearance of a `prepotential' for $v$.  Still, such
`obstructions' can be argued to be rather harmless. We will come back
to these issues in several examples below, especially when discussing
an explicit solution in \Sec\ref{sec:sdil-sol}.

This complication can be made to disappear by choosing $u = u_0 =
\const \neq 0$. Then the fermionic extension ($v'=\partial_\phi v$)
\begin{align}
  \Poisson^{ab} &= \left( v - \frac{1}{4u_0} \chi^2 v' \right)
  \epsilon^{ab}, \label{BDA-P-bb} \\
  \Poisson^{\alpha b} &= 0, \label{BDA-P-fb} \\
  \Poisson^{\alpha\beta} &= u_0 \gamma^3\^{\alpha\beta}
  \label{BDA-P-ff}
\end{align}
does not lead to restrictions on the purely bosonic part $v(\phi, Y)$
of the Poisson tensor, nor does it introduce additional singularities,
besides the ones which may already be present in the potential
$v(\phi,Y)$. But then no genuine supersymmetry survives (see
\Sec\ref{sec:BDS} below).

It should be noted that all dilaton models mentioned in the
introduction can be accommodated in a nontrivial version $u \neq
\const$ of this gPSM. We shall call the corresponding supergravity
actions their `diagonal extensions'.

\subsection{Nondegenerate Chiral Algebra}
\label{sec:SUSY/2-P}

Two further models follow by setting $u=u_0=\const \neq 0$ and
$\hat{u}=\pm \tilde{u}_0=\const$.  In this way a generalization with
full rank of the chiral $\N=(1,0)$ and $\N=(0,1)$ algebras is obtained
(\cf \App\ref{app:spinors})
\begin{equation}
  v^{\alpha\beta} = i \tilde{u}_0 X^c (\gamma_c P_\pm)^{\alpha\beta}
  + u_0 \gamma^3\^{\alpha\beta}.
\end{equation}
This particular choice for the coefficients of $X^c$ in
$v^{\alpha\beta}$ also has the advantage that $X^c$ drops out from
$\detv = -u_0^2/4$, thus its inverse exists everywhere.  Restricting
furthermore $f^c=0$ we arrive at
\begin{equation}
  F^c\_\alpha\^\beta = -\frac{i\tilde{u}_0 v}{2u_0}
  (\gamma^c P_\pm)_\alpha\^\beta, \qquad
  v_2\^{\alpha\beta} = 0, \qquad
  v_2 = -\frac{v'}{2u_0}.
\end{equation}
This yields another graded Poisson tensor for the arbitrary bosonic
potential $v(\phi,Y)$
\begin{align}
  \Poisson^{ab} &= \left( v - \frac{1}{4u_0} \chi^2 v' \right)
  \epsilon^{ab},
  \label{SUSY/2-P-bb} \\
  \Poisson^{\alpha b} &= -\frac{i\tilde{u}_0v}{2u_0}
  (\chi\gamma^b P_\pm)^\alpha, \label{SUSY/2-P-fb} \\
  \Poisson^{\alpha\beta} &= i \tilde{u}_0 X^c
  (\gamma_c P_\pm)^{\alpha\beta} + u_0 \gamma^3\^{\alpha\beta}.
  \label{SUSY/2-P-ff}
\end{align}
There also are no obstructions for such models corresponding to any
bosonic gravity model, given by a particular choice of $v(\phi,Y)$.

The Casimir function (\cf (\ref{c2})) reads
\begin{equation}
  \label{SUSY/2-C}
  \Casimir = \casimir - \frac{1}{4 u_0} \chi^2 c',
\end{equation}
where $\casimir$ must obey (\ref{c}).

\subsection{Deformed Rigid Supersymmetry}
\label{sec:SUSY-P}

The structure of rigid supersymmetry is encoded within the Poisson
tensor by means of the components $v=0$ and (\cf (\ref{matrix}))
\begin{equation}
  \label{SUSY-v-ff}
  v^{\alpha\beta} = i \tilde{u}_0 X^c \gamma_c\^{\alpha\beta} =
  \mtrx{\sqrt{2} \tilde{u}_0 X^{++}}{0}{0}{\sqrt{2} \tilde{u}_0 X^{--}},
\end{equation}
where again $\tilde{u}=\tilde{u}_0=\const \neq 0$. Here $\detv = 2Y
\tilde{u}_0^2$ and
\begin{equation}
  \frac{1}{\detv} v_{\alpha\beta} = \frac{i}{2Y \tilde{u}_0} X^c
  \gamma_{c\alpha\beta} = \mtrx{\frac{1}{\sqrt{2} \tilde{u}_0
      X^{++}}}{0}{0}{\frac{1}{\sqrt{2} \tilde{u}_0 X^{--}}}.
\end{equation}

Generalizing this ansatz to $v \neq 0$, the simplest choice $f^c = 0$
with an arbitrary function $v(\phi,Y)$ (deformed rigid supersymmetry,
DRS) yields
\begin{equation}
  F^c\_\alpha\^\beta = \frac{v}{4Y} X^a (\gamma_a \gamma^c
  \gamma^3)_\alpha\^\beta, \qquad
  v_2\^{\alpha\beta} = \frac{v}{4Y} \gamma^3\^{\alpha\beta},
  \qquad 
  v_2 = 0,
\end{equation}
and thus for $\Poisson^{IJ}$
\begin{align}
  \Poisson^{ab} &= v \epsilon^{ab}, \label{SUSY-P-bb} \\
  \Poisson^{\alpha b} &= \frac{v}{4Y} X^c
  (\chi\gamma_c\gamma^b\gamma^3)^\alpha, \label{SUSY-P-fb} \\
  \Poisson^{\alpha\beta} &= i \tilde{u}_0 X^c \gamma_c\^{\alpha\beta}
  + \frac{1}{2} \chi^2 \frac{v}{4Y} \gamma^3\^{\alpha\beta},
  \label{SUSY-P-ff}
\end{align}
and for the Casimir function $\Casimir = \casimir$ with (\ref{c}).

{}From (\ref{SUSY-P-bb})--(\ref{SUSY-P-ff}) it is clear---in contrast
to the algebras \ref{sec:BDA} and \ref{sec:SUSY/2-P}---that this
fermionic extension for a generic $v \neq 0$ introduces a possible
further singularity at $Y=0$, which cannot be cured by further
assumptions on functions which are still arbitrary.

Of course, in order to describe flat spacetime, corresponding to the
Poisson tensor of rigid supersymmetry, one has to set $v(\phi,Y)=0$.
Then the singularity at $Y=0$ in the extended Poisson tensor
disappears.

We remark already here that despite the fact that for $v \neq 0$ the
corresponding supersymmetrically extended action functional (in
contrast to its purely bosonic part) becomes singular at field values
$Y\equiv \half X^aX_a =0$, we expect that if solutions of the field
equations are singular there as well, such singularities will not be
relevant if suitable `physical' observables are considered. We have in
mind the analogy to curvature invariants which are not affected by
`coordinate singularities'. We do, however, not intend to prove this
statement in detail within the present thesis; in
\Sec\ref{sec:sdil-sol} below we shall only shortly discuss the similar
singularities, caused by the prepotential in an explicit solution of
the related field-theoretical model.

\subsection{Dilaton Prepotential Algebra}
\label{sec:Izq-P}

We now assume that the bosonic potential $v$ is restricted to be a
function of the dilaton $\phi$ only, $\dot{v} = \partial_Y v = 0$.
Many models of 2d supergravity, already known in the literature, are
contained within algebras of this type, one of which was described in
ref.~\cite{\bibIzq}. Let deformed rigid supersymmetry of
\Sec\ref{sec:SUSY-P} again be the key component of the Poisson tensor
(\ref{SUSY-v-ff}). Our attempt in \Sec\ref{sec:SUSY-P} to provide a
Poisson tensor for arbitrary $v$ built around that component produced
a new singularity at $Y=0$ in the fermionic extension. However, the
Poisson tensor underlying the model considered in \cite{\bibIzq} was
not singular in $Y$.  Indeed there exists a mechanism by which this
singularity can be cancelled in the general solution
(\ref{detv})--(\ref{sol-v2}), provided the arbitrary functions are
chosen in a specific manner.

For this purpose we add to (\ref{SUSY-v-ff}), keeping
$\tilde{u}=\tilde{u}_0=\const$, the fermionic potential $u(\phi)$,
\begin{equation}
  \label{Izq-v-ff}
  v^{\alpha\beta} = i \tilde{u}_0 X^c \gamma_c\^{\alpha\beta} + u
  \gamma^3\^{\alpha\beta} =
  \mtrx{\sqrt{2} \tilde{u}_0 X^{++}}{-u}{-u}{\sqrt{2} \tilde{u}_0 X^{--}},
\end{equation}
with determinant 
\begin{equation}
  \label{Izq-detv}
  \detv = 2Y \tilde{u}_0^2 - u^2.
\end{equation}

The Hamiltonian vector field $\rvec T^c$ in the solution
(\ref{sol-f-bff}) generates a factor $f_{(t)} \neq 0$ in
(\ref{ans-f-bb}). The independent vector field $f^c$ can be used to
cancel that factor provided one chooses
\begin{equation}
  \label{Izq-f-b}
  f^c = \half u' X^c.
\end{equation}
Then the disappearance of $f_{(t)}$ is in agreement with the solution
given in ref.~\cite{Izquierdo:1998hg}.  The remaining coefficient
functions then follow as
\begin{align}
  F^c\_\alpha\^\beta &= \frac{1}{2\detv} \left( \tilde{u}_0^2 v + u u'
  \right) X^a (\gamma_a \gamma^c \gamma^3)_\alpha\^\beta +
  \frac{i\tilde{u}_0}{2\detv} \left( u v + 2Y u' \right)
  \gamma^c\_\alpha\^\beta, \label{Izq-f-bff} \\
  v_2\^{\alpha\beta} &= \frac{1}{2\detv} \left( \tilde{u}_0^2 v + u u'
  \right) \gamma^3\^{\alpha\beta}, \label{Izq-v2-ff} \\
  v_2 &= \frac{uv}{2\detv^2} \left( \tilde{u}_0^2 v + u u' \right) +
  \frac{uu'}{2\detv^2} \left( u v + 2Y u' \right) + \frac{1}{2\detv} \left(u
    v' + 2Y u' \dot{v} + 2Y u'' \right). \label{Izq-v2}
\end{align}

Up to this point the bosonic potential $v$ and the potential $u$ have
been arbitrary functions of $\phi$. Demanding now that
\begin{equation}
  \label{Izq-v}
  \tilde{u}_0^2 v + u u' = 0,
\end{equation}
the singularity at $\detv=0$ is found to be cancelled not only in the
respective first terms of (\ref{Izq-f-bff})--(\ref{Izq-v2}), but also
in the rest:
\begin{alignat}{2}
  v &= -\frac{(u^2)'}{2\tilde{u}_0^2}, &\qquad
  F^c\_\alpha\^\beta &= \frac{iu'}{2\tilde{u}_0}
  \gamma^c\_\alpha\^\beta, \label{Izq-pot1} \\
  v_2\^{\alpha\beta} &= 0, &\qquad
  v_2 &= \frac{u''}{2\tilde{u}_0^2}. \label{Izq-pot2}
\end{alignat}
Furthermore the fermionic potential $u(\phi)$ is seen to be promoted
to a `prepotential' for $v(\phi)$.  A closer look at (\ref{Izq-v})
with (\ref{Izq-detv}) shows that this relation is equivalent to $\derv
\detv = 0$ which happens to be precisely the defining equation
(\ref{c}) of the Casimir function $\casimir(\phi,Y)$ of the bosonic
model. The complete Casimir function follows from (\ref{c2}):
\begin{equation}
  \label{Izq-c2}
  \casimir_2 = \frac{1}{2\detv} \left( u
    \partial_\phi + 2Y u' \partial_Y \right) \casimir
\end{equation}
so that
\begin{equation}
  \label{Izq-C}
  \Casimir = \detv + \half\chi^2 u'.
\end{equation}

Thus the Poisson tensor for $v=v(\phi)$, related to $u(\phi)$ by
(\ref{Izq-v}), becomes
\begin{align}
  \Poisson^{ab} &= \frac{1}{2\tilde{u}_0^2} \left( -(u^2)'
    + \half \chi^2 u'' \right) \epsilon^{ab}, \label{Izq-P-bb} \\
  \Poisson^{\alpha b} &= \frac{iu'}{2\tilde{u}_0} (\chi\gamma^b)^\alpha,
  \label{Izq-P-fb} \\
  \Poisson^{\alpha\beta} &= i \tilde{u}_0 X^c \gamma_c\^{\alpha\beta}
  + u \gamma^3\^{\alpha\beta}, \label{Izq-P-ff}
\end{align}
which is indeed free from singularities produced by the supersymmetric
extension. However, this does not eliminate all pitfalls: Given a
bosonic model described by a particular potential $v (\phi)$ where
$\phi$ is assumed to take values in the interval $I \subseteq \R$, we
have to solve (\ref{Izq-v}) for the prepotential $u(\phi)$, \ie the
quadratic equation
\begin{equation}
  \label{Izq-u}
  u^2 = -2 \tilde{u}_0^2 \int_{\phi_0}^{\phi} v(\varphi) d\varphi,
\end{equation}
which may possess a solution within the real numbers only for a
restricted range $\phi \in J$.  The interval $J$ may have a nontrivial
intersection with $I$ or even none at all. Clearly no restrictions
occur if $v$ contains a potential for the dilaton which happens to
lead to a negative definite integral on the \rhs of (\ref{Izq-u}) for
\emph{all} values of $\phi$ in $I$ (this happens \eg if $v$ contains
only odd powers of $\phi$ with negative prefactors).  On the other
hand, the domain of $\phi$ is always restricted if $v$ contains even
potentials, as becomes immediately clear when viewing the special
solutions given in \Tab\ref{tab:IzqMdls}.
\begin{table}[ht]
  \begin{center}
    \begin{tabular}{|l||l|l|} \hline
      Model & $v(\phi) = -\frac{(u^2)'}{2\tilde{u}_0^2}$ & $u(\phi)$
      \\
      \hline\hline
      & $0$ & $\tilde{u}_0 \lambda$ \\
      String & $-\Lambda$ & $\pm \tilde{u}_0 \sqrt{2\Lambda
        (\phi-\phi_0)}$ \\
      JT & $-\lambda^2 \phi$ & $\tilde{u}_0 \lambda \phi$ \\
      $R^2$ & $-\frac{\alpha}{2} \phi^2$ & $\pm \tilde{u}_0
      \sqrt{\frac{\alpha}{3} (\phi^3-\phi_0^3)}$ \\
      Howe & $-2 \lambda^2 \phi^3$ & $\tilde{u}_0 \lambda \phi^2$ \\
      \hline
      $\bar{\mathrm{SRG}}$ & $-\frac{\lambda^2}{\sqrt{\phi}}$ & $2
      \tilde{u}_0 \lambda \sqrt[4]{\phi}$ \\
      \hline
    \end{tabular}
  \end{center}
  \caption{Special Dilaton Prepotential Algebras} \label{tab:IzqMdls}
\end{table}
There the different potentials $v(\phi)$ are labelled according to the
models: The string model with $\Lambda=\const$ of
\cite{\bibSI,\bibDBHmatter}, JT is the Jackiw-Teitelboim model
(\ref{JT}), $\overline{\mathrm{SRG}}$ the spherically reduced black
hole (\ref{EBH}) in the conformal description (\cf
\Sec\ref{cha:intro}); the cubic potential appeared in
\cite{Howe:1979ia}, $R^2$ gravity is self-explaining. Note that in the
case of $\overline{\mathrm{SRG}}$ $I=J=\R_+$ ($\phi > 0$), there is
already a (harmless) restriction on allowed values of $\phi$ at the
purely bosonic level, \cf (\ref{EBH}).

So, as argued above, one may get rid of the singularities at $Y=0$ of
supersymmetric extensions obtained in the previous section. In some
cases, however, this leads to a restricted range for allowed values of
the dilaton, or, alternatively, to complex valued Poisson tensors.
Similarly to our expectation of the harmlessness of the above
mentioned $1/Y$-singularities on the level of the solutions (\cf also
\cite{Strobl:1999zz}), we expect that also complex-valued Poisson
tensors are no serious obstacle (both of these remarks apply to the
classical analysis only!). In fact, a similar scenario was seen to be
harmless (classically) also in the Poisson Sigma formulation of the
$G/G$ model for compact gauge groups like $SU(2)$, \cf
\cite{Schaller:1995xk,Alekseev:1995py}. We further illustrate these
remarks for the class of supergravity models considered in
\cite{\bibIzq} at the end of \Sec\ref{sec:sdil-sol}.

\subsection[Bosonic Potential Linear in $Y$]{Bosonic Potential Linear
in \mathversion{bold}$Y$}
\label{sec:cIzq-P}

In order to retain the $Y$-dependence and thus an algebra with bosonic
torsion, we take solution (\ref{Izq-v-ff})--(\ref{Izq-v2}) but instead
of (\ref{Izq-v}) we may also choose
\begin{equation} \label{cIzq-v} v = -\frac{(u^2)'}{2\tilde{u}_0^2} -
\frac{\detv}{2} f,
\end{equation} where $f$ is an arbitrary function of $\phi$ and
$Y$. Thanks to the factor $\detv$ also in this case the fermionic
extension does not introduce new singularities at
$\detv=0$.\footnote{Clearly also in (\ref{cIzq-v}) the replacement
$\detv f \rightarrow G(\detv,\phi,Y)$ with $G(\detv,\phi,Y)/\detv$
regular at $\detv=0$ has a similar effect. But linearity in $\detv$ is
sufficient for our purposes.}  Even if $f$ is a function of $\phi$
only ($\dot f = 0$), this model is quadratic in (bosonic) torsion,
because of (\ref{Izq-detv}).  A straightforward calculation using
(\ref{cIzq-v}) gives
\begin{align}
  F^c\_\alpha\^\beta &= -\frac{\tilde{u}_0^2 f}{4} X^a
  (\gamma_a \gamma^c \gamma^3)_\alpha\^\beta + i \left( \frac{u'}{2
      \tilde{u}_0} - \frac{\tilde{u}_0 u f}{4} \right)
  \gamma^c\_\alpha\^\beta, \label{cIzq-f-bff} \\
  v_2\^{\alpha\beta} &= -\frac{\tilde{u}_0^2 f}{4}
  \gamma^3\^{\alpha\beta}, \label{cIzq-v2-ff} \\
  v_2 &= \half \left( \frac{u''}{\tilde{u}_0^2} - u'f - \frac{uf'}{2}
    + \frac{\tilde{u}_0^2 u f^2}{4} - \frac{2Y u' \dot{f}}{2}
  \right). \label{cIzq-v2}
\end{align}

It is worthwhile to note that the present algebra, where the bosonic
potential $v$ is of the type (\ref{vdil}), can be reached from the
algebra of \Sec\ref{sec:Izq-P} with $\bar{v}=\bar{v}(\bar\phi)$ by a
conformal transformation, \ie a target space diffeomorphism in the
sense of \Sec\ref{sec:diffeo}. We use bars to denote quantities and
potentials of the algebra of \Sec\ref{sec:Izq-P}, but not for
$\tilde{u}_0$ because it remains unchanged, \ie $\bar{v} =
-\frac{(\bar{u}^2)'}{2\tilde{u}_0^2}$. By
\begin{equation}
  \label{conf-tr-X}
  \phi = \bar{\phi}, \qquad X^a = e^{\varphi(\phi)} \bar{X}^a, \qquad
  \chi^\alpha = e^{\half \varphi(\phi)} \bar{\chi}^\alpha,
\end{equation} the transformed Poisson tensor, expanded in terms of
unbarred coefficient functions (\cf \Sec\ref{sec:ans-gP}) becomes
\begin{alignat}{3}
  \tilde{u} &= \tilde{u}_0, &\qquad \tilde{u}_2 &= 0, \\
  u &= e^\varphi \bar{u}, &\qquad u_2 &= -\half \varphi', \\
  v &= e^{2 \varphi} \bar{v} - 2 Y \varphi', &\qquad v_2 &= e^\varphi
  \frac{\bar{u}''}{2\tilde{u}_0^2}, \\
  f_{(12)} &= \half \varphi', &\qquad f_{(31)} &= -\half \varphi', \\
  f_{(s)} &= e^\varphi \frac{\bar{u}'}{2\tilde{u}_0}
\end{alignat}
and $f_{(11)} = f_{(32)} = f_{(t)} = f_{(h)} = 0$. When $u(\phi)$ and
$\varphi(\phi)$ are taken as basic independent potentials we arrive at
\begin{align}
  v &= -\frac{1}{2\tilde{u}_0^2} e^{2\varphi} \left(
    e^{-2\varphi} u^2 \right)' - 2 Y \varphi', \label{cIzq-v-alt} \\
  v_2 &= \frac{1}{2\tilde{u}_0^2} e^{\varphi} \left( e^{-\varphi} u
  \right)'', \\
  f_{(s)} &= \frac{1}{2\tilde{u}_0} e^{\varphi} \left(
    e^{-\varphi} u \right)'.
\end{align}
If we set $\varphi' = \tilde{u}_0^2 f / 2$ we again obtain solution
(\ref{cIzq-v})--(\ref{cIzq-v2}) for $Y$-in\-de\-pend\-ent $f$.  The
components $\bar{\Poisson}^{a\phi}$ and $\bar{\Poisson}^{\alpha \phi}$
remain form invariant,
\begin{equation}
  \Poisson^{a\phi} = X^b \epsilon_b\^a, \qquad
  \Poisson^{\alpha \phi} = -\half \chi^\beta \gamma^3\_\beta\^\alpha,
\end{equation}
in agreement with the requirement determined for this case in
\Sec\ref{sec:ans-gP}. For completeness we list the transformation of
the 1-forms $A_I = (\omega, e_a, \psi_\alpha)$ according to
(\ref{tr-A})
\begin{equation}
  \label{conf-tr-A}
  \omega = \bar{\omega} - \varphi'
  \left( \bar{X}^b \bar{e}_b + \half \bar{\chi}^\beta \bar{\psi}_\beta
  \right), \qquad e_a = e^{-\varphi} \bar{e}_a, \qquad \psi_\alpha =
  e^{-\half \varphi} \bar{\psi}_\alpha.
\end{equation}
The second equation in (\ref{conf-tr-A}) provide the
justification for the name `conformal transformation'.

With the help of the scaling parameter $\varphi$ we can write
(\ref{cIzq-v-alt}), and also (\ref{cIzq-v}), in its equivalent form
$\derv(e^{-2\varphi} \detv) = 0$, thus exposing the Casimir function
to be $\casimir(\phi,Y) = e^{-2\varphi} \detv$.  Now $u(\phi)$ and
$\varphi(\phi)$ are to be viewed as two independent parameter
functions labelling specific types of Poisson tensors.  The solution
\begin{align}
  \Poisson^{ab} &= \left( -\frac{1}{2\tilde{u}_0^2} e^{2\varphi}
    \left( e^{-2\varphi} u^2 \right)' - 2 Y \varphi' +
    \frac{1}{4\tilde{u}_0^2} \chi^2 e^{\varphi} \left( e^{-\varphi} u
    \right)'' \right) \epsilon^{ab}, \label{cIzq-P-bb} \\
  \Poisson^{\alpha b} &= -\half \varphi' X^a (\chi\gamma_a \gamma^b
  \gamma^3)^\alpha + \frac{i}{2 \tilde{u}_0} e^{\varphi} \left(
    e^{-\varphi} u \right)' (\chi\gamma^b)^\alpha, \label{cIzq-P-fb}
  \\
  \Poisson^{\alpha\beta} &= i \tilde{u}_0 X^c \gamma_c\^{\alpha\beta}
  + \left( u - \frac{1}{4}\chi^2 \varphi' \right)
  \gamma^3\^{\alpha\beta} \label{cIzq-P-ff}
\end{align}
does not introduce a new singularity at $Y=0$, but in order to provide
the extension of the bosonic potential (\ref{vdil}) we have to solve
(\ref{cIzq-v-alt}) for the scaling parameter $\varphi(\phi)$ and the
fermionic potential $u(\phi)$, which may again lead to obstructions
similar to the ones described at the end of \Sec\ref{sec:Izq-P}. With
the integrals over $Z(\phi)$ and $V(\phi)$ introduced in (\ref{QW}) we
find
\begin{align}
  \varphi &= -\half Q(\phi), \label{cIzq-tr} \\
  u &= \pm \sqrt{-2 \tilde{u}_0^2 e^{-Q(\phi)}
    W(\phi)}. \label{cIzq-u}
\end{align}
Now we can read off the restriction to be $W(\phi) < 0$, yielding
singularities at the boundary $W(\phi) = 0$. The ansatz (\ref{cIzq-v})
can be rewritten in the equivalent form
\begin{equation}
  \label{cIzq-c} \derv(e^Q \detv) = 0 \Leftrightarrow
  \casimir(\phi,Y) = e^Q \detv = 2\tilde{u}_0^2 (Y e^Q + W).
\end{equation}
The complete Casimir from (\ref{c2}), which again exhibits the simpler
form (\ref{Izq-c2}), reads
\begin{equation}
  \label{cIzq-C} \Casimir = e^Q \left( \detv +
    \half\chi^2 e^{-\half Q} \left( e^{\half Q} u \right)' \right).
\end{equation}
As expected from ordinary 2d gravity $\Casimir$ is conformally
invariant.

Expressing the Poisson tensor in terms of the potentials $V(\phi)$ and
$Z(\phi)$ of the original bosonic theory, and with $u(\phi)$ as in
(\ref{cIzq-u}) we arrive at
\begin{align}
  \Poisson^{ab} &= \left( V + Y Z - \half \chi^2 \left[
      \frac{V Z + V'}{2u} + \frac{\tilde{u}_0^2 V^2}{2u^3} \right]
  \right) \epsilon^{ab}, \label{cIzq-P-bb-alt} \\
  \Poisson^{\alpha b} &=
  \frac{Z}{4} X^a (\chi\gamma_a \gamma^b \gamma^3)^\alpha -
  \frac{i\tilde{u}_0V}{2u} (\chi\gamma^b)^\alpha, \label{cIzq-P-fb-alt}
  \\
  \Poisson^{\alpha\beta} &= i \tilde{u}_0 X^c
  \gamma_c\^{\alpha\beta} + \left( u + \frac{Z}{8}\chi^2 \right)
  \gamma^3\^{\alpha\beta}.  \label{cIzq-P-ff-alt}
\end{align}
As will be shown in \Sec\ref{sec:sdil} this provides a
supersymmetrization for all the dilaton theories (\ref{dil}), because
it covers all theories (\ref{FOG}) with $v$ linear in $Y$. Among these
two explicit examples, namely SRG and the KV model, will be treated in
more detail now.

\subsubsection[SRG, Nondiagonal Extension I]{Spherically Reduced
Gravity (SRG), Nondiagonal Extension I}
\label{sec:SRG1}

In contrast to the KV-model below, no obstructions are found when
(\ref{cIzq-f-bff})--(\ref{cIzq-v2}) with $v$ given by (\ref{cIzq-v})
is used for SRG.  For simplicity we take in (\ref{SRG}) the case $d=4$
and obtain $Q(\phi) = -\half \ln(\phi)$, $W(\phi) = -2 \lambda^2
\sqrt{\phi}$ and
\begin{equation}
  \label{SRG1-u-and-f} u = 2 \tilde{u}_0 \lambda
  \sqrt{\phi}, \qquad \varphi = \frac{1}{4} \ln(\phi),
\end{equation}
where $u_0$ is a constant. Here already the bosonic theory is defined
for $\phi > 0$ only. From (\ref{cIzq-P-bb})--(\ref{cIzq-P-ff}) in the
Poisson tensor of SRG
\begin{align}
  P^{ab} &= \left( -\lambda^2 - \frac{Y}{2\phi} - \frac{3\lambda}{32
      \tilde{u}_0 \phi^{3/2}} \chi^2 \right)
  \epsilon^{ab}, \label{SRG1-P-bb} \\
  P^{\alpha b} &= -\frac{1}{8\phi} X^c (\chi \gamma_c \gamma^b
  \gamma^3)^\alpha + \frac{i\lambda}{4\sqrt{\phi}} \,
  (\chi\gamma^b)^\alpha, \label{SRG1-P-fb} \\
  P^{\alpha\beta} &= i \tilde{u}_0 X^c \gamma_c\^{\alpha\beta} +
  \left( 2 \tilde{u}_0 \lambda \sqrt{\phi} - \frac{1}{16\phi} \chi^2
  \right) \gamma^3\^{\alpha\beta} \label{SRG1-P-ff}
\end{align}
the singularity of the bosonic part simply carries over to the
extension, without introducing any new restriction for $\phi > 0$.

The bosonic part of the Casimir function (\ref{cIzq-C}) is
proportional to the ADM mass for SRG.

\subsubsection[KV, Nondiagonal Extension I]{Katanaev-Volovich Model
(KV), Nondiagonal Extension I}
\label{sec:SKV1}

The bosonic potential (\ref{KV}) leads to $Q(\phi) = \alpha \phi$,
thus $\varphi = -\frac{\alpha}{2} \phi$, and
\begin{equation}
  W(\phi) = \int_{\phi_0}^{\phi} e^{\alpha\eta} \left(
    \frac{\beta}{2} \eta^2 - \Lambda \right) d\eta = \left. e^{\alpha\eta}
    \left[ \frac{\beta}{2} \left( \frac{2}{\alpha^3} -
        \frac{2\eta}{\alpha^2} + \frac{\eta^2}{\alpha} \right) -
      \frac{\Lambda}{\alpha} \right]
  \right|_{\phi_0}^{\phi}. \label{SKV1-W}
\end{equation}
With $u(\phi)$ calculated according to (\ref{cIzq-u}) the Poisson
tensor is
\begin{align}
  \Poisson^{ab} &= \left( \frac{\beta}{2} \phi^2 - \Lambda
    + \alpha Y + \half \chi^2 v_2 \right) \epsilon^{ab},
  \label{SKV1-P-bb} \\
  \Poisson^{\alpha b} &= \frac{\alpha}{4} X^a (\chi\gamma_a \gamma^b
  \gamma^3)_\alpha\^\beta - \frac{i\tilde{u}_0}{2u} \left(
    \frac{\beta}{2} \phi^2 - \Lambda \right) (\chi\gamma^b)_\alpha\^\beta,
  \label{SKV1-P-fb} \\
  \Poisson^{\alpha\beta} &= i \tilde{u}_0 X^c
  \gamma_c\^{\alpha\beta} + \left( u + \frac{\alpha}{8}\chi^2 \right)
  \gamma^3\^{\alpha\beta}, \label{SKV1-P-ff}
\end{align}
with
\begin{equation}
  \label{SKV1-v2}
  v_2 = -\frac{\alpha \left( \frac{\beta}{2} \phi^2 - \Lambda \right)
    + \beta \phi}{2u} - \frac{\tilde{u}_0^2 \left( \frac{\beta}{2}
      \phi^2 - \Lambda \right)^2}{2u^3}.
\end{equation}

For general parameters $\alpha$, $\beta$, $\Lambda$ from
(\ref{cIzq-u}) restrictions upon the range of $\phi$ will in general
emerge, if we do not allow singular and complex Poisson tensors. It
may even happen that no allowed interval for $\phi$ exists.  In fact,
as we see from (\ref{SKV1-v2}), in the present case, the `problem' of
complex-valued Poisson tensors comes together with the
`singularity-problem'.

On the other hand, for $\beta \le 0$ and $\Lambda \ge 0$, where at
least one of these parameter does not vanish, the integrand in
(\ref{SKV1-W}) becomes negative definite, leading to the restriction
$\phi > \phi_0$ with singularities at $\phi = \phi_0$. If we further
assume $\alpha > 0$ we can set $\phi_0 = -\infty$. In contrast to the
torsionless $R^2$ model (see \Tab\ref{tab:IzqMdls}) the restriction
for this particular case disappears and the fermionic potential
becomes
\begin{equation}
  \label{SKV1-u}
  u = \pm \tilde{u}_0 \sqrt{-\frac{\beta}{\alpha^3} \left( \left(
        1-\alpha\phi \right)^2 + 1 \right) +
    \frac{2\Lambda}{\alpha}}.
\end{equation}

\subsection{General Prepotential Algebra}
\label{sec:GPA-P}

This algebra represents the immediate generalization of the
torsionless one of \Sec\ref{sec:Izq-P}, when (\ref{Izq-v-ff}) is taken
for $v^{\alpha\beta}$, but now with $u$ depending on both $\phi$ and
$Y$.  Here also $v=v(\phi,Y)$.  Again we have $\detv = 2Y
\tilde{u}_0^2 - u^2$. By analogy to the step in \Sec\ref{sec:Izq-P} we
again cancel the $f_{(t)}$ term (\cf (\ref{ans-f-bb})) by the choice
\begin{equation}
  f^c = \half (\derv u) X^c.
\end{equation}
This yields
\begin{align}
  F^c\_\alpha\^\beta &= \frac{1}{2\detv} \left( \tilde{u}_0^2 v + u
    \derv u \right) X^a (\gamma_a \gamma^c \gamma^3)_\alpha\^\beta +
  \frac{i\tilde{u}_0}{2\detv} \left( u v + 2Y \derv u \right)
  \gamma^c\_\alpha\^\beta, \label{GPA-f-bff} \\
  v_2\^{\alpha\beta} &= \frac{1}{2\detv} \left( \left( \tilde{u}_0^2 v
      + u \derv u \right) + \dot{u} \left( u v + 2Y \derv u \right)
  \right) \gamma^3\^{\alpha\beta}, \label{GPA-v2-ff} \\
  v_2 &= \frac{uv}{2\detv^2} \left( \tilde{u}_0^2 v + u \derv u
  \right) + \frac{u \derv u}{2\detv^2} \left( u v + 2Y \derv u \right)
  \eqnsplit \hspace{9em} + \frac{1}{2\detv} \left(u v' + 2Y \dot{v}
    \derv u + 2Y \derv^2 u \right). \label{GPA-v2}
\end{align}

The factors $\detv^{-1}$, $\detv^{-2}$ indicate the appearance of
action functional singularities, at values of the fields where $\detv$
vanishes. Again we have to this point kept $u$ independent of $v$.  In
this case, even when we relate $v$ and $u$ by imposing \eg
\begin{equation}
  \tilde{u}_0^2 v + u \derv u = 0 \Leftrightarrow
\derv\detv = 0 \Leftrightarrow \casimir(\phi,Y) = \detv,
\end{equation}
in order to cancel the first terms in
(\ref{GPA-f-bff})--(\ref{GPA-v2}), a generic singularity obstruction
is seen to persist (previous remarks on similar occasions should apply
here, too, however).

\subsection[Algebra with $u(\phi,Y)$ and $\tilde{u}(\phi,Y)$]{Algebra
with \mathversion{bold}$u(\phi,Y)$ and $\tilde{u}(\phi,Y)$}
\label{sec:uTu-P}

For $v^{\alpha\beta}$ we retain (\ref{Izq-v-ff}), but now with both
$u$ and $\tilde{u}$ depending on $\phi$ and $Y$.  Again the
determinant
\begin{equation}
  \label{uTu-detv}
  \detv = 2Y \tilde{u}^2 - u^2
\end{equation}
will introduce singularities. If we want to cancel the $f_{(t)}$ term
(\cf (\ref{ans-f-bb})) as we did in \Sec\ref{sec:Izq-P} and
\Sec\ref{sec:GPA-P}, we have to set here
\begin{equation}
  f^c = \frac{1}{2\tilde{u}} \left( \tilde{u} \derv u - u
    \derv\tilde{u} \right) X^c.
\end{equation}
This leads to
\begin{equation}
  \label{uTu-f-bff}
  F^c = -\frac{1}{4} (\derv\ln\detv) X^a \gamma_a \gamma^c \gamma^3 +
  \half (\derv\ln\tilde{u}) X^c \gamma^3 + \frac{i \tilde{u}
    u}{2\detv} \left[ v + 2Y \left( \derv\ln\frac{u}{\tilde{u}}
    \right) \right] \gamma^c.
\end{equation}

Again we could try to fix $u$ and $\tilde{u}$ suitably so as to cancel
\eg the first term in (\ref{uTu-f-bff}):
\begin{equation}
  \label{uTu-v}
  -\half \derv \detv = \tilde{u}^2 v + u \derv u - 2Y \tilde{u} \derv
  \tilde{u} = 0.
\end{equation}
But then the singularity obstruction resurfaces in (\cf (\ref{uTu-v}))
\begin{equation}
  v = \frac{\detv'}{\dot{\detv}} = \frac{-uu' + 2Y
    \tilde{u}\tilde{u}'}{\tilde{u}^2 - u\dot{u} + 2Y
    \tilde{u}\Dot{\Tilde{u}}}\; .
\end{equation}
Eq.\ (\ref{uTu-f-bff}) becomes
\begin{equation}
  F^c = \frac{1}{\dot{\detv}} \left( \tilde{u}
    \tilde{u}' - \frac{u}{\tilde{u}} \left( \tilde{u}' \dot{u} -
      \Dot{\Tilde{u}} u' \right) \right) X^c \gamma^3 +
  \frac{i}{\dot{\detv}} \left( \tilde{u} u' - 2Y \left( \tilde{u}'
      \dot{u} - \Dot{\Tilde{u}} u' \right) \right) \gamma^c.
\end{equation}

The general formulae for the Poisson tensor are not very illuminating.
Instead, we consider two special cases.

\subsubsection[SRG, Nondiagonal Extension II]{Spherically Reduced
Gravity (SRG), Nondiagonal Extension II}
\label{sec:SRG2}

For SRG also \eg the alternative
\begin{equation}
  \label{SRG2-v} v^{\mathrm{SRG}}(\phi,Y) = \frac{\detv'}{\dot{\detv}}
\end{equation}
exists, where $\tilde{u}$ and $u$ are given by
\begin{equation}
  \label{SRG2-uTu}
  \tilde{u} = \frac{\tilde{u}_0}{\sqrt[4]{\phi}}, \qquad u = 2
  \tilde{u}_0\lambda \sqrt[4]{\phi}\,,
\end{equation}
and $\tilde{u}_0 = \const$. The Poisson tensor is
\begin{align}
  P^{ab} &= \left( -\lambda^2 - \frac{Y}{2\phi} -
    \frac{3\lambda}{32\tilde{u}_0 \phi^{5/4}} \chi^2 \right)
  \epsilon^{ab}, \label{SRG2-P-bb} \\
  P^{\alpha b} &= -\frac{1}{8\phi} X^b (\chi\gamma^3)^\alpha +
  \frac{i\lambda}{4\sqrt{\phi}} (\chi\gamma^b)^\alpha,
  \label{SRG2-P-fb} \\
  P^{\alpha\beta} &= \frac{i\tilde{u}_0}{\sqrt[4]{\phi}} X^c
  \gamma_c\^{\alpha\beta} + 2\tilde{u}_0\lambda \sqrt[4]{\phi}\,
  \gamma^3\^{\alpha\beta}. \label{SRG2-P-ff}
\end{align}
Regarding the absence of obstructions this solution is as acceptable
as (and quite similar to) (\ref{SRG1-P-bb})--(\ref{SRG1-P-ff}).
Together with the diagonal extension implied by \Sec\ref{sec:BDA} and
the nondegenerate chiral extension of \Sec\ref{sec:SUSY/2-P}, these
four solutions for the extension of the physically motivated 2d
gravity theory in themselves represent a counterexample to the
eventual hope that the requirement for nonsingular, real extensions
might yield a unique answer, especially also for a supersymmetric
$\N=1$ extension of SRG.

\subsubsection[KV, Nondiagonal Extension II]{Katanaev-Volovich Model
(KV), Nondiagonal Extension II}
\label{sec:SKV2}

Within the fermionic extension treated now also another alternative
version of the KV case may be formulated. As for SRG in
\Sec\ref{sec:SRG2} we may identify the bosonic potential (\ref{KV})
with
\begin{equation}
  \label{SKV2-v} v^{\mathrm{KV}}(\phi,Y) = \frac{\detv'}{\dot{\detv}}.
\end{equation}
Then $\tilde{u}$ and $u$ must be chosen as
\begin{align}
  \tilde{u} &= \tilde{u}_0 e^{\frac{\alpha}{2} \phi},
  \label{SKV2-uT} \\
  u &= \pm \sqrt{-2\tilde{u}_0^2 W(\phi)}, \label{SKV2-u}
\end{align}
where $\tilde{u}_0 = \const$ and $W(\phi)$ has been defined in
(\ref{SKV1-W}). Instead of (\ref{SKV1-P-bb})--(\ref{SKV1-P-ff}) we
then obtain
\begin{align}
  \Poisson^{ab} &= \left( \frac{\beta}{2} \phi^2 - \Lambda
    + \alpha Y + \half \chi^2 v_2 \right) \epsilon^{ab},
  \label{SKV2-P-bb} \\
  \Poisson^{\alpha b} &= \frac{\alpha}{4} X^b (\chi
  \gamma^3)_\alpha\^\beta - \frac{i\tilde{u}}{2u} \left( \frac{\beta}{2}
    \phi^2 - \Lambda \right) (\chi\gamma^b)_\alpha\^\beta,
  \label{SKV2-P-fb} \\
  \Poisson^{\alpha\beta} &= i \tilde{u} X^c \gamma_c\^{\alpha\beta} +
  u \gamma^3\^{\alpha\beta}. \label{SKV2-P-ff}
\end{align}
with
\begin{equation}
  \label{SKV2-v2} v_2 = -\frac{\alpha \left(
      \frac{\beta}{2} \phi^2 - \Lambda \right) + \beta \phi}{2u} -
  \frac{\tilde{u}^2 \left( \frac{\beta}{2} \phi^2 - \Lambda
    \right)^2}{2u^3},
\end{equation}
which, however, is beset with the same obstruction problems as the
nondiagonal extension I.

\section{Supergravity Actions}
\label{sec:action}

The algebras discussed in the last section have been selected in view
of their application in specific gravitational actions.

\subsection{First Order Formulation}
\label{sec:gFOG}

With the notation introduced in \Sec\ref{sec:ans-gP}, the
identification (\ref{ident1}), and after a partial integration, the
action (\ref{gPSM}) takes the explicit form (remember $e_A = (e_a,
\psi_\alpha)$)
\begin{equation}
  \label{gFOG}
  \Action^{\mathrm{gFOG}} = \int_\BMf \phi d\omega + X^a De_a +
  \chi^\alpha D\psi_\alpha + \half \Poisson^{AB} e_B e_A,
\end{equation}
where the elements of the Poisson structure by expansion in Lorentz
covariant components in the notation of \Sec\ref{sec:ans-gP} can be
expressed explicitly as (\cf (\ref{ans-P-bb})--(\ref{ans-f-bb}))
\begin{align}
  \label{ans-L} \half \Poisson^{AB} e_B e_A &= - \half U
  (\psi \gamma^3 \psi) - \frac{i}{2} \widetilde U X^a (\psi \gamma_a
  \psi) - \frac{i}{2} \widehat U X^a \epsilon_a\^b (\psi \gamma_b \psi)
  \eqnsplit + (\chi F^a e_a \psi) + \half V \epsilon^{ba} e_a e_b.
\end{align}
Here $F^a \equiv F^a\_\beta\^\gamma$, the quantity of (\ref{ans-f-bff}),
provides the direct coupling of $\psi$ and $\chi$, and $D$ is the
Lorentz covariant exterior derivative,
\begin{equation}
  \label{CovExtDer} DX^a = dX^a + X^b \omega
  \epsilon_b\^a, \qquad D\chi^\alpha = d\chi^\alpha - \half \chi^\beta
  \omega \gamma^3\_\beta\^\alpha.
\end{equation}

Of course, at this point the Jacobi identity had not been used as yet
to relate the arbitrary functions; hence the action functional
(\ref{ans-L}) is not invariant under a local supersymmetry.  On the
other hand, when the Jacobi identities restrict those arbitrary
functions, the action (\ref{gFOG}) possesses the local symmetries
(\ref{gPSM-symms}), where the parameters $\epsilon_I = (l, \epsilon_a,
\epsilon_\alpha)$ correspond to Lorentz symmetry, diffeomorphism
invariance and, in addition, to supersymmetry, respectively.
Nevertheless, already at this point we may list the explicit
supersymmetry transformations with parameter $\epsilon_I = (0, 0,
\epsilon_\alpha)$ for the scalar fields,
\begin{align}
  \delta \phi &= \half (\chi \gamma^3 \epsilon), \label{symm-ph} \\
  \delta X^a &= - (\chi F^a \epsilon), \label{symm-X-b} \\
  \delta \chi^\alpha &= U (\gamma^3 \epsilon)^\alpha
  + i \widetilde{U} X^c (\gamma_c \epsilon)^\alpha + i \widehat{U} X^d
  \epsilon_d\^c (\gamma_c \epsilon)^\alpha, \label{symm-X-f}
\end{align}
and also for the gauge fields,
\begin{align}
  \delta \omega &= U' (\epsilon \gamma^3 \psi) + i \widetilde{U}' X^b
  (\epsilon \gamma_b \psi) + i \widehat{U}' X^a \epsilon_a\^b
  (\epsilon \gamma_b \psi) + (\chi \partial_\phi F^b
  \epsilon) e_b, \label{symm-om} \\
  \delta e_a &= i \widetilde{U} (\epsilon \gamma_a \psi) + i
  \widehat{U} \epsilon_a\^b (\epsilon \gamma_b \psi) + (\chi
  \partial_a F^b \epsilon) e_b \eqnsplit + X_a \left[ \Dot{U}
    (\epsilon \gamma^3 \psi) + i \Dot{\widetilde{U}} X^b (\epsilon
    \gamma_b \psi) + i \Dot{\widehat{U}} X^b \epsilon_b\^c (\epsilon
    \gamma_c \psi) \right], \label{symm-e-b} \\
  \delta \psi_\alpha &= -D\epsilon_\alpha + (F^b \epsilon)_\alpha e_b
  \eqnsplit + \chi_\alpha \left[ u_2 (\epsilon \gamma^3 \psi) + i
    \tilde{u}_2 X^b (\epsilon \gamma_b \psi) + i \hat{u}_2 X^a
    \epsilon_a\^b (\epsilon \gamma_b \psi) \right], \label{symm-ps-f}
\end{align}
with the understanding that they represent symmetries of the action
(\ref{gFOG}) only after the relations between the still arbitrary
functions for some specific algebra are implied. The only
transformation independent of those functions is (\ref{symm-ph}).

\subsection[Elimination of the Auxiliary Fields $X^I$]{Elimination of
the Auxiliary Fields \mathversion{bold}$X^I$}
\label{sec:Legendre}

We can eliminate the fields $X^I$ by a Legendre transformation.  To
sketch the procedure, we rewrite the action (\ref{gPSM}) in a
suggestive form as Hamiltonian action principle ($d^2\!x = dx^1 \wedge
dx^0$)
\begin{equation}
  \label{calHamilton-S}
  \Action = \int_{\BMf} d^2\!x
  \left( X^I \dot{\mathcal{A}}_I - \mathcal{H}(X,A) \right),
\end{equation}
where $X^I$ should be viewed as the `momenta' conjugate to the
`velocities' $\dot{\mathcal{A}}_I = \partial_0 A_{1I} - \partial_1
A_{0I}$ and $A_{mI}$ as the `coordinates'.  Velocities
$\dot{\mathcal{A}}_I$ and the `Hamiltonian' $\mathcal{H}(X,A) =
\Poisson^{JK} A_{0K} A_{1J}$ are densities in the present definition.
The second PSM field equation (\ref{gPSM-eomA}), in the form obtained
when varying $X^I$ in (\ref{calHamilton-S}), acts as a Legendre
transformation of $\mathcal{H}(X,A)$ with respect to the variables
$X^I$,
\begin{equation}
  \label{Legendre-calAdot}
  \dot{\mathcal{A}}_I = \frac{\rpartial \mathcal{H}(X,A)}{\partial
    X^I},
\end{equation}
also justifying the interpretation of $\dot{\mathcal{A}}_I$ as
conjugate to $X^I$. When (\ref{Legendre-calAdot}) can be solved for
all $X^I$, we get $X^I = X^I(\dot{A},A)$.  Otherwise, not all of the
$X^I$ can be eliminated and additional constraints
$\Phi(\dot{\mathcal{A}},A) = 0$ emerge. In the latter situation the
constraints may be used to eliminate some of the gauge fields $A_I$ in
favour of others. When all $X^I$ can be eliminated the Legendre
transformed density
\begin{equation}
  \label{Legendre-calF}
  \mathcal{F}(\dot{\mathcal{A}},A) = X^I(\dot{\mathcal{A}},A)
  \dot{\mathcal{A}}_I - \mathcal{H}(X(\dot{\mathcal{A}},A),A)
\end{equation}
follows, as well as the second order Lagrangian action principle
\begin{equation}
  \label{Lagrange-calS} \Action = \int_{\BMf}
  d^2\!x\,\mathcal{F}(\dot{\mathcal{A}},A),
\end{equation}
where the coordinates $A_{mI}$ must be varied
independently.

The formalism already presented applies to any (graded) PSM. If there
is an additional volume form $\epsilon$ on the base manifold $\BMf$ it
may be desirable to work with functions instead of densities. This is
also possible if the volume is dynamical as in gravity theories,
$\epsilon = \epsilon(A)$, because a redefinition of the velocities
$\dot\mathcal{A}_I$ containing coordinates $A_{mI}$ but not momenta
$X^I$ is always possible, as long as we can interpret the field
equations from varying $X^I$ as Legendre transformation. In particular
we use $\dot{A}_I = \star d A_I = \epsilon^{mn} \partial_n A_{mI}$ as
velocities and $H(X,A) = \star \left( -\half \Poisson^{JK} A_K A_J
\right) = \half P^{JK} \epsilon^{mn} A_{nK} A_{mJ}$ as Hamiltonian
function, yielding
\begin{equation}
  \label{Hamilton-S} \Action = \int_{\BMf} \epsilon
  \left( X^I \dot{A}_I - H(X,A) \right).
\end{equation}
Variation of $X^I$ leads to
\begin{equation}
  \label{Legendre-Adot} \dot{A}_I = \frac{\rpartial
    H(X,A)}{\partial X^I}.
\end{equation}
Solving this equation for $X^I = X^I(\dot{A},A)$ the Legendre
transformed function $F(\dot{A},A) = X^I(\dot{A},A) \dot{A}_I -
H(X(\dot{A},A),A)$ constitutes the action
\begin{equation}
  \label{Lagrange-S} \Action = \int_{\BMf} \epsilon
  F(\dot{A},A).
\end{equation}
If the Poisson tensor is linear in the coordinates $\Poisson^{JK} =
X^I f_I\^{JK}$, where $f_I\^{JK}$ are structure constants,
(\ref{Legendre-Adot}) cannot be used to solve for $X^I$, instead the
constraint $\dot{A}_I - \half f_I\^{JK} \epsilon^{mn} A_{nK} A_{mJ} =
0$ appears, implying that the field strength of ordinary gauge theory
is zero.  For a nonlinear Poisson tensor we always have the freedom to
move $X^I$-independent terms from the \rhs to the \lhs of
(\ref{Legendre-Adot}), thus using this particular type of covariant
derivatives as velocities conjugate to the momenta $X^I$ in the
Legendre transformation. This redefinition can already be done in the
initial action (\ref{Hamilton-S}) leading to a redefinition of the
Hamiltonian $H(X,A)$.

In order to bring 2d gravity theories into the form
(\ref{Hamilton-S}), but with covariant derivatives, it is desirable to
split off $\phi$-components of the Poisson tensor and to define the
`velocities' (\cf (\ref{ep-tensor}) and (\ref{HodgeDual}) in
\App\ref{app:gravity}) as
\begin{align}
  \rho &:= \star d\omega = \epsilon^{mn} (\partial_n
  \omega_{m}), \label{Ldre-ro} \\
  \tor_a &:= \star De_a = \epsilon^{mn}
  (\partial_n e_{ma}) - \omega_a, \label{Ldre-tau} \\
  \sigma_\alpha &:= \star D\psi_\alpha = \epsilon^{mn} \left(
    \partial_n \psi_{m\alpha} + \half \omega_n (\gamma^3\psi_m)_\alpha
  \right). \label{Ldre-si}
\end{align}
Here $\rho = R/2$ is proportional to the Ricci scalar; $\tor_a$ and
$\sigma_\alpha$ are the torsion vector and the spinor built from the
derivative of the Rarita-Schwinger field, respectively. As a
consequence the Lorentz connection $\omega_m$ is absent in the
Hamiltonian,
\begin{equation}
  V(\phi,X^A;e_{mA}) = \half \Poisson^{BC}
  \epsilon^{mn} e_{nC} e_{mB} = \frac{1}{e} \Poisson^{BC} e_{0C} e_{1B},
\end{equation}
of the supergravity action
\begin{equation}
  \label{Hamilton-sugra-S}
  \Action = \int_{\BMf} \epsilon \left( \phi \rho + X^a \tor_a +
    \chi^\alpha \sigma_\alpha - V(\phi,X^A;e_{mA}) \right).
\end{equation}

\subsection{Superdilaton Theory}
\label{sec:sdil}

As remarked already in \Sec\ref{cha:intro}, first order formulations
of (bosonic) 2d gravity (and hence PSMs) allow for an---at least on
the classical level---globally equivalent description of general
dilaton theories (\ref{dil}). Here we show that this statement remains
valid also in the case of additional supersymmetric partners (\ie for
gPSMs).  We simply have to eliminate the Lorentz connection $\omega_a$
and the auxiliary field $X^a$. Of course, also the validity of an
algebraic elimination procedure in the most general case should (and
can) be checked by verifying that the correct \eom{}s also follow from
the final action (\ref{sdil-S2}) or (\ref{sdil-S3}).  (Alternatively
to the procedure applied below one may also proceed as in
\cite{Strobl:1999Habil}, performing two `Gaussian integrals' to
eliminate $X^a$ and $\tor^a$ from the action).  In fact, in the
present section we will allow also for Poisson structures
characterized by a bosonic potential $v$ not necessarily linear in $Y
\equiv \half X^a X_a$ as in (\ref{vdil}).

Variation of $X^a$ in (\ref{gFOG}) yields the torsion equation
\begin{equation}
  \label{sdil-tau} \tor_a = \half (\partial_a
  \Poisson^{AB}) \epsilon^{mn} e_{nB} e_{mA}.
\end{equation}
{}From (\ref{sdil-tau}) using $\tilde{\omega}_a := \star de_a =
\epsilon^{mn} (\partial_n e_{ma})$ and $\tor_a = \tilde{\omega}_a -
\omega_a$ (\cf (\ref{Ldre-tau}))\footnote{For specific supergravities
  it may be useful to base this separation upon a different
  SUSY-covariant Lorentz connection $\susy{\omega}_a$ (\cf
  \Sec\ref{sec:Izq} and \ref{sec:cIzq}).} we get
\begin{equation}
  \label{sdil-om}
  \omega_a = \tilde{\omega}_a - \half
  (\partial_a \Poisson^{AB}) \epsilon^{mn} e_{nB} e_{mA}.
\end{equation}
Using (\ref{sdil-om}) to eliminate $\omega_a$ the separate terms of
(\ref{gFOG}) read
\begin{align}
  \phi d\omega &= \phi d\tilde{\omega} + \epsilon
  \epsilon^{an} (\partial_n \phi) \tor_a + \text{total div.},
  \label{sdil-phdom} \\
  X^a De_a &= \epsilon X^a \tor_a, \\
  \chi^\alpha D\psi_\alpha &= \chi^\alpha \tilde{D}\psi_\alpha + \half
  \epsilon \epsilon^{an} (\chi\gamma^3\psi_n) \tor_a,
\end{align}
where $\tor_a = \tor_a(X^I,e_{mA})$ is determined by (\ref{sdil-tau}).
The action, discarding the boundary term in (\ref{sdil-phdom}),
becomes
\begin{multline}
  \label{sdil-S1} \Action = \int_\BMf d^2\!x e \biggl[
  \phi \tilde{\rho} + \chi^\alpha \tilde{\sigma}_\alpha - \half
  \Poisson^{AB} \epsilon^{mn} e_{nB} e_{mA} \\
  + \left( X^a + \epsilon^{ar} \partial_r\phi + \half \epsilon^{ar}
    (\chi\gamma^3\psi_r) \right) \half (\partial_a \Poisson^{AB})
  \epsilon^{mn} e_{nB} e_{mA} \biggr].
\end{multline}
Here $\tilde{\rho}$ and $\tilde{\sigma}$ are defined in analogy to
(\ref{Ldre-ro}) and (\ref{Ldre-si}), but calculated with
$\tilde{\omega}_a$ instead of $\omega_a$.

To eliminate $X^a$ we vary once more with respect to $\delta X^b$:
\begin{equation}
  \left[ X^a + \epsilon^{an} (\partial_n \phi) + \half
    \epsilon^{an} (\chi\gamma^3\psi_n) \right] (\partial_b\partial_a
  \Poisson^{AB}) \epsilon^{mn} e_{nB} e_{mA} = 0.
\end{equation}
For $(\partial_b\partial_a \Poisson^{AB}) \epsilon^{mn} e_{nB} e_{mA}
\neq 0$ this leads to the (again \emph{algebraic}) equation
\begin{equation}
  \label{sdil-X-b} X^a = -\epsilon^{an} \left[
    (\partial_n \phi) + \half (\chi\gamma^3\psi_n) \right]
\end{equation}
for $X^a$. It determines $X^a$ in a way which does not depend on the
specific gPSM, because (\ref{sdil-X-b}) is nothing else than the \eom
for $\delta\omega$ in (\ref{gFOG}).

We thus arrive at the superdilaton action for an arbitrary gPSM
\begin{equation}
  \label{sdil-S2}
  \Action = \int_\BMf \phi d\tilde{\omega} + \chi^\alpha
  \tilde{D}\psi_\alpha + \half \Poisson^{AB} \bigg|_{X^a} e_B e_A,
\end{equation}
where $\vert_{X^a}$ means that $X^a$ has to be replaced by
(\ref{sdil-X-b}). The action (\ref{sdil-S2}) expressed in component
fields
\begin{equation}
  \label{sdil-S3} \Action = \int_\BMf d^2\!x e \left[
    \phi \frac{\tilde{R}}{2} + \chi^\alpha \tilde{\sigma}_\alpha - \half
    \Poisson^{AB}\bigg|_{X^a} \epsilon^{mn} e_{nB} e_{mA} \right]
\end{equation}
explicitly shows the fermionic generalization of the bosonic dilaton
theory (\ref{dil}) for any gPSM. Due to the quadratic term $X^2=X^a
X_a$ in $\Poisson^{AB}$ the first term in (\ref{sdil-X-b}) provides
the usual kinetic term of the $\phi$-field in (\ref{dil}) if we take
the special case (\ref{vdil}),
\begin{align}
  \Action &= \Action^{\mathrm{dil}} + \Action^{\mathrm{f}}
  \\
  \Action^{\mathrm{f}} &= \int_\BMf d^2\!x e \biggl[ \chi^\alpha
  \tilde{\sigma}_\alpha - \frac{Z}{2} (\partial^n \phi)
  (\chi\gamma^3\psi_n) + \frac{Z}{16} \chi^2 (\psi^n \psi_n) \eqnsplit
  +\half\chi^2 v_2\bigg|_{X^a} + (\chi F^a\bigg|_{X^a} \epsilon_a\^m
  \psi_m) - \half \Poisson^{\alpha\beta}\bigg|_{X^a} \epsilon^{mn}
  \psi_{n\beta} \psi_{m\alpha} \biggr].
\end{align}
However, (\ref{sdil-S3}) even allows an arbitrary dependence on $X^2$
and a corresponding dependence on higher powers of $(\partial^n \phi)
(\partial_n \phi)$. For the special case where $\Poisson^{AB}$ is
linear in $X^a$ (\ref{sdil-S1}) shows that $X^a$ drops out of that
action without further elimination. However, the final results
(\ref{sdil-S2}) and (\ref{sdil-S3}) are the same.

\section{Actions for Particular Models}
\label{sec:models}

Whereas in \Sec\ref{sec:Poisson} a broad range of solutions of graded
Poisson algebras has been constructed, we now discuss the related
actions and their (eventual) relation to a supersymmetrization of
(\ref{grav}) or (\ref{dil}). It will be found that, in contrast to the
transition from (\ref{FOG}) to (\ref{dil}) the form (\ref{grav}) of
the supersymmetric action requires that the different functions in the
gPSM solution obey certain conditions which are not always fulfilled.

For example, in order to obtain the supersymmetrization of
(\ref{grav}), $\phi$ and $X^a$ should be eliminated by a Legendre
transformation.  This is possible only if the Hessian determinant of
$v(\phi,Y)$ with respect to $X^i=(\phi,X^a)$ is regular,
\begin{equation}
  \label{HesseDet}
  \det\left( \frac{\partial^2 v}{\partial X^i \partial X^j} \right)
  \neq 0.
\end{equation}
Even in that case the generic situation will be that no closed
expression of the form (\ref{grav}) can be given.

In the following subsections for each algebra of \Sec\ref{sec:Poisson}
the corresponding FOG action (\ref{gFOG}), the related supersymmetry
and examples of the superdilaton version (\ref{sdil-S3}) will be
presented. In the formulae for the local supersymmetry we always drop
the (common) transformation law for $\delta\phi$ (\ref{symm-ph}).

\subsection{Block Diagonal Supergravity}
\label{sec:BDS}

The action functional can be read off from (\ref{gFOG}) and
(\ref{ans-L}) for the Poisson tensor of \Sec\ref{sec:BDA} (\cf
(\ref{BDA-U}) for $U$ and (\ref{BDA-v2}) together with (\ref{ans-V})
to determine $V$). It reads
\begin{equation}
  \label{BDS-S}
  \Action^{\mathrm{BDS}} = \int_\BMf \phi d\omega + X^a De_a +
  \chi^\alpha D\psi_\alpha - \half U (\psi \gamma^3 \psi) + \half V
  \epsilon^{ba} e_a e_b,
\end{equation}
and according to (\ref{symm-ph})--(\ref{symm-ps-f}) possesses the
local supersymmetry
\begin{alignat}{3}
  & &\qquad
  \delta X^a &= 0, &\qquad
  \delta \chi^\alpha &= U (\gamma^3 \epsilon)^\alpha,
  \label{BDS-symm-X} \\
  \delta \omega &= U' (\epsilon \gamma^3 \psi), &\qquad
  \delta e_a &= X_a \Dot{U} (\epsilon \gamma^3 \psi), &\qquad
  \delta \psi_\alpha &= -D\epsilon_\alpha + \chi_\alpha u_2 (\epsilon
  \gamma^3 \psi). \label{BDS-symm-A}
\end{alignat}
This transformation leads to a translation of $\chi^\alpha$ on the
hypersurface $\Casimir = \const$ if $u \neq 0$. Comparing with the
usual supergravity type symmetry (\ref{susytrafo}) we observe that the
first term in $\delta\psi_\alpha$ has the required form
(\ref{susytrafo}), but the variation $\delta e_a$ is quite different.

The fermionic extension (\ref{BDA-C}) of the Casimir function
(\ref{Casimir}) for this class of theories implies an absolute
conservation law $\Casimir = \casimir_0 = \const$.

Whether the supersymmetric extension of the action of type
(\ref{grav}) can be reached depends on the particular choice of the
bosonic potential $v$. An example where the elimination of all target
space coordinates $\phi$, $X^a$ and $\chi^\alpha$ is feasible and
actually quite simple is $R^2$-supergravity with $v =
-\frac{\alpha}{2} \phi^2$ and $U = u_0 = \const$.\footnote{Clearly
  then no genuine supersymmetry is implied by (\ref{BDS-symm-A}). But
  we use this example as an illustration for a complete elimination of
  different combinations of $\phi$, $X^a$ and $\chi^\alpha$.} The
result in this case is (\cf \Sec\ref{sec:Legendre} and especially
(\ref{Hamilton-sugra-S}))
\begin{equation}
  \Action^{R^2} = \int_\BMf d^2\!x e \left[ \frac{1}{8\alpha}
    \tilde{R}^2 - \frac{2 u_0}{\tilde{R}} \tilde{\sigma}^\alpha
    \tilde{\sigma}_\alpha + \frac{u_0}{2} \epsilon^{nm} (\psi_m
    \gamma^3 \psi_n) \right].
\end{equation}
Here the tilde in $\tilde{R}$ and $\tilde{\sigma}^\alpha$ indicates
that the torsion-free connection $\tilde{\omega}_a = \epsilon^{mn}
\partial_n e_{ma}$ is used to calculate the field strengths. In
supergravity it is not convenient to eliminate the field $\phi$.
Instead it should be viewed as the `auxiliary field' of supergravity
and therefore remain in the action. Thus eliminating only $X^a$ and
$\chi^\alpha$ yields
\begin{equation}
  \Action^{R^2} = \int_\BMf d^2\!x e \left[ \phi \frac{\tilde{R}}{2} -
    \frac{u_0}{\alpha \phi} \tilde{\sigma}^\alpha
    \tilde{\sigma}_\alpha - \frac{\alpha}{2} \phi^2 + \frac{u_0}{2}
    \epsilon^{nm} (\psi_m \gamma^3 \psi_n) \right].
\end{equation}

Also for SRG in $d$-dimensions (\ref{SRG}) such an elimination is
possible if $d \neq 4$, $d < \infty$. However, interestingly enough,
the Hessian determinant vanishes just for the physically most relevant
dimension four (SRG) and for the DBH (\ref{DBH}), preventing in this
case a transition to the form (\ref{grav}).

The formula for the equivalent superdilaton theories (\ref{sdil-S3})
is presented for the restriction (\ref{vdil}) to quadratic torsion
only, in order to have a direct comparison with (\ref{dil}). The
choice $U = u_0 = \const$ yields
\begin{align}
  \Action^{\mathrm{BDA}} &= \Action^{\mathrm{dil}} +
  \Action^{\mathrm{f}} \\
  \intertext{with the fermionic extension}
  \Action^{\mathrm{f}} &= \int_\BMf d^2\!x e \biggl[ \chi^\alpha
  \tilde{\sigma}_\alpha - \frac{Z}{2} (\partial^n \phi)
  (\chi\gamma^3\psi_n) + \frac{Z}{16} \chi^2 (\psi^n \psi_n) \eqnsplit
  -\frac{1}{4u_0} \chi^2 \left( V' - \frac{Z'}{2} (\partial^n \phi)
    (\partial_n \phi) \right) + \frac{u_0}{2} \epsilon^{mn} (\psi_n
  \gamma^3 \psi_m) \biggr].
\end{align}
It should be noted that this model represents a superdilaton theory
for arbitrary functions $V(\phi)$ and $Z(\phi)$ in (\ref{dil}).

\subsection{Parity Violating Supergravity}
\label{sec:PVS}

The action corresponding to the algebra of \Sec\ref{sec:SUSY/2-P}
inserted into (\ref{gFOG}) and (\ref{ans-L}) becomes
\begin{multline}
  \label{PVS-S}
  \Action^{\mathrm{PVS}} = \int_\BMf \phi d\omega + X^a De_a +
  \chi^\alpha D\psi_\alpha + \epsilon \left( v - \frac{1}{4 u_0}
    \chi^2 v' \right) \\
   - \frac{i \tilde{u}_0 v}{2 u_0} (\chi \gamma^a P_{\pm} e_a \psi)
  - \frac{i \tilde{u}_0}{2} X^a (\psi \gamma_a P_{\pm} \psi) -
  \frac{u_0}{2} (\psi \gamma^3 \psi),
\end{multline}
with local supersymmetry
\begin{equation}
  \delta X^a = \frac{i \tilde{u}_0 v}{2 u_0} (\chi \gamma^a P_{\pm}
  \epsilon), \qquad
  \delta \chi^\alpha = i \tilde{u}_0 X^b (\gamma_b P_{\pm}
  \epsilon)^\alpha + u_0 (\gamma^3 \epsilon)^\alpha,
\end{equation}
as well as
\begin{align}
  \delta \omega &= -\frac{i \tilde{u}_0 v'}{2 u_0} (\chi \gamma^b
  P_{\pm} \epsilon) e_b, \\
  \delta e_a &= i \tilde{u}_0 (\epsilon \gamma_a P_{\pm} \psi) -
  \frac{i \tilde{u}_0}{2 u_0} X_a \dot{v} (\chi \gamma^b P_{\pm}
  \epsilon) e_b, \\
  \delta \psi_\alpha &= -D\epsilon_\alpha - \frac{i \tilde{u}_0 v}{2
    u_0} (\gamma^b P_{\pm} \epsilon)_\alpha e_b,
\end{align}
and the absolute conservation law (\ref{SUSY/2-C}). Here, in contrast
to the model of \Sec\ref{sec:BDS}, the transformation law of $e_a$
essentially has the `canonical' form (\ref{susytrafo}).

As seen from the action (\ref{PVS-S}) and the symmetry transformations
the two chiralities are treated differently, but we do not have the
case of a genuine chiral supergravity (\cf \Sec\ref{sec:CPA} below).

\subsection{Deformed Rigid Supersymmetry}
\label{sec:SUSY}

In this case with the algebra (\ref{SUSY-P-bb})--(\ref{SUSY-P-ff}) we
obtain
\begin{multline}
  \label{SUSY-S}
  \Action^\mathrm{DRS} = \int_\BMf \phi d\omega + X^a De_a +
  \chi^\alpha D\psi_\alpha + \epsilon v \\
  + \frac{v}{4Y} X^a (\chi \gamma^3 \gamma_a \gamma^b e_b \psi) -
  \frac{i \tilde{u}_0}{2} X^a (\psi \gamma_a \psi) - \frac{v}{16Y}
  \chi^2 (\psi \gamma^3 \psi)
\end{multline}
with local supersymmetry (\ref{symm-ph}),
\begin{equation}
  \delta X^a = -\frac{v}{4Y} X^b (\chi \gamma_b \gamma^a \gamma^3
  \epsilon), \qquad
  \delta \chi^\alpha = i \tilde{u}_0 X^b (\gamma_b \epsilon)^\alpha +
  \half\chi^2 \frac{v}{4Y} (\gamma^3 \epsilon)^\alpha,
\end{equation}
and
\begin{align}
  \delta \omega &= \left( \frac{v}{4Y} \right)' \left[ X^c
    (\chi\gamma_c\gamma^b\gamma^3\epsilon) e_b + \half\chi^2
    (\epsilon\gamma^3\psi) \right], \\
  \delta e_a &= i \tilde{u}_0 (\epsilon \gamma_a \psi) + \frac{v}{4Y}
  (\chi\gamma_a\gamma^b\gamma^3\epsilon) e_b \eqnsplit
  + X_a \left( \frac{v}{4Y} \right)\spdot \left[ X^c
    (\chi\gamma_c\gamma^b\gamma^3\epsilon) e_b + \half\chi^2
    (\epsilon\gamma^3\psi) \right], \\
  \delta \psi_\alpha &= -D\epsilon_\alpha + \frac{v}{4Y} \left[ X^c
    (\gamma_c\gamma^b\gamma^3\epsilon)_\alpha e_b + \chi_\alpha
    (\epsilon\gamma^3\psi) \right].
\end{align}

Clearly, this model exhibits a `genuine' supergravity symmetry
(\ref{susytrafo}). As pointed out already in \Sec\ref{sec:SUSY-P} the
bosonic potential $v(\phi,Y)$ is not restricted in any way by the
super-extension. However, a new singularity of the action functional
occurs at $Y = 0$. The corresponding superdilaton theory can be
derived along the lines of \Sec\ref{sec:sdil}.

\subsection{Dilaton Prepotential Supergravities}
\label{sec:Izq}

In its FOG version the action from (\ref{gFOG}) with
(\ref{Izq-P-bb})--(\ref{Izq-P-ff}) reads
\begin{multline}
  \label{Izq-S}
  \Action^{\mathrm{DPA}} = \int_\BMf \phi d\omega + X^a De_a +
  \chi^\alpha D\psi_\alpha - \frac{1}{2\tilde{u}_0^2} \epsilon \left(
    (u^2)' - \half \chi^2 u'' \right) \\
   + \frac{iu'}{2\tilde{u}_0} (\chi \gamma^a e_a \psi) - \frac{i
     \tilde{u}_0}{2} X^a (\psi \gamma_a \psi) - \frac{u}{2} (\psi
   \gamma^3 \psi),
\end{multline}
where $\tilde{u}_0=\const$ and the `prepotential' $u$ depends on
$\phi$ only. The corresponding supersymmetry becomes (\ref{symm-ph}),
\begin{equation}
  \label{Izq-symm-X}
  \delta X^a = -\frac{iu'}{2\tilde{u}_0} (\chi\gamma^a\epsilon),
  \qquad
  \delta \chi^\alpha = i \tilde{u}_0 X^b (\gamma_b \epsilon)^\alpha +
  u (\gamma^3 \epsilon)^\alpha,
\end{equation}
and further
\begin{align}
  \delta \omega &= u' (\epsilon \gamma^3 \psi) +
  \frac{iu''}{2\tilde{u}_0} (\chi\gamma^b\epsilon) e_b,
  \label{Izq-symm-om} \\
  \delta e_a &= i \tilde{u}_0 (\epsilon \gamma_a \psi),
  \label{Izq-symm-e-b} \\
  \delta \psi_\alpha &= -D\epsilon_\alpha + \frac{iu'}{2\tilde{u}_0}
  (\gamma^b\epsilon)_\alpha e_b. \label{Izq-symm-ps-f}
\end{align}

Here we have the special situation of an action linear in $X^a$,
described at the end of \Sec\ref{sec:sdil}. Variation of $X^a$ in
(\ref{Izq-S}) leads to the constraint $De_a - \frac{i\tilde{u}_0}{2}
(\psi\gamma_a\psi) = 0$.  It can be used to eliminate the Lorentz
connection, \ie $\omega_a = \susy{\omega}_a$, where we introduced the
SUSY-covariant connection
\begin{equation}
  \label{Izq-om}
  \susy{\omega}_a := \epsilon^{mn} (\partial_n e_{ma}) +
  \frac{i\tilde{u}_0}{2} \epsilon^{mn} (\psi_n\gamma_a\psi_m).
\end{equation}
The action reads
\begin{multline}
  \label{Izq-S1}
  \Action^{\mathrm{DPA}} = \int_\BMf d^2\!x e \biggl[ \phi
  \frac{\susy{R}}{2} + \chi^\alpha \susy{\sigma}_\alpha -
  \frac{1}{2\tilde{u}_0^2} \left( (u^2)' - \half \chi^2 u'' \right) \\
  + \frac{iu'}{2\tilde{u}_0} (\chi \gamma^m \epsilon_m\^n \psi_n) +
  \frac{u}{2} \epsilon^{mn}(\psi_n \gamma^3 \psi_m) \biggr],
\end{multline}
where $\susy{R}$ and $\susy{\sigma}$ indicate that these covariant
quantities are built with the spinor dependent Lorentz connection
(\ref{Izq-om}).

The present model is precisely the one studied in \cite{\bibIzq}.  In
\Sec\ref{sec:sdil-sol} we give the explicit solution of the PSM field
equations for this model. The $R^2$ model and the model of Howe will
be treated in a little more detail now.

\subsubsection[$R^2$ Model]{\mathversion{bold}$R^2$ Model}
\label{sec:R2}


The supersymmetric extension of $R^2$-gravity, where $v =
-\frac{\alpha}{2} \phi^2$, is obtained with the general solution $u =
\pm \tilde{u}_0 \sqrt{\frac{\alpha}{3} (\phi^3 - \phi_0^3)}$ (\cf
(\ref{Izq-u})).

In order to simplify the analysis we choose $u = \tilde{u}_0
\sqrt{\frac{\alpha}{3} \phi^3}$.  The parameter $\alpha$ can have both
signs, implying the restriction on the range of the dilaton field such
that $\alpha \phi > 0$. Thus the superdilaton action (\ref{Izq-S1})
becomes
\begin{multline}
  \label{R2-S}
  \Action^{R^2} = \int_\BMf d^2\!x e \biggl[ \phi \frac{\susy{R}}{2} +
  \chi^\alpha \susy{\sigma}_\alpha - \frac{\alpha}{2} \phi^2 +
  \frac{1}{16\tilde{u}_0} \chi^2 \sqrt{\frac{3\alpha}{\phi}} \\
  + \frac{3i\alpha}{4} \sqrt{\frac{\phi}{3\alpha}}\,
  (\chi\gamma^m\epsilon_m\^n\psi_n) + \frac{\tilde{u}_0}{2}
  \sqrt{\frac{\alpha}{3} \phi^3}\, \epsilon^{mn}
  (\psi_n\gamma^3\psi_m) \biggr].
\end{multline}

The equation obtained when
varying $\chi^\alpha$ yields
\begin{equation}
  \label{R2-ch}
  \chi^\alpha = -8 \tilde{u}_0 \sqrt{\frac{\phi}{3\alpha}}\,
  \susy{\sigma}^\alpha - 2i \tilde{u}_0 (\psi^n \epsilon_n\^m
  \gamma_m)^\alpha.
\end{equation}
Eliminating the $\chi^\alpha$ field gives
\begin{multline}
  \Action^{R^2} = \int_\BMf d^2\!x e \biggl[ \phi \frac{\susy{R}}{2} -
  4 \tilde{u}_0 \sqrt{\frac{\phi}{3\alpha}}\, \susy{\sigma}^\alpha
  \susy{\sigma}_\alpha - 2i \tilde{u}_0 \phi (\susy{\sigma}
  \gamma^m \epsilon_m\^n \psi_n) - \frac{\alpha}{2} \phi^2 \\
  + \frac{\tilde{u}_0}{4} \sqrt{\frac{\alpha}{3} \phi^3}\, \left( 3
    (\psi^n \psi_n) - \epsilon^{mn} (\psi_n\gamma^3\psi_m) \right)
  \biggr].
\end{multline}
Further elimination of $\phi$ requires the solution of a cubic
equation for $\sqrt{\phi}$ with a complicated explicit solution,
leading to an equally complicated supergravity generalization of the
formulation (\ref{grav}) of this model.

\subsubsection{Model of Howe}
\label{sec:Howe}

The supergravity model of Howe \cite{Howe:1979ia}, originally derived
in terms of superfields, is just a special case of our generic model
in the graded PSM approach. Using for the various independent
potentials the particular values
\begin{equation}
  \label{Howe-pot1}
  \tilde{u}_0 = -2, \qquad u = -\phi^2
\end{equation}
we obtain $\detv = 8 Y - \phi^4$ and for the other nonzero potentials
(\cf (\ref{Izq-pot1}), (\ref{Izq-pot2}))
\begin{equation}
  \label{Howe-pot2}
  v = -\half \phi^3, \qquad
  v_2 = -\frac{1}{4}, \qquad
  f_{(s)} = \half \phi.
\end{equation}


The Lagrange density for this model in the formulation (\ref{FOG}) is
a special case of (\ref{Izq-S}):
\begin{multline}
  \label{Howe-S}
  \Action^{\mathrm{Howe}} = \int_\BMf \phi d\omega + X^a De_a +
  \chi^\alpha D\psi_\alpha \\ 
  + \half \phi^2 (\psi \gamma^3 \psi) + i X^a (\psi \gamma_a \psi) +
  \frac{i}{2} \phi (\chi \gamma^a e_a \psi) - \half \epsilon
  \left( \phi^3 + \frac{1}{4} \chi^2 \right).
\end{multline}

The local supersymmetry transformations from
(\ref{Izq-symm-X})--(\ref{Izq-symm-ps-f}) are
\begin{equation}
  \label{Howe-symm-X}
  \delta X^a = -\frac{i}{2} \phi (\chi \gamma^a \epsilon), \qquad
  \delta \chi^\alpha = -\phi^2 (\gamma^3 \epsilon)^\alpha - 2i X^b
  (\gamma_b \epsilon)^\alpha,
\end{equation}
and
\begin{align}
  \delta \omega &= -2 \phi (\epsilon \gamma^3 \psi) + \frac{i}{2} (\chi
  \gamma^b \epsilon) e_b, \label{Howe-symm-om} \\
  \delta e_a &= -2i (\epsilon \gamma_a \psi), \label{Howe-symm-e-b} \\
  \delta \psi_\alpha &= -D\epsilon_\alpha + \frac{i}{2} \phi (\gamma^b
  \epsilon)_\alpha e_b. \label{Howe-symm-ps-f}
\end{align}


Starting from the dilaton action (\ref{Izq-S1}) with (\ref{Howe-pot1})
and (\ref{Howe-pot2}), the remaining difference to the formulation of
Howe is the appearance of the fermionic coordinate $\chi^\alpha$. Due
to the quadratic term of $\chi^\alpha$ in (\ref{Howe-S}) we can use
its own algebraic field equations to eliminate it. Applying the
Hodge-dual yields
\begin{equation}
  \label{Howe-ch}
  \chi_\alpha = 4 \susy{\sigma}_\alpha + 2i \phi (\gamma^n
  \epsilon_n\^m \psi_m)_\alpha.
\end{equation}
Inserting this into the Lagrange density (\ref{Howe-S}) and into the
supersymmetry transformations (\ref{symm-ph}) as well as into
(\ref{Howe-symm-X})--(\ref{Howe-symm-ps-f}) and identifying $\phi$
with the scalar, usually interpreted as auxiliary field $A$, $\phi
\equiv A$, reveals precisely the supergravity model of Howe. That
model, in a notation almost identical to the one used here, is also
contained in \cite{Ertl:1997ib}, where a superfield approach was used.

\subsection{Supergravities with Quadratic Bosonic Torsion}
\label{sec:cIzq}

The algebra (\ref{cIzq-P-bb-alt})--(\ref{cIzq-P-ff-alt}) in
(\ref{gFOG}) leads to
\begin{multline}
  \label{cIzq-S}
  \Action^{\mathrm{QBT}} = \int_\BMf \phi d\omega + X^a De_a +
  \chi^\alpha D\psi_\alpha + \epsilon \left( V + \half X^a X_a Z +
    \half\chi^2 v_2 \right) \\
  + \frac{Z}{4} X^a (\chi\gamma^3\gamma_a\gamma^b e_b \psi) -
  \frac{i\tilde{u}_0 V}{2u} (\chi \gamma^a e_a \psi) \\
  - \frac{i \tilde{u}_0}{2} X^a (\psi \gamma_a \psi) - \frac{1}{2}
  \left( u + \frac{Z}{8} \chi^2 \right) (\psi \gamma^3 \psi),
\end{multline}
with $u(\phi)$ determined from $V(\phi)$ and $Z(\phi)$ according to
(\ref{cIzq-u}) and
\begin{equation}
  v_2 = -\frac{1}{2u} \left( V Z + V' + \frac{\tilde{u}_0^2 V^2}{u^2}
  \right).
\end{equation}

The special interest in models of this type derives from the fact that
because of their equivalence to the dilaton theories with dynamical
dilaton field (\cf \Sec\ref{sec:sdil}) they cover a large class of
physically interesting models. Also, as shown in \Sec\ref{sec:cIzq-P}
these models are connected by a simple conformal transformation to
theories without torsion, discussed in \Sec\ref{sec:Izq}.

Regarding the action (\ref{cIzq-S}) it should be kept in mind that
calculating the prepotential $u(\phi)$ we discovered the condition
$W(\phi) < 0$ (\cf (\ref{cIzq-u})). This excludes certain bosonic
theories from supersymmetrization with real actions, and it may lead
to restrictions on $\phi$, but there is even more information
contained in this inequality: It leads also to a restriction on $Y$.
Indeed, taking into account (\ref{c-sol}) we find
\begin{equation}
  \label{cIzq-resY}
  Y > \casimir(\phi,Y) e^{-Q(\phi)}.
\end{equation}

The local supersymmetry transformations of the action (\ref{cIzq-S})
become (\ref{symm-ph}),
\begin{align}
  \delta X^a &=  -\frac{Z}{4} X^b
  (\chi\gamma_b\gamma^a\gamma^3\epsilon) + \frac{i\tilde{u}_0 V}{2u}
  (\chi\gamma^a\epsilon), \label{cIzq-symm-X-b} \\
  \delta \chi^\alpha &= i \tilde{u}_0 X^b (\gamma_b \epsilon)^\alpha +
  \left( u + \frac{Z}{8} \chi^2 \right) (\gamma^3 \epsilon)^\alpha,
  \label{cIzq-symm-X-f}
\end{align}
and
\begin{align}
  \delta \omega &= \left( -\frac{\tilde{u}_0^2 V}{u} - \frac{Zu}{2} +
    \frac{Z'}{8} \chi^2 \right) (\epsilon \gamma^3 \psi) +
  \frac{Z'}{4} X^a (\chi\gamma^3\gamma_a\gamma^b\epsilon) e_b
  \eqnsplit\hspace{8.5em} - \frac{\tilde{u}_0}{2u} \left( V' +
    \frac{VZ}{2} + \frac{\tilde{u}_0^2 V^2}{u^2} \right)
  (\chi\gamma^b\epsilon) e_b, \label{cIzq-symm-om} \\
  \delta e_a &= i \tilde{u}_0 (\epsilon \gamma_a \psi) + \frac{Z}{4}
  (\chi\gamma^3\gamma_a\gamma^b\epsilon) e_b, \label{cIzq-symm-e-b} \\
  \delta \psi_\alpha &= -D\epsilon_\alpha + \frac{Z}{4} X^a
  (\gamma^3\gamma_a\gamma^b\epsilon)_\alpha e_b - \frac{i\tilde{u}_0
    V}{2u} (\gamma^b\epsilon)_\alpha e_b + \frac{Z}{4} \chi_\alpha
  (\epsilon\gamma^3\psi). \label{cIzq-symm-ps-f}
\end{align}

Finally, we take a closer look at the torsion condition. Variation of
$X^a$ in (\ref{cIzq-S}) yields
\begin{equation}
  \label{cIzq-tor}
  De_a - \frac{i\tilde{u}_0}{2} (\psi\gamma_a\psi) + \frac{Z}{4}
  (\chi\gamma^3\gamma_a\gamma^b e_b\psi) + \epsilon X_a Z = 0.
\end{equation}
For $Z \neq 0$ this can be used to eliminate $X^a$ directly, as
described in \Sec\ref{sec:Legendre} for a generic PSM.  The general
procedure to eliminate instead $\omega_a$ by this equation was
outlined in \Sec\ref{sec:sdil}. There, covariant derivatives were
expressed in terms of $\tilde{\omega}_a$.  For supergravity theories
it is desirable to use SUSY-covariant derivatives instead. The
standard covariant spinor dependent Lorentz connection
$\susy{\omega}_a$ was given in (\ref{Izq-om}).  However, that quantity
does not retain its SUSY-covariance if torsion is dynamical. Eq.\ 
(\ref{cIzq-tor}) provides such a quantity. Taking the Hodge dual,
using (\ref{Ldre-tau}) we find
\begin{align}
  \omega_a &= \susy{\omega}_a + X_a Z, \label{cIzq-om} \\
  \susy{\omega}_a &\equiv \tilde{\omega}_a + \frac{i\tilde{u}_0}{2}
  \epsilon^{mn} (\psi_n\gamma_a\psi_m) + \frac{Z}{4}
  (\chi\gamma^3\gamma_a\gamma^b\epsilon_b\^n\psi_n). \label{cIzq-s-om}
\end{align}
Clearly, $\omega_a$ possesses the desired properties (\cf
(\ref{gPSM-symms})), but it is not the minimal covariant connection.
The last term in (\ref{cIzq-om}) is a function of the target space
coordinates $X^I$ only, thus covariant by itself, which leads to the
conclusion that $\susy{\omega}_a$ is the required quantity.
Unfortunately, no generic prescription to construct $\susy{\omega}_a$
exists however.  The rest of the procedure of \Sec\ref{sec:sdil} for
the calculation of a superdilaton action starting with
(\ref{sdil-phdom}) still remains valid, but with $\susy{\omega}_a$ of
(\ref{cIzq-s-om}) replacing $\tilde{\omega}_a$.

We point out that it is essential to have the spinor field
$\chi^\alpha$ in the multiplet; just as $\phi$ has been identified
with the usual auxiliary field of supergravity in \Sec\ref{sec:Howe},
we observe that general supergravity (with torsion) requires an
additional auxiliary spinor field $\chi^\alpha$.

\subsubsection{Spherically Reduced Gravity (SRG)}
\label{sec:SRG}

The special case (\ref{SRG}) with $d=4$ for the potentials $V$ and $Z$
in (\ref{cIzq-S}) yields
\begin{multline}
  \label{SRG-S}
  \Action^{\mathrm{SRG}} = \int_\BMf \phi d\omega + X^a De_a +
  \chi^\alpha D\psi_\alpha - \epsilon \left( \lambda^2 +
    \frac{1}{4\phi} X^a X_a + \frac{3\lambda}{32\tilde{u}_0
      \phi^{3/2}} \chi^2 \right) \\
  - \frac{1}{8\phi} X^a (\chi\gamma^3\gamma_a\gamma^b e_b \psi) +
  \frac{i\lambda}{4 \sqrt{\phi}} (\chi \gamma^a e_a \psi) \\
  - \frac{i \tilde{u}_0}{2} X^a (\psi \gamma_a \psi) - \frac{1}{2}
  \left( 2\tilde{u}_0 \lambda \sqrt{\phi} - \frac{1}{16\phi} \chi^2
  \right) (\psi \gamma^3 \psi).
\end{multline}
We do not write down the supersymmetry transformations which follow
from (\ref{cIzq-symm-X-b})--(\ref{cIzq-symm-ps-f}).
We just note that our transformations $\delta e_a$ and $\delta
\psi_\alpha$ are different from the ones obtained in
\cite{Park:1993sd}. There, the supergravity multiplet is the same as
in the underlying model \cite{Howe:1979ia}, identical to the one of
\Sec\ref{sec:Howe}. The difference is related to the use of an
additional scalar superfield in \cite{Park:1993sd} to construct a
superdilaton action. Such an approach lies outside the scope of the
present thesis, where we remain within pure gPSM without additional
fields which, from the point of view of PSM are `matter' fields.

Here, according to the general derivation of \Sec\ref{sec:sdil}, we
arrive at the superdilaton action
\begin{align}
  \Action^\mathrm{SRG} &= \Action^{\mathrm{dil}} +
  \Action^{\mathrm{f}}, \label{SRG-dilS1} \\
  \intertext{with bosonic part (\ref{dil}) and the fermionic extension}
  \Action^{\mathrm{f}} &= \int_\BMf d^2\!x e \biggl[ \chi^\alpha
  \tilde{\sigma}_\alpha + i\tilde{u}_0 \left\{ (\partial^n \phi) +
    \half (\chi\gamma^3\psi^n) \right\} (\psi_n\gamma^m\psi_m)
  \eqnspl + \frac{1}{8\phi} (\partial^n \phi)
  (\chi\gamma^3\gamma^m\gamma_n\psi_m) - \frac{i\lambda}{4\sqrt{\phi}}
  (\chi\gamma^3\gamma^m\psi_m) + \tilde{u}_0 \lambda \sqrt{\phi}\,
  \epsilon^{mn} (\psi_n\gamma^3\psi_m) \eqnsplit - \frac{1}{32} \chi^2
  \left\{ \frac{1}{\phi} (\psi^n\psi_n) + \frac{1}{\phi}
    (\psi^n\gamma_n\gamma^m\psi_m) + \frac{3\lambda}{\tilde{u}_0
      \phi^{3/2}} + \epsilon^{mn} (\psi_n\gamma^3\psi_m) \right\}
  \biggr].
\end{align}

However, as already noted in the previous section, it may be
convenient to use the SUSY-covariant $\susy{\omega}_a$ (\cf
(\ref{cIzq-s-om})) instead of $\tilde{\omega}_a$, with the result:
\begin{multline}
  \label{SRG-dilS2}
  \Action^{\mathrm{SRG}} = \int_\BMf \phi d\susy{\omega} + \chi^\alpha
  \susy{D}\psi_\alpha + \frac{\epsilon}{4\phi}
  \biggl[ (\partial^n \phi) (\partial_n \phi) + (\partial^n \phi)
    (\chi\gamma^3\psi_n) + \frac{1}{8} \chi^2 (\psi^n \psi_n) \biggr] \\
  - \epsilon \biggl[ \lambda^2 + \frac{3\lambda}{32\tilde{u}_0
      \phi^{3/2}} \chi^2 \biggr]
  + \frac{i\lambda}{4 \sqrt{\phi}} (\chi \gamma^a e_a \psi) -
  \frac{1}{2} \biggl[ 2\tilde{u}_0 \lambda \sqrt{\phi} -
    \frac{1}{16\phi} \chi^2 \biggr] (\psi \gamma^3 \psi).
\end{multline}

\subsubsection{Katanaev-Volovich Model (KV)}
\label{sec:SKV}

The supergravity action (\ref{cIzq-S}) for the algebra of
\Sec\ref{sec:SKV1} reads
\begin{multline}
  \label{SKV-S}
  \Action^{\mathrm{KV}} = \int_\BMf d^2\!x e \biggl[ \phi R + X^a
  \tor_a + \chi^\alpha \sigma_\alpha + \frac{\beta}{2} \phi^2 -
  \Lambda + \frac{\alpha}{2} X^a X_a + \half\chi^2 v_2 \\
  + \frac{\alpha}{4} X^a (\chi \gamma_a \gamma^m \psi_m) -
  \frac{i\tilde{u}_0 \left( \frac{\beta}{2} \phi^2 - \Lambda
  \right)}{2u} (\chi\gamma^m\epsilon_m\^n\psi_n) \\
  + \frac{i\tilde{u}_0}{2} X^a \epsilon^{mn} (\psi_n\gamma_a\psi_m) +
  \half \left( u + \frac{\alpha}{8} \chi^2 \right) \epsilon^{mn}
  (\psi_n\gamma^3\psi_m) \biggr],
\end{multline}
where $v_2$ and $u$ were given in (\ref{SKV1-v2}) and (\ref{SKV1-u}).
It is invariant under the local supersymmetry transformations
(\ref{symm-ph}) and
\begin{align}
  \delta X^a &= -\frac{\alpha}{4} X^b
  (\chi\gamma_b\gamma^a\gamma^3\epsilon) + \frac{i\tilde{u}_0 \left(
      \frac{\beta}{2} \phi^2 - \Lambda \right)}{2u}
  (\chi\gamma^a\epsilon), \\
  \delta \chi^\alpha &= i \tilde{u}_0 X^b (\gamma_b \epsilon)^\alpha +
  \left( u + \frac{\alpha}{8} \chi^2 \right) (\gamma^3
  \epsilon)^\alpha,
\end{align}
in conjunction with the transformations
\begin{align}
  \delta \omega &= \left[ -\frac{\tilde{u}_0^2 \left(
      \frac{\beta}{2} \phi^2 - \Lambda \right)}{u} - \frac{\alpha u}{2}
  \right] (\epsilon \gamma^3 \psi) \eqnsplit - \frac{\tilde{u}_0}{2u}
  \left[ \beta\phi + \frac{\alpha \left( \frac{\beta}{2} \phi^2 -
      \Lambda \right)}{2} + \frac{\tilde{u}_0^2 \left(
      \frac{\beta}{2} \phi^2 - \Lambda \right)^2}{u^2} \right]
  (\chi\gamma^b\epsilon) e_b, \\
  \delta e_a &= i \tilde{u}_0 (\epsilon \gamma_a \psi) +
  \frac{\alpha}{4} (\chi\gamma^3\gamma_a\gamma^b\epsilon) e_b, \\
  \delta \psi_\alpha &= -D\epsilon_\alpha + \frac{\alpha}{4} X^a
  (\gamma^3\gamma_a\gamma^b\epsilon)_\alpha e_b
  -\frac{i\tilde{u}_0 \left( \frac{\beta}{2} \phi^2 - \Lambda
  \right)}{2u}  (\gamma^b\epsilon)_\alpha e_b + \frac{\alpha}{4} \chi_\alpha
  (\epsilon\gamma^3\psi)
\end{align}
of the gauge fields.

The explicit formula for the superdilaton action may easily be
obtained here, the complicated formulae, however, are not very
illuminating. It turns out that $\delta\omega$ and $\delta\psi_\alpha$
contain powers $u^k$, $k = -3,\ldots,1$. Therefore, singularities
related to the prepotential formulae (\ref{cIzq-u}) also affect these
transformations.

\subsection[$\N=(1,0)$ Dilaton Supergravity]%
{\mathversion{bold}$\N=(1,0)$ Dilaton Supergravity}
\label{sec:CPA}

The case where one chiral component in $\chi^\alpha$ (say $\chi^-$)
decouples from the theory is of special interest among the degenerate
algebras of \Sec\ref{sec:DFS-P} and \ref{sec:CPA-P}.

We adopt the solution (\ref{CPA-P-bb})--(\ref{CPA-F-b}) for the
Poisson algebra accordingly.  To avoid a cross term of the form
$(\chi^- \psi_+)$ appearing in $\Poisson^{a\beta} \psi_\beta e_a =
(\chi F^a e_a \psi)$ we have to set $f_{(s)} = f_{(t)} = 0$. Similarly
$f_{(12)} = 0$ cancels a $(\chi^-\chi^+)$-term in
$\half\Poisson^{\alpha\beta} \psi_\beta \psi_\alpha$. Furthermore we
choose $\tilde{u} = \tilde{u}_0 = \const$:
\begin{multline}
  \label{CPA-S}
  \Action = \int_\BMf \phi d\omega + X^{++} De_{++} + X^{--} De_{--} +
  \chi^+ D\psi_+ + \chi^- D\psi_- \\
  + \epsilon v + \frac{v}{2Y} X^{--} e_{--} (\chi^+ \psi_+ - \chi^-
  \psi_-) + \frac{\tilde{u}_0}{\sqrt{2}} X^{++} \psi_+ \psi_+
\end{multline}
The chiral components $\chi^-$ and $\psi_-$ can be set to zero
consistently. The remaining local supersymmetry has one parameter
$\epsilon_+$ only:
\begin{alignat}{2}
  \delta \phi &= \half (\chi^+ \epsilon_+), \label{CPA-symm-ph} \\
  \delta X^{++} &= 0, &\qquad
  \delta X^{--} &= -\frac{v}{2Y} X^{--} (\chi^+ \epsilon_+),
  \label{CPA-symm-X-b} \\
  \delta \chi^+ &= \sqrt{2} \tilde{u}_0 X^{++} \epsilon_+, &\qquad
  \delta \chi^- &= 0 \label{CPA-symm-X-f},
\end{alignat}
and
\begin{align}
  \delta \omega &= \frac{v'}{2Y} X^{--} e_{--} (\chi^+ \epsilon_+),
  \label{CPA-symm-om} \\
  \delta e_{++} &= -\sqrt{2} \tilde{u}_0 (\epsilon_+ \psi_+) + X_{++}
  \left( \frac{v}{2Y} \right)\spdot X^{--} e_{--} (\chi^+ \epsilon_+),
  \label{CPA-symm-e++} \\
  \delta e_{--} &= \frac{v}{2Y} e_{--} (\chi^+ \epsilon_+) + X_{--}
  \left( \frac{v}{2Y} \right)\spdot X^{--} e_{--} (\chi^+ \epsilon_+),
  \label{CPA-symm-e--} \\
  \delta \psi_+ &= -D\epsilon_+ + \frac{v}{2Y} X^{--} e_{--}, \qquad
  \delta \psi_- = 0. \label{CPA-symm-ps-f}
\end{align}

\section{Solution of the Dilaton Supergravity Model}
\label{sec:sdil-sol}

\newcommand{\chTp}{\tilde{\chi}^{(+)}}
\newcommand{\chTm}{\tilde{\chi}^{(-)}}
\newcommand{\chTpm}{\tilde{\chi}^{(\pm)}}
\newcommand{\chTT}{\Tilde{\Tilde{\chi}}^{(-)}}
\newcommand{\chSm}{\susy{\chi}^{(-)}}

For the dilaton prepotential supergravity model of \Sec\ref{sec:Izq}
\cite{\bibIzq}
the Poisson algebra was derived in \Sec\ref{sec:Izq-P}. The PSM field
equations (\ref{gPSM-eomX}) and (\ref{gPSM-eomA}) simplify
considerably in Casimir-Darboux coordinates which can be found
explicitly here, as in the pure gravity PSM. This is an
improvement as compared to \cite{Strobl:1999zz} where only the
\emph{existence} of such target coordinates was used.

We start with the Poisson tensor (\ref{Izq-P-bb})--(\ref{Izq-P-ff}) in
the coordinate system $X^I = (\phi,X^{++},X^{--},\chi^+,\chi^-)$. The
algebra under consideration has maximal rank $(2|2)$, implying that
there is one bosonic Casimir function $\Casimir$. Rescaling
(\ref{Izq-C}) we choose here
\begin{equation}
  \label{co-C}
  \Casimir = X^{++} X^{--} - \frac{u^2}{2\tilde{u}_0^2} + \half\chi^2
  \frac{u'}{2\tilde{u}_0^2}\,.
\end{equation}
In regions $X^{++} \neq 0$ we use $\Casimir$ instead of $X^{--}$ as
coordinate in target space ($X^{--} \rightarrow \Casimir$). In regions
$X^{--} \neq 0$ $X^{++} \rightarrow \Casimir$ is possible.

Treating the former case explicitly we replace $X^{++} \rightarrow
\lambda = -\ln|X^{++}|$ in each of the two patches $X^{++} > 0$ and
$X^{++} < 0$.  This function is conjugate to the generator of Lorentz
transformation $\phi$ (\cf (\ref{LorentzB}))
\begin{equation}
  \label{PB-la-ph}
  \{\lambda,\phi\} = 1.
\end{equation}
The functions $(\phi,\lambda,\Casimir)$ constitute a Casimir-Darboux
coordinate system for the bosonic sector \cite{\bibPSM}. Now our aim
is to decouple the bosonic sector from the fermionic one. The
coordinates $\chi^\alpha$ constitute a Lorentz spinor (\cf
(\ref{LorentzF})). With the help of $X^{++}$ they can be converted to
Lorentz scalars, \ie $\{\chTpm, \phi\} = 0$,
\begin{equation}
  \label{tr-ScalF}
  \chi^+ \rightarrow \chTp = \frac{1}{\sqrt{|X^{++}|}}
  \chi^+, \qquad
  \chi^- \rightarrow \chTm = \sqrt{|X^{++}|} \chi^-.
\end{equation}
A short calculation for the set of coordinates $(\phi, \lambda,
\Casimir, \chTp, \chTm)$ shows that $\{\lambda, \chTp\} = 0$ but
$\{\lambda, \chTm\} = -\frac{\sigma u'}{\sqrt{2} \tilde{u}_0} \chTp$,
where $\sigma := \sign(X^{++})$. The redefinition
\begin{equation}
  \label{tr-BD-F}
  \chTm \rightarrow \chTT = \chTm +
  \frac{\sigma u}{\sqrt{2}\tilde{u}_0} \chTp
\end{equation}
yields $\{\lambda, \chTT\} = 0$ and makes the algebra block diagonal.
For the fermionic sector
\begin{align}
  \{\chTp,\chTp\} &= \sigma \sqrt{2}
  \tilde{u}_0, \label{PB-F+F+} \\
  \{\chTT,\chTT\} &= \sigma \sqrt{2} \tilde{u}_0 \Casimir,
  \label{PB-F-F-} \\
  \{\chTp,\chTT\} &= 0 \label{PB-F+F-}
\end{align}
is obtained.  We first assume that the Casimir function $\Casimir$
appearing explicitly on the \rhs of (\ref{PB-F-F-}) is invertible.
Then the redefinition
\begin{equation}
  \label{tr-C-F-}
  \chTT \rightarrow \chSm = \frac{1}{\sqrt{|\Casimir|}}
  \chTT
\end{equation}
can be made. That we found the desired Casimir-Darboux system can be
seen from
\begin{equation}
  \label{PB-sF-sF-}
  \{\chSm,\chSm\} = s \sigma \sqrt{2}
  \tilde{u}_0.
\end{equation}
Here $s := \sign(\Casimir)$ denotes the sign of the Casimir function.
In fact we could rescale $\chSm$ and $\chTp$ so as to reduce the
respective right hand sides to $\pm 1$; the signature of the fermionic
$2 \times 2$ block cannot be changed however. In any case, we call the
local coordinates $\bar{X}^I := (\phi, \lambda, \Casimir, \chTp,
\chSm)$ Casimir-Darboux since the respective Poisson tensor is
constant, which is the relevant feature here.  Its non-vanishing
components can be read off from (\ref{PB-la-ph}), (\ref{PB-F+F+}) and
(\ref{PB-sF-sF-}).

Now it is straightforward to solve the PSM field equations.  Bars are
used to denote the gauge fields $\bar{A}_I = (\bar{A}_\phi,
\bar{A}_\lambda, \bar{A}_C, \bar{A}_{(+)}, \bar{A}_{(-)})$
corresponding to the coordinates $\bar{X}^I$.  The first PSM \eom{}s
(\ref{gPSM-eomX}) then read
\begin{align}
  d\phi - \bar{A}_\lambda &= 0, \label{Izq-eomPh} \\
  d\lambda + \bar{A}_\phi &= 0, \label{Izq-eomLa} \\
  d\Casimir &= 0, \label{Izq-eomC} \\
  d\chTp + \sigma \sqrt{2} \tilde{u}_0 \bar{A}_{(+)} &= 0,
  \label{Izq-eomCh+} \\
  d\chSm + s \sigma \sqrt{2} \tilde{u}_0 \bar{A}_{(-)} &=
  0. \label{Izq-eomCh-}
\end{align}
These equations decompose in two parts (which is true also in the case
of several Casimir functions). In regions where the Poisson tensor is
of constant rank we obtain the statement that any solution of
(\ref{gPSM-eomX}) and (\ref{gPSM-eomA}) lives on symplectic leaves
which is expressed here by the differential equation (\ref{Izq-eomC})
with the one-parameter solution $\Casimir = \casimir_0 = \const$.
Eqs.\ (\ref{Izq-eomPh})--(\ref{Izq-eomCh-}) without (\ref{Izq-eomC})
are to be used to solve for all gauge fields excluding the ones which
correspond to the Casimir functions, thus excluding $\bar{A}_C$ in our
case. Note that this solution is purely algebraic.  The second set of
PSM equations (\ref{gPSM-eomA}) again split in two parts. The
equations $d\bar{A}_\phi = d\bar{A}_\lambda = d\bar{A}_{(+)} =
d\bar{A}_{(-)} = 0$ are identically fulfilled, as can be seen from
(\ref{Izq-eomPh})--(\ref{Izq-eomCh-}). To show this property in the
generic case the first PSM equations (\ref{gPSM-eomX}) in conjunction
with the Jacobi identity of the Poisson tensor have to be used. The
remainder of the second PSM equations are the equations for the gauge
fields corresponding to the Casimir functions. In a case as simple as
ours we find, together with the local solution in terms of an
integration function $F(x)$ (taking values in the commuting
supernumbers),
\begin{equation}
  \label{Izq-A-C}
  d\bar{A}_C = 0 \quad \Rightarrow \quad \bar{A}_C = -dF.
\end{equation}

The explicit solution for the original gauge fields $A_I = (\omega,
e_a, \psi_\alpha)$ is derived from the target space transformation
$A_I = (\partial_I \bar{X}^J) \bar{A}_J$, but in order to compare with
the case $\Casimir = 0$ we introduce an intermediate step and give the
solution in coordinates $\tilde{X}^I = (\phi, \lambda, \Casimir,
\chTp, \chTT)$ first. With $\tilde{A}_I = (\tilde{A}_\phi,
\tilde{A}_\lambda, \tilde{A}_C, \tilde{A}_{(+)}, \tilde{A}_{(-)})$ the
calculation $\tilde{A}_I = (\tilde{\partial}_I \bar{X}^J) \bar{A}_J$
yields
\begin{equation}
  \label{Izq-solA1}
  \tilde{A}_\phi = -d\lambda, \qquad
  \tilde{A}_\lambda = d\phi, \qquad
  \tilde{A}_{(+)} = -\frac{\sigma}{\sqrt{2}\tilde{u}_0} d\chTp,
\end{equation}
and
\begin{equation}
  \label{Izq-solA2a}
  \tilde{A}_C = -dF + \frac{\sigma}{2 \sqrt{2}\tilde{u}_0 \Casimir^2}
  \chTT d\chTT, \qquad
  \tilde{A}_{(-)} = -\frac{\sigma}{\sqrt{2}\tilde{u}_0 \Casimir}
  d\chTT.
\end{equation}

This has to be compared with the case $\Casimir = 0$. Obviously, the
fermionic sector is no longer of full rank, and $\chTT$ is an
additional, fermionic Casimir function as seen from (\ref{PB-F-F-})
and also from the corresponding field equation
\begin{equation}
  \label{Izq-eomChBD}
  d\chTT = 0.
\end{equation}
Thus, the $\tilde{X}^I$ are Casimir-Darboux coordinates on the
subspace $\Casimir = 0$. The PSM \eom{}s in barred coordinates still
are of the form (\ref{Izq-eomPh})--(\ref{Izq-eomCh+}).  Therefore, the
solution (\ref{Izq-solA1}) remains unchanged, but (\ref{Izq-solA2a})
has to be replaced by the solution of $d\tilde{A}_C = 0$ and
$d\tilde{A}_{(-)} = 0$. In terms of the bosonic function $F(x)$ and an
additional fermionic function $\rho(x)$ the solution is
\begin{equation}
  \label{Izq-solA2b}
  \tilde{A}_C = -dF, \qquad
  \tilde{A}_{(-)} = -d\rho.
\end{equation}

Collecting all formulae the explicit solution for the original gauge
fields $A_I = (\omega, e_a, \psi_\alpha)$ calculated with $A_I =
(\partial_I \tilde{X}^J) \tilde{A}_J$ reads (\cf
\App\ref{app:spin-comp} for the definition of $++$ and $--$ components
of Lorentz vectors)
\begin{align}
  \omega &= \frac{d X^{++}}{X^{++}} + V \tilde{A}_C +
  \frac{\sigma u'}{\sqrt{2}\tilde{u}_0} \chTp \tilde{A}_{(-)},
  \label{Izq-solOm} \\
  e_{++} &= -\frac{d\phi}{X^{++}} + X^{--}
  \tilde{A}_C \eqnsplit + \frac{1}{2 X^{++}} \left[
    \frac{\sigma}{\sqrt{2}\tilde{u}_0} \chTp d\chTp + \left( \chTm -
      \frac{\sigma u}{\sqrt{2}\tilde{u}_0} \chTp \right) \tilde{A}_{(-)}
  \right], \label{Izq-solE++} \\
  e_{--} &= X^{++} \tilde{A}_C, \label{Izq-solE--} \\
  \psi_+ &= -\frac{u'}{2\tilde{u}_0^2} \chi^-
  \tilde{A}_C - \frac{\sigma}{\sqrt{2}\tilde{u}_0}
  \frac{1}{\sqrt{|X^{++}|}} \left( d\chTp - u \tilde{A}_{(-)} \right),
  \label{Izq-solPs+} \\
  \psi_- &= \frac{u'}{2\tilde{u}_0^2} \chi^+
  \tilde{A}_C + \sqrt{|X^{++}|} \tilde{A}_{(-)}. \label{Izq-solPs-}
\end{align}
$\tilde{A}_C$ and $\tilde{A}_{(-)}$ are given by (\ref{Izq-solA2a})
for $\Casimir \neq 0$ and by (\ref{Izq-solA2b}) for $\Casimir = 0$.

For $\Casimir \neq 0$ our solution depends on the free function $F$
and the coordinate functions $(\phi, X^{++}, X^{--}, \chi^+, \chi^-)$
which, however, are constrained by $\Casimir = \casimir_0 = \const$
according to (\ref{co-C}). For $\Casimir = 0$ the free functions are
$F$ and $\rho$. The coordinate functions $(\phi, X^{++}, X^{--},
\chi^+, \chi^-)$ here are restricted by $\Casimir = 0$ in (\ref{co-C})
and by $\chTT = \const$.

This solution holds for $X^{++} \neq 0$; an analogous set of relations
can be derived exchanging the role of $X^{++}$ and $X^{--}$.

The solution (\cf (\ref{Izq-solE++}) and (\ref{Izq-solE--})) is free
from coordinate singularities in the line element, exhibiting a sort
of `super Eddington-Finkelstein' form. For special choices of the
potentials $v(\phi)$ or the related prepotential $u(\phi)$ we refer to
\Tab\ref{tab:IzqMdls}.

This provides also the solution for the models with quadratic bosonic
torsion of \Sec\ref{sec:cIzq} by a further change of variables
(\ref{conf-tr-X}) with parameter (\ref{cIzq-tr}). Its explicit form is
calculated using (\ref{conf-tr-A}).

As of now we did not use the gauge freedom. Actually, in supergravity
theories this is generically not that easy, since the fermionic part
of the symmetries are known only in their \emph{infinitesimal} form
(the bosonic part corresponds on a global level to diffeomorphisms and
local Lorentz transformations). This changes in the present context
for the case that Casimir-Darboux coordinates are available. Indeed,
for a constant Poisson tensor the otherwise field-dependent, nonlinear
symmetries (\ref{gPSM-symms}) can be integrated easily: Within the
range of applicability of the target coordinates, $\bar{X}^I$ may be
shifted by some arbitrary function (except for the Casimir function
$\Casimir$, which, however, was found to be constant over $\BMf$), and
$\bar{A}_C$ may be redefined by the addition of some exact part. The
only restrictions to these symmetries come from nondegeneracy of the
metric (thus \eg $\tilde{A}_C$ should not be put to zero, \cf
(\ref{Izq-solE--})). In particular we are thus allowed to put both
$\chTp$ and $\chSm$ to zero in the present patch, and thus, if one
follows back the transformations introduced, also the original fields
$\chi^\pm$. Next, in the patches with $X^{++} \neq 0$, we may fix the
local Lorentz invariance by $X^{++} := 1$ and the diffeomorphism
invariance by choosing $\phi$ and $F$ as local coordinates on the
spacetime manifold $\BMf$. The resulting gauge fixed solution agrees
with the one found in the original bosonic theory, \cf \eg
\cite{Klosch:1996fi}. This is in agreement with the general
considerations of \cite{Strobl:1999zz}, here however made explicit.

A final remark concerns the discussion following (\ref{Izq-u}): As
noted there, for some choices of the bosonic potential $v$ the
potential $u$ becomes complex valued if the range of $\phi$ is not
restricted appropriately. It is straightforward to convince oneself
that the above formulae are still valid in the case of complex valued
potentials $u$ (\ie complex valued Poisson tensors). Just in
intermediary steps, such as (\ref{tr-BD-F}), one uses complex valued
fields (with some reality constraints). The final gauge fixed
solution, however, is not affected by this, being real as it should
be.


\chapter{PSM with Superfields}
\label{cha:SPSM}


So far the question is still open how the supergravity models derived
using the graded PSM approach (\cf \Cha\ref{cha:gPSM} above) fit
within the superfield context. The former approach relies on a Poisson
tensor of a target space, but in the superfield context of
\Cha\ref{cha:sfield} there was no such structure, even worse, there
was no target space at all. The connection can be established by
extending the gPSM to superspace.

\section{Outline of the Approach}
\label{sec:sPSM-outline}

\subsection{Base Supermanifold}
\label{sec:BSMf}

It is suggestive to enlarge the two-dimensional base manifold $\BMf$
with coordinates $x^m$ to become a four-dimensional supermanifold
$S_\BMf$ parametrized by coordinates $z^M = (x^m, \theta^\mu)$.  We do
not modify the target space $\TSp$, where we retain still the
coordinates $X^I = (\phi, X^a, \chi^\alpha)$ and the Poisson tensor
$\Poisson^{IJ}(X)$ as in \Cha\ref{cha:gPSM}. When we consider the map
$S_\BMf \to \TSp$ in coordinate representation, we obtain the
superfields $X^I(z) = X^I(x, \theta)$, which we expand as usual
\begin{equation}
  \label{eq:sX}
  X^I(z) = X^I(x) + \theta^\mu X_\mu\^I(x) + \half \theta^2
  X_{2}\^I(x).
\end{equation}
It is important to note that there is no change of the Poisson tensor
of the model. If we want to calculate the equations of the PSM
apparatus the only thing we have to do with the Poisson tensor is to
evaluate it at $X(z)$, \ie $\Poisson^{IJ}(X(z))$. 

In the PSM with base manifold $\BMf$ we had the gauge fields $A_I$.
These former 1-forms on $\BMf$ are now promoted to 1-superforms on
$S_\BMf$,
\begin{equation}
  \label{eq:sAI}
  A_I(z) = dz^M A_{MI}(z).
\end{equation}
The enlargement of the base manifold thus added further gauge fields
to the model. Beside the already known ones,
\begin{equation}
  \label{eq:sAmI}
  A_{mI} = (\omega_m, e_{ma}, \RS_{m \alpha}),
\end{equation}
the new fields
\begin{equation}
  \label{eq:sAmuI}
  A_{\mu I} = (\omega_\mu, e_{\mu a}, \RS_{\mu \alpha})
\end{equation}
are obtained.

Because we don't know the PSM action for base manifolds of dimension
other than two, nor do we know the action for base supermanifolds, the
further analysis rests solely on the properties of the PSM equations
of motion. The observation that neither the closure of the PSM field
equations nor its symmetries depend on the underlying base manifold
but only on the Jacobi identities of the Poisson tensor makes this
possible. The field equations in superspace read
\begin{gather}
  dX^I + \Poisson^{IJ} A_J = 0, \label{eq:sPSM-eom1} \\
  dA_I + \half (\partial_I \Poisson^{JK}) A_K A_J = 0,
  \label{eq:sPSM-eom2}
\end{gather}
and the symmetry transformations with parameters $(\epsilon_I) = (l,
\epsilon_a, \epsilon_\alpha)$, which are now functions on $S_\BMf$,
$\epsilon_I = \epsilon_I(z)$, are given by
\begin{align}
  \delta X^I &= \Poisson^{IJ} \epsilon_J, \label{eq:sPSM-symm1} \\
  \delta A_I &= -d\epsilon_I - (\partial_I \Poisson^{JK}) \epsilon_K
  A_J. \label{eq:sPSM-symm2}
\end{align}

Collecting indices by introducing a superindex $A = (a,\alpha)$ the
supervielbein
\begin{equation}
  \label{eq:svielbein}
  E_{MA} = \mtrx{e_{ma}}{\RS_{m\alpha}}{e_{\mu a}}{\RS_{\mu\alpha}}
\end{equation}
can be formed. The vielbein $e_m\^a$ therein is invertible per
definition and we assume further that $\RS_\mu\^\alpha$ possesses
this property too.  Lorentz indices are raised and lowered with metric
$\eta_{ab}$, spinor indices with $\epsilon_{\alpha\beta}$ so that the
space spanned by the coordinates $X^A$ is equipped with the direct
product supermetric
\begin{equation}
  \label{eq:supermetric}
  \eta_{AB} = \mtrx{\eta_{ab}}{0}{0}{\epsilon_{\alpha\beta}}.
\end{equation}
This allows to raise and lower $A$-type indices, $E_M\^A = \eta^{AB}
E_{MB}$. The inverse supervielbein fulfills $E_A\^M E_M\^B =
\delta_A\^B$.

\subsection{Superdiffeomorphism}
\label{sec:sPSM-diffeo}

Our next task is to recast superdiffeomorphism of the base manifold
$S_\BMf$ into PSM symmetry transformations with parameter
$\epsilon_I$. Let
\begin{equation}
  \label{eq:sdiffeo}
  \delta z^M = -\xi^M(z)
\end{equation}
be the infinitesimal displacement of points on $S_\BMf$ by a
vector superfield
\begin{equation}
  \label{eq:svectorfield}
  \xi^M(z) = \xi^M(x) + \theta^\mu \xi_\mu\^M(x) + \half \theta^2
  \xi_2\^M(x).
\end{equation}
The transformations of the fields $X^I(z)$ under (\ref{eq:sdiffeo}),
\begin{equation}
  \label{eq:diffeo-XI}
  \delta X^I(z) = -\xi^M(z) \partial_M X^I(z),
\end{equation}
can be brought to the form (\ref{eq:sPSM-symm1}) by taking the field
equations (\ref{eq:sPSM-eom1}) into account, in components they read
$\partial_M X^I = -\Poisson^{IJ} A_{MJ} (-1)^{M(I+J)}$, and by the
choice
\begin{equation}
  \label{eq:transf-XI}
  \epsilon_I(z) = \xi^M(z) A_{MI}(z)
\end{equation}
for the PSM symmetry parameters. To confirm (\ref{eq:transf-XI}) the
transformations $\delta A_I$ have to be considered in order to rule
out the arbitrariness $\epsilon_I \rightarrow \epsilon_I + \partial_I
\Casimir$ which does not change the transformation rules
(\ref{eq:sPSM-symm1}) of $X^I$.

If we want to go the other way and choose $\epsilon_A(z)$ as the
primary transformation parameters, the corresponding $\xi^M(z)$ can be
derived from $\epsilon_A(z) = \xi^M(z) E_{MA}(z)$ due to the
invertibility of the supervielbein, but note that we have to
accommodate for an additional $\epsilon_A(z)$-dependent Lorentz
transformation in order to recast superdiffeomorphism:
\begin{align}
  \xi^M(\epsilon_A(z); z) &= \epsilon^A(z) E_A\^M(z),
  \label{eq:diffeo-from-symm1} \\
  l(\epsilon_A(z); z) &= \epsilon^A(z) E_A\^M(z) \omega_M(z).
  \label{eq:diffeo-from-symm2}
\end{align}

\subsection{Component Fields}
\label{sec:sPSM-comp}

The component fields of supergravity in the superfield formulation can
be restored by solving the PSM field equations partly. This can be
achieved by separating the $\partial_\nu$ derivatives in the PSM field
equations (\ref{eq:sPSM-eom1}) and (\ref{eq:sPSM-eom2}). From
(\ref{eq:sPSM-eom1})
\begin{equation}
  \label{eq:sPSM-eom1-f-der}
  \partial_\nu X^I(z) + \Poisson^{IJ}(X(z)) A_{\nu J}(z) (-1)^{I+J} = 0
\end{equation}
is obtained. To zeroth order in $\theta$ this equation expresses
$X_\mu\^I(x)$ (\cf (\ref{eq:sX})) in terms of $X^I(x)$ and $A_{\nu
  J}(x) = A_{\nu J}(x, 0)$. For the same purpose we choose
\begin{gather}
  \partial_\nu A_{\mu I}(z) + \partial_\mu A_{\nu I}(z) - \partial_I
  \Poisson^{JK}|_{X(z)} A_{\mu K}(z) A_{\nu J}(z) (-1)^K = 0,
  \label{eq:sPSM-eom2-f-der1} \\
  \partial_\nu A_{mI}(z) + \partial_m A_{\nu I}(z) + \partial_I
  \Poisson^{JK}|_{X(z)} A_{mK}(z) A_{\nu J}(z) (-1)^{I+J} = 0
  \label{eq:sPSM-eom2-f-der2}
\end{gather}
from the field equations (\ref{eq:sPSM-eom2}). The equations
(\ref{eq:sPSM-eom1-f-der}), (\ref{eq:sPSM-eom2-f-der1}) and
(\ref{eq:sPSM-eom2-f-der2}) can be solved simultaneously order by
order in $\theta$, therefore expressing higher order
$\theta$-components of $X^I(z)$, $A_{mI}(z)$ and $A_{\mu I}(z)$ in
terms of $X^I(x)$, $A_{mI}(x)$, $A_{\mu I}(x)$, $(\partial_{[\nu}
A_{\mu]I})(x)$ and $(\partial_m A_{\mu I})(x)$, $(\partial_m
\partial_{[\nu} A_{\mu]I})(x)$.

Counting one $x$-space field as one degree of freedom there are $16$
independent symmetry parameters in the vector superfield $\xi^M(z)$,
the same gauge degree of freedom as in $\epsilon_A(z) =
(\epsilon_a(z), \epsilon_\alpha(z))$, and there are $4$ independent
parameters in the Lorentz supertransformation $l(z) =
\epsilon_\phi(z)$. In summary $20$ gauge degrees of freedom are found
in the set of superfield parameters $\epsilon_I(z) = (l(z),
\epsilon_a(z), \epsilon_\alpha(z))$. This huge amount of $x$-space
symmetries can be broken by gauging some component fields of the newly
added fields (\ref{eq:sAmuI}). In particular we impose the conditions
\begin{alignat}{3}
  \omega_{\mu}|_{\theta=0} &= 0, &\qquad e_{\mu a}|_{\theta=0} &= 0,
  &\qquad \RS_{\mu\alpha}|_{\theta=0} &= \epsilon_{\mu\alpha},
  \label{eq:gfAmuI0} \\
  \partial_{[\nu} \omega_{\mu]}|_{\theta=0} &= 0, &\qquad
  \partial_{[\nu} e_{\mu] a}|_{\theta=0} &= 0, &\qquad \partial_{[\nu}
  \RS_{\mu] \alpha}|_{\theta=0} &= 0.
  \label{eq:gfAmuI1}
\end{alignat}
This gauge conditions turn out to be identical to the ones of the
superfield formulation \cf\Sec\ref{sec:Howe-symms} and
\ref{sec:ertl-gf}.  

Counting the number of gauged $x$-space fields we obtain $10$ for
(\ref{eq:gfAmuI0}) and $5$ for (\ref{eq:gfAmuI1}), thus reducing the
number of independent $x$-field symmetry parameters to $5$. These
transformations are given by $\epsilon_I(x) = \epsilon_I(x,
\theta=0)$, where local Lorentz rotation, local translation and local
supersymmetry can be identified within the parameter set
$\epsilon_I(x) = (l(x), \epsilon_a(x), \epsilon_\alpha(x))$.
Alternatively, one could use $\xi^M(x)$ instead of $\epsilon_A(x)$
(\cf \Sec\ref{sec:sPSM-diffeo} before) to account for translation and
supersymmetry. All $\theta$-components of $\epsilon_I(z)$ can be
determined from the PSM symmetries (\ref{eq:sPSM-symm2}),
\begin{equation}
  \label{eq:sPSM-symm2-f-der}
  \delta A_{\mu I}(z) = -\partial_\mu \epsilon_I(z) - \partial_I
  \Poisson^{JK}|_{X(z)} \epsilon_K A_{\mu J} (-1)^{I+J},
\end{equation}
when the gauge fixing conditions (\ref{eq:gfAmuI0}) and
(\ref{eq:gfAmuI1}) for $A_{\mu I}$ are taken into account.

\section{Rigid Supersymmetry}
\label{sec:rigid-SUSY}

The Poisson tensor of rigid supersymmetry is
\begin{alignat}{2}
  \Poisson^{a\phi} &= X^b \epsilon_b\^a, &\qquad \Poisson^{\alpha
    \phi} &= -\half \chi^\beta \gamma^3\_\beta\^\alpha, \\
  \Poisson^{\alpha\beta} &= 2i \tilde{c} X^a \gamma_a\^{\alpha\beta}
  + 2 c \gamma^3\^{\alpha\beta}, \\
  \Poisson^{ab} &= 0, &\qquad \Poisson^{a\beta} &= 0,
\end{alignat}
where $c$ and $\tilde{c}$ are constant.

Choosing the gauge conditions (\ref{eq:gfAmuI0}) and
(\ref{eq:gfAmuI1}) we can solve (\ref{eq:sPSM-eom1-f-der}) for
$X^I(z)$ at once. In order to make notations more transparent we drop
the arguments of the functions and place a hat above a symbol, if it
is meant to be a function of $z$, otherwise with no hat it is a
function of $x$ only. We get
\begin{align}
  \hat{\phi} &= \phi + \half (\theta \gamma^3 \chi) - \half \theta^2 c, \\
  \hat{X}^a &= X^a, \\
  \hat{\chi}^\alpha &= \chi^\alpha + 2i \tilde{c} X^a (\theta
  \gamma_a)^\alpha + 2 c (\theta \gamma^3)^\alpha.
\end{align}
Solving (\ref{eq:sPSM-eom2-f-der1}) yields $A_{\mu I}(z)$,
\begin{align}
  \hat{\omega}_\mu &= 0, \\
  \hat{e}_{\mu a} &= -i \tilde{c} (\theta \gamma_a)_\mu, \\
  \hat{\RS}_{\mu\alpha} &= \epsilon_{\mu\alpha},
\end{align}
and finally we get the component fields of $A_{mI}(z)$ using
(\ref{eq:sPSM-eom2-f-der2}),
\begin{align}
  \hat{\omega}_m &= \omega_m, \\
  \hat{e}_{ma} &= e_{ma} - 2i \tilde{c} (\theta \gamma_a \RS_m), \\
  \hat{\RS}_{m\alpha} &= \RS_{m\alpha} - \half \omega_m (\theta
  \gamma^3)_\alpha.
\end{align}

The inverse of the supervielbein $\hat{E}_M\^A$, \cf
(\ref{eq:svielbein}), is given by $\hat{E}_A\^M$ with the components
\begin{align}
  \hat{E}_a\^m &= e_a\^m + i \tilde{c} (\theta \gamma^m \RS_a) -
  \half \theta^2\, \tilde{c}^2 (\RS_a \gamma^n \gamma^m \RS_n), \\
  \hat{E}_a\^\mu &= -\RS_a\^\mu -i \tilde{c} (\theta \gamma^n \RS_a)
  \RS_n\^\mu + \half \omega_a (\theta \gamma^3)^\mu \eqnsplit + \half
  \theta^2 \left[ \tilde{c}^2 (\RS_a \gamma^n \gamma^m \RS_n)
    \RS_m\^\mu + \frac{i}{2} \tilde{c} \omega_n (\RS_a \gamma^n
    \gamma^3)^\mu \right], \\
  \hat{E}_\alpha\^m &= i \tilde{c} (\theta \gamma^m)_\alpha - \half
  \theta^2\, \tilde{c}^2 (\gamma^n \gamma^m \RS_n)_\alpha, \\
  \hat{E}_\alpha\^\mu &= \delta_\alpha\^\mu -i \tilde{c} (\theta
  \gamma^n)_\alpha \RS_n\^\mu + \half \theta^2 \left[ \tilde{c}^2
    (\gamma^n \gamma^m \RS_n)_\alpha \RS_m\^\mu + \frac{i}{2}
    \tilde{c} \omega_n (\gamma^n \gamma^3)_\alpha\^\mu \right].
\end{align}

The chosen gauge restricts the symmetry parameters $\epsilon_I(z)$ due
to the variations (\ref{eq:sPSM-symm2-f-der}), from which we get
\begin{align}
  \hat{l} &= l, \\
  \hat{\epsilon}_a &= \epsilon_a + 2i \tilde{c} (\epsilon \gamma_a
  \theta), \\
  \hat{\epsilon}_\alpha &= \epsilon_\alpha - \half l (\theta
  \gamma^3)_\alpha.
\end{align}

Using (\ref{eq:diffeo-from-symm1}) and (\ref{eq:diffeo-from-symm2}) we
calculate the world sheet transformations corresponding to above
parameters. For parameter $\epsilon_a(x)$ only, we get
\begin{equation}
  \hat{\xi}^m = \epsilon^a \hat{E}_a\^m, \qquad
  \hat{\xi}^\mu = \epsilon^a \hat{E}_a\^\mu, \qquad
  \hat{l} = \epsilon^a \hat{E}_a\^m \omega_m.
\end{equation}
\begin{align}
  \hat{\xi}^m &= \epsilon^a e_a\^m - i \tilde{c} \epsilon^a (\RS_a
  \gamma^m \theta) - \half \theta^2\, \tilde{c}^2 \epsilon^a (\RS_a
  \gamma^n \gamma^m \RS_n), \label{eq:diffeo-a-m} \\
  \hat{\xi}^\mu &= \half \epsilon^a \omega_a (\theta \gamma^3)^\mu -
  \epsilon^a \RS_a\^\mu + i \tilde{c} \epsilon^a (\RS_a \gamma^n
  \theta) \RS_n\^\mu \eqnsplit + \half \theta^2 \left[ \tilde{c}^2
    \epsilon^a (\RS_a \gamma^n \gamma^m \RS_n) \RS_m\^\mu +
    \frac{i}{2} \tilde{c} \omega_n \epsilon^a (\RS_a \gamma^n
    \gamma^3)^\mu \right], \label{eq:diffeo-a-mu} \\
  \hat{l} &= \epsilon^a \omega_a - i \tilde{c} \epsilon^a (\RS_a
  \gamma^m \theta) \omega_m - \half \theta^2\, \tilde{c}^2 \epsilon^a
  (\RS_a \gamma^n \gamma^m \RS_n) \omega_m. \label{eq:diffeo-a-l}
\end{align}
Making a supersymmetry transformation with parameter
$\epsilon_\alpha(x)$ yields
\begin{align}
  \hat{\xi}^m &= i \tilde{c} (\epsilon \gamma^m \theta) + \half
  \theta^2\, \tilde{c}^2 (\epsilon \gamma^n \gamma^m \RS_n), \\
  \hat{\xi}^\mu &= \epsilon^\mu - i \tilde{c} (\epsilon \gamma^n
  \theta) \RS_n\^\mu + \half \theta^2 \left[ -\tilde{c}^2 (\epsilon
    \gamma^n \gamma^m \RS_n) \RS_m\^\mu - \frac{i}{2} \tilde{c}
    \omega_n (\epsilon \gamma^n \gamma^3)^\mu \right], \\
  \hat{l} &= i \tilde{c} (\epsilon \gamma^m \theta) \omega_m + \half
  \theta^2\, \tilde{c}^2 (\epsilon \gamma^n \gamma^m \RS_n) \omega_m.
\end{align}
Finally, for Lorentz transformations with parameter $l(x)$ the correct
representation in superspace is
\begin{equation}
  \hat{\xi}^m = 0, \qquad
  \hat{\xi}^\mu = -\half l (\theta \gamma^3)^\mu, \qquad
  \hat{l} = l.
\end{equation}

When analyzing (\ref{eq:diffeo-a-m})--(\ref{eq:diffeo-a-l}) we
recognize that a $\epsilon_a(x)$-transformation creates a pure bosonic
diffeomorphism on the world sheet, $\delta x^m = -\eta^m(x)$, with
parameter $\eta^m(x) = \epsilon^a(x) e_a\^m(x)$, a Lorentz
transformation with parameter $l(x) = \epsilon^a(x) \omega_a(x)$ and a
supersymmetry transformation with parameter $\epsilon^\alpha(x) =
-\epsilon^a(x) \RS_a\^\alpha(x)$. We can go the other way and impose a
bosonic diffeomorphism $\delta x^m = -\eta^m(x)$. Then using
$\hat{\epsilon}_I = \eta^m \hat{A}_{mI}$ (\cf (\ref{eq:transf-XI})) we
obtain
\begin{align}
  \hat{l} &= \eta^m \omega_m, \\
  \hat{\epsilon}_a &= \eta^m e_{ma} + 2i \tilde{c} \eta^m (\RS_m
  \gamma_a \theta), \\
  \hat{\epsilon}_\alpha &= \eta^m \RS_{m\alpha} - \half \eta^m
  \omega_m (\theta \gamma^3)_\alpha,
\end{align}
where we see that we not only get an $\epsilon_a$-transformation but
also a Lorentz and a supersymmetry transformation.

Finally we note the transformations of the scalars $X^I(x)$,
\begin{align}
  \delta \phi &= -X^b \epsilon_b\^a \epsilon_a - \half (\epsilon
  \gamma^3
  \chi), \\
  \delta X^a &= l X^b \epsilon_b\^a, \\
  \delta \chi^\alpha &= -\half l (\chi \gamma^3)^\alpha - 2i \tilde{c}
  X^a (\epsilon \gamma_a)^\alpha - 2 c (\epsilon \gamma^3)^\alpha,
\end{align}
and the transformations of the gauge fields $A_{mI}(x)$,
\begin{align}
  \delta \omega_m &= -\partial_m l, \\
  \delta e_{ma} &= -D_m \epsilon_a - l \epsilon_a\^b e_{mb} + 2i
  \tilde{c} (\epsilon \gamma_a \RS_m), \\
  \delta \RS_{m\alpha} &= -D_m \epsilon_\alpha + \half l (\gamma^3
  \RS_m)_\alpha.
\end{align}
We can see that there is nothing new, because this is just the result
that we had for a PSM with the purely bosonic base manifold $\BMf$.
The same is true for the remaining superspace PSM equations. They are
just the same as in the simpler model with world sheet $\BMf$.

\section{Supergravity Model of Howe}
\label{sec:sPSM-Howe}

In order to establish the connection between the superfield method of
\Cha\ref{cha:sfield} and the graded PSM approach of \Cha\ref{cha:gPSM}
explicitly we choose the well-known supergravity model of Howe
\cite{Howe:1979ia}. It was derived with superfield methods, further
developments can be found in \cite{\bibHoweOther}. For the purposes
needed here a summary of the superfield expressions for that model was
given in \Sec\ref{sec:sugra-howe}.

The corresponding Poisson tensor for the Howe model has been derived
in \Cha\ref{cha:gPSM}:
\begin{align}
  \Poisson^{\alpha\beta} &= i \tilde{u}_0 X^a \gamma_a\^{\alpha\beta}
  + \tilde{u}_0 \lambda \phi^2 \gamma^3\^{\alpha\beta} \\
  \Poisson^{\alpha b} &= i \lambda \phi (\chi \gamma^b)^\alpha \\
  \Poisson^{ab} &= \left( -2 \lambda^2 \phi^3 + \frac{\lambda}{2
      \tilde{u}_0} \chi^2 \right) \epsilon^{ab}
\end{align}
Here $\lambda$ and $\tilde{u}_0$ are constant. For $\Poisson^{a\phi}$
and $\Poisson^{\alpha \phi}$ we refer to (\ref{P-Lorentz}) in
\Sec\ref{sec:ans-gP}, where also the ansatz of the Poisson tensor
(\ref{ans-P-bb})--(\ref{v-ff}) defining the various potentials can be
found. The Poisson tensor was derived in \Sec\ref{sec:Izq-P}, in
particular the prepotential $u(\phi)$ was given in
\Tab\ref{tab:IzqMdls}. The corresponding PSM action and its symmetries
are to be found in \Sec\ref{sec:Howe}.

With the Wess-Zumino type gauge conditions (\ref{eq:gfAmuI0}) and
(\ref{eq:gfAmuI1}) the PSM field equation (\ref{eq:sPSM-eom1-f-der})
can be solved for $X^I(z)$ at once. In order to make notations more
concise the arguments of the functions are dropped and replaced by a
hat above a symbol to signalize that it is a function of $z$,
otherwise with no hat it is a function of $x$ only. We obtain
\begin{align}
  \hat{\phi} &= \phi + \half (\theta \gamma^3 \chi) -
  \frac{\lambda\tilde{u}_0}{4} \theta^2 \phi^2, \\
  \hat{X}^a &= X^a  - i \lambda \phi (\theta \gamma^a \chi) -
  \frac{\lambda\tilde{u}_0}{2} \theta^2 \phi X^a, \\
  \hat{\chi}^\alpha &= \chi^\alpha + i \tilde{u}_0 X^a (\theta
  \gamma_a)^\alpha + \lambda\tilde{u}_0 \phi^2 (\theta
  \gamma^3)^\alpha + \frac{3 \lambda\tilde{u}_0}{4} \theta^2 \phi
  \chi^\alpha.
\end{align}
Solving (\ref{eq:sPSM-eom2-f-der1}) yields $A_{\mu I}(z)$,
\begin{align}
  \hat{\omega}_\mu &= -\lambda\tilde{u}_0 \phi (\theta \gamma^3)_\mu,
  \label{sPSM-Howe-om-f} \\
  \hat{e}_{\mu a} &= -\frac{i \tilde{u}_0}{2} (\theta \gamma_a)_\mu,
  \label{sPSM-Howe-e-fb} \\
  \hat{\RS}_{\mu\alpha} &= \left( 1 + \frac{\lambda\tilde{u}_0}{4}
    \theta^2 \phi \right) \epsilon_{\mu\alpha},
  \label{sPSM-Howe-ps-ff}
\end{align}
and $A_{mI}(z)$ derived from (\ref{eq:sPSM-eom2-f-der2}) reads,
\begin{align}
  \hat{\omega}_m &= \omega_m + i \lambda (\theta \gamma_m \chi) - 2
  \lambda\tilde{u}_0 \phi (\theta \gamma^3 \RS_m) \eqnsplit
  \hspace{4em} +
  \frac{\lambda\tilde{u}_0}{2} \theta^2 \left[ X_m - \phi \omega_m
    + \half (\chi \RS_m) \right], \label{sPSM-Howe-om-b} \\
  \hat{e}_{ma} &= e_{ma} - i \tilde{u}_0 (\theta \gamma_a \RS_m) -
  \frac{\lambda\tilde{u}_0}{2} \theta^2 \phi e_{ma},
  \label{sPSM-Howe-e-bb} \\
  \hat{\RS}_{m\alpha} &= \RS_{m\alpha} - \half \omega_m (\theta
  \gamma^3)_\alpha + i \lambda \phi (\theta \gamma_m)_\alpha \eqnsplit
  \hspace{6em} + \half\theta^2 \left[ \frac{3 \lambda\tilde{u}_0}{2}
    \phi \RS_{m\alpha} - \frac{i \lambda}{2} (\chi \gamma_m
    \gamma^3)_\alpha \right]. \label{sPSM-Howe-ps-bf}
\end{align}

The choice $\lambda = \half$ and $\tilde{u}_0 = -2$ together with the
identification of the dilaton as the supergravity auxiliary field
$\phi = A$ (\cf also \Sec\ref{sec:Howe}) should lead to the superfield
expressions obtained in \Sec\ref{sec:Howe-comp}. Indeed,
(\ref{sPSM-Howe-om-f}), (\ref{sPSM-Howe-e-fb}) and
(\ref{sPSM-Howe-ps-ff}) are already identical to the formulae
(\ref{Howe-Om-f}), (\ref{Howe-E-fb}) and (\ref{Howe-E-ff}),
respectively. The same is true for (\ref{sPSM-Howe-e-bb}) as a
comparison to (\ref{Howe-E-bb}) confirms, but $X_m$ and $\chi^\alpha$
are present in (\ref{sPSM-Howe-om-b}) and (\ref{sPSM-Howe-ps-bf})
which do not show up directly in (\ref{Howe-Om-b}) and
(\ref{Howe-E-bf}). Furthermore $\omega_m$ in (\ref{sPSM-Howe-om-b})
and (\ref{sPSM-Howe-ps-bf}) is an independent $x$-space field, but in
supergravity from superspace only the dependent connection
$\susy{\omega}_m$ (\cf (\ref{Howe-om})) is left over.

First we take a closer look at $\omega_m$ and $X_m$: To zeroth order
in $\theta$ the superspace field equations (\ref{eq:sPSM-eom1}) and
(\ref{eq:sPSM-eom2}) reduce to the ones of graded PSM where the
underlying base manifold is two-dimensional, so that the
considerations of \Cha\ref{cha:gPSM} apply. Especially we refer to
\Sec\ref{sec:sdil} where a uniform way to eliminate $\omega_m$ and
$X^a$ at once was offered. From (\ref{sdil-om}) for the class of
dilaton prepotential supergravities (\cf \Sec\ref{sec:Izq}) to which
the Poisson tensor considered here belongs the result was already
derived and given in (\ref{Izq-om}). This is identical to
(\ref{Howe-om}) as required. For $X^a$ the expression (\ref{sdil-X-b})
was obtained, which does not rely on a particular Poisson tensor at
all.

Once the equivalence $\omega \equiv \susy{\omega}$ is established and
after replacing $X^a$ in (\ref{sPSM-Howe-om-b}) by (\ref{sdil-X-b}) it
remains to show that
\begin{equation}
  \label{sPSM-Howe-ch}
  \chi_\alpha = -\frac{\tilde{u}_0}{\lambda} \susy{\sigma}_\alpha,
\end{equation}
where $\susy{\sigma}_\alpha$ was given by (\ref{Howe-si}). The reader
should be aware of the fact that $\susy{\sigma}_\alpha$ was defined
without the auxiliary scalar field in \Cha\ref{cha:gPSM}, \cf also
(\ref{Ldre-si}) and (\ref{Howe-ch}). Actually (\ref{sPSM-Howe-ch}) is
a PSM field equation. To zeroth order in $\theta$ in
(\ref{eq:sPSM-eom2}) the $x$-space differential form equation
\begin{equation}
  D\RS_\alpha + i \lambda \phi (\gamma^a e_a \RS)_\alpha +
  \frac{\lambda}{\tilde{u}_0} \epsilon \chi_\alpha = 0
\end{equation}
is contained, from which (\ref{sPSM-Howe-ch}) is derived.

To summarize: with $\omega \equiv \susy{\omega}$ (\cf (\ref{Izq-om})),
$X^a$ given by (\ref{sdil-X-b}), and with $\chi_\alpha$ as in
(\ref{sPSM-Howe-ch}) the superconnection (\ref{sPSM-Howe-om-b}) and
the Rarita-Schwinger superfield (\ref{sPSM-Howe-ps-bf}) become to be
identical to the ones obtained from superspace constraints namely
(\ref{Howe-Om-b}) and (\ref{Howe-E-bf}).


\chapter{Conclusion}
\label{cha:concl}



In \Cha\ref{cha:sfield} a new formulation of the superfield approach
to general supergravity has been established which allows nonvanishing
bosonic torsion. It is based upon the new minimal set of constraints
(\ref{constr-new}) for $\N = (1,1)$ superspace. The computational
problems which would occur for nonvanishing torsion following the
approach of the seminal work of Howe \cite{Howe:1979ia} are greatly
reduced by working in terms of a special decomposition of the
supervielbein in terms of superfields $B_m\^a$, $B_\mu\^\alpha$,
$\Phi_\mu\^m$ and $\Psi_m\^\mu$ (\cf (\ref{SVdecomp}) and
(\ref{ISVdecomp})).

In \Sec\ref{sec:ertl-cf} the component fields in the Wess-Zumino type
gauge (\ref{gf-0}) and (\ref{gf-1}) were derived explicitly for the
decompositon superfields as well as for the supervielbein and the
Lorenz superconnection. They also turned out to be expressed by an
additional, new multiplet $\{k^a, \varphi_m\^\alpha, \omega_m\}$
consisting of a vector field, a spin-vector and the Lorentz
connection, besides the well-known supergravity multiplet $\{e_m\^a,
\RS_m\^\alpha, A\}$ consisting of the zweibein, the Rarita-Schwinger
field and a scalar $A$.

The Bianchi identities have shown that the scalar superfield $S$,
already present in the Howe model, is accompanied by a vector
superfield $K^a$.  Together they represent the independent and
unconstrained superfields of supertorsion and supercurvature (\cf
(\ref{ertl-F-ff-T})--(\ref{ertl-T-bb})).  Since the component fields
transform with respect to the correct local diffeomorphisms, local
Lorentz and supersymmetry transformations, a generic superfield
Lagrangian is a superscalar built from $S$ and $K^a$, multiplied by
the superdeterminant of $E_M\^A$. With that knowledge one could
produce the immediate generalization of two-dimensional gravity with
torsion \cite{\bibKV}, but the explicit derivation of the component
field content of supertorsion and supercurvature turned out to be
extremely cumbersome.  That goal may be attained with the help of a
computer algebra program \cite{Ertl:Index-0.14.2} in future
investigations.


In the course of introducing the PSM concept we also presented
(\Cha\ref{cha:PSM}) a new extension of that approach, where the
singular Poisson structure of the PSM is embedded into a symplectic
one. In this connection we also showed how the general solution of the
field equations can be obtained in a systematic manner.


In \Cha\ref{cha:gPSM} the extension of the concept of Poisson Sigma
Models (PSM) to the graded case \cite{\bibNLSGT,Strobl:1999zz} has
been explored in some detail for the application in general
two-dimensional supergravity theories, when a dilaton field is
present. Adding one ($N=1$) or more ($N>1$) pairs of Majorana fields
representing respectively a target space (spinor) variable
$\chi^\alpha$ and a related `gravitino' $\psi_m\^\alpha$,
automatically leads to a supergravity with local supersymmetry closing
on-shell. Our approach yields the minimal supermultiplets, avoiding
the imposition and evaluation of constraints which is necessary in the
superfield formalism. Instead we have to solve Jacobi identities,
which the (degenerate) Poisson structure $\Poisson^{AB}$ of a PSM must
obey. In our present work we have performed this task for the full
$N=1$ problem. The solution for the algebras turns out to be quite
different according to the rank (defined in \Sec\ref{sec:remJac}) of
the fermionic extension, but could be reduced essentially to an
algebraic problem---despite the fact that the Jacobi identities
represent a set of nonlinear first order differential equations in
terms of the target space coordinates.

In this argument the Casimir functions are found to play a key role.
If the fermionic extension is of full rank that function of the
corresponding bosonic PSM simply generalizes to a quantity $\Casimir$,
taking values in the (commuting) supernumbers, because a quadratic
contribution of $\chi^\alpha$ is included (\cf (\ref{Casimir})). If
the extension is not of full rank, apart from the commuting $\Casimir$
also anticommuting Casimir functions of the form (\ref{DFS-Cf+}) and
(\ref{DFS-Cf-}) appear.

In certain cases, but not in general, the use of target space
diffeomorphisms (\cf \Sec\ref{sec:diffeo}) was found to be a useful
tool for the construction of the specific algebras and ensuing
supergravity models.  The study of `stabilisators', target space
transformations which leave an initially given bosonic algebra
invariant, also clarified the large arbitrariness (dependence of the
solution on arbitrary functions) found for the Poisson superalgebras
and the respective supergravity actions.

Because of this we have found it advisable to study explicit
specialized algebras and supergravity theories of increasing
complexity (\Sec\ref{sec:Poisson} and \ref{sec:models}). Our examples
are chosen in such a way that the extension of known bosonic 2d models
of gravity, like the Jackiw-Teitelboim model \cite{\bibJT}, the
dilaton black-hole \cite{\bibDil}, spherically symmetric gravity, the
Katanaev-Volovich model \cite{\bibKV}, $R^2$-gravity and others could
be covered (\cf \Sec\ref{sec:models}).  The arbitrariness referred to
above has the consequence that in all cases examples of several
possible extensions can be given. For a generic supergravity, obtained
in this manner, obstructions for the allowed values of the bosonic
target space coordinates emerge. Certain extensions are even found to
be not viable within real extensions of the bosonic algebra. We
identified two sources of these problems: the division by a certain
determinant (\cf (\ref{detv})) in the course of the (algebraic)
solution of the Jacobi identities and the appearance of a
`prepotential' which may be nontrivially related to the potential in
the original PSM.  Our hopes for the existence of an eventual criterion
for a reduction of the inherent arbitrariness, following from the
requirement that such obstructions should be absent, unfortunately did
not materialize: \eg for the physically interesting case of an
extension of spherically reduced Einstein gravity no less than four
different obstruction-free supergravities are among the examples
discussed here, and there exist infinitely more.

The PSM approach for 2d gravities contains a preferred formulation of
gravity as a `first order' (in derivatives) action (\ref{FOG}) in the
bosonic, as well as in the supergravity case (\Sec\ref{sec:action}).

In this formulation the target space coordinates $X^I = (\phi, X^a,
\chi^\alpha)$ of the gPSM are seen to coincide with the momenta in a
Hamiltonian action. A Hamiltonian analysis is not pursued in our
present work. Instead we discuss the possibility to eliminate $X^I$ in
part or completely.

The elimination of $X^a$ is possible in the action of a generic
supergravity PSM together with a torsion dependent part of the spin
connection. We show that in this way the most general superdilaton
theory with usual bosonic part (\ref{dil}) and minimal content of
fermionic fields in its extension (the Majorana spinor $\chi^\alpha$
as partner of the dilaton field $\phi$, and the 1-form `gravitino'
$\psi_\alpha$) is produced.

By contrast, the elimination of the dilaton field $\phi$ and/or the
related spinor $\chi^\alpha$ can only be achieved in particular cases.
Therefore, these fields should be regarded as substantial ingredients
when extending a bosonic 2d gravity action of the form (\ref{grav}),
depending on curvature and torsion.

The supergravity models whose bosonic part is torsion-free ($Z=0$ in
(\ref{FOG}) with (\ref{vdil}), or in (\ref{dil})) have been studied
before \cite{\bibIzq}. Specializing the potential $v(\phi)$ \emph{and}
the extension appropriately, one arrives at the supersymmetric
extension of the $R^2$-model ($v = -\frac{\alpha}{2} \phi^2$) and the
model of Howe ($v = -\half\phi^3$) \cite{Howe:1979ia} originally
derived in terms of superfields (\cf also \cite{Ertl:1997ib}).  In the
latter case the `auxiliary field' $A$ is found to coincide with the
dilaton field $\phi$. It must be emphasized, though, that in the PSM
approach all these models are obtained by introducing a cancellation
mechanism for singularities and ensuing obstructions (for actions and
solutions in the real numbers).

When the bosonic model already contains torsion in the PSM form
(\ref{FOG}) or when, equivalently, $Z \neq 0$ in (\ref{dil}) an
extension of a conformal transformation to a target space
diffeomorphism in the gPSM allows an appropriate generalization of the
models with $Z=0$. Our discussion of spherically symmetric gravity
(\ref{SRG}) and of the Katanaev-Volovich model (\ref{KV}) show basic
differences. Whereas no new obstruction appears for the former
`physical' theory ($\phi > 0$), as already required by (\ref{SRG}),
the latter model develops a problem with real actions, except when the
parameters $\alpha$, $\beta$ and $\Lambda$ are chosen in a very
specific manner.

We also present a field theoretic model for a gPSM with rank $(2|1)$,
when only one component of the target space spinor $\chi^\alpha$ is
involved.  Its supersymmetry only contains one anticommuting function
so that this class of models can be interpreted as $\N = (1,0)$
supergravity.

Finally (\Sec\ref{sec:sdil-sol}) we make the general considerations of
\cite{Strobl:1999zz} more explicit by giving the full (analytic)
solution to the class of models summarized in the two preceding
paragraphs. It turns out to be sufficient to discuss the case $Z=0$,
because $Z \neq 0$ can be obtained by conformal transformation. Our
formulation in terms of Casimir-Darboux coordinates (including the
fermionic extension) allows the integration of the infinitesimal
supersymmetry transformation to finite ones. Within the range of
applicability for the target space coordinates $X^I$ this permits a
gauging of the target space spinors to zero. In this sense
supergravities (without matter) are `trivial'.  However, as stressed
in the introduction, such arguments break down when (supersymmetric)
matter is coupled to the model.


Finally, in \Cha\ref{cha:SPSM} the relations between the superfield
approach and the gPSM one are discussed. Taking the Howe model as an
explicit example the auxiliary field $A$ in Howe's model is identified
with the dilaton field $\phi$ which appears already in the bosonic
theory. In this manner also the relation to the first oder formulation
\cite{\bibFOG,Strobl:1994eu,\bibDilKKL,Kummer:1997jj,Strobl:1999Habil}
is clarified.

\bigskip


This leads us to an outlook on possible further applications.  There
is a multitude of directions for future work suggested by our present
results. We only give a few examples.

Clearly starting from any of the models described here, its
supertransformations could be used---at least in a trial-and-error
manner as in the original $d=4$ supergravity \cite{\bibSUGRA}---to
extend the corresponding bosonic action \cite{Izquierdo:1998hg}.

The even simpler introduction of matter in the form of a scalar
`testparticle' in gravity explicitly (or implicitly) is a necessary
prerequisite for defining the global manifold geometrically to its
geodesics (including null directions). We believe that a (properly
defined) spinning testparticle would be the appropriate instrument for
2d supergravity (\cf \eg \cite{Knutt-Wehlau:1998gq}).  `Trivial'
supergravity should be without influence on its (`super')-geodesics.
This should work in the same way as coordinate singularities are not
felt in bosonic gravity.

Another line of investigation concerns the reduction of $d \geq 4$
supergravities to a $d=2$ effective superdilaton theory. In this way
it should perhaps be possible to nail down the large arbitrariness of
superdilaton models, when---as in our present work---this problem is
regarded from a strictly $d=2$ point of view. It can be verified in
different ways that the introduction of Killing spinors within such an
approach inevitably leads to complex fermionic (Dirac) components.
Thus 2d (dilatonic) supergravities with $\N \geq 2$ must be
considered. As explained in \Sec\ref{sec:gPSM} also for this purpose
the gPSM approach seems to be the method of choice. Of course, the
increase in the number of fields, together with the restrictions of
the additional $SO(\N)$ symmetry will provide an even more complicated
structure.  Already for $\N=1$ we had to rely to a large extent on
computer-aided techniques.

Preliminary computations show that the `minimal' supergravity actions,
provided by the PSM approach, also seem to be most appropriate for a
Hamiltonian analysis leading eventually to a quantum 2d supergravity,
extending the analogous result for a purely bosonic case
\cite{Schaller:1994es,\bibQGr,Strobl:1999Habil}. The role of the
obstruction for real supersymmetric extensions, encountered for some
of the models within this thesis, will have to be reconsidered
carefully in this context.


\begin{appendix}
\renewcommand{\chaptermark}[1]{
  \markboth{\appendixname\ \thechapter.\ #1}{}}


\chapter{Forms and Gravity}
\label{app:gravity}

We use the characters $a,b,c,\ldots$ to denote Lorentz indices and
$m,n,l,\ldots$ to denote world indices, both types take the values
$(0,1)$. For the Kronecker symbol we write $\delta_a\^b \equiv
\delta_a^b$ and for later convenience we also define the generalized
Kronecker symbol
\begin{equation}
  \label{genKronecker}
  \hat\epsilon_{ab}^{cd} := \delta_a^c \delta_b^d - \delta_a^d
  \delta_b^c = \hat\epsilon_{ab} \hat\epsilon^{cd},
\end{equation}
where $\hat\epsilon_{01} \equiv \hat\epsilon^{01} := 1$ is the
alternating $\epsilon$-symbol.

In $d=2$ our Minkowski metric is
\begin{equation}
  \label{lor-metric}
  \eta_{ab} = \eta^{ab} = \mtrx{1}{0}{0}{-1},
\end{equation}
and for the antisymmetric $\epsilon$-tensor we set $\epsilon_{ab} :=
\hat\epsilon_{ab}$ and consistently $\epsilon^{ab} =
-\hat\epsilon^{ab}$, so that
\begin{equation}
  \label{lor-ep}
  \epsilon_{ab} = -\epsilon^{ab} = \mtrx{0}{1}{-1}{0}.
\end{equation}
It obeys
\begin{equation}
  \epsilon_{ab} \epsilon^{cd} = \delta_a\^d \delta_b\^c - \delta_a\^c
  \delta_b\^d, \qquad
  \epsilon_a\^b \epsilon_b\^c = \delta_a\^c, \qquad
  \epsilon^{ab} \epsilon_{ba} = 2,
\end{equation}
and $\epsilon_a\^b$ is also the generator of Lorentz transformations
in $d=2$.  The totally antisymmetric tensor and the Minkowskian metric
satisfy the Fierz-type identity
\begin{equation}
  \eta_{ab} \epsilon_{cd} + \eta_{da} \epsilon_{bc} + \eta_{cd}
  \epsilon_{ab} + \eta_{bc} \epsilon_{da} = 0,
\end{equation}
which allows to make rearrangements in third and higher order
monomials of Lorentz vectors. 

Let $x^m = (x^0, x^1)$ be local coordinates on a two-dimensional
manifold.  We define the components of a 1-form $\lambda$ and a 2-form
$\epsilon$ according to
\begin{equation}
  \label{form}
  \lambda = dx^m \lambda_m, \qquad
  \epsilon = \half dx^m \wedge dx^n \epsilon_{nm}.
\end{equation}
The exterior derivative of a function $f$ is $df = dx^m \partial_m f$,
as customary, but for a 1-form $\lambda$ we choose
\begin{equation}
  \label{form-d}
  d\lambda := dx^{m} \wedge dx^{n} \partial_n \lambda_m.
\end{equation}
As a consequence $d$ acts from the \emph{right}, \ie for a $q$-form
$\psi$ and a $p$-form $\phi$ the Leibniz rule is
\begin{equation}
  \label{form-Leibnitz}
  d(\psi \wedge \phi) = \psi \wedge d\phi + (-1)^{p} d\psi
  \wedge \phi.
\end{equation}
This convention is advantageous for the extension to spinors and
superspace (\cf \Sec\ref{sec:sfield} and \App\ref{app:spinors}) where
we assume the same summation convention of superindices.

Purely bosonic two-dimensional gravity is described in terms of the
anholonomic orthonormal basis in tangent or equivalently in cotangent
space,
\begin{equation}
  \label{zweibein}
  \partial_a = e_a\^m \partial_m, \qquad e^a = dx^m e_m\^a,
\end{equation}
where $e_m\^a$ is the zweibein and $e_a\^m$ its inverse. We use
$\partial_a = e_a\^m \partial_m$ to denote the moving frame, because
$e_a$ is used in the PSM context for the 1-forms $e_a = e^b
\eta_{ba}$.  The zweibein is an isomorphism which transforms world
indices into tangent indices and vice versa, therefore $e :=
\det(e_m\^a) = \det(e_a\^m)^{-1} \neq 0$.
In terms of the metric $g_{mn} = e_n\^b e_m\^a \eta_{ab}$ and its
determinant $g = \det(g_{mn})$ we have $e = \sqrt{-g}$. The
relationship $e_a\^m e_m\^b = \delta_a\^b$ can be expressed in $d=2$
by the closed formulae
\begin{equation}
  e_m\^a = \frac{1}{\det(e_a\^m)}\, \hat\epsilon_{mn}^{ab} e_b\^n, \qquad
  e_a\^m = \frac{1}{\det(e_m\^a)}\, \hat\epsilon_{ab}^{mn} e_n\^b,
\end{equation}
where $\hat\epsilon_{mn}^{ab}$ and $\hat\epsilon_{ab}^{mn}$ are the
generalized Kronecker symbols derived from (\ref{genKronecker}) by
simply rewriting that formula with appropriate indices.

The transition to world indices of the $\epsilon$-tensor
(\ref{lor-ep}) yields
\begin{equation}
  \label{ep-tensor}
  \epsilon_{mn} = e\, \hat\epsilon_{mn}, \qquad
  \epsilon^{mn} = - \frac{1}{e}\, \hat\epsilon^{mn}.
\end{equation}
The induced volume form
\begin{equation}
  \label{volume-form}
  \epsilon = \half e^a \wedge e^b \epsilon_{ba} = e^1 \wedge e^0
  = e\, dx^1 \wedge dx^0
\end{equation}
enables us to derive the useful relation $dx^m \wedge dx^n =
\epsilon\, \epsilon^{mn}$, and to define the Hodge dual according to
\begin{equation}
  \label{HodgeDual}
  \star 1 = \epsilon, \qquad
  \star dx^m = dx^n \epsilon_n\^m, \qquad
  \star\epsilon = 1. \qquad
\end{equation}
It is a linear map, \ie $\star(\phi f) = (\star\phi) f$ for functions
$f$ and forms $\phi$, and bijective $\star\star = \id$.

The anholonomicity coefficients $c_{ab}\^c$ from
\begin{equation}
  \label{anhol}
  [\partial_a, \partial_b] = c_{ab}\^c \partial_c, \qquad
  de^a = -\half dx^m \wedge dx^n c_{nm}\^a
\end{equation}
can be expressed by the zweibein as well as by its inverse according to
\begin{equation}
  c_{ab}\^c = (\partial_a e_b\^n - \partial_b e_a\^n) e_n\^c, \qquad
  c_{mn}\^a = -(\partial_m e_n\^a - \partial_n e_m\^a).
\end{equation}
In two dimensions the anholonomicity coefficients are in one-to-one
correspondence with their own trace
\begin{equation}
  c_b = c_{ab}\^a, \qquad
  c_{ab}\^c = \delta_a\^c c_b - \delta_b\^c c_a.
\end{equation}
With
\begin{equation}
  \label{omT}
  \tilde{\omega}^b := \epsilon^{nm} (\partial_m e_n\^b)
\end{equation}
the useful relations
\begin{equation}
  \tilde{\omega}^c = \epsilon^{cb} c_b = -\half \epsilon^{ba}
  c_{ab}\^c, \qquad
  c_{ab}\^c = -\epsilon_{ab} \tilde{\omega}^c
\end{equation}
and
\begin{equation}
  \epsilon^{nm} \partial_m v_n = \epsilon^{ba} (\partial_a v_b
  - \tilde{\omega}_a \epsilon_b\^c v_c) =
  \epsilon^{ba} \partial_a v_b + \tilde{\omega}^c v_c
\end{equation}
are obtained.

Apart from the zweibein a two-dimensional spacetime is characterized
by the Lorentz connection $\omega_{ma}\^b = \omega_m \epsilon_a\^b$
providing the connection 1-form $\omega = dx^m \omega_{m}$ and the
exterior covariant derivative of vector and covector valued
differential forms
\begin{equation}
  \label{extCovDer}
  \D \phi^a := d \phi^a + \phi^b \wedge \omega \epsilon_b\^a, \qquad
  \D \phi_a := d \phi_a -  \epsilon_a\^b \phi_b \wedge \omega.
\end{equation}
The exterior covariant derivative $\D_m$ acts exclusively on Lorentz
indices, whereas the full covariant derivative $\nabla_m$ includes
also world indices. The latter is expressed with the help of the
linear connection $\Gamma_{mn}\^l$, \eg for vectors and covectors the
component expressions
\begin{alignat}{2}
  \D_m v^b &= \partial_m v^b + v^c \omega_{mc}\^b, &\qquad
  \nabla_m v^n &= \partial_m v^n + v^l \Gamma_{ml}\^n, \\
  \D_m v_b &= \partial_m v_b - \omega_{mb}\^c v_c, &\qquad
  \nabla_m v_n &= \partial_m v_n - \Gamma_{mn}\^l v_l.  
\end{alignat}
are significant.
Consistency for interchanging both types of indices leads to the
assertion
\begin{equation}
  (\D_m v^b) e_b\^n = \nabla_m v^n \quad \Leftrightarrow \quad
  \nabla_m e_b\^n = 0.
\end{equation}
From the last equation we obtain
\begin{align}
  \label{linCon}
  \Gamma_{mn}\^l = (\D_m e_n\^b) e_b\^l
  = (\partial_m e_n\^b + e_n\^c \omega_{mc}\^b) e_b\^l.
\end{align}

The torsion 2-form and its Hodge dual (\cf (\ref{HodgeDual})) are
defined by
\begin{equation}
  \label{tor}
  t^a := \D e^a, \qquad
  \tor^a := \Hodge t^a.
\end{equation}
For the components of $t^a = \half dx^m \wedge dx^n t_{nm}\^a$ and for
$\tor^a$ the expressions
\begin{equation}
  t_{ab}\^c = -c_{ab}\^c + \omega_a \epsilon_b\^c - \omega_b
  \epsilon_a\^c, \qquad
  \tor^a = \half \epsilon^{mn} t_{nm}\^a
\end{equation}
are obtained, and due to condition (\ref{linCon})
\begin{equation}
  t_{mn}\^l = \Gamma_{mn}\^l - \Gamma_{nm}\^l.
\end{equation}
The vector $\tor^a$ is frequently encountered in \Cha\ref{cha:gPSM}.
It immediately provides a simple formula relating torsion and Lorentz
connection:
\begin{equation}
  \omega_a = \tilde{\omega}_a - \tor_a
\end{equation}
Here $\tilde{\omega}_a$ is the quantity already defined by (\ref{omT})
which turns out to be the torsion free connection of Einstein gravity.

The curvature tensor is quite simple in two dimensions:
\begin{equation}
  \label{cur}
  r_{mna}\^b = f_{mn} \epsilon_a\^b, \qquad
  f_{mn} = \partial_m \omega_n - \partial_n \omega_m
\end{equation}
Expressed in terms of $\omega_a$ one obtains
\begin{equation}
  f_{ab} = \partial_a \omega_b - \partial_b \omega_a - c_{ab}\^c
  \omega_c = \nabla_a \omega_b - \nabla_b \omega_a + t_{ab}\^c
  \omega_c.
\end{equation}
Ricci tensor and scalar read
\begin{equation}
  \label{ricci}
  r_{ab} = r_{cab}\^c = \epsilon_a\^c f_{cb}, \qquad
  r = r_a\^a = \epsilon^{ba} f_{ab},
\end{equation}
leading to the relation
\begin{equation}
  \label{ricci-r}
  \frac{r}{2} = \Hodge d\omega = \epsilon^{nm} \partial_m \omega_n,
\end{equation}
which is valid exclusively in the two-dimensional case.

\chapter{Conventions of Spinor-Space and Spinors}
\label{app:spinors}
  
Of course, the properties of Clifford algebras and spinors in any
number of dimensions (including $d = 1+1$) are well-known, but in view
of the tedious calculations required in our present work we include
this appendix in order to prevent any misunderstandings of our results
and facilitate the task of the intrepid reader who wants to redo
derivations.
  

The $\gamma$ matrices which are the elements of the Clifford algebra
defined by the relation
\begin{equation}
  \label{edefgm}
  \gamma^a \gamma^b + \gamma^b \gamma^a = 2 \eta^{ab} \1, \qquad
  \eta_{ab} = \eta^{ab} = \diag(+-),
\end{equation}
are represented by two-dimensional matrices
\begin{equation}
  \label{egamma}
  \gamma^0\_\alpha\^\beta = \mtrx{0}{1}{1}{0}, \qquad
  \gamma^1\_\alpha\^\beta = \mtrx{0}{1}{-1}{0}.
\end{equation}
As indicated, the lower index is assumed to be the first one.  The
spinor indices are often suppressed assuming the summation from `ten
to four'.  The generator of a Lorentz boost (hyperbolic rotation) has
the form
\begin{equation}
  \sigma^{ab} = \frac12 (\gamma^a \gamma^b - \gamma^b \gamma^a) =
  \epsilon^{ab} \gamma^3,
\end{equation}
where
\begin{equation}
  \gamma^3 = \gamma^1 \gamma^0 = \mtrx{1}{0}{0}{-1}, \qquad
  (\gamma^3)^2 = \1.
\end{equation}
In two dimensions the $\gamma$-matrices satisfy the relation
\begin{equation}
  \label{egmpro1}
  \gamma^a \gamma^b = \eta^{ab} \1 + \epsilon^{ab} \gamma^3,
\end{equation}
which is equivalent to the definition (\ref{edefgm}). The following
formulae are frequently used in our calculations:
\begin{equation}
  \label{egmpro2}
  \begin{aligned}[t]
    \gamma^a \gamma_a &=2, \\
    \gamma^a \gamma^3 + \gamma^3 \gamma^a &= 0, \\
    \tr(\gamma^a \gamma^b) &= 2 \eta^{ab}.
  \end{aligned} \qquad
  \begin{aligned}[t]
    \gamma^a \gamma^b \gamma_a &= 0, \\
    \gamma^a \gamma^3 &= \gamma^b \epsilon_b\^a,
  \end{aligned}
\end{equation}
As usual the trace of the product of an odd number of
$\gamma$-matrices vanishes.

In two dimensions the $\gamma$-matrices satisfy the Fierz identity
\begin{multline}
  \label{efierz}
  2 \gamma^a\_\alpha\^\gamma \gamma^b\_\beta\^\delta =
  \gamma^a\_\alpha\^\delta
  \gamma^b\_\beta\^\gamma + \gamma^b\_\alpha\^\delta
  \gamma^a\_\beta\^\gamma + \\
  + \eta^{ab} (\delta_\alpha\^\delta \delta_\beta\^\gamma -
  \gamma^3\_\alpha\^\delta \gamma^3\_\beta\^\gamma -
  \gamma^c\_\alpha\^\delta \gamma_{c\beta}\^\gamma) + \epsilon^{ab}
  (\gamma^3\_\alpha\^\delta \delta_\beta\^\gamma -
  \delta_\alpha\^\delta \gamma^3\_\beta\^\gamma),
\end{multline}
which can be checked by direct calculation. Different contractions of
it with $\gamma$-matrices then yield different but equivalent versions
\begin{align}
  2 \delta_\alpha\^\gamma \delta_\beta\^\delta &=
  \delta_\alpha\^\delta \delta_\beta\^\gamma +\gamma^3\_\alpha\^\delta
  \gamma^3\_\beta\^\gamma + \gamma^a\_\alpha\^\delta
  \gamma_{a\beta}\^\gamma,
  \label{efier1} \\
  2 \gamma^3\_\alpha\^\gamma \gamma^3\_\beta\^\delta &=
  \delta_\alpha\^\delta \delta_\beta\^\gamma +
  \gamma^3\_\alpha\^\delta \gamma^3\_\beta\^\gamma -
  \gamma^a\_\alpha\^\delta \gamma_{a\beta}\^\gamma,
  \label{efier2} \\
  \gamma^a\_\alpha\^\gamma \gamma_{a\beta}\^\delta &=
  \delta_\alpha\^\delta \delta_\beta\^\gamma -
  \gamma^3\_\alpha\^\delta \gamma^3\_\beta\^\gamma, \label{efier3}
\end{align}
which allow to manipulate third and higher order monomials in spinors.
Notice that equation (\ref{egmpro1}) is also the consequence of
(\ref{efierz}). An other manifestation of the Fierz identity is the
completeness relation
\begin{equation}
  \label{ga-compl}
  \Gamma_\alpha\^\beta = \half \Gamma_\gamma\^\gamma\,
  \delta_\alpha\^\beta + \half (\Gamma\gamma_a)_\gamma\^\gamma\,
  \gamma^a\_\alpha\^\beta + \half (\Gamma\gamma^3)_\gamma\^\gamma\,
  \gamma^3\_\alpha\^\beta.
\end{equation}

A Dirac spinor in two dimensions, forming an irreducible
representation for the full Lorentz group including space and time
reflections, has two complex components. We write it---in contrast
to the usual convention in field theory, but in agreement with
conventional superspace notations---as a row
\begin{equation}
  \chi^\alpha = (\chi^+, \chi^-).
\end{equation}
In our notation the first and second components of a Dirac spinor
correspond to right and left chiral Weyl spinors $\chi^{(\pm)}$,
respectively,
\begin{equation}
  \chi^{(\pm)} = \chi P_{\pm}, \qquad
  \chi^{(\pm)} \gamma^3 = \pm \chi^{(\pm)},
\end{equation}
where the chiral projectors are given by
\begin{equation}
  \label{ChiralP}
  P_{\pm} = \half (\1 \pm \gamma^3)
\end{equation}
so that
\begin{equation}
  \chi^{(+)} = (\chi^+, 0), \qquad
  \chi^{(-)} = (0, \chi^-).
\end{equation}
Here matrices act on spinors form the right according to the usual
multiplication law. All spinors are always assumed to be anticommuting
variables.  The notation with upper indices is a consequence of our
convention to contract indices, together with the usual multiplication
rule for matrices. Under the Lorentz boost by the parameter $\omega$
spinors transform as
\begin{equation}
  \chi'\^\alpha = \chi^\beta S_\beta\^\alpha,
\end{equation}
where
\begin{equation}
  \label{erotsp}
  S_\beta\^\alpha = \delta_\beta\^\alpha \cosh{\frac{\omega}{2}} -
  \gamma^3\_\beta\^\alpha \sinh{\frac{\omega}{2}} =
  \mtrx{e^{-\omega/2}}{0}{0}{e^{+\omega/2}},
\end{equation}
when the Lorentz boost of a vector is given by the matrix
\begin{equation}
  \label{erotve}
  S_b\^a = \delta_b\^a \cosh{\omega} + \epsilon_b\^a \sinh{\omega} =
  \mtrx{\cosh{\omega}}{\sinh{\omega}}{\sinh{\omega}}{\cosh{\omega}}.
\end{equation}
By (\ref{erotsp}), (\ref{erotve}) the $\gamma$-matrices are invariant
under simultaneous transformation of Latin and Greek indices. This
requirement fixes the relative factors in the bosonic and fermionic
sectors of the Lorentz generator.

For a spinor
\begin{equation}
  \chi_\alpha = \clmn{\chi_+}{\chi_-}
\end{equation}
the Dirac conjugation $\bar\chi^\alpha = \chi^\dagger\^{\dot\alpha}
A_{\dot\alpha}\^\alpha$ depends on the matrix $A$, which obeys
\begin{equation}
  \label{DiracAProp}
  A \gamma^a A^{-1} = (\gamma^a)^\dagger, \qquad
  A^\dagger = A.
\end{equation}
We make the usual choice
\begin{equation}
  \label{DiracA}
  A = \mtrx{0}{1}{1}{0} = \gamma^0.
\end{equation}

The charge conjugation of a spinor using complex conjugation is
\begin{equation}
  \label{ChargeConj}
  \chi^c = B \chi^*
\end{equation}
\begin{equation}
  \label{ChargeBProp}
  B^{-1} \gamma^a B = -(\gamma^a)^*, \qquad B B^* = \1.
\end{equation}
For our choice of $\gamma^a$ (\ref{egamma})
\begin{equation}
  \label{ChargeB}
  B = \mtrx{-1}{0}{0}{1}.
\end{equation}
Alternatively, one can define the charge conjugated spinor with the
help of the Dirac conjugation matrix (\ref{DiracA}),
\begin{equation}
  \chi^c = (\bar\chi C)^T = C^T A^T \chi^*, \qquad
  \chi^c\_\alpha = \bar\chi^\beta C_{\beta\alpha},
\end{equation}
\begin{equation}
  \label{ChargeCProp}
  C^{-1} \gamma^a C = - (\gamma^a)^T, \qquad C^T = -C,
\end{equation}
\begin{equation}
  \label{ChargeC}
  C = (C_{\beta\alpha}) = \mtrx{0}{1}{-1}{0}.
\end{equation}

By means of
\begin{equation}
  \label{spin-metric}
  \epsilon_{\alpha\beta} = \epsilon^{\alpha\beta} = \mtrx{0}{1}{-1}{0}
\end{equation}
indices of Majorana spinors $\chi^c = \chi$, in components $\chi_+ =
-(\chi_+)^*, \chi_- = (\chi_-)^*$, can be raised and lowered as
$\chi^\alpha = \epsilon^{\alpha\beta} \chi_\beta$ and $\chi_\alpha =
\chi^\beta \epsilon_{\beta\alpha}$. In components we get
\begin{equation}
  \label{spin-down}
  \chi^+ = \chi_-, \qquad \chi^- = -\chi_+.
\end{equation}
This yields $\psi^\alpha \chi_\alpha = -\psi_\alpha \chi^\alpha
= \chi^\alpha \psi_\alpha = \psi^- \chi^+ - \psi^+ \chi^-$
for two anticommuting Majorana spinors $\psi$ and $\chi$. For
bilinear forms we use the shorthand
\begin{equation}
  (\psi\chi) = \psi^\alpha \chi_\alpha, \qquad
  (\psi\gamma^a\chi) = \psi^\alpha \gamma^a\_\alpha\^\beta \chi_\beta,
  \qquad
  (\psi\gamma^3\chi) = \psi^\alpha \gamma^3\_\alpha\^\beta \chi_\beta.
\end{equation}
A useful property is
\begin{equation}
  \label{spin-metric-prop}
  \epsilon_{\alpha\beta} \epsilon^{\gamma\delta} =
  \delta_\alpha\^\gamma \delta_\beta\^\delta - \delta_\alpha\^\delta
  \delta_\beta\^\gamma.
\end{equation}
The Fierz identity (\ref{efier1}) yields
\begin{equation}
   \label{ecompl}
  \chi_\alpha \psi^\beta = -\half (\psi \chi) \delta_\alpha\^\beta -
  \half (\psi \gamma_a \chi) \gamma^a\_\alpha\^\beta - \half (\psi
  \gamma^3 \chi) \gamma^3\_\alpha\^\beta.
\end{equation}

Among the spinor matrices $\gamma^a\^{\alpha\beta}$ and
$\gamma^3\^{\alpha\beta}$ are symmetric in $\alpha \leftrightarrow
\beta$, whereas $\epsilon^{\alpha\beta}$ is antisymmetric.

\section{Spin-Components of Lorentz Tensors}
\label{app:spin-comp}
  
\newcommand{\<}{\langle}
\renewcommand{\>}{\rangle}

The light cone components of a Lorentz vector $v^a = (v^0,v^1)$ are
defined according to
\begin{equation}
  \label{LC-comp}
  v^\oplus = \frac{1}{\sqrt{2}} \left( v^0 + v^1 \right), \qquad
  v^\ominus = \frac{1}{\sqrt{2}} \left( v^0 - v^1 \right).
\end{equation}
We use the somewhat unusual symbols $\oplus$ and $\ominus$ to denote
light cone indices, in order to avoid confusion with spinor indices.
The Lorentz metric is off-diagonal in this basis,
$\eta_{\oplus\ominus} = \eta_{\ominus\oplus} = 1$, leading to the
usual property for raising and lowering indices $v^\oplus = v_\ominus$
and $v^\ominus = v_\oplus$. The nonzero components of the epsilon
tensor are $\epsilon_{\oplus\ominus} = -1$ and
$\epsilon_{\ominus\oplus} = 1$.  When raising the last index the
generator of Lorentz transformation $\epsilon_a\^b$ is found to be
diagonal, $\epsilon_\oplus\^\oplus = -1$ and
$\epsilon_\ominus\^\ominus = 1$.  For quadratic forms we obtain
\begin{equation}
  v^a w^b \eta_{ba} = v^\oplus w^\ominus + v^\ominus w^\oplus, \qquad
  v^a w^b \epsilon_{ba} = v^\oplus w^\ominus - v^\ominus w^\oplus.
\end{equation}


In theories with spinors, and thus in the presence of
$\gamma$-matrices, it is convenient to associate to every Lorentz
vector $v^a$ a spin-tensor according to $v^{\alpha\beta} = a v^c
\gamma_c\^{\alpha\beta}$ with the convenient value of the constant $a$
to be determined below. In $d=2$ this spin-tensor is symmetric
($v^{\alpha\beta} = v^{\beta\alpha}$) and $\gamma^3$-traceless
($v^{\alpha\beta} \gamma^3\_{\beta\alpha} = 0$). In order to make
these properties manifest we define the projectors
\begin{equation}
  T^{\<\alpha\beta\>} := -\half T^{\gamma\delta}
  \gamma^a\_{\delta\gamma} \gamma_a\^{\alpha\beta} =
  T^{(\alpha\beta)} + \half T^{\gamma\delta}
  \gamma^3\_{\delta\gamma} \gamma^3\^{\alpha\beta},
\end{equation}
thus $v^{\alpha\beta} = v^{\<\alpha\beta\>}$. We use angle brackets to
express explicitly the fact that a spin-pair originates from a Lorentz
vector and write $v^{\<\alpha\beta\>}$ from now on. In the chosen
representation of the $\gamma$-matrices mixed components vanish,
$v^{\<+-\>} = v^{\<-+\>} = 0$. Furthermore, as a consequence of the
spinor metric (\ref{spin-metric}) the spin pairs $\<++\>$ and $\<--\>$
behave exactly as light cone indices; these indices are lowered
according to
\begin{equation}
  \label{spin-pair-down}
  v^{\<++\>} = v_{\<--\>}, \qquad v^{\<--\>} = v_{\<++\>}.
\end{equation}

In order to be consistent the Lorentz metric in spinor components has
to be of light cone form too: By definition we have
$\eta_{\<\alpha\beta\>\<\delta\gamma\>} = a^2 \gamma^a\_{\alpha\beta}
\gamma^b\_{\delta\gamma} \eta_{ab}$, which is indeed off diagonal and
symmetric when interchanging the spin-pairs. Demanding
$\eta_{\<++\>\<--\>} = 1$ fixes the parameter $a$ up to a sign,
yielding $a = \pm \frac{i}{\sqrt{2}}$. We choose the positive sign
yielding for the injection of vectors into spin-tensors, and for the
extraction of vectors from spin-tensors
\begin{equation}
  \label{spin-vector}
  v^{\<\alpha\beta\>} := \frac{i}{\sqrt{2}} v^a
  \gamma_a\^{\alpha\beta}, \qquad
  T^a := \frac{i}{\sqrt{2}} T^{\alpha\beta} \gamma^a\_{\beta\alpha},
\end{equation}
respectively. This definition leads to
\begin{align}
  \eta_{\<\alpha\beta\>\<\delta\gamma\>} &= -\half
  \gamma^a\_{\alpha\beta} \gamma_{a\delta\gamma}, \\
  \gamma_{\<\alpha\beta\>}\_{\delta\gamma} &= \frac{i}{\sqrt{2}}
  \gamma^a\_{\alpha\beta} \gamma_{a\delta\gamma} = -i \sqrt{2}
  \eta_{\<\alpha\beta\>\<\delta\gamma\>}
\end{align}
for the conversion of Lorentz indices in $\eta_{ab}$ and
$\gamma_{a\delta\gamma}$ to spin-pairs, yielding for the latter also a
spin-tensor which is symmetric when interchanging the spin pairs,
$\gamma_{\<\alpha\beta\>\<\delta\gamma\>} =
\gamma_{\<\delta\gamma\>\<\alpha\beta\>}$.

The spin pair components are not identical to the light cone ones as
defined above in (\ref{LC-comp}).  The former are purely imaginary and
related to the latter according to
\begin{equation}
  v^{\<++\>} = i v^\oplus, \qquad v^{\<--\>} = -i v^\ominus.
\end{equation}
The components of the epsilon tensor are $\epsilon_{\<++\>\<--\>} = -1$
and $\epsilon_{\<--\>\<++\>} = 1$ and the quadratic forms read
\begin{align}
  v^a w_a &= v^{\<\alpha\beta\>} w_{\<\beta\alpha\>} = v^{\<++\>}
  w^{\<--\>} + v^{\<--\>} w^{\<++\>}, \\
  v^a w^b \epsilon_{ba} &= v^{\<\alpha\beta\>} w^{\<\gamma\delta\>}
  \epsilon_{\<\delta\gamma\>\<\beta\alpha\>} = v^{\<++\>} w^{\<--\>} -
  v^{\<--\>} w^{\<++\>}.
\end{align}
Note that we can exchange contracted Lorentz indices with spin pair
indices in angle brackets---a consequence of our choice for parameter
$a$.  

The metric $\eta_{\<\alpha\beta\>\<\gamma\delta\>}$ is only applicable
to lower spin-pair indices which correspond to Lorentz vectors. It can
be extended to a full metric $\eta_{\alpha\beta}\^{\delta\gamma}$,
demanding $T^{\beta\alpha} \eta_{\alpha\beta}\^{\delta\gamma} =
T^{\delta\gamma}$ for any spin-tensor $T^{\beta\alpha}$. This implies
the $\gamma$-matrix expansion
\begin{equation}
  \eta_{\alpha\beta}\^{\delta\gamma} = -\half \gamma^a\_{\alpha\beta}
  \gamma_a\^{\delta\gamma} - \half \gamma^3\_{\alpha\beta}
  \gamma^3\^{\delta\gamma} - \half \epsilon_{\alpha\beta}
  \epsilon^{\delta\gamma}.
\end{equation}
Using Fierz identity (\ref{efier1}) we find the simple formula
\begin{equation}
  \eta_{\alpha\beta}\^{\delta\gamma} = \delta_\alpha\^\gamma
  \delta_\beta\^\delta.
\end{equation}

For easier writing we omit in the main text of this work the brackets
when adequate ($v^{\<++\>} \rightarrow v^{++}$, $v^{\<--\>}
\rightarrow v^{--}$, \etc).

\section{Decompositions of Spin-Tensors}
\label{app:spin-tensor}
  
Any symmetric spinor $W_{\alpha\beta} = W_{\beta\alpha}$ can be
decomposed in a vector and a pseudoscalar
\begin{equation}
  W_{\alpha\beta} = W^a \gamma_{a\alpha\beta} + W^3
  \gamma^3\_{\alpha\beta}
\end{equation}
given by
\begin{equation}
  W^a = \half (W \gamma^a)_\gamma\^\gamma, \qquad W^3 = \half (W
  \gamma^3)_\gamma\^\gamma.
\end{equation}
The symmetric tensor product of two spinors is therefore
\begin{equation}
  \chi_\alpha \psi_\beta + \chi_\beta \psi_\alpha = (\chi \gamma_a
  \psi)\, \gamma^a\_{\alpha\beta} + (\chi \gamma^3 \psi)\,
  \gamma^3\_{\alpha\beta}.
\end{equation}

The field $\RS^{a\alpha}$ has one vector and one spinor index. We
assume that for each $a$ it is a Majorana spinor.  Therefore it has
two real and two purely imaginary components forming a reducible
representation of the Lorentz group. In many applications it becomes
extremely useful to work with its Lorentz covariant decomposition
\begin{equation}
  \label{ederas}
  \RS^a = \RS\gamma^a + \lambda^a,
\end{equation}
where
\begin{equation}
  \label{RS-decomp}
  \RS = \frac12 (\RS^a \gamma_a), \qquad
  \lambda^a = \frac12 (\RS^b \gamma^a \gamma_b) = 
  \frac12 (\gamma_b \gamma^a \RS^b).
\end{equation}
The spinor $\RS^\alpha$ and the spin-vector $\lambda_a\^\alpha$ form
irreducible representations of the Lorentz group and each of them has
two independent components. The spin-vector $\lambda_a$ satisfies the
Rarita-Schwinger condition
\begin{equation}
  \label{equlam}
  \lambda_a \gamma^a = 0
\end{equation}
valid for such a field. In two dimensions equation (\ref{equlam}) may
be written in equivalent forms
\begin{equation}
  \label{eculef}
  \lambda_a \gamma_b = \lambda_b \gamma_a
  \quad \text{or} \quad
  \epsilon^{ab} \lambda_a \gamma_b = 0.
\end{equation}
If one chooses the $\lambda^{0\alpha}$ components as independent ones
then the components of $\lambda^{1\alpha}$ can be found from
(\ref{equlam}) to be
\begin{displaymath}
  \lambda^{0\alpha} = (\lambda^{0+}, \lambda^{0-}), \qquad
  \lambda^{1\alpha} = (\lambda^{0+}, -\lambda^{0-}).
\end{displaymath}
It is important to note that as a consequence any cubic or higher
monomial of the (anticommuting) $\RS$ or $\lambda_a$ vanishes
identically. Furthermore, the field $\lambda_a$ satisfies the useful
relation
\begin{equation}
  \epsilon_a\^b \lambda_b = \lambda_a \gamma^3
\end{equation}
which together with (\ref{egmpro2}) yields
\begin{equation}
  \begin{aligned}
    \epsilon_a\^b \RS_b &= -\RS \gamma_a \gamma^3 + \lambda_a
    \gamma^3.
  \end{aligned}
\end{equation} 
For the sake of brevity we often introduce the obvious notations
\begin{equation}
  \RS^2 = \RS^\alpha \RS_\alpha, \qquad
  \lambda^2 = \lambda^{a\alpha} \lambda_{a\alpha}.
\end{equation}
Other convenient identities used for $\lambda_a$ in our present work
are
\begin{align}
  (\lambda_a \lambda_b) &= \frac12 \eta_{ab} \lambda^2, \\
  (\lambda_a \gamma^3 \lambda_b) &= \frac12 \epsilon_{ab} \lambda^2, \\
  (\lambda_a \gamma_c \lambda_b) &= 0.
\end{align}
The first of these identities can be proved by inserting the unit
matrix $\gamma_a\gamma^a/2$ inside the product and interchanging the
indices due to equation (\ref{eculef}).  The second and third equation
is antisymmetric in indices $a$, $b$, and therefore to be calculated
easily because they are proportional to $\epsilon_{ab}$.

Quadratic combinations of the vector-spinor field can be decomposed in
terms of irreducible components:
\begin{align}
  (\RS_a \RS_b) &= \eta_{ab} \left( -\RS^2 + \frac12
    \lambda^2
  \right) + 2 (\RS \gamma_a \lambda_b) \\
  (\RS_a \gamma^3 \RS_b) &= \epsilon_{ab} \left( \RS^2 +
    \frac12 \lambda^2
  \right) \\
  (\RS_a \gamma_c \RS_b) &= 2 \epsilon_{ab} (\RS \gamma^3
  \lambda_c) \label{rs-ga-rs} \\
  (\RS_a \gamma_c \gamma^3 \RS_b) &= 2 \epsilon_{ab} (\RS
  \lambda_c)
\end{align}

Any totally symmetric spinor $W_{\alpha\beta\gamma} =
W_{(\alpha\beta\gamma)}$ can be written as
\begin{equation}
  \label{W-fff}
  W_{\alpha\beta\gamma} = \gamma^a\_{\alpha\beta} W_{a\gamma} +
  \gamma^a\_{\beta\gamma} W_{a\alpha} + \gamma^a\_{\gamma\alpha}
  W_{a\beta},
\end{equation}
where the decomposition of the vector-spinor $W_{a\gamma} =
W^+\_{a\gamma} + W^-\_{a\gamma}$ is given by
\begin{align}
  W^+\_{a\delta} &= -\frac{1}{4} (\gamma_a \gamma^b)_\delta\^\gamma
  \gamma_b\^{\beta\alpha} W_{\alpha\beta\gamma}, \\
  W^-\_{a\delta} &= -\frac{1}{12} (\gamma^b \gamma_a)_\delta\^\gamma
  \gamma_b\^{\beta\alpha} W_{\alpha\beta\gamma}.
\end{align}
The part $W^+\_{a\gamma} = -\gamma_{a\gamma}\^\delta W^+\_\delta$ can
be calculated, with help of the identity
\begin{equation}
  \gamma^b\_\delta\^{(\gamma} \gamma_b\^{\beta\alpha)} =
  -\gamma^3\_\delta\^{(\gamma} \gamma^3\^{\beta\alpha)},
\end{equation}
by the formula
\begin{equation}
  W^+\_\delta = \frac{1}{4} \gamma^b\_\delta\^\gamma
  \gamma_b\^{\beta\alpha} W_{\alpha\beta\gamma}
  = -\frac{1}{4} \gamma^3\_\delta\^\gamma \gamma^3\^{\beta\alpha}
  W_{\alpha\beta\gamma}.
\end{equation}
If a totally symmetric spinor $V_{\alpha\beta\gamma} =
V_{(\alpha\beta\gamma)}$ is given in the form
\begin{equation}
  V_{\alpha\beta\gamma} = \gamma^3\_{\alpha\beta} V_\gamma +
  \gamma^3\_{\beta\gamma} V_\alpha + \gamma^3\_{\gamma\alpha} V_\beta,
\end{equation}
the representation of $V_{\alpha\beta\gamma}$ in the form
(\ref{W-fff}) is given by
\begin{equation}
  W^+\_\delta = (\gamma^3 V)_\delta, \qquad W^-\_{a\delta} = 0,
\end{equation}
therefore
\begin{equation}
  \label{V-fff}
  V_{\alpha\beta\gamma} = -\gamma^a\_{\alpha\beta} (\gamma_a \gamma^3
  V)_\gamma - \gamma^a\_{\beta\gamma} (\gamma_a \gamma^3
  V)_\alpha -  \gamma^a\_{\gamma\alpha} (\gamma_a \gamma^3 V)_\beta.
\end{equation}

\section[Properties of the $\Gamma$-Matrices]{Properties of the
  \mathversion{bold}$\Gamma$-Matrices}
\label{app:Gamma}

There are several useful brackets the of $\gamma$-matrices with
$\Gamma^a$ and $\Gamma^3$ which have been used in
\Sec\ref{sec:ertl-cf}:
\begin{alignat}{2}
  \{\Gamma^3, \gamma^3\} &= 2 \cdot \1 &\qquad [\Gamma^3, \gamma^3] &=
  -2i k^a \epsilon_a\^b \gamma_b \\
  \{\Gamma^3, \gamma_a\} &= 2i k_a \1 &\qquad [\Gamma^3, \gamma_a] &=
  2
  \epsilon_a\^b \Gamma_b \\
  \{\Gamma^a, \gamma^3\} &= -2i k^a \1 &\qquad [\Gamma^a, \gamma^3] &=
  [\gamma^a, \gamma^3] = 2 \gamma^a \gamma^3 \\
  \{\Gamma^a, \gamma^b\} &= 2 \eta^{ab} \1 &\qquad [\Gamma^a,
  \gamma^b] &= 2 \epsilon^{ab} \gamma^3 + 2i k^a (\gamma^b \gamma^3)
\end{alignat}
The algebra of the field dependent $\Gamma$-matrices reads
\begin{alignat}{2}
  \{\Gamma^3, \Gamma^3\} &= 2 (1-k^2) \1, &\qquad
  [\Gamma^3, \Gamma^3] &= 0, \\
  \{\Gamma^3, \Gamma_a\} &= 0, &\qquad
  [\Gamma^3, \Gamma_a] &= 2 (\delta_a\^b - k_a k^b) \epsilon_b\^c
  \Gamma_c, \\
  \{\Gamma^a, \Gamma^b\} &= 2 \left( \eta^{ab} - k^a k^b \right) \1,
  &\qquad
  [\Gamma^a, \Gamma^b] &= 2 \epsilon^{ab} \Gamma^3, \label{GammaBr}
\end{alignat}
where the abbreviation $k^2 = k^a k_a$ was introduced. The
anticommutator in (\ref{GammaBr}) suggests the interpretation of
$\eta^{ab} - k^a k^b$ as a metric in tangent space (\cf for a similar
feature \cite{Hehl:1995ue}). From the algebra
\begin{align}
  \Gamma^3 \Gamma^3 &= (1-k^2) \1, \\
  \Gamma^3 \Gamma_a &= (\delta_a\^b - k_a k^b) \epsilon_b\^c \Gamma_c,
  \\
  \Gamma^a \Gamma^b &= (\eta^{ab} - k^a k^b) \1 + \epsilon^{ab}
  \Gamma^3
\end{align}
is derived. Note that because of $(\delta_a\^b - k_a k^b)
\epsilon_b\^c = \epsilon_a\^b \left( (1-k^2) \delta_b\^c + k_b k^c
\right)$ we can also write
\begin{equation}
  \Gamma^3 \Gamma_a
  = (1-k^2) \epsilon_a\^b \Gamma_b + k_b k^c \Gamma_c
  = (1-k^2) \epsilon_a\^b \gamma_b - i \epsilon_a\^b k_b \Gamma^3.
\end{equation}
For three products of $\Gamma^a$ the symmetry property $\Gamma^a
\Gamma^b \Gamma^c = \Gamma^c \Gamma^b \Gamma^a$ is valid.

Contractions of the Rarita-Schwinger field with $\Gamma^a$ similar to
(\ref{RS-decomp}) yield
\begin{align}
  \half (\RS_b \Gamma_c \Gamma^b)^\alpha &= \lambda_b\^\alpha -
  \frac{i}{2} k^b (\RS_b \Gamma_c \gamma^3)^\alpha + i k^c (\RS
  \gamma^3)^\alpha, \\
  \half (\RS_b \Gamma^b)^\alpha &= \RS^\alpha - \frac{i}{2} k^b
  (\RS_b \gamma^3)^\alpha.
\end{align}
Some further formulae used in the calculations for terms quadratic in
the Rarita-Schwinger field are
\begin{gather}
  (\delta_a\^d - k_a k^d) (\RS_d \Gamma^c \Gamma_b \RS_c) =
  (\delta_b\^d - k_b k^d) (\RS_d \Gamma^c \Gamma_a \RS_c), \\
  (\RS_c \Gamma^c \Gamma^3 \Gamma^b \Gamma^a \RS_b) = (\RS_c
  \Gamma^a \Gamma^3 \Gamma^b \Gamma^c \RS_b) = (1-k^2) \epsilon^{cb}
  (\RS_b \Gamma^a \RS_c), \\
  \epsilon^{cb} (\RS_b \Gamma^a \RS_c) = (\RS_b \gamma^a \gamma^3
  \Gamma^c \Gamma^b \RS_c) = 4 (\RS \gamma^3 \lambda^a)
  - i k^a (2 \RS^2 + \lambda^2),
\end{gather}
and in the emergence of two different spin-vectors
\begin{equation}
  (\RS_b \Gamma^a \gamma_m \varphi^b) = (\varphi^b \gamma_m \Gamma^a
  \RS_b), \qquad
  (\RS_b \Gamma^a \gamma^3 \varphi^b) = (\varphi^b \gamma^3 \Gamma^a
  \RS_b).
\end{equation}


\end{appendix}

\backmatter

\lhead{\slshape \bibname}                             \rhead{\thepage}
\bibliographystyle{utphys}
\bibliography{master}

\providecommand{\href}[2]{#2}\begingroup\raggedright\begin{thebibliography}{10%
0}

\bibitem{jordan55}
P.~Jordan, {\em Schwerkraft und Weltall}.
\newblock Braunschweig, second~ed., 1955.

\bibitem{Dicke:1957RM}
R.~H. Dicke {\em Rev. Mod. Phys.} {\bf 29} (1957) 29.

\bibitem{Jordan:1959eg}
P.~Jordan, ``The present state of {D}irac's cosmological hypothesis,'' {\em Z.
  Phys.} {\bf 157} (1959)
112--121.

\bibitem{Fierz:1956}
M.~Fierz, ``{\"U}ber die physikalische {D}eutung der erweiterten
  {G}ravitationstheorie {P.} {J}ordans,'' {\em Helv. Phys. Acta} {\bf 29}
  (1956) 128.

\bibitem{Brans:1961sx}
C.~Brans and R.~H. Dicke, ``{M}ach's principle and a relativistic theory of
  gravitation,'' {\em Phys. Rev.} {\bf 124} (1961)
925--935.

\bibitem{Wang:1999fa}
L.~min Wang, R.~R. Caldwell, J.~P. Ostriker, and P.~J. Steinhardt, ``Cosmic
  concordance and quintessence,'' {\em Astrophys. J.} {\bf 530} (2000) 17--35,
\href{http://www.arXiv.org/abs/astro-ph/9901388}{{\tt astro-ph/9901388}}.

\bibitem{Diaz-Rivera:1999wd}
L.~M. Diaz-Rivera and L.~O. Pimentel, ``Cosmological models with dynamical
  {$\Lambda$} in scalar-tensor theories,'' {\em Phys. Rev.} {\bf D60} (1999)
  123501,
\href{http://www.arXiv.org/abs/gr-qc/9907016}{{\tt gr-qc/9907016}}.

\bibitem{Matos:1999et}
T.~Matos, F.~S. Guzman, and L.~A. Urena-Lopez, ``Scalar field as dark matter in
  the universe,'' {\em Class. Quant. Grav.} {\bf 17} (2000) 1707,
\href{http://www.arXiv.org/abs/astro-ph/9908152}{{\tt astro-ph/9908152}}.

\bibitem{Coley:1999yq}
A.~A. Coley, ``Qualitative properties of scalar-tensor theories of gravity,''
  {\em Gen. Rel. Grav.} {\bf 31} (1999) 1295,
\href{http://www.arXiv.org/abs/astro-ph/9910395}{{\tt astro-ph/9910395}}.

\bibitem{Bertolami:1999dp}
O.~Bertolami and P.~J. Martins, ``Non-minimal coupling and quintessence,'' {\em
  Phys. Rev.} {\bf D61} (2000) 064007,
\href{http://www.arXiv.org/abs/gr-qc/9910056}{{\tt gr-qc/9910056}}.

\bibitem{Sokolowski:1995dk}
L.~M. Sokolowski, ``Universality of {E}instein's general relativity,''
\href{http://www.arXiv.org/abs/gr-qc/9511073}{{\tt gr-qc/9511073}}.

\bibitem{Perlmutter:1997hx}
{\bf The Supernova Cosmology Project} Collaboration, S.~Perlmutter {\em et
  al.}, ``Cosmology from type {Ia} supernovae,'' {\em Bull. Am. Astron. Soc.}
  {\bf 29} (1997) 1351,
\href{http://www.arXiv.org/abs/astro-ph/9812473}{{\tt astro-ph/9812473}}.

\bibitem{Perlmutter:1998np}
{\bf Supernova Cosmology Project} Collaboration, S.~Perlmutter {\em et al.},
  ``Measurements of {$\Omega$} and {$\Lambda$} from 42 high-redshift
  supernovae,''
\href{http://www.arXiv.org/abs/astro-ph/9812133}{{\tt astro-ph/9812133}}.

\bibitem{Riess:1998cb}
A.~G. Riess {\em et al.}, ``Observational evidence from supernovae for an
  accelerating universe and a cosmological constant,'' {\em Astron. J.} {\bf
  116} (1998) 1009--1038,
\href{http://www.arXiv.org/abs/astro-ph/9805201}{{\tt astro-ph/9805201}}.

\bibitem{Garnavich:1998nb}
P.~M. Garnavich {\em et al.}, ``Constraints on cosmological models from hubble
  space telescope observations of high-{$z$} supernovae,'' {\em Astrophys. J.}
  {\bf 493} (1998) L53--57,
\href{http://www.arXiv.org/abs/astro-ph/9710123}{{\tt astro-ph/9710123}}.

\bibitem{Garnavich:1998th}
P.~M. Garnavich {\em et al.}, ``Supernova limits on the cosmic equation of
  state,'' {\em Astrophys. J.} {\bf 509} (1998) 74,
\href{http://www.arXiv.org/abs/astro-ph/9806396}{{\tt astro-ph/9806396}}.

\bibitem{Schmidt:1998}
B.~Schmidt {\em et al.}, ``The high-{$z$} supernova search,'' {\em Astrophys.
  J.} {\bf 507} (1998) 46,
  \href{http://www.arXiv.org/abs/astro-ph/9805200}{{\tt astro-ph/9805200}}.

\bibitem{Hayashi:1977jd}
K.~Hayashi, ``The gauge theory of the translation group and underlying
  geometry,'' {\em Phys. Lett.} {\bf B69} (1977)
441.

\bibitem{Hayashi:1979qx}
K.~Hayashi and T.~Shirafuji, ``New general relativity,'' {\em Phys. Rev.} {\bf
  D19} (1979) 3524--3553.
Addendum-ibid. \textbf{D24} (1981) 3312--3314.

\bibitem{hehl79}
F.~W. Hehl, ``Four lectures on {P}oincare gauge field theory,'' in {\em
  Proceedings of the 6th School of Cosmology and Gravitation on Spin, Torsion,
  Rotation and Supergravity}, P.~G. Bergmann and V.~de~Sabbata, eds., vol.~58
  of {\em Series B: Physics}.
\newblock NATO Advanced Study Institute, Erice, May 6--18 1979, 1980.

\bibitem{kopczynski82}
W.~Kopczy{\'n}ski, ``Problems with metric-teleparallel theories of
  gravitation,'' {\em J. Phys.} {\bf A15} (1982) 493--506.

\bibitem{Kopczynski:1990af}
W.~Kopczy{\'n}ski, ``Variational principles for gravity and fluids,'' {\em
  Annals Phys.} {\bf 203} (1990)
308.

\bibitem{Muller-Hoissen:1983vc}
F.~M{\"u}ller-Hoissen and J.~Nitsch, ``Teleparallelism - a viable theory of
  gravity?,'' {\em Phys. Rev.} {\bf D28} (1983)
718.

\bibitem{mueller-hoissen85}
F.~M{\"u}ller-Hoissen and J.~Nitsch, ``On the tetrad theory of gravity,'' {\em
  Gen. Rel. Grav.} {\bf 17} (1985) 747--760.

\bibitem{Bakler:1988ub}
P.~Baekler and E.~W. Mielke, ``{H}amiltonian structure of {P}oincare gauge
  theory and separation of nondynamical variables in exact torsion solutions,''
  {\em Fortschr. Phys.} {\bf 36} (1988)
549.

\bibitem{Mielke:1990my}
E.~W. Mielke, ``Positive gravitational energy proof from complex variables?,''
  {\em Phys. Rev.} {\bf D42} (1990)
3388--3394.

\bibitem{mielke92}
E.~W. Mielke, ``{A}shtekar's complex variables in general relativity and its
  teleparallelism equivalent,'' {\em Ann. Phys.} {\bf 219} (1992) 78.

\bibitem{nester89}
J.~M. Nester, ``Positive energy via the teleparallel {H}amiltonian,'' {\em Int.
  J. Mod. Phys.} {\bf A4} (1989) 1755--1772.

\bibitem{DeAndrade:2000sf}
V.~C. {De Andrade}, L.~C.~T. Guillen, and J.~G. Pereira, ``Teleparallel
  gravity: An overview,''
\href{http://www.arXiv.org/abs/gr-qc/0011087}{{\tt gr-qc/0011087}}.

\bibitem{Hehl:1995ue}
F.~W. Hehl, J.~D. McCrea, E.~W. Mielke, and Y.~Ne{'e}man, ``Metric affine gauge
  theory of gravity: Field equations, {N}oether identities, world spinors, and
  breaking of dilation invariance,'' {\em Phys. Rept.} {\bf 258} (1995) 1--171,
\href{http://www.arXiv.org/abs/gr-qc/9402012}{{\tt gr-qc/9402012}}.

\bibitem{Freedman:1976xh}
D.~Z. Freedman, P.~{van Nieuwenhuizen}, and S.~Ferrara, ``Progress toward a
  theory of supergravity,'' {\em Phys. Rev.} {\bf D13} (1976)
3214--3218.

\bibitem{Freedman:1976py}
D.~Z. Freedman and P.~{van Nieuwenhuizen}, ``Properties of supergravity
  theory,'' {\em Phys. Rev.} {\bf D14} (1976)
912.

\bibitem{Deser:1976eh}
S.~Deser and B.~Zumino, ``Consistent supergravity,'' {\em Phys. Lett.} {\bf
  B62} (1976)
335.

\bibitem{Deser:1976rb}
S.~Deser and B.~Zumino, ``A complete action for the spinning string,'' {\em
  Phys. Lett.} {\bf B65} (1976)
369--373.

\bibitem{Grimm:1978kp}
R.~Grimm, J.~Wess, and B.~Zumino, ``Consistency checks on the superspace
  formulation of supergravity,'' {\em Phys. Lett.} {\bf B73} (1978)
415.

\bibitem{green87}
M.~B. Green, J.~H. Schwarz, and E.~Witten, {\em Superstring Theory}, vol.~1 and
  2.
\newblock Cambridge University Press, Cambridge, 1987.

\bibitem{luest89}
D.~L{\"u}st and S.~Theisen, {\em Lectures on String Theory}.
\newblock Springer-Verlag, Berlin, 1989.

\bibitem{polchinski98}
J.~Polchinski, {\em String Theory}, vol.~I and II.
\newblock Cambridge University Press, Cambridge, 1998.

\bibitem{Wess:1977fn}
J.~Wess and B.~Zumino, ``Superspace formulation of supergravity,'' {\em Phys.
  Lett.} {\bf B66} (1977)
361--364.

\bibitem{Wess:1978bu}
J.~Wess and B.~Zumino, ``Superfield {L}agrangian for supergravity,'' {\em Phys.
  Lett.} {\bf B74} (1978)
51.

\bibitem{Golfand:1971iw}
Y.~A. Golfand and E.~P. Likhtman, ``Extension of the algebra of {P}oincare
  group generators and violation of {P} invariance,'' {\em JETP Lett.} {\bf 13}
  (1971)
323--326.

\bibitem{Volkov:1973ix}
D.~V. Volkov and V.~P. Akulov, ``Is the neutrino a {G}oldstone particle?,''
  {\em Phys. Lett.} {\bf B46} (1973)
109--110.

\bibitem{Vafa:1996xn}
C.~Vafa, ``Evidence for {F}-theory,'' {\em Nucl. Phys.} {\bf B469} (1996)
  403--418,
\href{http://www.arXiv.org/abs/hep-th/9602022}{{\tt hep-th/9602022}}.

\bibitem{Witten:1995ex}
E.~Witten, ``String theory dynamics in various dimensions,'' {\em Nucl. Phys.}
  {\bf B443} (1995) 85--126,
\href{http://www.arXiv.org/abs/hep-th/9503124}{{\tt hep-th/9503124}}.

\bibitem{Sezgin:1997cj}
E.~Sezgin, ``The {M}-algebra,'' {\em Phys. Lett.} {\bf B392} (1997) 323--331,
\href{http://www.arXiv.org/abs/hep-th/9609086}{{\tt hep-th/9609086}}.

\bibitem{Howe:1979ia}
P.~S. Howe, ``Super {W}eyl transformations in two dimensions,'' {\em J. Phys.}
  {\bf A12} (1979)
393--402.

\bibitem{Fayet:1977yc}
P.~Fayet, ``Spontaneously broken supersymmetric theories of weak,
  electromagnetic and strong interactions,'' {\em Phys. Lett.} {\bf B69} (1977)
489.

\bibitem{Hindawi:1996fy}
A.~Hindawi, B.~A. Ovrut, and D.~Waldram, ``Two-dimensional higher-derivative
  supergravity and a new mechanism for supersymmetry breaking,'' {\em Nucl.
  Phys.} {\bf B471} (1996) 409--429,
\href{http://www.arXiv.org/abs/hep-th/9509174}{{\tt hep-th/9509174}}.

\bibitem{Aichelburg:1978fn}
P.~C. Aichelburg and T.~Dereli, ``Exact plane wave solutions of supergravity
  field equations,'' {\em Phys. Rev.} {\bf D18} (1978)
1754.

\bibitem{Aichelburg:1980eb}
P.~C. Aichelburg, ``Identification of trivial solutions in supergravity,'' {\em
  Phys. Lett.} {\bf B91} (1980)
382.

\bibitem{Aichelburg:1983ux}
P.~C. Aichelburg and R.~Gueven, ``Supersymmetric black holes in {$N=2$}
  supergravity theory,'' {\em Phys. Rev. Lett.} {\bf 51} (1983)
1613.

\bibitem{Rosenbaum:1986cn}
M.~Rosenbaum, M.~Ryan, L.~F. Urrutia, and R.~A. Matzner, ``Colliding plane
  waves in {$N=1$} classical supergravity,'' {\em Phys. Rev.} {\bf D34} (1986)
409--415.

\bibitem{Knutt-Wehlau:1998gq}
M.~E. Knutt-Wehlau and R.~B. Mann, ``Supergravity from a massive superparticle
  and the simplest super black hole,'' {\em Nucl. Phys.} {\bf B514} (1998)
  355--378,
\href{http://www.arXiv.org/abs/hep-th/9708126}{{\tt hep-th/9708126}}.

\bibitem{Katanaev:1986wk}
M.~O. Katanaev and I.~V. Volovich, ``String model with dynamical geometry and
  torsion,'' {\em Phys. Lett.} {\bf B175} (1986)
413--416.

\bibitem{Katanaev:1990qm}
M.~O. Katanaev and I.~V. Volovich, ``Two-dimensional gravity with dynamical
  torsion and strings,'' {\em Ann. Phys.} {\bf 197} (1990)
1.

\bibitem{Schaller:1994np}
P.~Schaller and T.~Strobl, ``Canonical quantization of non-{E}insteinian
  gravity and the problem of time,'' {\em Class. Quant. Grav.} {\bf 11} (1994)
  331--346,
\href{http://www.arXiv.org/abs/hep-th/9211054}{{\tt hep-th/9211054}}.

\bibitem{Schaller:1994es}
P.~Schaller and T.~Strobl, ``Poisson structure induced (topological) field
  theories,'' {\em Mod. Phys. Lett.} {\bf A9} (1994) 3129--3136,
\href{http://www.arXiv.org/abs/hep-th/9405110}{{\tt hep-th/9405110}}.

\bibitem{Strobl:1994eu}
T.~Strobl, ``{D}irac quantization of gravity {Y}ang-{M}ills systems in
  (1+1)-dimensions,'' {\em Phys. Rev.} {\bf D50} (1994) 7346--7350,
\href{http://www.arXiv.org/abs/hep-th/9403121}{{\tt hep-th/9403121}}.

\bibitem{Klosch:1996fi}
T.~Kl{\"o}sch and T.~Strobl, ``Classical and quantum gravity in
  (1+1)-dimensions. {P}art 1: {A} unifying approach,'' {\em Class. Quant.
  Grav.} {\bf 13} (1996) 965--984,
  \href{http://www.arXiv.org/abs/gr-qc/9508020}{{\tt gr-qc/9508020}}.
Erratum ibid. 14 (1997) 825.

\bibitem{Klosch:1996qv}
T.~Kl{\"o}sch and T.~Strobl, ``Classical and quantum gravity in 1+1 dimensions.
  {P}art 2: {T}he universal coverings,'' {\em Class. Quant. Grav.} {\bf 13}
  (1996) 2395--2422,
\href{http://www.arXiv.org/abs/gr-qc/9511081}{{\tt gr-qc/9511081}}.

\bibitem{Kloesch:1997fi}
T.~Kl{\"o}sch and T.~Strobl, ``Classical and quantum gravity in
  (1+1)-dimensions. {P}art 3: {S}olutions of arbitrary topology,'' {\em Class.
  Quant. Grav.} {\bf 14} (1997)
1689--1723.

\bibitem{Katanaev:1996bh}
M.~O. Katanaev, W.~Kummer, and H.~Liebl, ``Geometric interpretation and
  classification of global solutions in generalized dilaton gravity,'' {\em
  Phys. Rev.} {\bf D53} (1996) 5609--5618,
\href{http://www.arXiv.org/abs/gr-qc/9511009}{{\tt gr-qc/9511009}}.

\bibitem{Katanaev:1997ni}
M.~O. Katanaev, W.~Kummer, and H.~Liebl, ``On the completeness of the black
  hole singularity in 2d dilaton theories,'' {\em Nucl. Phys.} {\bf B486}
  (1997) 353--370,
\href{http://www.arXiv.org/abs/gr-qc/9602040}{{\tt gr-qc/9602040}}.

\bibitem{Katanaev:1993fu}
M.~O. Katanaev, ``All universal coverings of two-dimensional gravity with
  torsion,'' {\em J. Math. Phys.} {\bf 34} (1993)
700.

\bibitem{Katanaev:1997je}
M.~O. Katanaev, ``{E}uclidean two-dimensional gravity with torsion,'' {\em J.
  Math. Phys.} {\bf 38} (1997)
946--980.

\bibitem{kummer92}
W.~Kummer, ``Deformed iso(2,1)-symmetry and non-{E}insteinian 2d-gravity with
  matter,'' in {\em Hadron Structure '92}, D.~Brunsko and J.~Urb\'an, eds.,
  pp.~48--56.
\newblock Ko\v sice University, 1992.

\bibitem{Solodukhin:1993xs}
S.~N. Solodukhin, ``Two-dimensional black hole with torsion,'' {\em Phys.
  Lett.} {\bf B319} (1993) 87--95,
\href{http://www.arXiv.org/abs/hep-th/9302040}{{\tt hep-th/9302040}}.

\bibitem{Solodukhin:1995ux}
S.~Solodukhin, ``Exact solution of 2-d {P}oincar\'e gravity coupled to fermion
  matter,'' {\em Phys. Rev.} {\bf D51} (1995) 603--608,
\href{http://www.arXiv.org/abs/hep-th/9404045}{{\tt hep-th/9404045}}.

\bibitem{Kummer:1992rt}
W.~Kummer and D.~J. Schwarz, ``Renormalization of {$R^2$} gravity with
  dynamical torsion in {$d = 2$},'' {\em Nucl. Phys.} {\bf B382} (1992)
171--186.

\bibitem{Haider:1992cw}
F.~Haider and W.~Kummer, ``Quantum functional integration of non-{E}insteinian
  gravity in {$d = 2$},'' {\em Int. J. Mod. Phys.} {\bf A9} (1994)
207.

\bibitem{Kummer:1997jj}
W.~Kummer, H.~Liebl, and D.~V. Vassilevich, ``Non-perturbative path integral of
  2d dilaton gravity and two-loop effects from scalar matter,'' {\em Nucl.
  Phys.} {\bf B513} (1998) 723,
\href{http://www.arXiv.org/abs/hep-th/9707115}{{\tt hep-th/9707115}}.

\bibitem{Mandal:1991tz}
G.~Mandal, A.~M. Sengupta, and S.~R. Wadia, ``Classical solutions of
  two-dimensional string theory,'' {\em Mod. Phys. Lett.} {\bf A6} (1991)
1685--1692.

\bibitem{Elitzur:1991cb}
S.~Elitzur, A.~Forge, and E.~Rabinovici, ``Some global aspects of string
  compactifications,'' {\em Nucl. Phys.} {\bf B359} (1991)
581--610.

\bibitem{Witten:1991yr}
E.~Witten, ``On string theory and black holes,'' {\em Phys. Rev.} {\bf D44}
  (1991)
314--324.

\bibitem{Callan:1992rs}
C.~G. {Callan Jr.}, S.~B. Giddings, J.~A. Harvey, and A.~Strominger,
  ``Evanescent black holes,'' {\em Phys. Rev.} {\bf D45} (1992) 1005--1009,
\href{http://www.arXiv.org/abs/hep-th/9111056}{{\tt hep-th/9111056}}.

\bibitem{Fabbri:1996bz}
A.~Fabbri and J.~G. Russo, ``Soluble models in 2d dilaton gravity,'' {\em Phys.
  Rev.} {\bf D53} (1996) 6995--7002,
\href{http://www.arXiv.org/abs/hep-th/9510109}{{\tt hep-th/9510109}}.

\bibitem{Schmidt:1999wb}
H.~J. Schmidt, ``The classical solutions of two-dimensional gravity,'' {\em
  Gen. Rel. Grav.} {\bf 31} (1999) 1187,
\href{http://www.arXiv.org/abs/gr-qc/9905051}{{\tt gr-qc/9905051}}.

\bibitem{Obukhov:1997uc}
Y.~N. Obukhov and F.~W. Hehl, ``Black holes in two dimensions,'' in {\em Black
  Holes: Theory and Observation}, F.~W. Hehl, C.~Kiefer, and R.~J. Metzler,
  eds., pp.~289--316.
\newblock Springer 1998, Proceedings of the 179.\ WE-Heraeus Seminar in Bad
  Honnef, Germany, 1997.
\newblock
\href{http://www.arXiv.org/abs/hep-th/9807101}{{\tt hep-th/9807101}}.
\newblock

\bibitem{Strobl:1999Habil}
T.~Strobl, {\em Gravity in Two Spacetime Dimensions}.
\newblock Habilitationsschrift, Rheinisch-Westf\"alische Technische Hochschule
  Aachen, 1999.

\bibitem{Banks:1991mk}
T.~Banks and M.~O'Loughlin, ``Two-dimensional quantum gravity in {M}inkowski
  space,'' {\em Nucl. Phys.} {\bf B362} (1991)
649--664.

\bibitem{Odintsov:1991qu}
S.~D. Odintsov and I.~L. Shapiro, ``One loop renormalization of two-dimensional
  induced quantum gravity,'' {\em Phys. Lett.} {\bf B263} (1991)
183--189.

\bibitem{Louis-Martinez:1994eh}
D.~Louis-Martinez, J.~Gegenberg, and G.~Kunstatter, ``Exact {D}irac
  quantization of all 2-d dilaton gravity theories,'' {\em Phys. Lett.} {\bf
  B321} (1994) 193--198,
\href{http://www.arXiv.org/abs/gr-qc/9309018}{{\tt gr-qc/9309018}}.

\bibitem{Gegenberg:1995pv}
J.~Gegenberg, G.~Kunstatter, and D.~Louis-Martinez, ``Observables for
  two-dimensional black holes,'' {\em Phys. Rev.} {\bf D51} (1995) 1781--1786,
\href{http://www.arXiv.org/abs/gr-qc/9408015}{{\tt gr-qc/9408015}}.

\bibitem{Klosch:1998yh}
T.~Klosch and T.~Strobl, ``A global view of kinks in 1+1 gravity,'' {\em Phys.
  Rev.} {\bf D57} (1998) 1034--1044,
\href{http://www.arXiv.org/abs/gr-qc/9707053}{{\tt gr-qc/9707053}}.

\bibitem{Verlinde:1991rf}
H.~Verlinde, ``Black holes and strings in two dimensions,'' in {\em The Sixth
  Marcel Grossmann Meeting on General Relativity, Kyoto, Japan 1991}, H.~Sato,
  ed., pp.~813--831.
\newblock World Scientific, Singapore, 1992.

\bibitem{Kummer:1997hy}
W.~Kummer, H.~Liebl, and D.~V. Vassilevich, ``Exact path integral quantization
  of generic 2-d dilaton gravity,'' {\em Nucl. Phys.} {\bf B493} (1997)
  491--502,
\href{http://www.arXiv.org/abs/gr-qc/9612012}{{\tt gr-qc/9612012}}.

\bibitem{Kummer:1998zs}
W.~Kummer, H.~Liebl, and D.~V. Vassilevich, ``Integrating geometry in general
  2d dilaton gravity with matter,'' {\em Nucl. Phys.} {\bf B544} (1999) 403,
\href{http://www.arXiv.org/abs/hep-th/9809168}{{\tt hep-th/9809168}}.

\bibitem{Thomi:1984na}
P.~Thomi, B.~Isaak, and P.~Hajicek, ``Spherically symmetric systems of fields
  and black holes. 1. definition and properties of apparent horizon,'' {\em
  Phys. Rev.} {\bf D30} (1984)
1168.

\bibitem{Hajicek:1984mz}
P.~Hajicek, ``Spherically symmetric systems of fields and black holes. 2.
  apparent horizon in canonical formalism,'' {\em Phys. Rev.} {\bf D30} (1984)
1178.

\bibitem{Schmidt:1997mq}
H.~J. Schmidt, ``A new proof of {B}irkhoff's theorem,'' {\em Grav. Cosmol.}
  {\bf 3} (1997) 185,
\href{http://www.arXiv.org/abs/gr-qc/9709071}{{\tt gr-qc/9709071}}.

\bibitem{Schmidt:1998ih}
H.~J. Schmidt, ``A two-dimensional representation of four-dimensional
  gravitational waves,'' {\em Int. J. Mod. Phys.} {\bf D7} (1998) 215--224,
\href{http://www.arXiv.org/abs/gr-qc/9712034}{{\tt gr-qc/9712034}}.

\bibitem{Barbashov:1979bm}
B.~M. Barbashov, V.~V. Nesterenko, and A.~M. Chervyakov, ``The solitons in some
  geometrical field theories,'' {\em Theor. Math. Phys.} {\bf 40} (1979)
  572--581.
Teor. Mat. Fiz. 40 (1979) 15--27, J. Phys. A13 (1980) 301--312.

\bibitem{Teitelboim:1983ux}
C.~Teitelboim, ``Gravitation and hamiltonian structure in two space-time
  dimensions,'' {\em Phys. Lett.} {\bf B126} (1983)
41.

\bibitem{teitelboim84}
C.~Teitelboim, ``The {H}amiltonian structure of two-dimensional space-time and
  its relation with the conformal anomaly,'' in {\em Quantum Theory of Gravity.
  Essays in Honor of the 60th Birthday of Bryce S. DeWitt}, S.~Christensen,
  ed., pp.~327--344.
\newblock Hilger, Bristol, 1984.

\bibitem{jackiw84}
R.~Jackiw, ``{L}iouville field theory: a two-dimensional model for gravity,''
  in {\em Quantum Theory of Gravity. Essays in Honor of the 60th Birthday of
  Bryce S. DeWitt}, S.~Christensen, ed., pp.~403--420.
\newblock Hilger, Bristol, 1984.

\bibitem{Jackiw:1985je}
R.~Jackiw, ``Lower dimensional gravity,'' {\em Nucl. Phys.} {\bf B252} (1985)
343--356.

\bibitem{Mikovic:1992id}
A.~Mikovi{\'c}, ``Exactly solvable models of 2-d dilaton quantum gravity,''
  {\em Phys. Lett.} {\bf B291} (1992) 19--25,
\href{http://www.arXiv.org/abs/hep-th/9207006}{{\tt hep-th/9207006}}.

\bibitem{Mikovic:1993vy}
A.~Mikovi{\'c}, ``Two-dimensional dilaton gravity in a unitary gauge,'' {\em
  Phys. Lett.} {\bf B304} (1993) 70--76,
\href{http://www.arXiv.org/abs/hep-th/9211082}{{\tt hep-th/9211082}}.

\bibitem{Mikovic:1995ub}
A.~Mikovi{\'c}, ``{H}awking radiation and back reaction in a unitary theory of
  2-d quantum gravity,'' {\em Phys. Lett.} {\bf B355} (1995) 85--91,
\href{http://www.arXiv.org/abs/hep-th/9407104}{{\tt hep-th/9407104}}.

\bibitem{Mikovic:1997de}
A.~Mikovi{\'c} and V.~Radovanovi{\'c}, ``Loop corrections in the spectrum of 2d
  {H}awking radiation,'' {\em Class. Quant. Grav.} {\bf 14} (1997) 2647--2661,
\href{http://www.arXiv.org/abs/gr-qc/9703035}{{\tt gr-qc/9703035}}.

\bibitem{Kuchar:1997zm}
K.~V. Kucha{\v r}, J.~D. Romano, and M.~Varadarajan, ``{D}irac constraint
  quantization of a dilatonic model of gravitational collapse,'' {\em Phys.
  Rev.} {\bf D55} (1997) 795--808,
\href{http://www.arXiv.org/abs/gr-qc/9608011}{{\tt gr-qc/9608011}}.

\bibitem{Varadarajan:1998qz}
M.~Varadarajan, ``Quantum gravity effects in the {CGHS} model of collapse to a
  black hole,'' {\em Phys. Rev.} {\bf D57} (1998) 3463--3473,
\href{http://www.arXiv.org/abs/gr-qc/9801058}{{\tt gr-qc/9801058}}.

\bibitem{Cangemi:1996yz}
D.~Cangemi, R.~Jackiw, and B.~Zwiebach, ``Physical states in matter coupled
  dilaton gravity,'' {\em Ann. Phys.} {\bf 245} (1996) 408--444,
\href{http://www.arXiv.org/abs/hep-th/9505161}{{\tt hep-th/9505161}}.

\bibitem{Benedict:1996qy}
E.~Benedict, R.~Jackiw, and H.~J. Lee, ``Functional {S}chr{\"o}dinger and
  {BRST} quantization of (1+1)-dimensional gravity,'' {\em Phys. Rev.} {\bf
  D54} (1996) 6213--6225,
\href{http://www.arXiv.org/abs/hep-th/9607062}{{\tt hep-th/9607062}}.

\bibitem{Schaller:1994uj}
P.~Schaller and T.~Strobl, ``{P}oisson sigma models: {A} generalization of 2d
  gravity {Y}ang-{M}ills systems,''
\href{http://www.arXiv.org/abs/hep-th/9411163}{{\tt hep-th/9411163}}.

\bibitem{Schaller:1995xk}
P.~Schaller and T.~Strobl, ``Introduction to {Poisson-$\sigma$} models,'' in
  {\em Low-Dimensional Models in Statistical Physics and Quantum Field Theory},
  H.~Grosse and L.~Pittner, eds., vol.~469 of {\em Lecture Notes in Physics},
  p.~321.
\newblock Springer, Berlin, 1996.
\newblock
\href{http://www.arXiv.org/abs/hep-th/9507020}{{\tt hep-th/9507020}}.
\newblock

\bibitem{Ikeda:1993qz}
N.~Ikeda and K.~I. Izawa, ``Quantum gravity with dynamical torsion in
  two-dimensions,'' {\em Prog. Theor. Phys.} {\bf 89} (1993)
223--230.

\bibitem{Ikeda:1993nk}
N.~Ikeda and K.~I. Izawa, ``Gauge theory based on quadratic {L}ie algebras and
  2-d gravity with dynamical torsion,'' {\em Prog. Theor. Phys.} {\bf 89}
  (1993)
1077--1086.

\bibitem{Ikeda:1993aj}
N.~Ikeda and K.~I. Izawa, ``General form of dilaton gravity and nonlinear gauge
  theory,'' {\em Prog. Theor. Phys.} {\bf 90} (1993) 237--246,
\href{http://www.arXiv.org/abs/hep-th/9304012}{{\tt hep-th/9304012}}.

\bibitem{Cattaneo:1999fm}
A.~S. Cattaneo and G.~Felder, ``A path integral approach to the {K}ontsevich
  quantization formula,'' {\em Commun. Math. Phys.} {\bf 212} (2000) 591,
\href{http://www.arXiv.org/abs/math.qa/9902090}{{\tt math.qa/9902090}}.

\bibitem{Cattaneo:2000iw}
A.~S. Cattaneo and G.~Felder, ``{P}oisson sigma models and symplectic
  groupoids,''
\href{http://www.arXiv.org/abs/math.sg/0003023}{{\tt math.sg/0003023}}.

\bibitem{Strobl:1994PhD}
T.~Strobl, {\em {P}oisson Structure Induced Field Thoeries and Models of 1+1
  Dimensional Gravity}.
\newblock PhD thesis, Vienna University of Technology, May, 1994.

\bibitem{Weinstein:1983MM}
A.~Weinstein, ``The local structure of {P}oisson manifolds,'' {\em J. Diff.
  Geom.} {\bf 18} (1983) 523--557.

\bibitem{choquet-bruhat89}
Y.~Choquet-Bruhat and C.~DeWitt-Morette, {\em Analysis, Manifolds and Physics.
  Part II: 92 Applications}.
\newblock North-Holland, Amsterdam, 1989.

\bibitem{Kummer:1995qv}
W.~Kummer and P.~Widerin, ``Conserved quasilocal quantities and general
  covariant theories in two-dimensions,'' {\em Phys. Rev.} {\bf D52} (1995)
  6965--6975,
\href{http://www.arXiv.org/abs/gr-qc/9502031}{{\tt gr-qc/9502031}}.

\bibitem{Grumiller:1999rz}
D.~Grumiller and W.~Kummer, ``Absolute conservation law for black holes,'' {\em
  Phys. Rev.} {\bf D61} (2000) 064006,
\href{http://www.arXiv.org/abs/gr-qc/9902074}{{\tt gr-qc/9902074}}.

\bibitem{berezin66}
F.~A. Berezin, {\em The Method of Second Quantization}.
\newblock Academic Press, New York and London, 1966.

\bibitem{gates83}
S.~J. {Gates Jr.}, M.~T. Grisaru, M.~Ro\v{c}ek, and W.~Siegel, {\em
  {SUPERSPACE} \textit{or} One Thousand and One Lessons in Supersymmetry},
  vol.~58 of {\em Frontiers in Physics}.
\newblock The Benjamin/Cummings Publishing Company, London, 1983.

\bibitem{dewitt84}
B.~DeWitt, {\em Supermanifolds}.
\newblock Cambridge Monographs on Mathematical Physics. Cambridge University
  Press, 1984.

\bibitem{Vladimirov:1984zj}
V.~S. Vladimirov and I.~V. Volovich, ``Superanalysis. {I}. {D}ifferential
  calculus,'' {\em Theor. Math. Phys.} {\bf 59} (1984)
317--335.

\bibitem{Vladimirov:1985ZJ}
V.~S. Vladimirov and I.~V. Volovich, ``Superanalysis. {II}. {I}ntegral
  calculus,'' {\em Theor. Math. Phys.} {\bf 60} (1985) 743--765.

\bibitem{constantinescu94}
F.~Constantinescu and H.~F. de~Groote, {\em Geometrische und algebraische
  {M}ethoden der {P}hysik: {S}upermannigfaltigkeiten und
  {V}irasoro-{A}lgebren}.
\newblock Teubner Studienbücher Mathematik. Teubner, Stuttgart, 1994.

\bibitem{Strobl:1993xt}
T.~Strobl, ``Comment on gravity and the {P}oincar{\'e} group,'' {\em Phys.
  Rev.} {\bf D48} (1993) 5029--5031,
\href{http://www.arXiv.org/abs/hep-th/9302041}{{\tt hep-th/9302041}}.

\bibitem{VanNieuwenhuizen:1981ae}
P.~{van Nieuwenhuizen}, ``Supergravity,'' {\em Phys. Rept.} {\bf 68} (1981)
189--398.

\bibitem{Park:1993sd}
Y.~Park and A.~Strominger, ``Supersymmetry and positive energy in classical and
  quantum two-dimensional dilaton gravity,'' {\em Phys. Rev.} {\bf D47} (1993)
  1569--1575,
\href{http://www.arXiv.org/abs/hep-th/9210017}{{\tt hep-th/9210017}}.

\bibitem{Chamseddine:1991fg}
A.~H. Chamseddine, ``Superstrings in arbitrary dimensions,'' {\em Phys. Lett.}
  {\bf B258} (1991)
97--103.

\bibitem{Rivelles:1994xs}
V.~O. Rivelles, ``Topological two-dimensional dilaton supergravity,'' {\em
  Phys. Lett.} {\bf B321} (1994) 189--192,
\href{http://www.arXiv.org/abs/hep-th/9301029}{{\tt hep-th/9301029}}.

\bibitem{Cangemi:1994mj}
D.~Cangemi and M.~Leblanc, ``Two-dimensional gauge theoretic supergravities,''
  {\em Nucl. Phys.} {\bf B420} (1994) 363--378,
\href{http://www.arXiv.org/abs/hep-th/9307160}{{\tt hep-th/9307160}}.

\bibitem{Ikeda:1994dr}
N.~Ikeda, ``Gauge theory based on nonlinear {L}ie superalgebras and structure
  of 2-d dilaton supergravity,'' {\em Int. J. Mod. Phys.} {\bf A9} (1994)
1137--1152.

\bibitem{Ikeda:1994fh}
N.~Ikeda, ``Two-dimensional gravity and nonlinear gauge theory,'' {\em Ann.
  Phys.} {\bf 235} (1994) 435--464,
\href{http://www.arXiv.org/abs/hep-th/9312059}{{\tt hep-th/9312059}}.

\bibitem{Izquierdo:1998hg}
J.~M. Izquierdo, ``Free differential algebras and generic 2d dilatonic
  (super)gravities,'' {\em Phys. Rev.} {\bf D59} (1999) 084017,
\href{http://www.arXiv.org/abs/hep-th/9807007}{{\tt hep-th/9807007}}.

\bibitem{Strobl:1999zz}
T.~Strobl, ``Target-superspace in 2d dilatonic supergravity,'' {\em Phys.
  Lett.} {\bf B460} (1999) 87,
\href{http://www.arXiv.org/abs/hep-th/9906230}{{\tt hep-th/9906230}}.

\bibitem{Kac:1977qb}
V.~G. Kac, ``A sketch of {L}ie superalgebra theory,'' {\em Commun. Math. Phys.}
  {\bf 53} (1977)
31--64.

\bibitem{Scheunert:1976uf}
M.~Scheunert, W.~Nahm, and V.~Rittenberg, ``Classification of all simple graded
  {L}ie algebras whose {L}ie algebra is reductive. 1,'' {\em J. Math. Phys.}
  {\bf 17} (1976)
1626.

\bibitem{Scheunert:1976ug}
M.~Scheunert, W.~Nahm, and V.~Rittenberg, ``Classification of all simple graded
  {L}ie algebras whose {L}ie algebra is reductive. 2. (construction of the
  exceptional algebras),'' {\em J. Math. Phys.} {\bf 17} (1976)
1640.

\bibitem{Frappat:1996pb}
L.~Frappat, P.~Sorba, and A.~Sciarrino, ``Dictionary on {L}ie superalgebras,''
\href{http://www.arXiv.org/abs/hep-th/9607161}{{\tt hep-th/9607161}}.

\bibitem{deBoer:1996nu}
J.~{de Boer}, F.~Harmsze, and T.~Tjin, ``Nonlinear finite {$W$}-symmetries and
  applications in elementary systems,'' {\em Phys. Rept.} {\bf 272} (1996)
  139--214,
\href{http://www.arXiv.org/abs/hep-th/9503161}{{\tt hep-th/9503161}}.

\bibitem{Brown:1979ma}
M.~Brown and S.~J. {Gates, Jr.}, ``Superspace {B}ianchi identities and the
  supercovariant derivative,'' {\em Ann. Phys.} {\bf 122} (1979)
443.

\bibitem{Martinec:1983um}
E.~Martinec, ``Superspace geometry of fermionic strings,'' {\em Phys. Rev.}
  {\bf D28} (1983)
2604--2613.

\bibitem{Rocek:1986iz}
M.~Ro{\v c}ek, P.~{van Nieuwenhuizen}, and S.~C. Zhang, ``Superspace path
  integral measure of the {$N=1$} spinning string,'' {\em Ann. Phys.} {\bf 172}
  (1986)
348--370.

\bibitem{Ertl:1997ib}
M.~F. Ertl, M.~O. Katanaev, and W.~Kummer, ``Generalized supergravity in two
  dimensions,'' {\em Nucl. Phys.} {\bf B530} (1998) 457--486,
\href{http://www.arXiv.org/abs/hep-th/9710051}{{\tt hep-th/9710051}}.

\bibitem{Ertl:Index-0.14.2}
M.~Ertl, ``\textsl{Mathematica} {I}ndex {P}ackage.'' A \textsl{Mathematica}
  Package for index manipulation, endowed with an anticommutative product, as
  well as left and right derivatives suitable for superspace calculations.
  Version~0.14.2. Unpublished. Information about the program can be obtained
  from the author (e-mail: \texttt{ertl@tph.tuwien.ac.at}).

\bibitem{Kummer:1994ur}
W.~Kummer and P.~Widerin, ``Non{E}insteinian gravity in {$d=2$}: {S}ymmetry and
  current algebra,'' {\em Mod. Phys. Lett.} {\bf A9} (1994)
1407--1414.

\bibitem{Brink:1977vg}
L.~Brink and J.~H. Schwarz, ``Local complex supersymmetry in two-dimensions,''
  {\em Nucl. Phys.} {\bf B121} (1977)
285.

\bibitem{Cangemi:1992bj}
D.~Cangemi and R.~Jackiw, ``Gauge invariant formulations of lineal gravity,''
  {\em Phys. Rev. Lett.} {\bf 69} (1992) 233--236,
\href{http://www.arXiv.org/abs/hep-th/9203056}{{\tt hep-th/9203056}}.

\bibitem{Cangemi:1993sd}
D.~Cangemi and R.~Jackiw, ``{P}oincare gauge theory for gravitational forces in
  (1+1)- dimensions,'' {\em Ann. Phys.} {\bf 225} (1993) 229--263,
\href{http://www.arXiv.org/abs/hep-th/9302026}{{\tt hep-th/9302026}}.

\bibitem{Alekseev:1995py}
A.~Y. Alekseev, P.~Schaller, and T.~Strobl, ``The topological {$G/G$} {WZW}
  model in the generalized momentum representation,'' {\em Phys. Rev.} {\bf
  D52} (1995) 7146--7160,
\href{http://www.arXiv.org/abs/hep-th/9505012}{{\tt hep-th/9505012}}.

\end{thebibliography}\endgroup

\end{document}